%% file: 2f-Chapt.tex
\documentclass[11pt]{cernrep}   

\usepackage{graphicx}
\usepackage{here}
\usepackage{epsfig}
\usepackage{cite}
\usepackage{color}

\usepackage{axodraw}
\usepackage{epsfig}                            
\usepackage{graphicx}
\usepackage{latexsym}

\begin{document}    
\marginparwidth 3cm
\include{2f-Chapt-Defs}

\include{2f-Chapt-Title}

\tableofcontents 
\include{2f-Chapt-Part1}

\include{2f-Chapt-Part1a}

\include{2f-Chapt-Part2}
\include{2f-Chapt-Part4}

\include{2f-Chapt-Part5}

\include{2f-Chapt-Part6}

\include{2f-Chapt-Part7}

\include{2f-Chapt-Summary}

%
\bibliographystyle{utphys_spires}
\include{2f-Chapt-Biblio}
\end{document}

%% file: 2f-Chapt-Defs.tex

\def\Was{W\c as}
\def\Order#1{${\cal O}(#1$)}
\def\KKMC{${\cal{KK}}$MC}
\def\lint{\int\limits}
\def\bbeta{\bar{\beta}}
\def\tbeta{\tilde{\beta}}
\def\talpha{\tilde{\alpha}}
\def\tomega{\tilde{\omega}}
\def\fpfp {\mathrm{f'}\overline{\mathrm{f'}}}
\def\ffbar{\relax\ifmmode{\mathrm{f}\overline{\mathrm{f}}}%
          \else${\mathrm{f}\overline{\mathrm{f}}}$\fi}  
\def\qqbar{\relax\ifmmode{\mathrm{q}\overline{\mathrm{q}}}%
          \else${\mathrm{q}\overline{\mathrm{q}}}$\fi}  
\def\vvbar{\relax\ifmmode{\nu\overline{\nu}}%
          \else${\nu\overline{\nu}}$\fi}  
\def\llbar{\relax\ifmmode{\ell\overline{\ell}}%
          \else${\ell\overline{\ell}}$\fi}  
\def\reff#1{(\ref{#1})}
\def\citobs#1{\mbox{\texttt{\textbf{#1}}}}
\def\labobs#1{\mbox{\texttt{\textbf{#1}}}}

\let\at=@
\catcode`\@=\active


\newcommand{\ffo}{\ifmmode {\rm f}_1\overline{\rm f}_1 
                                                       \else f$_1\overline{\rm f}_1$ \fi}
\newcommand{\fft}{\ifmmode {\rm f}_2\overline{\rm f}_2 
                                                       \else f$_2\overline{\rm f}_2$ \fi}
\newcommand{\ffp}{\ifmmode {\rm f}^\prime\overline{{\rm f}^\prime}
                                                       \else f$^\prime\overline{{\rm f}^\prime}$ \fi}
\newcommand{\ff}{\ifmmode {\rm f}\overline{\rm f} 
                                                       \else f$\overline{\rm f}$ \fi}

\newcommand{\eps}{\varepsilon}
\newcommand{\Li}{\mathrm{Li}_2}
\newcommand{\Ri}{R(\infty)}
\newcommand{\dd}{{\mathrm d}}
\newcommand{\matrm}[1]{\mbox{\scriptsize #1}}
\newcommand{\vecc}[1]{\mbox{\boldmath $#1$}}

\def\la{\mathrel{\mathpalette\fun <}}
\def\ga{\mathrel{\mathpalette\fun >}}
\def\fun#1#2{\lower3.6pt\vbox{\baselineskip0pt\lineskip.9pt
  \ialign{$\mathsurround=0pt#1\hfil##\hfil$\crcr#2\crcr\sim\crcr}}}  

\newcommand{\zf}{{ZFITTER}}
\newcommand{\ee}{$e^+e^-$}
\newcommand{\sss}[1]{\scriptscriptstyle{#1}}

\newcommand{\mt}{M_t}
\newcommand{\mh}{M_{H}}
\newcommand{\mb}{M_b}
\newcommand{\mc}{M_c}
\newcommand{\dal}{\Delta\alpha^{(5)}_{had}}

\def\Order#1{${\cal O}(#1$)}

%
\newcommand{\nl}{\nonumber\\}
\newcommand{\nn}{\nonumber}
\newcommand{\ds}{\displaystyle}
\newcommand{\mpar}[1]{{\marginpar{\hbadness10000%
                      \sloppy\hfuzz10pt\boldmath\bf#1}}%
                      \typeout{marginpar: #1}\ignorespaces}
\def\mnew{\mpar{\hfil NEW \hfil}\ignorespaces}
\newcommand{\lpar}{\left(}                            
\newcommand{\rpar}{\right)} 
\newcommand{\lrbr}{\left[}
\newcommand{\rrbr}{\right]}
\newcommand{\lcbr}{\left\{}
\newcommand{\rcbr}{\right\}} 
\newcommand{\rbrak}[1]{\lrbr#1\rrbr}
\newcommand{\bq}{\begin{equation}}                    
\newcommand{\eq}{\end{equation}}
\newcommand{\bqa}{\begin{eqnarray}}
\newcommand{\eqa}{\end{eqnarray}}
\newcommand{\ba}[1]{\begin{array}{#1}}
\newcommand{\ea}{\end{array}}
\newcommand{\ben}{\begin{enumerate}}
\newcommand{\een}{\end{enumerate}}
\newcommand{\bei}{\begin{itemize}}
\newcommand{\eei}{\end{itemize}}
\newcommand{\eqn}[1]{Eq.(\ref{#1})}
\newcommand{\eqns}[2]{Eqs.(\ref{#1})--(\ref{#2})}
\newcommand{\eqnss}[1]{Eqs.(\ref{#1})}
\newcommand{\eqnsc}[2]{Eqs.(\ref{#1}) and (\ref{#2})}
\newcommand{\eqnst}[3]{Eqs.(\ref{#1}), (\ref{#2}) and (\ref{#3})}
\newcommand{\eqnsf}[4]{Eqs.(\ref{#1}), (\ref{#2}), (\ref{#3}) and (\ref{#4})}
\newcommand{\eqnsv}[5]{Eqs.(\ref{#1}), (\ref{#2}), (\ref{#3}), (\ref{#4}) and (\ref{#5})}
\newcommand{\tbn}[1]{Tab.~\ref{#1}}
\newcommand{\tabn}[1]{Tab.~\ref{#1}}
\newcommand{\tbns}[2]{Tabs.~\ref{#1}--\ref{#2}}
\newcommand{\tabns}[2]{Tabs.~\ref{#1}--\ref{#2}}
\newcommand{\tbnsc}[2]{Tabs.~\ref{#1} and \ref{#2}}
\newcommand{\fig}[1]{Fig.~\ref{#1}}
\newcommand{\figs}[2]{Figs.~\ref{#1}--\ref{#2}}
\newcommand{\sect}[1]{Section~\ref{#1}}
\newcommand{\sects}[2]{Section~\ref{#1} and \ref{#2}}
\newcommand{\subsect}[1]{Subsection~\ref{#1}}
\newcommand{\appendx}[1]{Appendix~\ref{#1}}
\newcommand{\hsp}{\hspace{.5mm}}
\def\negs{\hspace{-0.26in}}
\def\negsh{\hspace{-0.13in}}
%
%
\newcommand{\TeV}{\mathrm{TeV}}                     
\newcommand{\GeV}{\mathrm{GeV}}
\newcommand{\MeV}{\mathrm{MeV}}
\newcommand{\nb}{\mathrm{nb}}
\newcommand{\pb}{\mathrm{pb}}
\newcommand{\fb}{\mathrm{fb}}
\def\Re{\mathop{\operator@font Re}\nolimits}
\def\Im{\mathop{\operator@font Im}\nolimits}
\newcommand{\ord}[1]{{\cal O}\lpar#1\rpar}
\newcommand{\group}{SU(2)\otimes U(1)}
\newcommand{\ib}{i}
\newcommand{\asums}[1]{\sum_{#1}}
\newcommand{\asumt}[2]{\sum_{#1}^{#2}}
\newcommand{\asum}[3]{\sum_{#1=#2}^{#3}}
%
%
\newcommand{\tmi}{\times 10^{-1}}
\newcommand{\tmii}{\times 10^{-2}}
\newcommand{\tmiii}{\times 10^{-3}}
\newcommand{\tmiv}{\times 10^{-4}}
\newcommand{\tmfv}{\times 10^{-5}}
\newcommand{\tmfvi}{\times 10^{-6}}
\newcommand{\tmfvii}{\times 10^{-7}}
\newcommand{\tmfviii}{\times 10^{-8}}
\newcommand{\tmfix}{\times 10^{-9}}
\newcommand{\tmfx}{\times 10^{-10}}
%
%
\newcommand{\fer}{{\rm{fer}}}
\newcommand{\bos}{{\rm{bos}}}
\newcommand{\lep}{{l}}
\newcommand{\had}{{h}}
\newcommand{\gen}{\rm{g}}
\newcommand{\dbl}{\rm{d}}
\newcommand{\philone}{\phi}
\newcommand{\philoneb}{\phi_{0}}
\newcommand{\phiind}[1]{\phi_{#1}}
\newcommand{\gBi}[2]{B_{#1}^{#2}}
\newcommand{\gBn}[1]{B^{#1}}
%
%
\newcommand{\ph}{\gamma}
\newcommand{\ab}{A}
\newcommand{\abp}{A'}
\newcommand{\abr}{A^r}
\newcommand{\abb}{A^{0}}
\newcommand{\abi}[1]{A_{#1}}
\newcommand{\abpi}[1]{A'_{#1}}
\newcommand{\abri}[1]{A^r_{#1}}
\newcommand{\abbi}[1]{A^{0}_{#1}}
\newcommand{\wb}{W}            
\newcommand{\wbi}[1]{W_{#1}}           
\newcommand{\wbp}{W^{+}}
\newcommand{\wbm}{W^{-}}
\newcommand{\wbpm}{W^{\pm}}
\newcommand{\wbpi}[1]{W^{+}_{#1}}
\newcommand{\wbmi}[1]{W^{-}_{#1}}
\newcommand{\wbpmi}[1]{W^{\pm}_{#1}}
\newcommand{\wbmpi}[1]{W^{\mp}_{#1}}
\newcommand{\wbli}[1]{W^{[+}_{#1}}
\newcommand{\wbri}[1]{W^{-]}_{#1}}
\newcommand{\zb}{Z}
\newcommand{\zbi}[1]{Z_{#1}}
\newcommand{\vb}{V}
\newcommand{\vbi}[1]{V_{#1}}      
\newcommand{\vbiv}[1]{V^{*}_{#1}}      
\newcommand{\Pb}{P}
\newcommand{\Sb}{S}
\newcommand{\Bb}{B}
%
%
\newcommand{\vph}{\varphi}
\newcommand{\hk}{K}
\newcommand{\hKi}[1]{K_{#1}}
\newcommand{\hkg}{\phi}
\newcommand{\hkn}{\phi^{0}}                 
\newcommand{\hkp}{\phi^{+}}
\newcommand{\hkm}{\phi^{-}}
\newcommand{\hkpm}{\phi^{\pm}}
\newcommand{\hkmp}{\phi^{\mp}}
\newcommand{\hki}[1]{\phi^{#1}}
\newcommand{\hb}{H}
\newcommand{\hbi}[1]{H_{#1}}
\newcommand{\hkl}{\phi^{[+\cgfi\cgfi}}
\newcommand{\hkr}{\phi^{-]}}
%
%
\newcommand{\fpx}{X}
\newcommand{\fpy}{Y}
\newcommand{\fpxp}{X^+}
\newcommand{\fpxm}{X^-}
\newcommand{\fpxpm}{X^{\pm}}
\newcommand{\fpxi}[1]{X^{#1}}
\newcommand{\fpyZ}{Y^{\ssZ}}
\newcommand{\fpyA}{Y^{\ssA}}
\newcommand{\fpyG}{Y^{\ssG}}
\newcommand{\fpyZA}{Y_{\ssZ,\ssA}}
\newcommand{\fpbxi}[1]{{\overline{X}}^{#1}}
\newcommand{\fpbyZ}{{\overline{Y}}^{\ssZ}}
\newcommand{\fpbyA}{{\overline{Y}}^{\ssA}}
\newcommand{\fpbyZA}{{\overline{Y}}^{\ssZ,\ssA}}
%
%
\newcommand{\Flone}{F}
\newcommand{\fpsi}{\psi}
\newcommand{\fpsif}{\psi_f}
\newcommand{\fpsifp}{\psi'_f}
\newcommand{\fpsii}[1]{\psi^{#1}}
\newcommand{\fbpsif}{{\overline{\psi}_f}}
\newcommand{\fbpsifp}{{\overline{\psi}'_f}}
\newcommand{\fpsib}{\psi_{\sszero}}
\newcommand{\fpsir}{\psi^r}
\newcommand{\fpsiL}{\psi_{_L}}
\newcommand{\fpsiR}{\psi_{_R}}
\newcommand{\fpsiLi}[1]{\psi_{_L}^{#1}}
\newcommand{\fpsiRi}[1]{\psi_{_R}^{#1}}
\newcommand{\fpsiLbi}[1]{\psi_{_{0L}}^{#1}}
\newcommand{\fpsiRbi}[1]{\psi_{_{0R}}^{#1}}
\newcommand{\fpsiLR}{\psi_{_{L,R}}}
\newcommand{\fbpsi}{{\overline{\psi}}}
\newcommand{\fbpsii}[1]{{\overline{\psi}}^{#1}}
\newcommand{\fbpsir}{{\overline{\psi}}^r}
\newcommand{\fbpsiL}{{\overline{\psi}}_{_L}}
\newcommand{\fbpsiR}{{\overline{\psi}}_{_R}}
\newcommand{\fbpsiLi}[1]{\overline{\psi_{_L}}^{#1}}
\newcommand{\fbpsiRi}[1]{\overline{\psi_{_R}}^{#1}}
\newcommand{\FQED}[2]{F_{#1#2}}
\newcommand{\FQEDp}[2]{F'_{#1#2}}
\newcommand{\fe}{e}
\newcommand{\fep}{e^{+}}
\newcommand{\fem}{e^{-}}
\newcommand{\fepm}{e^{\pm}}
\newcommand{\fp}{f^{+}}
\newcommand{\fm}{f^{-}}
\newcommand{\fhp}{h^{+}}
\newcommand{\fhm}{h^{-}}
\newcommand{\fh}{h}
\newcommand{\flm}{\mu}
\newcommand{\flmp}{\mu^{+}}
\newcommand{\flmm}{\mu^{-}}
\newcommand{\fll}{l}
\newcommand{\fllp}{l^{+}}
\newcommand{\fllm}{l^{-}}
\newcommand{\flt}{\tau}
\newcommand{\fltp}{\tau^{+}}
\newcommand{\fltm}{\tau^{-}}
\newcommand{\fq}{q}
\newcommand{\fqi}[1]{\fq_{#1}}
\newcommand{\bfqi}[1]{\barq_{#1}}
\newcommand{\ffQ}{Q}
\newcommand{\fu}{u}
\newcommand{\fd}{d}
\newcommand{\fc}{c}
\newcommand{\fs}{s}
\newcommand{\fqp}{q'}
\newcommand{\fup}{u'}
\newcommand{\fdp}{d'}
\newcommand{\fcp}{c'}
\newcommand{\fsp}{s'}
\newcommand{\fdpp}{d''}
\newcommand{\ffi}[1]{f_{#1}}
\newcommand{\bffi}[1]{{\overline{f}}_{#1}}
\newcommand{\ffpi}[1]{f'_{#1}}
\newcommand{\bffpi}[1]{{\overline{f}}'_{#1}}
\newcommand{\ft}{t}
\newcommand{\ffb}{b}
\newcommand{\fl}{l}
\newcommand{\fli}[1]{\fl_{#1}}
\newcommand{\fnu}{\nu}
\newcommand{\fU}{U}
\newcommand{\fD}{D}
\newcommand{\fUc}{\overline{U}}
\newcommand{\fDc}{\overline{D}}
\newcommand{\fnul}{\nu_l}
\newcommand{\fnue}{\nu_e}
\newcommand{\fnum}{\nu_{\mu}}
\newcommand{\fnut}{\nu_{\tau}}
\newcommand{\fbe}{{\overline{e}}}
\newcommand{\fblm}{{\overline{\mu}}}
\newcommand{\fblt}{{\overline{\tau}}}
\newcommand{\fbu}{{\overline{u}}}
\newcommand{\fbd}{{\overline{d}}}
\newcommand{\fbf}{{\overline{f}}}
\newcommand{\fbfp}{{\overline{f}}'}
\newcommand{\fbl}{{\overline{l}}}
\newcommand{\fbnu}{{\overline{\nu}}}
\newcommand{\fbnul}{{\overline{\nu}}_{\fl}}
\newcommand{\fbnue}{{\overline{\nu}}_{\fe}}
\newcommand{\fbnum}{{\overline{\nu}}_{\flm}}
\newcommand{\fbnut}{{\overline{\nu}}_{\flt}}
\newcommand{\fuL}{u_{_L}}
\newcommand{\fdL}{d_{_L}}
\newcommand{\ffL}{f_{_L}}
\newcommand{\fbuL}{{\overline{u}}_{_L}}
\newcommand{\fbdL}{{\overline{d}}_{_L}}
\newcommand{\fbfL}{{\overline{f}}_{_L}}
\newcommand{\fuR}{u_{_R}}
\newcommand{\fdR}{d_{_R}}
\newcommand{\ffR}{f_{_R}}
\newcommand{\fbuR}{{\overline{u}}_{_R}}
\newcommand{\fbdR}{{\overline{d}}_{_R}}
\newcommand{\fbfR}{{\overline{f}}_{_R}}
%
%
\newcommand{\barf}{\overline f}                
\newcommand{\barl}{\overline l}
\newcommand{\barq}{\overline q}
\newcommand{\barqp}{\overline{q}'}
\newcommand{\barb}{\overline b}
\newcommand{\bart}{\overline t}
\newcommand{\barc}{\overline c}
\newcommand{\baru}{\overline u}
\newcommand{\bard}{\overline d}
\newcommand{\bars}{\overline s}
\newcommand{\barv}{\overline v}
\newcommand{\barnu}{\overline{\nu}}
\newcommand{\barne}{\overline{\nu}_{\fe}}
\newcommand{\barnm}{\overline{\nu}_{\flm}}
\newcommand{\barnt}{\overline{\nu}_{\flt}}
%
%
\newcommand{\glu}{g}
%
%
\newcommand{\prot}{p}
\newcommand{\aprot}{{\bar{p}}}
\newcommand{\Nucln}{N}
%
%
\newcommand{\tM}{{\tilde M}}
\newcommand{\tMs}{{\tilde M}^2}
\newcommand{\tW}{{\tilde \Gamma}}
\newcommand{\tWs}{{\tilde\Gamma}^2}
\newcommand{\fphi}{\phi}
\newcommand{\fJpsi}{J/\psi}
\newcommand{\fgpsi}{\psi}
\newcommand{\Glone}{\Gamma}
\newcommand{\Gloni}[1]{\Gamma_{#1}}
\newcommand{\Glones}{\Gamma^2}
\newcommand{\Glonec}{\Gamma^3}
\newcommand{\glone}{\gamma}
\newcommand{\glones}{\gamma^2}
\newcommand{\gloneq}{\gamma^4}
\newcommand{\gloni}[1]{\gamma_{#1}}
\newcommand{\glonis}[1]{\gamma^2_{#1}}
\newcommand{\Grest}[2]{\Gamma_{#1}^{#2}}
\newcommand{\grest}[2]{\gamma_{#1}^{#2}}
\newcommand{\resampl}{A_{_{\rm{R}}}}
\newcommand{\resasyi}[1]{{\cal{A}}_{#1}}
\newcommand{\sSrest}[1]{\sigma_{#1}}
\newcommand{\Srest}[2]{\sigma_{#1}\lpar{#2}\rpar}
\newcommand{\Gdist}[1]{{\cal{G}}\lpar{#1}\rpar}
\newcommand{\sGdist}{{\cal{G}}}
\newcommand{\Aarea}{A_{0}}
\newcommand{\Aareai}[1]{{\cal{A}}\lpar{#1}\rpar}
\newcommand{\sAarea}{{\cal{A}}}
\newcommand{\resolw}{\sigma_{\ssE}}
\newcommand{\resolws}{\sigma^2_{\ssE}}
\newcommand{\chizer}{\chi_{0}}
\newcommand{\ini}{\rm{in}}
\newcommand{\fin}{\rm{fin}}
\newcommand{\ifi}{\rm{if}}
\newcommand{\ipf}{\rm{i+f}}
\newcommand{\tot}{\rm{T}}
\newcommand{\FB}{\rm{FB}}
\newcommand{\Nn}{\rm{N}}
\newcommand{\Bac}{\rm{Q}}
\newcommand{\Res}{\rm{R}}
\newcommand{\Int}{\rm{I}}
\newcommand{\NRe}{\rm{NR}}
\newcommand{\ratoe}{\delta}
\newcommand{\ratoes}{\delta^2}
%
%
\newcommand{\Fbox}[2]{f^{\rm{box}}_{#1}\lpar{#2}\rpar}
\newcommand{\Dbox}[2]{\delta^{\rm{box}}_{#1}\lpar{#2}\rpar}
\newcommand{\Bbox}[3]{{\cal{B}}_{#1}^{#2}\lpar{#3}\rpar}
%
%
\newcommand{\phm}{\lambda}
\newcommand{\phms}{\lambda^2}
\newcommand{\mV}{M_{_V}}
\newcommand{\mw}{M_{_W}}
\newcommand{\mX}{M_{_X}}
\newcommand{\mY}{M_{_Y}}
\newcommand{\LM}{M}
\newcommand{\mz}{M_{_Z}}
\newcommand{\bzm}{M_{_0}}
\newcommand{\bhm}{M_{_{0H}}}
\newcommand{\mf}{m_f}
\newcommand{\mfp}{m_{f'}}
\newcommand{\mfh}{m_{h}}
\newcommand{\me}{m_e}
\newcommand{\mm}{m_{\mu}}
\newcommand{\mtau}{m_{\tau}}
\newcommand{\muq}{m_u}
\newcommand{\md}{m_d}
\newcommand{\muqp}{m'_u}
\newcommand{\mdqp}{m'_d}
\newcommand{\ms}{m_s}
\newcommand{\mup}{M_u}                              
\newcommand{\mdp}{M_d}
\newcommand{\mcp}{M_c}
\newcommand{\msp}{M_s}
\newcommand{\mbp}{M_b}
%
%
\newcommand{\mls}{m^2_l}
\newcommand{\mVs}{M^2_{_V}}
\newcommand{\mVq}{M^4_{_V}}
\newcommand{\mws}{M^2_{_W}}
\newcommand{\mwc}{M^3_{_W}}
\newcommand{\LMs}{M^2}
\newcommand{\LMc}{M^3}
\newcommand{\mzs}{M^2_{_Z}}
\newcommand{\mzc}{M^3_{_Z}}
\newcommand{\bzms}{M^2_{_0}}
\newcommand{\bzmc}{M^3_{_0}}
\newcommand{\bhms}{M^2_{_{0H}}}
\newcommand{\mhs}{M^2_{_H}}
\newcommand{\mfs}{m^2_f}
\newcommand{\mfc}{m^3_f}
\newcommand{\mfps}{m^2_{f'}}
\newcommand{\mfhs}{m^2_{h}}
\newcommand{\mfpc}{m^3_{f'}}
\newcommand{\mts}{m^2_t}
\newcommand{\mes}{m^2_e}
\newcommand{\mms}{m^2_{\mu}}
\newcommand{\mmc}{m^3_{\mu}}
\newcommand{\mmfour}{m^4_{\mu}}
\newcommand{\mmf}{m^5_{\mu}}
\newcommand{\mmfive}{m^5_{\mu}}
\newcommand{\mmsix}{m^6_{\mu}}
\newcommand{\mtsix}{m^6_{\ft}}
\newcommand{\mminv}{\frac{1}{m_{\mu}}}
\newcommand{\mtaus}{m^2_{\tau}}
\newcommand{\mus}{m^2_u}
\newcommand{\mds}{m^2_d}
\newcommand{\muqps}{m'^2_u}
\newcommand{\mdqps}{m'^2_d}
\newcommand{\mcs}{m^2_c}
\newcommand{\mss}{m^2_s}
\newcommand{\mbs}{m^2_b}
\newcommand{\mups}{M^2_u}
\newcommand{\mdps}{M^2_d}
\newcommand{\mcps}{M^2_c}
\newcommand{\msps}{M^2_s}
\newcommand{\mbps}{M^2_b}
%
%
\newcommand{\muf}{\mu_{\ff}}
\newcommand{\mufs}{\mu^2_{\ff}}
\newcommand{\mufq}{\mu^4_{\ff}}
\newcommand{\mufx}{\mu^6_{\ff}}
\newcommand{\muz}{\mu_{_{\zb}}}
\newcommand{\muw}{\mu_{_{\wb}}}
\newcommand{\mut}{\mu_{\ft}}
\newcommand{\muzs}{\mu^2_{_{\zb}}}
\newcommand{\muws}{\mu^2_{_{\wb}}}
\newcommand{\muts}{\mu^2_{\ft}}
\newcommand{\mubs}{\mu^2_{\ffb}}
\newcommand{\muSW}{\mu^2_{_{\wb}}}
\newcommand{\muwq}{\mu^4_{_{\wb}}}
\newcommand{\muwsx}{\mu^6_{_{\wb}}}
\newcommand{\muwms}{\mu^{-2}_{_{\wb}}}
\newcommand{\muhs}{\mu^2_{_{\hb}}}
\newcommand{\muhq}{\mu^4_{_{\hb}}}
\newcommand{\muhsx}{\mu^6_{_{\hb}}}
\newcommand{\mutq}{\mu^4_{_{\hb}}}   
\newcommand{\mutsx}{\mu^6_{_{\hb}}}  
\newcommand{\muL}{\mu}
\newcommand{\muS}{\mu^2}
\newcommand{\muQ}{\mu^4}
\newcommand{\muizs}{\mu^2_{0}}
\newcommand{\muizq}{\mu^4_{0}}
\newcommand{\muis}{\mu^2_{1}}
\newcommand{\muiis}{\mu^2_{2}}
\newcommand{\muiiis}{\mu^2_{3}}
\newcommand{\muii}[1]{\mu_{#1}}
\newcommand{\muisi}[1]{\mu^2_{#1}}
\newcommand{\muiqi}[1]{\mu^4_{#1}}
\newcommand{\muixi}[1]{\mu^6_{#1}}
\newcommand{\zm}{z_m}
\newcommand{\mwzs}{M^2_{\ssW,\ssZ}}
\newcommand{\ri}[1]{r_{#1}}
\newcommand{\xw}{x_w}
\newcommand{\xws}{x^2_w}
\newcommand{\xwc}{x^3_w}
\newcommand{\xth}{x_t}
\newcommand{\xths}{x^2_t}
\newcommand{\xthc}{x^3_t}
\newcommand{\xthf}{x^4_t}
\newcommand{\xthv}{x^5_t}
\newcommand{\xthx}{x^6_t}
\newcommand{\xh}{x_h}
\newcommand{\xhs}{x^2_h}
\newcommand{\xhc}{x^3_h}
\newcommand{\Rl}{R_{\fl}}
\newcommand{\Rb}{R_{\ffb}}
\newcommand{\Rc}{R_{\fc}}
%
%
\newcommand{\mwq}{M^4_{_\wb}}
\newcommand{\mwf}{M^4_{_\wb}}
\newcommand{\LMq}{M^4}
\newcommand{\mzq}{M^4_{_Z}}
\newcommand{\bzmq}{M^4_{_0}}
\newcommand{\mhq}{M^4_{_H}}
\newcommand{\mfq}{m^4_f}
\newcommand{\mfpq}{m^4_{f'}}
\newcommand{\mtq}{m^4_t}
\newcommand{\meq}{m^4_e}
\newcommand{\mmq}{m^4_{\mu}}
\newcommand{\mtauq}{m^4_{\tau}}
\newcommand{\muqq}{m^4_u}
\newcommand{\mdq}{m^4_d}
\newcommand{\mcq}{m^4_c}
\newcommand{\msq}{m^4_s}
\newcommand{\mbq}{m^4_b}
\newcommand{\mupq}{M^4_u}
\newcommand{\mdpq}{M^4_d}
\newcommand{\mcpq}{M^4_c}
\newcommand{\mspq}{M^4_s}
\newcommand{\mbpq}{M^4_b}
%
%
\newcommand{\mwx}{M^6_{_W}}
\newcommand{\mzx}{M^6_{_Z}}
\newcommand{\mfx}{m^6_f}
\newcommand{\mfpx}{m^6_{f'}}
\newcommand{\LMx}{M^6}
%
%
\newcommand{\mer}{m_{er}}
\newcommand{\mlep}{m_l}
\newcommand{\mleps}{m^2_l}
\newcommand{\mzer}{m_0}
\newcommand{\mone}{m_1}
\newcommand{\mtwo}{m_2}
\newcommand{\mtre}{m_3}
\newcommand{\mfor}{m_4}
\newcommand{\mlone}{m}
\newcommand{\mloneb}{\bar{m}}
\newcommand{\mind}[1]{m_{#1}}
\newcommand{\mzers}{m^2_0}
\newcommand{\mones}{m^2_1}
\newcommand{\mtwos}{m^2_2}
\newcommand{\mtres}{m^2_3}
\newcommand{\mfors}{m^2_4}
\newcommand{\mlones}{m^2}
\newcommand{\minds}[1]{m^2_{#1}}
\newcommand{\mlonec}{m^3}
\newcommand{\moneq}{m^4_1}
\newcommand{\mtwoq}{m^4_2}
\newcommand{\mtreq}{m^4_3}
\newcommand{\mforq}{m^4_4}
\newcommand{\mloneq}{m^4}
\newcommand{\mindq}[1]{m^4_{#1}}
\newcommand{\mlonev}{m^5}
\newcommand{\mindv}[1]{m^5_{#1}}
\newcommand{\monex}{m^6_1}
\newcommand{\mtwox}{m^6_2}
\newcommand{\mtrex}{m^6_3}
\newcommand{\mforx}{m^6_4}
\newcommand{\mlonex}{m^6}
\newcommand{\mindx}[1]{m^6_{#1}}
\newcommand{\Mone}{M_1}
\newcommand{\Mtwo}{M_2}
\newcommand{\Mtre}{M_3}
\newcommand{\Mfor}{M_4}
\newcommand{\Mlone}{M}
\newcommand{\Mlonep}{M'}
\newcommand{\Miind}{M_i}
\newcommand{\Mind}[1]{M_{#1}}
\newcommand{\Minds}[1]{M^2_{#1}}
\newcommand{\Mindc}[1]{M^3_{#1}}
\newcommand{\Mindf}[1]{M^4_{#1}}
\newcommand{\Mones}{M^2_1}
\newcommand{\Mtwos}{M^2_2}
\newcommand{\Mtres}{M^2_3}
\newcommand{\Mfors}{M^2_4}
\newcommand{\Mlones}{M^2}
\newcommand{\Mloneps}{M'^2}
\newcommand{\Miinds}{M^2_i}
\newcommand{\Mlonec}{M^3}
\newcommand{\Monec}{M^3_1}
\newcommand{\Mtwoc}{M^3_2}
\newcommand{\Moneq}{M^4_1}
\newcommand{\Mtwoq}{M^4_2}
\newcommand{\Mtreq}{M^4_3}
\newcommand{\Mforq}{M^4_4}
\newcommand{\Mloneq}{M^4}
\newcommand{\Miindq}{M^4_i}
\newcommand{\Monex}{M^6_1}
\newcommand{\Mtwox}{M^6_2}
\newcommand{\Mtrex}{M^6_3}
\newcommand{\Mforx}{M^6_4}
\newcommand{\Mlonex}{M^6}
\newcommand{\Miindx}{M^6_i}
\newcommand{\meb}{m_{\sszero}}
\newcommand{\mebs}{m^2_{\sszero}}
%
%
\newcommand{\Mq }{M_q  }
\newcommand{\MqS}{M^2_q}
\newcommand{\Ms }{M_s  }
\newcommand{\MsS}{M^2_s}
\newcommand{\Mc }{M_c  }
\newcommand{\McS}{M^2_c}
\newcommand{\Mb }{M_b  }
\newcommand{\MbS}{M^2_b}
\newcommand{\Mt }{M_t  }
\newcommand{\MtS}{M^2_t}
%
%
\newcommand{\mq}{m_q}
\newcommand{\mqs}{m^2_q}
\newcommand{\mqS}{m^2_q}
\newcommand{\mqQ}{m^4_q}
\newcommand{\mqX}{m^6_q}
\newcommand{\mqp}{m'_q }
\newcommand{\mqpS}{m'^2_q}
\newcommand{\mqpQ}{m'^4_q}
%
%
\newcommand{\lL}{l}
\newcommand{\lLi}[1]{l_{#1}}
\newcommand{\ls}{l^2}
\newcommand{\LL}{L}
\newcommand{\LcalL}{\cal{L}}
\newcommand{\LS}{L^2}
\newcommand{\LC}{L^3}
\newcommand{\LQ}{L^4}
\newcommand{\lw}{l_w}
\newcommand{\Lw}{L_w}
\newcommand{\Lws}{L^2_w}
\newcommand{\Lz}{L_z}
\newcommand{\Lzs}{L^2_z}
\newcommand{\Lis}[1]{L^2_{#1}}
\newcommand{\Lic}[1]{L^3_{#1}}
%
%
\newcommand{\sman}{s}
\newcommand{\tman}{t}
\newcommand{\uman}{u}
\newcommand{\smani}[1]{s_{#1}}
\newcommand{\bsmani}[1]{{\bar{s}}_{#1}}
\newcommand{\smans}{s^2}
\newcommand{\tmans}{t^2}
\newcommand{\umans}{u^2}
\newcommand{\shat}{{\hat s}}
\newcommand{\that}{{\hat t}}
\newcommand{\uhat}{{\hat u}}
%
%
\newcommand{\smanp}{s'}
\newcommand{\smanpi}[1]{s'_{#1}}
\newcommand{\tmanp}{t'}
\newcommand{\umanp}{u'}
\newcommand{\kappi}[1]{\kappa_{#1}}
\newcommand{\zetai}[1]{\zeta_{#1}}
%
%
%
\newcommand{\Phaspi}[1]{\Gamma_{#1}}
\newcommand{\rbetai}[1]{\beta_{#1}}
\newcommand{\ralphai}[1]{\alpha_{#1}}
\newcommand{\rbetais}[1]{\beta^2_{#1}}
\newcommand{\Lambdi}[1]{\Lambda_{#1}}
\newcommand{\Nomini}[1]{N_{#1}}
\newcommand{\smlone}{\frac{-\sman-\ib\ep}{\mlones}}
%
%
\newcommand{\theti}[1]{\theta_{#1}}
\newcommand{\delti}[1]{\delta_{#1}}
\newcommand{\phigi}[1]{\phi_{#1}}
\newcommand{\acoli}[1]{\xi_{#1}}
\newcommand{\scats}{s}
\newcommand{\scatss}{s^2}
\newcommand{\scatsi}[1]{s_{#1}}
\newcommand{\scatsis}[1]{s^2_{#1}}
\newcommand{\scatst}[2]{s_{#1}^{#2}}
\newcommand{\scatc}{c}
\newcommand{\scatcs}{c^2}
\newcommand{\scatci}[1]{c_{#1}}
\newcommand{\scatcis}[1]{c^2_{#1}}
\newcommand{\scatct}[2]{c_{#1}^{#2}}
\newcommand{\angamt}[2]{\gamma_{#1}^{#2}}
\newcommand{\scatsin}{\sin\theta}
\newcommand{\scatsins}{\sin^2\theta}
\newcommand{\scatcos}{\cos\theta}
\newcommand{\scatcoss}{\cos^2\theta}
%
%
\newcommand{\Regia}{{\cal{R}}}
\newcommand{\Iconi}[2]{{\cal{I}}_{#1}\lpar{#2}\rpar}
\newcommand{\sIcon}[1]{{\cal{I}}_{#1}}
\newcommand{\betaf}{\beta_{\ff}}
\newcommand{\betafs}{\beta^2_{\ff}}
\newcommand{\betafc}{\beta^3_{\ff}}
\newcommand{\Kfact}[2]{{\cal{K}}_{#1}\lpar{#2}\rpar}
%
%
\newcommand{\Struf}[4]{{\cal D}^{#1}_{#2}\lpar{#3;#4}\rpar}
\newcommand{\sStruf}[2]{{\cal D}^{#1}_{#2}}
\newcommand{\Fluxf}[2]{H\lpar{#1;#2}\rpar}
\newcommand{\Fluxfi}[4]{H_{#1}^{#2}\lpar{#3;#4}\rpar}
\newcommand{\sFluxf}{H}
\newcommand{\Bflux}[2]{{\cal{B}}_{#1}\lpar{#2}\rpar}
\newcommand{\bflux}[2]{{\cal{B}}_{#1}\lpar{#2}\rpar}
\newcommand{\Fluxd}[2]{D_{#1}\lpar{#2}\rpar}
\newcommand{\fluxd}[2]{C_{#1}\lpar{#2}\rpar}
\newcommand{\Fluxh}[4]{{\cal{H}}_{#1}^{#2}\lpar{#3;#4}\rpar}
\newcommand{\Sluxh}[4]{{\cal{S}}_{#1}^{#2}\lpar{#3;#4}\rpar}
\newcommand{\Fluxhb}[4]{{\overline{{\cal{H}}}}_{#1}^{#2}\lpar{#3;#4}\rpar}
\newcommand{\sFluxhb}{{\overline{{\cal{H}}}}}
\newcommand{\Sluxhb}[4]{{\overline{{\cal{S}}}}_{#1}^{#2}\lpar{#3;#4}\rpar}
\newcommand{\sSluxhb}[2]{{\overline{{\cal{S}}}}_{#1}^{#2}}
\newcommand{\fluxh}[4]{h_{#1}^{#2}\lpar{#3;#4}\rpar}
\newcommand{\fluxhs}[3]{h_{#1}^{#2}\lpar{#3}\rpar}
\newcommand{\sfluxhs}[2]{h_{#1}^{#2}}
\newcommand{\fluxhb}[4]{{\overline{h}}_{#1}^{#2}\lpar{#3;#4}\rpar}
\newcommand{\Strufd}[2]{D\lpar{#1;#2}\rpar}
%
%
\newcommand{\rMQ}[1]{r^2_{#1}}
\newcommand{\rMQs}[1]{r^4_{#1}}
\newcommand{\hf}{h_{\ff}}
\newcommand{\rf}{w_{\ff}}
\newcommand{\rfs}{w^2_{\ff}}
\newcommand{\zfs}{z^2_{\ff}}
\newcommand{\rfc}{w^3_{\ff}}
\newcommand{\zfc}{z^3_{\ff}}
\newcommand{\df}{d_{\ff}}
\newcommand{\rfp}{w_{\ffp}}
\newcommand{\rfps}{w^2_{\ffp}}
\newcommand{\rfpc}{w^3_{\ffp}}
\newcommand{\Lrfp}{L_{w}}
\newcommand{\rt}{w_{\ft}}
\newcommand{\rts}{w^2_{\ft}}
\newcommand{\rb}{w_{\ffb}}
\newcommand{\rbs}{w^2_{\ffb}}
\newcommand{\dt}{d_{\ft}}
\newcommand{\dts}{d^2_{\ft}}
\newcommand{\rh}{r_{h}}
\newcommand{\Lnrt}{\ln{\rt}}
\newcommand{\Rw}{R_{_{\wb}}}
\newcommand{\Rws}{R^2_{_{\wb}}}
\newcommand{\Rz}{R_{_{\zb}}}
\newcommand{\Rzp}{R^{+}_{_{\zb}}}
\newcommand{\Rzm}{R^{-}_{_{\zb}}}
\newcommand{\Rzs}{R^2_{_{\zb}}}
\newcommand{\Rzc}{R^3_{_{\zb}}}
\newcommand{\Rv}{R_{_{\vb}}}
\newcommand{\rhw}{w_h}
\newcommand{\rhz}{z_h}
\newcommand{\rhws}{w^2_h}
\newcommand{\rhzs}{z^2_h}
%
%
\newcommand{\vqrato}{z}
\newcommand{\vqrats}{w}
\newcommand{\vqratq}{w^2}
\newcommand{\seyrat}{z}
\newcommand{\sexrat}{w}
\newcommand{\sexrats}{w^2}
\newcommand{\sehrat}{h}
\newcommand{\sewrat}{w}
\newcommand{\sezrat}{z}
\newcommand{\zetav}{\zeta}
\newcommand{\zetavi}[1]{\zeta_{#1}}
\newcommand{\bpo}{\beta^2}
\newcommand{\bpos}{\beta^4}
\newcommand{\bpt}{{\tilde\beta}^2}
\newcommand{\lap}{\kappa}
\newcommand{\hw}{w_h}
\newcommand{\hz}{z_h}
%
%
\newcommand{\ec}{e}
\newcommand{\ecs}{e^2}
\newcommand{\ect}{e^3}
\newcommand{\ecq}{e^4}
\newcommand{\ecb}{e_{\sszero}}
\newcommand{\ecbs}{e^2_{_0}}
\newcommand{\ecbq}{e^4_{_0}}
\newcommand{\eci}[1]{e_{#1}}
\newcommand{\ecis}[1]{e^2_{#1}}
\newcommand{\hate}{{\hat e}}
\newcommand{\gss}{g_{_S}}
\newcommand{\gsss}{g^2_{_S}}
\newcommand{\gssb}{g^2_{_{S_0}}}
\newcommand{\als}{\alpha_{_S}}
\newcommand{\as}{a_{_S}}
\newcommand{\ass}{a^2_{_S}}
\newcommand{\gf}{G_{\ssF}}
\newcommand{\gfs}{G^2_{\ssF}}
\newcommand{\gb}{g} 
\newcommand{\gbi}[1]{g_{#1}}
\newcommand{\gbb}{g_{0}}
\newcommand{\gbs}{g^2}
\newcommand{\gbc}{g^3}
\newcommand{\gbf}{g^4}
\newcommand{\gpb}{g'}
\newcommand{\gpbs}{g'^2}
\newcommand{\vc}[1]{v_{#1}}
\newcommand{\ac}[1]{a_{#1}}
\newcommand{\vcc}[1]{v^*_{#1}}
\newcommand{\acc}[1]{a^*_{#1}}
\newcommand{\hatv}[1]{{\hat v}_{#1}}
\newcommand{\vcs}[1]{v^2_{#1}}
\newcommand{\acs}[1]{a^2_{#1}}
\newcommand{\gcv}[1]{g^{#1}_{\ssV}}
\newcommand{\gca}[1]{g^{#1}_{\ssA}}
\newcommand{\gcp}[1]{g^{+}_{#1}}
\newcommand{\gcm}[1]{g^{-}_{#1}}
\newcommand{\gcpm}[1]{g^{\pm}_{#1}}
\newcommand{\vci}[2]{v^{#2}_{#1}}
\newcommand{\aci}[2]{a^{#2}_{#1}}
\newcommand{\vceff}[1]{v^{#1}_{\rm{eff}}}
\newcommand{\hvc}[1]{\hat{v}_{#1}}
\newcommand{\hvcs}[1]{\hat{v}^2_{#1}}
\newcommand{\Vc}[1]{V_{#1}}
\newcommand{\Ac}[1]{A_{#1}}
\newcommand{\Vcs}[1]{V^2_{#1}}
\newcommand{\Acs}[1]{A^2_{#1}}
\newcommand{\vpa}[2]{\sigma_{#1}^{#2}}
\newcommand{\vma}[2]{\delta_{#1}^{#2}}
\newcommand{\vfw}{\sigma^{a}_{\ff}}
\newcommand{\vfpw}{\sigma^{a}_{\ffp}}
\newcommand{\vfwi}[1]{\sigma^{a}_{#1}}
\newcommand{\vfwsi}[1]{\lpar\sigma^{a}_{#1}\rpar^2}
\newcommand{\vvfw}{v^{a}_{\ff}}
\newcommand{\vvew}{v^{a}_{\fe}}
\newcommand{\vzm}{v^{-}_{\ssZ}}
\newcommand{\vzp}{v^{+}_{\ssZ}}
\newcommand{\vzpm}{v^{\pm}_{\ssZ}}
\newcommand{\vzmp}{v^{\mp}_{\ssZ}}
\newcommand{\vam}{v^{-}_{\ssA}}
\newcommand{\vap}{v^{+}_{\ssA}}
\newcommand{\vapm}{v^{\pm}_{\ssA}}
\newcommand{\gv}{g_{_V}}
\newcommand{\gve}{g^{\fe}_{_{V}}}
\newcommand{\gae}{g^{\fe}_{_{A}}}
\newcommand{\gvf}{g^{\ff}_{_{V}}}
\newcommand{\gaf}{g^{\ff}_{_{A}}}
\newcommand{\gva}{g_{_{V,A}}}
\newcommand{\gvae}{g^{\fe}_{_{V,A}}}
\newcommand{\gvaf}{g^{\ff}_{_{V,A}}}
\newcommand{\sGv}{{\cal{G}}_{_V}}
\newcommand{\cGa}{{\cal{G}}^{*}_{_A}}
\newcommand{\cGv}{{\cal{G}}^{*}_{_V}}
\newcommand{\sGa}{{\cal{G}}_{_A}}
\newcommand{\Gvf}{{\cal{G}}^{\ff}_{_{V}}}
\newcommand{\Gaf}{{\cal{G}}^{\ff}_{_{A}}}
\newcommand{\Gvaf}{{\cal{G}}^{\ff}_{_{V,A}}}
\newcommand{\Gve}{{\cal{G}}^{\fe}_{_{V}}}
\newcommand{\Gae}{{\cal{G}}^{\fe}_{_{A}}}
\newcommand{\Gvae}{{\cal{G}}^{\fe}_{_{V,A}}}
\newcommand{\gvl}{g^{\fl}_{_{V}}}
\newcommand{\gal}{g^{\fl}_{_{A}}}
\newcommand{\gval}{g^{\fl}_{_{V,A}}}
\newcommand{\gvb}{g^{\ffb}_{_{V}}}
\newcommand{\gab}{g^{\ffb}_{_{A}}}
\newcommand{\fvf}{F_{_V}^{\ff}}
\newcommand{\faf}{F_{_A}^{\ff}}
\newcommand{\fvl}{F_{_V}^{\fl}}
\newcommand{\fal}{F_{_A}^{\fl}}
\newcommand{\corat}{\kappa}
\newcommand{\corats}{\kappa^2}
%
%
\newcommand{\dr}{\Delta r}
\newcommand{\drl}{\Delta r_{_L}}
\newcommand{\drh}{\Delta{\hat r}}
\newcommand{\drhw}{\Delta{\hat r}_{_W}}
\newcommand{\rhou}{\rho_{_U}}
\newcommand{\rhoz}{\rho_{_\zb}}
\newcommand{\rZ}{\rho_{_\zb}}
\newcommand{\rhob}{\rho_{_0}}
\newcommand{\rZf}{\rho^{f}_{_\zb}}
\newcommand{\rhoe}{\rho_{\fe}}
\newcommand{\rhof}{\rho_{\ff}}
\newcommand{\rhoi}[1]{\rho_{#1}}
\newcommand{\kZf}{\kappa^{f}_{_\zb}}
\newcommand{\rWf}{\rho^{\ff}_{_\wb}}
\newcommand{\brWf}{{\bar{\rho}}^{\ff}_{_\wb}}
\newcommand{\rHf}{\rho^{\ff}_{_\hb}}
\newcommand{\brHf}{{\bar{\rho}}^{\ff}_{_\hb}}
\newcommand{\rhoR}{\rho^{\ssR}_{_{\zb}}}
\newcommand{\hatrh}{{\hat\rho}}
\newcommand{\ku}{\kappa_{\ssU}}
\newcommand{\rZdf}[1]{\rho^{#1}_{_\zb}}
\newcommand{\kZdf}[1]{\kappa^{#1}_{_\zb}}
\newcommand{\rdfL}[1]{\rho^{#1}_{_L}}
\newcommand{\kdfL}[1]{\kappa^{#1}_{_L}}
\newcommand{\rdfR}[1]{\rho^{#1}_{\rm{rem}}}
\newcommand{\kdfR}[1]{\kappa^{#1}_{\rm{rem}}}
\newcommand{\bark}{\overline\kappa}
%
%
\newcommand{\stw}{s_{\theta}}             
\newcommand{\ctw}{c_{\theta}}
\newcommand{\stws}{s_{\theta}^2}
\newcommand{\stwc}{s_{\theta}^3}
\newcommand{\stwf}{s_{\theta}^4}
\newcommand{\stwx}{s_{\theta}^6}
\newcommand{\ctws}{c_{\theta}^2}
\newcommand{\ctwc}{c_{\theta}^3}
\newcommand{\ctwf}{c_{\theta}^4}
\newcommand{\ctwx}{c_{\theta}^6}
\newcommand{\stwfiv}{s_{\theta}^5}
\newcommand{\ctwfiv}{c_{\theta}^5}
\newcommand{\stwsix}{s_{\theta}^6}
\newcommand{\ctwsix}{c_{\theta}^6}
%
%
\newcommand{\siw}{s_{_W}}           
\newcommand{\cow}{c_{_W}}
\newcommand{\siws}{s^2_{_W}}
\newcommand{\cows}{c^2_{_W}}
\newcommand{\siwc}{s^3_{_W}}
\newcommand{\cowc}{c^3_{_W}}
\newcommand{\siwf}{s^4_{_W}}
\newcommand{\cowf}{c^4_{_W}}
\newcommand{\siwx}{s^6_{_W}}
\newcommand{\cowx}{c^6_{_W}}
\newcommand{\sons}{s_{_W}}
\newcommand{\sonss}{s^2_{_W}}
\newcommand{\cons}{c_{_W}}
\newcommand{\cooss}{c^2_{_W}}
%
%
\newcommand{\szs}{{\overline s}^2}
\newcommand{\szq}{{\overline s}^4}
\newcommand{\czs}{{\overline c}^2}
\newcommand{\sbs}{s_{_0}^2}
\newcommand{\cbs}{c_{_0}^2}
\newcommand{\dss}{\Delta s^2}
\newcommand{\snes}{s_{\nu e}^2}
\newcommand{\cnes}{c_{\nu e}^2}
\newcommand{\shs}{{\hat s}^2}
\newcommand{\chs}{{\hat c}^2}
\newcommand{\chl}{{\hat c}}
\newcommand{\seffs}{s^2_{\rm{eff}}}
\newcommand{\seffsf}[1]{\sin^2\theta^{#1}_{\rm{eff}}}
\newcommand{\sress}{s^2_{\rm res}}                
\newcommand{\sR}{s_{_R}}
\newcommand{\sRs}{s^2_{_R}}
\newcommand{\ctwe}{c_{\theta}^6}
\newcommand{\sany}{s}
\newcommand{\cany}{c}
\newcommand{\sanys}{s^2}
\newcommand{\canys}{c^2}
%
%
\newcommand{\sip}{u}                             
\newcommand{\siap}{{\bar{v}}}                    
\newcommand{\sop}{{\bar{u}}}                     
\newcommand{\soap}{v}                            
\newcommand{\ip}[1]{u\lpar{#1}\rpar}             
\newcommand{\iap}[1]{{\bar{v}}\lpar{#1}\rpar}    
\newcommand{\op}[1]{{\bar{u}}\lpar{#1}\rpar}     
\newcommand{\oap}[1]{v\lpar{#1}\rpar}            
%
%
\newcommand{\ipp}[2]{u\lpar{#1,#2}\rpar}         
\newcommand{\ipap}[2]{{\bar v}\lpar{#1,#2}\rpar} 
\newcommand{\opp}[2]{{\bar u}\lpar{#1,#2}\rpar}  
\newcommand{\opap}[2]{v\lpar{#1,#2}\rpar}        
\newcommand{\upspi}[1]{u\lpar{#1}\rpar}
\newcommand{\vpspi}[1]{v\lpar{#1}\rpar}
\newcommand{\wpspi}[1]{w\lpar{#1}\rpar}
\newcommand{\ubpspi}[1]{{\bar{u}}\lpar{#1}\rpar}
\newcommand{\vbpspi}[1]{{\bar{v}}\lpar{#1}\rpar}
\newcommand{\wbpspi}[1]{{\bar{w}}\lpar{#1}\rpar}
\newcommand{\udpspi}[1]{u^{\dagger}\lpar{#1}\rpar}
\newcommand{\vdpspi}[1]{v^{\dagger}\lpar{#1}\rpar}
\newcommand{\wdpspi}[1]{w^{\dagger}\lpar{#1}\rpar}
\newcommand{\Ubilin}[1]{U\lpar{#1}\rpar}
\newcommand{\Vbilin}[1]{V\lpar{#1}\rpar}
\newcommand{\Xbilin}[1]{X\lpar{#1}\rpar}
\newcommand{\Ybilin}[1]{Y\lpar{#1}\rpar}
\newcommand{\up}[2]{u_{#1}\lpar #2\rpar}
\newcommand{\vp}[2]{v_{#1}\lpar #2\rpar}
\newcommand{\ubp}[2]{{\overline u}_{#1}\lpar #2\rpar}
\newcommand{\vbp}[2]{{\overline v}_{#1}\lpar #2\rpar}
\newcommand{\Pje}[1]{\frac{1}{2}\lpar 1 + #1\,\gfd\rpar}
\newcommand{\Pj}[1]{\Pi_{#1}}
\newcommand{\trace}{\mbox{Tr}}
%
%
\newcommand{\Poper}[2]{P_{#1}\lpar{#2}\rpar}
\newcommand{\Loper}[2]{\Lambda_{#1}\lpar{#2}\rpar}
\newcommand{\proj}[3]{P_{#1}\lpar{#2,#3}\rpar}
\newcommand{\sproj}[1]{P_{#1}}
\newcommand{\Nden}[3]{N_{#1}^{#2}\lpar{#3}\rpar}
\newcommand{\sNden}[1]{N_{#1}}
\newcommand{\nden}[2]{n_{#1}^{#2}}
%
%
\newcommand{\vwf}[2]{e_{#1}\lpar#2\rpar}             
\newcommand{\vwfb}[2]{{\overline e}_{#1}\lpar#2\rpar}
\newcommand{\pwf}[2]{\epsilon_{#1}\lpar#2\rpar}      
\newcommand{\sla}[1]{/\!\!\!#1}
\newcommand{\slac}[1]{/\!\!\!\!#1}
%
%
\newcommand{\iemom}{p_{_-}}                    
\newcommand{\ipmom}{p_{_+}}
\newcommand{\oemom}{q_{_-}}                    
\newcommand{\opmom}{q_{_+}}
%
%
\newcommand{\spro}[2]{{#1}\cdot{#2}}
%
%
\newcommand{\gfour}{\gamma_4}                    
\newcommand{\gfd}{\gamma_5}                    
\newcommand{\gap}{\lpar 1+\gamma_5\rpar}
\newcommand{\gam}{\lpar 1-\gamma_5\rpar}
\newcommand{\gdp}{\gamma_+}
\newcommand{\gdm}{\gamma_-}
\newcommand{\gdpm}{\gamma_{\pm}}
\newcommand{\gad}{\gamma}
\newcommand{\gapm}{\lpar 1\pm\gamma_5\rpar}
\newcommand{\gadi}[1]{\gamma_{#1}}
\newcommand{\gadu}[1]{\gamma_{#1}}
\newcommand{\gaduc}[1]{\gamma^{*}_{#1}}
\newcommand{\gaduh}[1]{\gamma^{+}_{#1}}
\newcommand{\gadut}[1]{\gamma^{\ssT}_{#1}}
\newcommand{\gapu}[1]{\gamma^{#1}}
\newcommand{\sigd}[2]{\sigma_{#1#2}}
\newcommand{\sumsp}{\overline{\sum_{\mbox{\tiny{spins}}}}}
%
%
\newcommand{\li}[2]{\mathrm{Li}_{#1}\lpar\displaystyle{#2}\rpar} 
\newcommand{\sli}[1]{\mathrm{Li}_{#1}} 
\newcommand{\etaf}[2]{\eta\lpar#1,#2\rpar}
\newcommand{\lkall}[3]{\lambda\lpar#1,#2,#3\rpar}       
\newcommand{\slkall}[3]{\lambda^{1/2}\lpar#1,#2,#3\rpar}
\newcommand{\segam}{\Gamma}                             
\newcommand{\egam}[1]{\Gamma\lpar#1\rpar}               
\newcommand{\egams}[1]{\Gamma^2\lpar#1\rpar}            
\newcommand{\ebe}[2]{B\lpar#1,#2\rpar}                  
\newcommand{\ddel}[1]{\delta\lpar#1\rpar}               
\newcommand{\ddeln}[1]{\delta^{(n)}\lpar#1\rpar}        
\newcommand{\drii}[2]{\delta_{#1#2}}                    
\newcommand{\driv}[4]{\delta_{#1#2#3#4}}                
\newcommand{\intmomi}[2]{\int\,d^{#1}#2}
\newcommand{\intmomii}[3]{\int\,d^{#1}#2\,\int\,d^{#1}#3}
\newcommand{\intfx}[1]{\int_{\scriptstyle 0}^{\scriptstyle 1}\,d#1}
\newcommand{\intfxy}[2]{\int_{\scriptstyle 0}^{\scriptstyle 1}\,d#1\,
                        \int_{\scriptstyle 0}^{\scriptstyle #1}\,d#2}
\newcommand{\intfxyz}[3]{\int_{\scriptstyle 0}^{\scriptstyle 1}\,d#1\,
                         \int_{\scriptstyle 0}^{\scriptstyle #1}\,d#2\,
                         \int_{\scriptstyle 0}^{\scriptstyle #2}\,d#3}
\newcommand{\Beta}[2]{{\rm{B}}\lpar #1,#2\rpar}
\newcommand{\sBeta}{\rm{B}}
\newcommand{\sign}[1]{{\rm{sign}}\lpar{#1}\rpar}
%
%
\newcommand{\gn}{\Gamma_{\nu}}
\newcommand{\gel}{\Gamma_{\fe}}
\newcommand{\gmu}{\Gamma_{\mu}}
\newcommand{\gff}{\Gamma_{\ff}}
\newcommand{\gt}{\Gamma_{\tau}}
\newcommand{\gl}{\Gamma_{\fl}}
\newcommand{\gq}{\Gamma_{\fq}}
\newcommand{\gu}{\Gamma_{\fu}}
\newcommand{\gd}{\Gamma_{\fd}}
\newcommand{\gc}{\Gamma_{\fc}}
\newcommand{\gs}{\Gamma_{\fs}}
\newcommand{\gbq}{\Gamma_{\ffb}}
\newcommand{\gz}{\Gamma_{_{\zb}}}
\newcommand{\gw}{\Gamma_{_{\wb}}}
\newcommand{\gh}{\Gamma_{\had}}
\newcommand{\ghb}{\Gamma_{_{\hb}}}
\newcommand{\gi}{\Gamma_{\rm{inv}}}
\newcommand{\gzs}{\Gamma^2_{_{\zb}}}
%
%
\newcommand{\tcie}{I^{(3)}_{\fe}}
\newcommand{\tcim}{I^{(3)}_{\flm}}
\newcommand{\tcif}{I^{(3)}_{f}}
\newcommand{\tciq}{I^{(3)}_{\fq}}
\newcommand{\tcib}{I^{(3)}_{\ffb}}
\newcommand{\tcih}{I^{(3)}_h}
\newcommand{\tcii}{I^{(3)}_i}
\newcommand{\tcift}{I^{(3)}_{\tilde f}}
\newcommand{\tcifp}{I^{(3)}_{f'}}
\newcommand{\wispt}[1]{I^{(3)}_{#1}}
\newcommand{\ql}{Q_l}
\newcommand{\qe}{Q_e}
\newcommand{\qu}{Q_u}
\newcommand{\qd}{Q_d}
\newcommand{\qb}{Q_b}
\newcommand{\qt}{Q_t}
\newcommand{\qup}{Q'_u}
\newcommand{\qdp}{Q'_d}
\newcommand{\qmu}{Q_{\mu}}
\newcommand{\qes}{Q^2_e}
\newcommand{\qec}{Q^3_e}
\newcommand{\qus}{Q^2_u}
\newcommand{\qds}{Q^2_d}
\newcommand{\qbs}{Q^2_b}
\newcommand{\qts}{Q^2_t}
\newcommand{\qbc}{Q^3_b}
\newcommand{\qf}{Q_f}
\newcommand{\qfs}{Q^2_f}
\newcommand{\qfc}{Q^3_f}
\newcommand{\qff}{Q^4_f}
\newcommand{\qep}{Q_{e'}}
\newcommand{\qfp}{Q_{f'}}
\newcommand{\qfps}{Q^2_{f'}}
\newcommand{\qfpc}{Q^3_{f'}}
\newcommand{\qq}{Q_q}
\newcommand{\qqs}{Q^2_q}
\newcommand{\qi}{Q_i}
\newcommand{\qis}{Q^2_i}
\newcommand{\qj}{Q_j}
\newcommand{\qjs}{Q^2_j}
\newcommand{\QW}{Q_{_\wb}}
\newcommand{\QWs}{Q^2_{_\wb}}
\newcommand{\Qd}{Q_d}
\newcommand{\Qds}{Q^2_d}
\newcommand{\Qu}{Q_u}
\newcommand{\Qus}{Q^2_u}
\newcommand{\vi}{v_i}
\newcommand{\vis}{v^2_i}
\newcommand{\ai}{a_i}
\newcommand{\ais}{a^2_i}
%
%
\newcommand{\piv}{\Pi_{_V}}
\newcommand{\pia}{\Pi_{_A}}
\newcommand{\piva}{\Pi_{_{V,A}}}
\newcommand{\pivi}[1]{\Pi^{({#1})}_{_V}}
\newcommand{\piai}[1]{\Pi^{({#1})}_{_A}}
\newcommand{\pivai}[1]{\Pi^{({#1})}_{_{V,A}}}
\newcommand{\pih}{{\hat\Pi}}
\newcommand{\sgh}{{\hat\Sigma}}
\newcommand{\Pgg}{\Pi_{\ph\ph}}
\newcommand{\Ptg}{\Pi_{_{3Q}}}
\newcommand{\Ptt}{\Pi_{_{33}}}
\newcommand{\Pzg}{\Pi_{_{\zb\ab}}}
\newcommand{\Pzga}[2]{\Pi^{#1}_{_{\zb\ab}}\lpar#2\rpar}
\newcommand{\Pf}{\Pi^{\ssF}}
\newcommand{\Sgg}{\Sigma_{_{\ab\ab}}}
\newcommand{\Szg}{\Sigma_{_{\zb\ab}}}
\newcommand{\SVV}{\Sigma_{_{\vb\vb}}}
\newcommand{\USvv}{{\hat\Sigma}_{_{\vb\vb}}}
\newcommand{\Sww}{\Sigma_{_{\wb\wb}}}
\newcommand{\Swwg}{\Sigma^{_G}_{_{\wb\wb}}}
\newcommand{\Szz}{\Sigma_{_{\zb\zb}}}
\newcommand{\Shh}{\Sigma_{_{\hb\hb}}}
\newcommand{\Spzz}{\Sigma'_{_{\zb\zb}}}
\newcommand{\Stg}{\Sigma_{_{3Q}}}
\newcommand{\Stt}{\Sigma_{_{33}}}
\newcommand{\bSww}{{\overline\Sigma}_{_{WW}}}
\newcommand{\bStg}{{\overline\Sigma}_{_{3Q}}}
\newcommand{\bStt}{{\overline\Sigma}_{_{33}}}
\newcommand{\Sssn}{\Sigma_{_{\hkn\hkn}}}
\newcommand{\Sssc}{\Sigma_{_{\phi\phi}}}
\newcommand{\Szn}{\Sigma_{_{\zb\hkn}}}
\newcommand{\Swc}{\Sigma_{_{\wb\hkg}}}
\newcommand{\mix}[2]{{\cal{M}}^{#1}\lpar{#2}\rpar}
\newcommand{\bmix}[2]{\Pi^{{#1},\ssF}_{_{\zb\ab}}\lpar{#2}\rpar}
\newcommand{\hPgg}[2]{{\hat{\Pi}^{{#1},\ssF}}_{\ph\ph}\lpar{#2}\rpar}
\newcommand{\hmix}[2]{{\hat{\Pi}^{{#1},\ssF}}_{_{\zb\ab}}\lpar{#2}\rpar}
\newcommand{\Dz}[2]{{\cal{D}}_{_{\zb}}^{#1}\lpar{#2}\rpar}
\newcommand{\bDz}[2]{{\cal{D}}^{{#1},\ssF}_{_{\zb}}\lpar{#2}\rpar}
\newcommand{\hDz}[2]{{\hat{\cal{D}}}^{{#1},\ssF}_{_{\zb}}\lpar{#2}\rpar}
\newcommand{\Szzd}[2]{\Sigma'^{#1}_{_{\zb\zb}}\lpar{#2}\rpar}
\newcommand{\Swwd}[2]{\Sigma'^{#1}_{_{\wb\wb}}\lpar{#2}\rpar}
\newcommand{\Shhd}[2]{\Sigma'^{#1}_{_{\hb\hb}}\lpar{#2}\rpar}
\newcommand{\ZFren}[2]{{\cal{Z}}^{#1}\lpar{#2}\rpar}
\newcommand{\WFren}[2]{{\cal{W}}^{#1}\lpar{#2}\rpar}
\newcommand{\HFren}[2]{{\cal{H}}^{#1}\lpar{#2}\rpar}
\newcommand{\WI}{\cal{W}}
%
%
\newcommand{\cf}{c_f}
\newcommand{\Cf}{C_{_F}}
\newcommand{\Nf}{N_f}
\newcommand{\Nc}{N_c}
\newcommand{\Ncs}{N^2_c}
\newcommand{\nf }{n_f}
\newcommand{\nfs}{n^2_f}
\newcommand{\nfc}{n^3_f}
\newcommand{\MSB}{\overline{MS}}
\newcommand{\LMSB}{\Lambda_{\overline{\mathrm{MS}}}}
\newcommand{\LMSBp}{\Lambda'_{\overline{\mathrm{MS}}}}
\newcommand{\LMSBS}{\Lambda^2_{\overline{\mathrm{MS}}}}
\newcommand{\LMSBv }{\mbox{$\Lambda^{(5)}_{\overline{\mathrm{MS}}}$}}
\newcommand{\LMSBvS}{\mbox{$\left(\Lambda^{(5)}_{\overline{\mathrm{MS}}}\right)^2$}}
\newcommand{\LMSBt }{\mbox{$\Lambda^{(3)}_{\overline{\mathrm{MS}}}$}}
\newcommand{\LMSBtS}{\mbox{$\left(\Lambda^{(3)}_{\overline{\mathrm{MS}}}\right)^2$}}
\newcommand{\LMSBf }{\mbox{$\Lambda^{(4)}_{\overline{\mathrm{MS}}}$}}
\newcommand{\LMSBfS}{\mbox{$\left(\Lambda^{(4)}_{\overline{\mathrm{MS}}}\right)^2$}}
\newcommand{\LMSBn }{\mbox{$\Lambda^{(\nf)}_{\overline{\mathrm{MS}}}$}}
\newcommand{\LMSBnS}{\mbox{$\left(\Lambda^{(\nf)}_{\overline{\mathrm{MS}}}\right)^2$}}
\newcommand{\LMSBnml }{\mbox{$\Lambda^{(\nf-1)}_{\overline{\mathrm{MS}}}$}}
\newcommand{\LMSBnmlS}{\mbox{$\left(\Lambda^{(\nf-1)}_{\overline{\mathrm{MS}}}\right)^2$}}
\newcommand{\Bnf}{\lpar\nf \rpar}
\newcommand{\Bnfm}{\lpar\nf-1 \rpar}
\newcommand{\LuM}{L_{_M}}
\newcommand{\bef}{\beta_{\ff}}
\newcommand{\befs}{\beta^2_{\ff}}
\newcommand{\befc}{\beta^3_{f}}
\newcommand{\alsp}{\alpha'_{_S}}
\newcommand{\api}{\displaystyle \frac{\als(s)}{\pi}}
\newcommand{\alss}{\alpha^2_{_S}}
\newcommand{\ztwo}{\zeta(2)}
\newcommand{\ztri}{\zeta(3)}
\newcommand{\zfor}{\zeta(4)}
\newcommand{\zfiv}{\zeta(5)}
\newcommand{\bi}[1]{b_{#1}}
\newcommand{\ci}[1]{c_{#1}}
\newcommand{\Ci}[1]{C_{#1}}
\newcommand{\bip}[1]{b'_{#1}}
\newcommand{\cip}[1]{c'_{#1}}
%
%
\newcommand{\osps}{16\,\pi^2}
\newcommand{\srt}{\sqrt{2}}
\newcommand{\ospsi}{\displaystyle{\frac{i}{16\,\pi^2}}}
%
%
\newcommand{\tfpromu}{\mbox{$e^+e^-\to \mu^+\mu^-$}}
\newcommand{\tfprotau}{\mbox{$e^+e^-\to \tau^+\tau^-$}}
\newcommand{\tfproe}{\mbox{$e^+e^-\to e^+e^-$}}
\newcommand{\tfpronu}{\mbox{$e^+e^-\to \barnu\nu$}}
\newcommand{\tfproqq}{\mbox{$e^+e^-\to \barq q$}}
\newcommand{\tfprohad}{\mbox{$e^+e^-\to\,$} hadrons}
%
%
\newcommand{\bpromu}{\mbox{$e^+e^-\to \mu^+\mu^-\ph$}}
\newcommand{\bprotau}{\mbox{$e^+e^-\to \tau^+\tau^-\ph$}}
\newcommand{\bproe}{\mbox{$e^+e^-\to e^+e^-\ph$}}
\newcommand{\bpronu}{\mbox{$e^+e^-\to \barnu\nu\ph$}}
\newcommand{\bproqq}{\mbox{$e^+e^-\to \barq q \ph$}}
%
%
\newcommand{\tbprow} {\mbox{$e^+e^-\to \wbp \wbm $}}
\newcommand{\tbproz} {\mbox{$e^+e^-\to \zb  \zb  $}}
\newcommand{\tbproh} {\mbox{$e^+e^-\to \zb  \hb  $}}
\newcommand{\tbprozg}{\mbox{$e^+e^-\to \zb  \ph  $}}
\newcommand{\tbprog} {\mbox{$e^+e^-\to \ph  \ph  $}}
%
%
\newcommand{\Fermionline}[1]{
\vcenter{\hbox{
  \begin{picture}(60,20)(0,{#1})
  \SetScale{2.}
    \ArrowLine(0,5)(30,5)
  \end{picture}}}
}
\newcommand{\AntiFermionline}[1]{
\vcenter{\hbox{
  \begin{picture}(60,20)(0,{#1})
  \SetScale{2.}
    \ArrowLine(30,5)(0,5)
  \end{picture}}}
}
\newcommand{\Photonline}[1]{
\vcenter{\hbox{
  \begin{picture}(60,20)(0,{#1})
  \SetScale{2.}
    \Photon(0,5)(30,5){2}{6.5}
  \end{picture}}}
}
\newcommand{\Gluonline}[1]{
\vcenter{\hbox{
  \begin{picture}(60,20)(0,{#1})
  \SetScale{2.}
    \Gluon(0,5)(30,5){2}{6.5}
  \end{picture}}}
}
\newcommand{\Wbosline}[1]{
\vcenter{\hbox{
  \begin{picture}(60,20)(0,{#1})
  \SetScale{2.}
    \Photon(0,5)(30,5){2}{4}
    \ArrowLine(13.3,3.1)(16.9,7.2)
  \end{picture}}}
}
\newcommand{\Zbosline}[1]{
\vcenter{\hbox{
  \begin{picture}(60,20)(0,{#1})
  \SetScale{2.}
    \Photon(0,5)(30,5){2}{4}
  \end{picture}}}
}
\newcommand{\Philine}[1]{
\vcenter{\hbox{
  \begin{picture}(60,20)(0,{#1})
  \SetScale{2.}
    \DashLine(0,5)(30,5){2}
  \end{picture}}}
}
\newcommand{\Phicline}[1]{
\vcenter{\hbox{
  \begin{picture}(60,20)(0,{#1})
  \SetScale{2.}
    \DashLine(0,5)(30,5){2}
    \ArrowLine(14,5)(16,5)
  \end{picture}}}
}
\newcommand{\Ghostline}[1]{
\vcenter{\hbox{
  \begin{picture}(60,20)(0,{#1})
  \SetScale{2.}
    \DashLine(0,5)(30,5){.5}
    \ArrowLine(14,5)(16,5)
  \end{picture}}}
}
%
%
\newcommand{\gauge}{g}
\newcommand{\gpar}{\xi}
\newcommand{\gparA}{\xi_{_A}}
\newcommand{\gparZ}{\xi_{_Z}}
\newcommand{\gpari}[1]{\gpar_{#1}}
\newcommand{\gparis}[1]{\gpar^2_{#1}}
\newcommand{\gpariq}[1]{\gpar^4_{#1}}
\newcommand{\gpars}{\xi^2}
\newcommand{\dgpar}{\Delta\gpar}
\newcommand{\dgparA}{\Delta\gparA}
\newcommand{\dgparZ}{\Delta\gparZ}
\newcommand{\gparq}{\xi^4}
\newcommand{\gparAs}{\xi^2_{_A}}
\newcommand{\gparAq}{\xi^4_{_A}}
\newcommand{\gparZs}{\xi^2_{_Z}}
\newcommand{\gparZq}{\xi^4_{_Z}}
\newcommand{\Rxi}{R_{\gpar}}
\newcommand{\UG}{U}
\newcommand{\UGi}{\ssU}
\newcommand{\hxi}{\chi}
%
%
\newcommand{\LSM}{{\cal{L}}_{_{\rm{SM}}}}
\newcommand{\LSMr}{{\cal{L}}^{\rm{\ssR}}_{_{\rm{SM}}}}
\newcommand{\LYM}{{\cal{L}}_{_{\rm{YM}}}}
\newcommand{\Lzer}{{\cal{L}}_{0}}
\newcommand{\Lone}{{\cal{L}}^{{\bos},{\rm{I}}}}
\newcommand{\Lpro}{{\cal{L}}_{\rm{prop}}}
\newcommand{\Ls  }{{\cal{L}}_{_{\rm{S}}}}
\newcommand{\Lsi }{{\cal{L}}^{\rm{I}}_{_{\rm{S}}}}
\newcommand{\Lgf }{{\cal{L}}_{\rm{gf}}}
\newcommand{\Lgfi}{{\cal{L}}^{\rm{I}}_{\rm{gf}}}
\newcommand{\Lf  }{{\cal{L}}^{{\fer},{\rm{I}}}_{\ssV}}
\newcommand{\LHf }{{\cal{L}}^{\fer}_{\ssS}}
\newcommand{\LHfm}{{\cal{L}}^{{\fer},m}_{\ssS}}
\newcommand{\LHfi}{{\cal{L}}^{{\fer},{\rm{I}}}_{\ssS}}
\newcommand{\Lren}{{\cal{L}}_{\rm{\ssR}}}
\newcommand{\Lct}{{\cal{L}}_{\rm{ct}}}
\newcommand{\Lcti}[1]{{\cal{L}}^{#1}_{\rm{ct}}}
\newcommand{\LctI}{{\cal{L}}^{(2)}_{\rm{ct}}}
\newcommand{\Llone}{{\cal{L}}}
\newcommand{\LQED}{{\cal{L}}_{_{\rm{QED}}}}
\newcommand{\LQEDz}{{\cal{L}}^{0}_{_{\rm{QED}}}}
\newcommand{\LQEDi}{{\cal{L}}^{\rm{\ssI}}_{_{\rm{QED}}}}
\newcommand{\LQEDr}{{\cal{L}}^{\rm{\ssR}}_{_{\rm{QED}}}}
\newcommand{\Greenf}{G}
\newcommand{\Greenfa}[1]{G\lpar{#1}\rpar}
\newcommand{\Greenft}[2]{G\lpar{#1,#2}\rpar}
\newcommand{\FST}[3]{F_{#1#2}^{#3}}
\newcommand{\cD}[1]{D_{#1}}
\newcommand{\pd}[1]{\partial_{#1}}
\newcommand{\tgen}[1]{\tau^{#1}}
\newcommand{\gbl}{g_1}
\newcommand{\lctt}[3]{\varepsilon_{#1#2#3}}
\newcommand{\lctf}[4]{\varepsilon_{#1#2#3#4}}
\newcommand{\lctfb}[4]{\varepsilon\lpar{#1#2#3#4}\rpar}
\newcommand{\slct}{\varepsilon}
\newcommand{\cgfi}[1]{{\cal{C}}^{#1}}
\newcommand{\cgfZ}{{\cal{C}}^{\ssZ}}
\newcommand{\cgfA}{{\cal{C}}^{\ssA}}
\newcommand{\hpms}{\mu^2}
\newcommand{\hpal}{\alpha_{_H}}
\newcommand{\hpals}{\alpha^2_{_H}}
\newcommand{\hpbe}{\beta_{_H}}
\newcommand{\hpbep}{\beta^{'}_{_H}}
\newcommand{\hpla}{\lambda}
\newcommand{\hpalf}{\alpha_{f}}
\newcommand{\hpbef}{\beta_{f}}
\newcommand{\tpar}[1]{\Lambda^{#1}}
\newcommand{\Mop}[2]{{\rm{M}}^{#1#2}}
\newcommand{\Lop}[2]{{\rm{L}}^{#1#2}}
\newcommand{\Lgen}[1]{T^{#1}}
\newcommand{\Rgen}[1]{t^{#1}}
\newcommand{\fpari}[1]{\lambda_{#1}}
\newcommand{\fQ}[1]{Q_{#1}}
\newcommand{\unm}{I}
\newcommand{\cDsla}{/\!\!\!\!D}
%
%
\newcommand{\saff}[1]{A_{#1}}                    
\newcommand{\aff}[2]{A_{#1}\lpar #2\rpar}                   
\newcommand{\sbff}[1]{B_{#1}}                    
\newcommand{\sfbff}[1]{B^{\ssF}_{#1}}
\newcommand{\bff}[4]{B_{#1}\lpar #2;#3,#4\rpar}             
\newcommand{\bfft}[3]{B_{#1}\lpar #2,#3\rpar}             
\newcommand{\fbff}[4]{B^{\ssF}_{#1}\lpar #2;#3,#4\rpar}        
\newcommand{\cdbff}[4]{\Delta B_{#1}\lpar #2;#3,#4\rpar}             
\newcommand{\sdbff}[4]{\delta B_{#1}\lpar #2;#3,#4\rpar}             
\newcommand{\cdbfft}[3]{\Delta B_{#1}\lpar #2,#3\rpar}             
\newcommand{\sdbfft}[3]{\delta B_{#1}\lpar #2,#3\rpar}             
\newcommand{\scff}[1]{C_{#1}}                    
\newcommand{\scffo}[2]{C_{#1}\lpar{#2}\rpar}                
\newcommand{\cff}[7]{C_{#1}\lpar #2,#3,#4;#5,#6,#7\rpar}    
\newcommand{\sccff}[5]{c_{#1}\lpar #2;#3,#4,#5\rpar} 
\newcommand{\sdff}[1]{D_{#1}}                    
\newcommand{\dffp}[7]{D_{#1}\lpar #2,#3,#4,#5,#6,#7;}       
\newcommand{\dffm}[4]{#1,#2,#3,#4\rpar}                     
\newcommand{\bzfa}[2]{B^{\ssF}_{_{#2}}\lpar{#1}\rpar}
\newcommand{\bzfaa}[3]{B^{\ssF}_{_{#2#3}}\lpar{#1}\rpar}
\newcommand{\shcff}[4]{C_{_{#2#3#4}}\lpar{#1}\rpar}
\newcommand{\shdff}[6]{D_{_{#3#4#5#6}}\lpar{#1,#2}\rpar}
\newcommand{\scdff}[3]{d_{#1}\lpar #2,#3\rpar} 
\newcommand{\scaldff}[1]{{\cal{D}}^{#1}}
\newcommand{\caldff}[2]{{\cal{D}}^{#1}\lpar{#2}\rpar}
\newcommand{\caldfft}[3]{{\cal{D}}_{#1}^{#2}\lpar{#3}\rpar}
%
%
\newcommand{\slaff}[1]{a_{#1}}                        
\newcommand{\slbff}[1]{b_{#1}}                        
\newcommand{\slbffh}[1]{{\hat{b}}_{#1}}    
\newcommand{\ssldff}[1]{d_{#1}}                        
\newcommand{\sslcff}[1]{c_{#1}}                        
\newcommand{\slcff}[2]{c_{#1}^{(#2)}}                        
\newcommand{\sldff}[2]{d_{#1}^{(#2)}}                        
\newcommand{\lbff}[3]{b_{#1}\lpar #2;#3\rpar}         
\newcommand{\lbffh}[2]{{\hat{b}}_{#1}\lpar #2\rpar}   
\newcommand{\lcff}[8]{c_{#1}^{(#2)}\lpar  #3,#4,#5;#6,#7,#8\rpar}         
\newcommand{\ldffp}[8]{d_{#1}^{(#2)}\lpar #3,#4,#5,#6,#7,#8;}
\newcommand{\ldffm}[4]{#1,#2,#3,#4\rpar}                   
%
%
\newcommand{\Iff}[4]{I_{#1}\lpar #2;#3,#4 \rpar}
\newcommand{\Jff}[4]{J_{#1}\lpar #2;#3,#4 \rpar}
\newcommand{\Jds}[5]{{\bar{J}}_{#1}\lpar #2,#3;#4,#5 \rpar}
\newcommand{\sJds}[1]{{\bar{J}}_{#1}}
%
\newcommand{\nhmt}{\frac{n}{2}-2}
\newcommand{\nhmts}{{n}/{2}-2}
\newcommand{\omnh}{1-\frac{n}{2}}
\newcommand{\nhmo}{\frac{n}{2}-1}
\newcommand{\fmon}{4-n}
\newcommand{\lpi}{\ln\pi}
\newcommand{\lmass}[1]{\ln #1}
\newcommand{\egnh}{\egam{\frac{n}{2}}}
\newcommand{\egomnh}{\egam{1-\frac{n}{2}}}
\newcommand{\egtmnh}{\egam{2-\frac{n}{2}}}
\newcommand{\Ddr}{{\ds\frac{1}{{\bar{\varepsilon}}}}}
\newcommand{\Ddrs}{{\ds\frac{1}{{\bar{\varepsilon}^2}}}}
\newcommand{\Ddrd}{{\bar{\varepsilon}}}
\newcommand{\ept}{\hat\varepsilon}
\newcommand{\Ddrh}{{\ds\frac{1}{\hat{\varepsilon}}}}
\newcommand{\Ddrp}{{\ds\frac{1}{\varepsilon'}}}
\newcommand{\Ddrps}{\lpar{\ds{\frac{1}{\varepsilon'}}}\rpar^2}
\newcommand{\dre}{\varepsilon}
\newcommand{\drei}[1]{\varepsilon_{#1}}
\newcommand{\epp}{\varepsilon'}
\newcommand{\ep}{\epsilon}
\newcommand{\propbt}[6]{{{#1_{#2}#1_{#3}}\over{\lpar #1^2 + #4 
-\ib\ep\rpar\lpar\lpar #5\rpar^2 + #6 -\ib\ep\rpar}}}
\newcommand{\propbo}[5]{{{#1_{#2}}\over{\lpar #1^2 + #3 - \ib\ep\rpar
\lpar\lpar #4\rpar^2 + #5 -\ib\ep\rpar}}}
\newcommand{\propc}[6]{{1\over{\lpar #1^2 + #2 - \ib\ep\rpar
\lpar\lpar #3\rpar^2 + #4 -\ib\ep\rpar
\lpar\lpar #5\rpar^2 + #6 -\ib\ep\rpar}}}
\newcommand{\propa}[2]{{1\over {#1^2 + #2^2 - \ib\ep}}}
\newcommand{\propb}[4]{{1\over {\lpar #1^2 + #2 - \ib\ep\rpar
\lpar\lpar #3\rpar^2 + #4 -\ib\ep\rpar}}}
\newcommand{\propbs}[4]{{1\over {\lpar\lpar #1\rpar^2 + #2 - \ib\ep\rpar
\lpar\lpar #3\rpar^2 + #4 -\ib\ep\rpar}}}
\newcommand{\propat}[4]{{#3_{#1}#3_{#2}\over {#3^2 + #4^2 - \ib\ep}}}
\newcommand{\propaf}[6]{{#5_{#1}#5_{#2}#5_{#3}#5_{#4}\over 
{#5^2 + #6^2 -\ib\ep}}}
\newcommand{\momeps}[1]{#1^2 - \ib\ep}
\newcommand{\mopeps}[1]{#1^2 + \ib\ep}
\newcommand{\propz}[1]{{1\over{#1^2 + \mzs - \ib\ep}}}
\newcommand{\propw}[1]{{1\over{#1^2 + \mws - \ib\ep}}}
\newcommand{\proph}[1]{{1\over{#1^2 + \mhs - \ib\ep}}}
\newcommand{\propf}[2]{{1\over{#1^2 + #2}}}
\newcommand{\propzrg}[3]{{{\delta_{#1#2}}\over{{#3}^2 + \mzs - \ib\ep}}}
\newcommand{\propwrg}[3]{{{\delta_{#1#2}}\over{{#3}^2 + \mws - \ib\ep}}}
\newcommand{\propzug}[3]{{
      {\delta_{#1#2} + \displaystyle{{{#3}^{#1}{#3}^{#2}}\over{\mzs}}}
                         \over{{#3}^2 + \mzs - \ib\ep}}}
\newcommand{\propwug}[3]{{
      {\delta_{#1#2} + \displaystyle{{{#3}^{#1}{#3}^{#2}}\over{\mws}}}
                        \over{{#3}^2 + \mws - \ib\ep}}}
\newcommand{\thf}[1]{\theta\lpar #1\rpar}
\newcommand{\epf}[1]{\varepsilon\lpar #1\rpar}
\newcommand{\singp}{\stackrel{\rm{sing}}{\rightarrow}}
\newcommand{\aint}[3]{\int_{#1}^{#2}\,d #3}
\newcommand{\aroot}[1]{\sqrt{#1}}
\newcommand{\gramc}{\Delta_3}
\newcommand{\gramd}{\Delta_4}
\newcommand{\pinch}[2]{P^{(#1)}\lpar #2\rpar}
\newcommand{\pinchc}[2]{C^{(#1)}_{#2}}
\newcommand{\pinchd}[2]{D^{(#1)}_{#2}}
\newcommand{\loarg}[1]{\ln\lpar #1\rpar}
\newcommand{\loargr}[1]{\ln\lrbr #1\rrbr}
\newcommand{\lsoarg}[1]{\ln^2\lpar #1\rpar}
\newcommand{\ltarg}[2]{\ln\lpar #1\rpar\lpar #2\rpar}
\newcommand{\rfun}[2]{R\lpar #1,#2\rpar}
\newcommand{\pinchb}[3]{B_{#1}\lpar #2,#3\rpar}
\newcommand{\lga}{\ph}
\newcommand{\lzga}{\ssZ\ph}
%
%
\newcommand{\afa}[5]{A_{#1}^{#2}\lpar #3;#4,#5\rpar}
\newcommand{\bfa}[5]{B_{#1}^{#2}\lpar #3;#4,#5\rpar} 
\newcommand{\hfa}[5]{H_{#1}^{#2}\lpar #3;#4,#5\rpar}
\newcommand{\rfa}[5]{R_{#1}^{#2}\lpar #3;#4,#5\rpar}
\newcommand{\afao}[3]{A_{#1}^{#2}\lpar #3\rpar}
\newcommand{\bfao}[3]{B_{#1}^{#2}\lpar #3\rpar}
\newcommand{\hfao}[3]{H_{#1}^{#2}\lpar #3\rpar}
\newcommand{\rfao}[3]{R_{#1}^{#2}\lpar #3\rpar}
\newcommand{\afax}[6]{A_{#1}^{#2}\lpar #3;#4,#5,#6\rpar}
\newcommand{\afas}[2]{A_{#1}^{#2}}
\newcommand{\bfas}[2]{B_{#1}^{#2}}
\newcommand{\hfas}[2]{H_{#1}^{#2}}
\newcommand{\rfas}[2]{R_{#1}^{#2}}
\newcommand{\tfas}[2]{T_{#1}^{#2}}
\newcommand{\afaR}[6]{A_{#1}^{\gpar}\lpar #2;#3,#4,#5,#6 \rpar}
\newcommand{\bfaR}[6]{B_{#1}^{\gpar}\lpar #2;#3,#4,#5,#6 \rpar}
\newcommand{\hfaR}[6]{H_{#1}^{\gpar}\lpar #2;#3,#4,#5,#6 \rpar}
\newcommand{\shfaR}[1]{H_{#1}^{\gpar}}
\newcommand{\rfaR}[6]{R_{#1}^{\gpar}\lpar #2;#3,#4,#5,#6 \rpar}
\newcommand{\srfaR}[1]{R_{#1}^{\gpar}}
\newcommand{\afaRg}[5]{A_{#1 \lga}^{\gpar}\lpar #2;#3,#4,#5 \rpar}
\newcommand{\bfaRg}[5]{B_{#1 \lga}^{\gpar}\lpar #2;#3,#4,#5 \rpar}
\newcommand{\hfaRg}[5]{H_{#1 \lga}^{\gpar}\lpar #2;#3,#4,#5 \rpar}
\newcommand{\shfaRg}[1]{H_{#1\lga}^{\gpar}}
\newcommand{\rfaRg}[5]{R_{#1 \lga}^{\gpar}\lpar #2;#3,#4,#5 \rpar}
\newcommand{\srfaRg}[1]{R_{#1\lga}^{\gpar}}
\newcommand{\afaRt}[3]{A_{#1}^{\gpar}\lpar #2,#3 \rpar}
\newcommand{\hfaRt}[3]{H_{#1}^{\gpar}\lpar #2,#3 \rpar}
\newcommand{\hfaRf}[4]{H_{#1}^{\gpar}\lpar #2,#3,#4 \rpar}
\newcommand{\afasm}[4]{A_{#1}^{\lpar #2,#3,#4 \rpar}}
\newcommand{\bfasm}[4]{B_{#1}^{\lpar #2,#3,#4 \rpar}}
\newcommand{\htf}[2]{H_2\lpar #1,#2\rpar}
\newcommand{\rof}[2]{R_1\lpar #1,#2\rpar}
\newcommand{\rtf}[2]{R_3\lpar #1,#2\rpar}
\newcommand{\rtrans}[2]{R_{#1}^{#2}}
\newcommand{\momf}[2]{#1^2_{#2}}
\newcommand{\Scalvert}[8][70]{
  \vcenter{\hbox{
  \SetScale{0.8}
  \begin{picture}(#1,50)(15,15)
    \Line(25,25)(50,50)      \Text(34,20)[lc]{#6} \Text(11,20)[lc]{#3}
    \Line(50,50)(25,75)      \Text(34,60)[lc]{#7} \Text(11,60)[lc]{#4}
    \Line(50,50)(90,50)      \Text(11,40)[lc]{#2} \Text(55,33)[lc]{#8}
    \GCirc(50,50){10}{1}          \Text(60,48)[lc]{#5} 
  \end{picture}}}
  }
%
%
\newcommand{\tHs}{\mu}
\newcommand{\tHsz}{\mu_{_0}}
\newcommand{\tHss}{\mu^2}
\newcommand{\Reb}{{\rm{Re}}}
\newcommand{\Imb}{{\rm{Im}}}
%
%
\newcommand{\spd}{\partial}
\newcommand{\ffun}[2]{F_{#1}\lpar #2\rpar}
\newcommand{\gfun}[2]{G_{#1}\lpar #2\rpar}
\newcommand{\sffun}[1]{F_{#1}}
\newcommand{\csffun}[1]{{\cal{F}}_{#1}}
\newcommand{\sgfun}[1]{G_{#1}}
\newcommand{\tpfi}{\lpar 2\pi\rpar^4\ib}
\newcommand{\ffv}{F_{_V}}
\newcommand{\fga}{G_{_A}}
\newcommand{\ffm}{F_{_M}}
\newcommand{\ffs}{F_{_S}}
\newcommand{\fgp}{G_{_P}}
\newcommand{\fge}{G_{_E}}
\newcommand{\ffa}{F_{_A}}
\newcommand{\ffps}{F_{_P}}
\newcommand{\ffe}{F_{_E}}
\newcommand{\gacom}[2]{\lpar #1 + #2\gfd\rpar}
\newcommand{\mft}{m_{\tilde f}}
\newcommand{\qft}{Q_{f'}}
\newcommand{\vft}{v_{\tilde f}}
\newcommand{\subb}[2]{b_{#1}\lpar #2 \rpar}
\newcommand{\fwfr}[5]{\Sigma\lpar #1,#2,#3;#4,#5 \rpar}
\newcommand{\slim}[2]{\lim_{#1 \to #2}}
\newcommand{\sprop}[3]{
{#1\over {\lpar q^2\rpar^2\lpar \lpar q+ #2\rpar^2+#3^2\rpar }}}
%
%
\newcommand{\xroot}[1]{x_{#1}}
\newcommand{\yroot}[1]{y_{#1}}
\newcommand{\zroot}[1]{z_{#1}}
\newcommand{\lvar}{l}
\newcommand{\rvar}{r}
\newcommand{\tvar}{t}
\newcommand{\uvar}{u}
\newcommand{\vvar}{v}
\newcommand{\xvar}{x}
\newcommand{\yvar}{y}
\newcommand{\zvar}{z}
\newcommand{\xvarp}{x'}
\newcommand{\yvarp}{y'}
\newcommand{\zvarp}{z'}
\newcommand{\rvars}{r^2}
\newcommand{\vvars}{v^2}
\newcommand{\xvars}{x^2}
\newcommand{\yvars}{y^2}
\newcommand{\zvars}{z^2}
\newcommand{\rvarc}{r^3}
\newcommand{\xvarc}{x^3}
\newcommand{\yvarc}{y^3}
\newcommand{\zvarc}{z^3}
\newcommand{\rvarq}{r^4}
\newcommand{\xvarq}{x^4}
\newcommand{\yvarq}{y^4}
\newcommand{\zvarq}{z^4}
\newcommand{\avar}{a}
\newcommand{\avars}{a^2}
\newcommand{\avarc}{a^3}
\newcommand{\avari}[1]{a_{#1}}
\newcommand{\avart}[2]{a_{#1}^{#2}}
\newcommand{\delvari}[1]{\delta_{#1}}
\newcommand{\rvari}[1]{r_{#1}}
\newcommand{\xvari}[1]{x_{#1}}
\newcommand{\yvari}[1]{y_{#1}}
\newcommand{\zvari}[1]{z_{#1}}
\newcommand{\rvart}[2]{r_{#1}^{#2}}
\newcommand{\xvart}[2]{x_{#1}^{#2}}
\newcommand{\yvart}[2]{y_{#1}^{#2}}
\newcommand{\zvart}[2]{z_{#1}^{#2}}
\newcommand{\rvaris}[1]{r^2_{#1}}
\newcommand{\xvaris}[1]{x^2_{#1}}
\newcommand{\yvaris}[1]{y^2_{#1}}
\newcommand{\zvaris}[1]{z^2_{#1}}
\newcommand{\Xvar}{X}
\newcommand{\Xvars}{X^2}
\newcommand{\Xvari}[1]{X_{#1}}
\newcommand{\Xvaris}[1]{X^2_{#1}}
\newcommand{\Yvar}{Y}
\newcommand{\Yvars}{Y^2}
\newcommand{\Yvari}[1]{Y_{#1}}
\newcommand{\Yvaris}[1]{Y^2_{#1}}
\newcommand{\lnx}{\ln\xvar}
\newcommand{\lnz}{\ln\zvar}
\newcommand{\lnsx}{\ln^2\xvar}
\newcommand{\lnsz}{\ln^2\zvar}
\newcommand{\lncz}{\ln^3\zvar}
\newcommand{\lnomz}{\ln\lpar 1-\zvar\rpar}
\newcommand{\lnsomz}{\ln^2\lpar 1-\zvar\rpar}
\newcommand{\ccoefi}[1]{c_{#1}}
\newcommand{\ccoeft}[2]{c^{#1}_{#2}}
%
%
\newcommand{\Smat}{{\cal{S}}}
\newcommand{\Mmat}{{\cal{M}}}
\newcommand{\Rmat}{{\cal{R}}}
\newcommand{\Xmat}[1]{X_{#1}}
\newcommand{\XmatI}[1]{X^{-1}_{#1}}
\newcommand{\unitmat}{I}
\newcommand{\zeromat}{O}
\newcommand{\paulimat}[1]{\tau_{#1}}
\newcommand{\Umat  }{U}
\newcommand{\Umath }{U^{+}}
\newcommand{\UmatL }{{\cal{U}}_{\ssL}}
\newcommand{\UmatLh}{{\cal{U}}^{+}_{\ssL}}
\newcommand{\UmatR }{{\cal{U}}_{\ssR}}
\newcommand{\UmatRh}{{\cal{U}}^{+}_{\ssR}}
\newcommand{\Vmat  }{V}
\newcommand{\Vmath }{V^{+}}
\newcommand{\VmatL }{{\cal{D}}_{\ssL}}
\newcommand{\VmatLh}{{\cal{D}}^{+}_{\ssL}}
\newcommand{\VmatR }{{\cal{D}}_{\ssR}}
\newcommand{\VmatRh}{{\cal{D}}^{+}_{\ssR}}
\newcommand{\Kmat}{{C}}
\newcommand{\Kmatc}{{C}^{\dagger}}
\newcommand{\Kmati}[1]{{C}_{#1}}
\newcommand{\Kmatci}[1]{{C}^{\dagger}_{#1}}
\newcommand{\ffac}[2]{f_{#1}^{#2}}
\newcommand{\Ffac}[1]{F_{#1}}
\newcommand{\Rvec}[2]{R^{(#1)}_{#2}}
\newcommand{\momfl}[2]{#1_{#2}}
\newcommand{\momfs}[2]{#1^2_{#2}}
\newcommand{\fpseZ}{A^{\ssF\ssP,\ssZ}}
\newcommand{\fpseA}{A^{\ssF\ssP,\ssA}}
\newcommand{\fptZ}{T^{\ssF\ssP,\ssZ}}
\newcommand{\fptA}{T^{\ssF\ssP,\ssA}}
\newcommand{\dprop}{\overline\Delta}
\newcommand{\dpropi}[1]{d_{#1}}
\newcommand{\dpropic}[1]{d^{c}_{#1}}
\newcommand{\dpropii}[2]{d_{#1}\lpar #2\rpar}
\newcommand{\dpropis}[1]{d^2_{#1}}
\newcommand{\dproppi}[1]{d'_{#1}}
\newcommand{\psf}[4]{P\lpar #1;#2,#3,#4\rpar}
\newcommand{\ssf}[5]{S^{(#1)}\lpar #2;#3,#4,#5\rpar}
\newcommand{\csf}[5]{C_{_S}^{(#1)}\lpar #2;#3,#4,#5\rpar}
%
%
\newcommand{\lvec}{l}
\newcommand{\lvecs}{l^2}
\newcommand{\lveci}[1]{l_{#1}}
\newcommand{\mvec}{m}
\newcommand{\mvecs}{m^2}
\newcommand{\mveci}[1]{m_{#1}}
\newcommand{\nvec}{n}
\newcommand{\nvecs}{n^2}
\newcommand{\nveci}[1]{n_{#1}}
\newcommand{\epi}[1]{\epsilon_{#1}}
\newcommand{\phep}[1]{\ep_{#1}}
\newcommand{\php}[3]{\ep^{#1}_{#2}\lpar #3 \rpar}
\newcommand{\sphep}{\ep}
\newcommand{\vbep}[1]{e_{#1}}
\newcommand{\vbepp}[1]{e^{+}_{#1}}
\newcommand{\vbepm}[1]{e^{-}_{#1}}
\newcommand{\svbep}{e}
%
%
\newcommand{\lpol}{\lambda}
\newcommand{\spol}{\sigma}
\newcommand{\rpol}{\rho  }
\newcommand{\kpol}{\kappa}
\newcommand{\lpols}{\lambda^2}
\newcommand{\spols}{\sigma^2}
\newcommand{\rpols}{\rho^2}
\newcommand{\kpols}{\kappa^2}
\newcommand{\lpoli}[1]{\lambda_{#1}}
\newcommand{\spoli}[1]{\sigma_{#1}}
\newcommand{\rpoli}[1]{\rho_{#1}}
\newcommand{\kpoli}[1]{\kappa_{#1}}
%
%
\newcommand{\uvec}{u}
\newcommand{\uveci}[1]{u_{#1}}
%
%
\newcommand{\imom}{q}
\newcommand{\imomi}[1]{q_{#1}}
\newcommand{\imoms}{q^2}
\newcommand{\pmom}{p}
\newcommand{\pmomp}{p'}
\newcommand{\pmoms}{p^2}
\newcommand{\pmomq}{p^4}
\newcommand{\pmomx}{p^6}
\newcommand{\pmomi}[1]{p_{#1}}
\newcommand{\pmompi}[1]{p'_{#1}}
\newcommand{\pmomis}[1]{p^2_{#1}}
\newcommand{\Pmom}{P}
\newcommand{\Pmoms}{P^2}
\newcommand{\Pmomi}[1]{P_{#1}}
\newcommand{\Pmomis}[1]{P^2_{#1}}
\newcommand{\Pmomp}{P'}
\newcommand{\Pmomps}{{P'}^2}
\newcommand{\Pmompi}[1]{P'_{#1}}
\newcommand{\Pmompis}[1]{{P'}^2_{#1}}
\newcommand{\Kmom}{K}
\newcommand{\Kmoms}{K^2}
\newcommand{\Kmomi}[1]{K_{#1}}
\newcommand{\Kmomis}[1]{K^2_{#1}}
\newcommand{\kmom}{k}
\newcommand{\kmoms}{k^2}
\newcommand{\kmomi}[1]{k_{#1}}
\newcommand{\lmom}{l}
\newcommand{\lmoms}{l^2}
\newcommand{\lmomi}[1]{l_{#1}}
\newcommand{\qmom}{q}
\newcommand{\qmoms}{q^2}
\newcommand{\qmomi}[1]{q_{#1}}
\newcommand{\qmomis}[1]{q^2_{#1}}
\newcommand{\smom}{s}
\newcommand{\smoms}{s^2}
\newcommand{\smomi}[1]{s_{#1}}
\newcommand{\tmom}{t}
\newcommand{\tmoms}{t^2}
\newcommand{\tmomi}[1]{t_{#1}}
\newcommand{\Trmom}{Q}
\newcommand{\Prmom}{P}
\newcommand{\gmv}{Q^2}
\newcommand{\Trmoms}{Q^2}
\newcommand{\Prmoms}{P^2}
\newcommand{\Ptmoms}{T^2}
\newcommand{\Pumoms}{U^2}
\newcommand{\Trmomq}{Q^4}
\newcommand{\Prmomq}{P^4}
\newcommand{\Ptmomq}{T^4}
\newcommand{\Pumomq}{U^4}
\newcommand{\Trmomx}{Q^6}
\newcommand{\Trmomi}[1]{Q_{#1}}
\newcommand{\Trmomis}[1]{Q^2_{#1}}
\newcommand{\Prmomi}[1]{P_{#1}}
\newcommand{\pone}{p_1}
\newcommand{\ptwo}{p_2}
\newcommand{\ptre}{p_3}
\newcommand{\pfor}{p_4}
\newcommand{\pones}{p_1^2}
\newcommand{\ptwos}{p_2^2}
\newcommand{\ptres}{p_3^2}
\newcommand{\pfors}{p_4^2}
\newcommand{\poneq}{p_1^4}
\newcommand{\ptwoq}{p_2^4}
\newcommand{\ptreq}{p_3^4}
\newcommand{\pforq}{p_4^4}
\newcommand{\modmom}[1]{\mid{\vec{#1}}\mid}
\newcommand{\modmoms}[1]{\mid{\vec{#1}}\mid^2}
\newcommand{\modmomi}[2]{\mid{\vec{#1}}_{#2}\mid}
\newcommand{\vect}[1]{{\vec{#1}}}
\newcommand{\Energ}{E}
\newcommand{\Energp}{E'}
\newcommand{\Energpp}{E''}
\newcommand{\Energs}{E^2}
\newcommand{\Energc}{E^3}
\newcommand{\Energf}{E^4}
\newcommand{\Energv}{E^5}
\newcommand{\Energx}{E^6}
\newcommand{\Energi}[1]{E_{#1}}
\newcommand{\Energt}[2]{E_{#1}^{#2}}
\newcommand{\Energis}[1]{E^2_{#1}}
\newcommand{\energ}{e}
\newcommand{\energp}{e'}
\newcommand{\energpp}{e''}
\newcommand{\energs}{e^2}
\newcommand{\energi}[1]{e_{#1}}
\newcommand{\energt}[2]{e_{#1}^{#2}}
\newcommand{\energis}[1]{e^2_{#1}}
\newcommand{\wenerg}{w}
\newcommand{\wenergs}{w^2}
\newcommand{\wenergi}[1]{w_{#1}}
\newcommand{\wenergp}{w'}
\newcommand{\wenergpp}{w''}
%
%
\newcommand{\ecut}{e}
\newcommand{\ecuts}{e^2}
\newcommand{\ecuti}[1]{e^{#1}}
\newcommand{\ccut}{c_m}
\newcommand{\ccuti}[1]{c_{#1}}
\newcommand{\ccuts}{c^2_m}
\newcommand{\scuts}{s^2_m}
\newcommand{\ccutis}[1]{c^2_{#1}}
\newcommand{\ccutic}[1]{c^3_{#1}}
\newcommand{\ccutc}{c^3_m}
\newcommand{\rcut}{\varrho}
\newcommand{\rcuts}{\varrho^2}
\newcommand{\rcuti}[1]{\varrho_{#1}}
\newcommand{\rcutu}[1]{\varrho^{#1}}
\newcommand{\Dcut}{\Delta}
%
\newcommand{\dwf}{\delta_{_{\rm{WF}}}}
\newcommand{\gbar}{\overline g}
\newcommand{\PP}{\mbox{PP}}
\newcommand{\mv}{m_{_V}}
\newcommand{\bGv}{{\overline\Gamma}_{_V}}
\newcommand{\Umuv}{\hat{\mu}_\ssV}
\newcommand{\Svv}{{\Sigma}_\ssV}
\newcommand{\muv}{p_\ssV}
\newcommand{\muvb}{\mu_{\ssV_{0}}}
\newcommand{\URPvv}{{P}_\ssV}
\newcommand{\RPvv}{{P}_\ssV}
\newcommand{\Svvrem}{{\Sigma}_\ssV^{\mathrm{rem}}}
\newcommand{\USvvrem}{\hat{\Sigma}_\ssV^{\mathrm{rem}}}
\newcommand{\Gv}{\Gamma_{_V}}
%
%
\newcommand{\param}{p}
\newcommand{\parami}[1]{p^{#1}}
\newcommand{\paramb}{p_{0}}
\newcommand{\Zcon}{Z}
\newcommand{\Zconi}[1]{Z_{#1}}
\newcommand{\zconi}[1]{z_{#1}}
\newcommand{\Zconim}[1]{{Z^-_{#1}}}
\newcommand{\zconim}[1]{{z^-_{#1}}}
\newcommand{\Zcont}[2]{Z_{#1}^{#2}}
\newcommand{\zcont}[2]{z_{#1}^{#2}}
\newcommand{\zcontm}[2]{z_{#1}^{{#2}-}}
\newcommand{\sZconi}[2]{\sqrt{Z_{#1}}^{\;#2}}
\newcommand{\gacome}[1]{\lpar #1 - \gfd\rpar}
\newcommand{\sPj}[2]{\Lambda^{#1}_{#2}}
\newcommand{\sPjs}[2]{\Lambda_{#1,#2}}
\newcommand{\amos}{\mbox{$M^2_{_1}$}}
\newcommand{\amts}{\mbox{$M^2_{_2}$}}
\newcommand{\er}{e_{_{R}}}
\newcommand{\epr}{e'_{_{R}}}
\newcommand{\ers}{e^2_{_{R}}}
\newcommand{\erc}{e^3_{_{R}}}
\newcommand{\erq}{e^4_{_{R}}}
\newcommand{\erf}{e^5_{_{R}}}
\newcommand{\sour}{J}
\newcommand{\sourb}{\overline J}
\newcommand{\lrm}{M_{_R}}
%
%
\newcommand{\vlami}[1]{\lambda_{#1}}
\newcommand{\vlamis}[1]{\lambda^2_{#1}}
\newcommand{\Vvert}{V}
\newcommand{\Avert}{A}
\newcommand{\Svert}{S}
\newcommand{\Pvert}{P}
\newcommand{\vvert}{F}
\newcommand{\Cvert}{\cal{V}}
\newcommand{\Bvert}{\cal{B}}
\newcommand{\Vveri}[2]{V_{#1}^{#2}}
\newcommand{\Fveri}[1]{{\cal{F}}^{#1}}
\newcommand{\Cveri}[1]{{\cal{V}}\lpar{#1}\rpar}
\newcommand{\Bveri}[1]{{\cal{B}}\lpar{#1}\rpar}
\newcommand{\Vverti}[3]{V_{#1}^{#2}\lpar{#3}\rpar}
\newcommand{\Averti}[3]{A_{#1}^{#2}\lpar{#3}\rpar}
\newcommand{\Gverti}[3]{G_{#1}^{#2}\lpar{#3}\rpar}
\newcommand{\Zverti}[3]{Z_{#1}^{#2}\lpar{#3}\rpar}
\newcommand{\Hverti}[2]{H^{#1}\lpar{#2}\rpar}
\newcommand{\Wverti}[3]{W_{#1}^{#2}\lpar{#3}\rpar}
\newcommand{\Cverti}[2]{{\cal{V}}_{#1}^{#2}}
\newcommand{\vverti}[3]{F^{#1}_{#2}\lpar{#3}\rpar}
\newcommand{\averti}[3]{{\overline{F}}^{#1}_{#2}\lpar{#3}\rpar}
\newcommand{\fveone}[1]{f_{#1}}
\newcommand{\fvetri}[3]{f^{#1}_{#2}\lpar{#3}\rpar}
\newcommand{\gvetri}[3]{g^{#1}_{#2}\lpar{#3}\rpar}
\newcommand{\cvetri}[3]{{\cal{F}}^{#1}_{#2}\lpar{#3}\rpar}
\newcommand{\hvetri}[3]{{\hat{\cal{F}}}^{#1}_{#2}\lpar{#3}\rpar}
\newcommand{\avetri}[3]{{\overline{\cal{F}}}^{#1}_{#2}\lpar{#3}\rpar}
\newcommand{\fverti}[2]{F^{#1}_{#2}}
\newcommand{\cverti}[2]{{\cal{F}}_{#1}^{#2}}
\newcommand{\fV}{f_{_{\Vvert}}}
\newcommand{\gA}{g_{_{\Avert}}}
\newcommand{\fVi}[1]{f^{#1}_{_{\Vvert}}}
\newcommand{\seai}[1]{a_{#1}}
\newcommand{\seapi}[1]{a'_{#1}}
\newcommand{\seAi}[2]{A_{#1}^{#2}}
\newcommand{\sewi}[1]{w_{#1}}
\newcommand{\seWi}[1]{W_{#1}}
\newcommand{\seWsi}[1]{W^{*}_{#1}}
\newcommand{\seWti}[2]{W_{#1}^{#2}}
\newcommand{\sewti}[2]{w_{#1}^{#2}}
\newcommand{\seSig}[1]{\Sigma_{#1}\lpar\sla{\pmom}\rpar}
\newcommand{\ww}{{\rm{w}}}
\newcommand{\ew}{{\rm{ew}}}
\newcommand{\ct}{{\rm{ct}}}
\newcommand{\nonSE}{\rm{non-SE}}
\newcommand{\leading}{\rm{\ssL}}
%
%
\newcommand{\bbff}[1]{{\overline B}_{#1}}
\newcommand{\sW}{p_{_W}}
\newcommand{\sZ}{p_{_Z}}
\newcommand{\ssp}{s_p}
\newcommand{\fW}{f_{_W}}
\newcommand{\fZ}{f_{_Z}}
\newcommand{\subMSB}[1]{{#1}_{\mbox{$\overline{\scriptscriptstyle MS}$}}}
\newcommand{\supMSB}[1]{{#1}^{\mbox{$\overline{\scriptscriptstyle MS}$}}}
\newcommand{\redMSB}{{\mbox{$\overline{\scriptscriptstyle MS}$}}}
\newcommand{\gpbb}{g'_{0}}
\newcommand{\Zconip}[1]{Z'_{#1}}
\newcommand{\bpff}[4]{B'_{#1}\lpar #2;#3,#4\rpar}             
\newcommand{\xidf}{\xi^2-1}
\newcommand{\tDdr}{1/{\bar{\varepsilon}}}
\newcommand{\cRz}{{\cal R}_{_Z}}
\newcommand{\cRg}{{\cal R}_{\gamma}}
\newcommand{\Sz}{\Sigma_{_Z}}
\newcommand{\alh}{{\hat\alpha}}
\newcommand{\alhz}{\alpha_{\ssZ}}
\newcommand{\Phzg}{{\hat\Pi}_{_{\zb\ab}}}
\newcommand{\fvvert}{F^{\rm vert}_{_V}}
\newcommand{\gavert}{G^{\rm vert}_{_A}}
\newcommand{\bmv}{{\overline m}_{_V}}
\newcommand{\Sgn}{\Sigma_{\gamma\hkn}}
\newcommand{\rmboxd}{{\rm Box}_d\lpar s,t,u;M_1,M_2,M_3,M_4\rpar}
\newcommand{\rmboxc}{{\rm Box}_c\lpar s,t,u;M_1,M_2,M_3,M_4\rpar}
%
%
\newcommand{\Afaci}[1]{A_{#1}}
\newcommand{\Afacis}[1]{A^2_{#1}}
\newcommand{\upar}[1]{u}
\newcommand{\upari}[1]{u_{#1}}
\newcommand{\vpari}[1]{v_{#1}}
\newcommand{\lpari}[1]{l_{#1}}
\newcommand{\Lpari}[1]{l_{#1}}
\newcommand{\Nff}[2]{N^{(#1)}_{#2}}
\newcommand{\Sff}[2]{S^{(#1)}_{#2}}
\newcommand{\sSff}{S}
\newcommand{\etafd}[2]{\eta_d\lpar#1,#2\rpar}
\newcommand{\sigdu}[2]{\sigma_{#1#2}}
\newcommand{\scalc}[4]{c_{0}\lpar #1;#2,#3,#4\rpar}
\newcommand{\scald}[2]{d_{0}\lpar #1,#2\rpar}
\newcommand{\scaldi}[3]{d_{0}^{#1}\lpar #2,#3\rpar}
\newcommand{\pir}[1]{\Pi^{\rm{\ssR}}\lpar #1\rpar}
\newcommand{\sigh}{\sigma_{\rm had}}
\newcommand{\dah}{\Delta\alpha^{(5)}_{\rm had}}
\newcommand{\dat}{\Delta\alpha_{\rm top}}
\newcommand{\Vqed}[3]{V_1^{\rm sub}\lpar#1;#2,#3\rpar}
\newcommand{\thetah}{{\hat\theta}}
\newcommand{\smlon}{\frac{\mlones}{s}}
\newcommand{\lntwo}{\ln 2}
\newcommand{\wmin}{w_{\rm min}}
\newcommand{\kmin}{k_{\rm min}}
\newcommand{\mdls}{\Big|}
\newcommand{\smf}{\frac{\mfs}{s}}
\newcommand{\bint}{\beta_{\rm int}}
\newcommand{\IRv}{V_{_{\rm IR}}}
\newcommand{\IRr}{R_{_{\rm IR}}}
\newcommand{\fssts}{\frac{s^2}{t^2}}
\newcommand{\fssus}{\frac{s^2}{u^2}}
\newcommand{\optM}{1+\frac{t}{M^2}}
\newcommand{\opuM}{1+\frac{u}{M^2}}
\newcommand{\ftM}{\lpar -\frac{t}{M^2}\rpar}
\newcommand{\fuM}{\lpar -\frac{u}{M^2}\rpar}
\newcommand{\omsM}{1-\frac{s}{M^2}}
\newcommand{\xsf}{\sigma_{_{\rm F}}}
\newcommand{\xsb}{\sigma_{_{\rm B}}}
\newcommand{\afb}{A_{_{\rm FB}}}
\newcommand{\rsoft}{\rm soft}
\newcommand{\rms}{\rm s}
\newcommand{\rsmx}{\sqrt{s_{\rm max}}}
\newcommand{\rspm}{\sqrt{s_{\pm}}}
\newcommand{\rsp}{\sqrt{s_{+}}}
\newcommand{\rsm}{\sqrt{s_{-}}}
\newcommand{\sigmx}{\sigma_{\rm max}}
\newcommand{\gG}[2]{G_{#1}^{#2}}
\newcommand{\gacomm}[2]{\lpar #1 - #2\gfd\rpar}
\newcommand{\fcsx}{\frac{1}{\ctwsix}}
\newcommand{\fcq}{\frac{1}{\ctwf}}
\newcommand{\fcs}{\frac{1}{\ctws}}
\newcommand{\affs}[2]{{\cal A}_{#1}\lpar #2\rpar}                   
\newcommand{\stwei}{s_{\theta}^8}
\def\mdan{\vspace{1mm}\mpar{\hfil$\downarrow$new\hfil}\vspace{-1mm}
          \ignorespaces}
\def\muan{\vspace{-1mm}\mpar{\hfil$\uparrow$new\hfil}\vspace{1mm}\ignorespaces}
\def\mlan{\vspace{-1mm}\mpar{\hfil$\rightarrow$new\hfil}\vspace{1mm}\ignorespaces}
\def\mnnew{\mpar{\hfil NEWNEW \hfil}\ignorespaces}
%
%
\newcommand{\boxc}[2]{{\cal{B}}_{#1}^{#2}}
\newcommand{\boxct}[3]{{\cal{B}}_{#1}^{#2}\lpar{#3}\rpar}
\newcommand{\hboxc}[3]{\hat{{\cal{B}}}_{#1}^{#2}\lpar{#3}\rpar}
\newcommand{\vev}{\langle v \rangle}
\newcommand{\vevi}[1]{\langle v_{#1}\rangle}
\newcommand{\vevs}{\langle v^2   \rangle}
\newcommand{\fwfrV}[5]{\Sigma_{_V}\lpar #1,#2,#3;#4,#5 \rpar}
\newcommand{\fwfrS}[7]{\Sigma_{_S}\lpar #1,#2,#3;#4,#5;#6,#7 \rpar}
\newcommand{\fSi}[1]{f^{#1}_{_{\Svert}}}
\newcommand{\fPi}[1]{f^{#1}_{_{\Pvert}}}
\newcommand{\mXs}{m_{_X}}
\newcommand{\mXss}{m^2_{_X}}
\newcommand{\mYs}{M^2_{_Y}}
\newcommand{\xik}{\xi_k}
\newcommand{\xiks}{\xi^2_k}
\newcommand{\mpls}{m^2_+}
\newcommand{\mmis}{m^2_-}
%
\newcommand{\SN}{\Sigma_{_N}}
\newcommand{\SC}{\Sigma_{_C}}
\newcommand{\SPN}{\Sigma'_{_N}}
\newcommand{\PFf}{\Pi^{\fer}_{_F}}
\newcommand{\PFb}{\Pi^{\bos}_{_F}}
\newcommand{\dPZ}{\Delta{\hat\Pi}_{_Z}}
\newcommand{\Sfin}{\Sigma_{_F}}
\newcommand{\Sfir}{\Sigma_{_R}}
\newcommand{\Sfinh}{{\hat\Sigma}_{_F}}
\newcommand{\Sfinf}{\Sigma^{\fer}_{_F}}
\newcommand{\Sfinbh}{\Sigma^{\bos}_{_F}}
\newcommand{\alf}{\alpha^{\fer}}
\newcommand{\alhfz}{\alpha^{\fer}\lpar{\mzs}\rpar}
\newcommand{\alhfs}{\alpha^{\fer}\lpar{\sman}\rpar}
\newcommand{\gfQ}{g^f_{_{Q}}}
\newcommand{\gfL}{g^f_{_{L}}}
\newcommand{\ccf}{\frac{\gbs}{16\,\pi^2}}
\newcommand{\chq}{{\hat c}^4}
\newcommand{\muuq}{m_{u'}}
\newcommand{\muus}{m^2_{u'}}
\newcommand{\mdd}{m_{d'}}
\newcommand{\clf}[2]{\mathrm{Cli}_{_#1}\lpar\displaystyle{#2}\rpar}
\def\stes{\sin^2\theta}
\def\acal{\cal A}
\def\alr{A_{_{\rm{LR}}}}
\newcommand{\barQ}{\overline Q}
\newcommand{\Sptg}{\Sigma'_{_{3Q}}}
\newcommand{\Sptt}{\Sigma'_{_{33}}}
\newcommand{\Ppgg}{\Pi'_{\ph\ph}}
\newcommand{\Pww}{\Pi_{_{\wb\wb}}}
\newcommand{\capV}[2]{{\cal F}^{#2}_{_{#1}}}
\newcommand{\bt}{\beta_t}
\newcommand{\btp}{\beta'_t}
\newcommand{\mhsix}{M^6_{_H}}
\newcommand{\topt}{{\cal T}_{33}}
\newcommand{\topq}{{\cal T}_4}
\newcommand{\Phzgf}{{\hat\Pi}^{\fer}_{_{\zb\ab}}}
\newcommand{\Phzgb}{{\hat\Pi}^{\bos}_{_{\zb\ab}}}
\newcommand{\Sfirh}{{\hat\Sigma}_{_R}}
\newcommand{\Szgh}{{\hat\Sigma}_{_{\zb\ab}}}
\newcommand{\Szghb}{{\hat\Sigma}^{\bos}_{_{\zb\ab}}}
\newcommand{\Szghf}{{\hat\Sigma}^{\fer}_{_{\zb\ab}}}
\newcommand{\Szgb}{\Sigma^{\bos}_{_{\zb\ab}}}
\newcommand{\Szgf}{\Sigma^{\fer}_{_{\zb\ab}}}
\newcommand{\chig}{\chi_{\ph}}
\newcommand{\chiz}{\chi_{\ssZ}}
\newcommand{\Sfih}{{\hat\Sigma}}
\newcommand{\Szzh}{\hat{\Sigma}_{_{\zb\zb}}}
\newcommand{\dPZf}{\Delta{\hat\Pi}^f_{_{\zb}}}
\newcommand{\khZdf}[1]{{\hat\kappa}^{#1}_{_{\zb}}}
\newcommand{\chf}{{\hat c}^4}
\newcommand{\amp}[2]{{\cal{A}}_{_{#1}}^{\rm{#2}}}
\newcommand{\hatvm}[1]{{\hat v}^-_{#1}}
\newcommand{\hatvp}[1]{{\hat v}^+_{#1}}
\newcommand{\hatvpm}[1]{{\hat v}^{\pm}_{#1}}
\newcommand{\kvz}[1]{\kappa^{\zb #1}_{_V}}
\newcommand{\barp}{\overline p}                
\newcommand{\delw}{\Delta_{_{\wb}}}
\newcommand{\bdelw}{{\bar{\Delta}}_{_{\wb}}}
\newcommand{\bdelf}{{\bar{\Delta}}_{\ff}}
\newcommand{\delz}{\Delta_{_\zb}}
\newcommand{\deli}[1]{\Delta\lpar{#1}\rpar}
\newcommand{\hdeli}[1]{{\hat{\Delta}}\lpar{#1}\rpar}
\newcommand{\chizb}{\chi_{_\zb}}
\newcommand{\Swwp}{\Sigma'_{_{\wb\wb}}}
\newcommand{\epph}{\varepsilon'/2}
\newcommand{\sbffp}[1]{B'_{#1}}                    
\newcommand{\epss}{\varepsilon^*}
\newcommand{\Ddrhs}{{\ds\frac{1}{\hat{\varepsilon}^2}}}
\newcommand{\lnmsb}{L_{_\wb}}
\newcommand{\lnsmsb}{L^2_{_\wb}}
\newcommand{\tpni}{\lpar 2\pi\rpar^n\ib}
\newcommand{\tpn}{2^n\,\pi^{n-2}}
\newcommand{\cmf}{M_f}
\newcommand{\cmfs}{M^2_f}
\newcommand{\toDdr}{{\ds\frac{2}{{\bar{\varepsilon}}}}}
\newcommand{\troDdr}{{\ds\frac{3}{{\bar{\varepsilon}}}}}
\newcommand{\totDdr}{{\ds\frac{3}{{2\,\bar{\varepsilon}}}}}
\newcommand{\foDdr}{{\ds\frac{4}{{\bar{\varepsilon}}}}}
\newcommand{\smh}{m_{_H}}
\newcommand{\smhs}{m^2_{_H}}
\newcommand{\Ph}{\Pi_{_\hb}}
\newcommand{\Sphh}{\Sigma'_{_{\hb\hb}}}
\newcommand{\bh}{\beta}
\newcommand{\alsn}{\alpha^{(n_f)}_{_S}}
\newcommand{\smq}{m_q}
\newcommand{\smqp}{m_{q'}}
\newcommand{\shb}{h}
\newcommand{\hab}{A}
\newcommand{\hbpm}{H^{\pm}}
\newcommand{\hbp}{H^{+}}
\newcommand{\hbm}{H^{-}}
\newcommand{\msh}{M_h}
\newcommand{\mha}{M_{_A}}
\newcommand{\mhc}{M_{_{H^{\pm}}}}
\newcommand{\mshs}{M^2_h}
\newcommand{\mhas}{M^2_{_A}}
\newcommand{\barfp}{\overline{f'}}                
\newcommand{\chiii}{{\hat c}^3}
\newcommand{\chiv}{{\hat c}^4}
\newcommand{\chv}{{\hat c}^5}
\newcommand{\chvi}{{\hat c}^6}
\newcommand{\alsvi}{\alpha^{6}_{_S}}
\newcommand{\tww}{t_{_W}}
\newcommand{\ti}{t_{1}}
\newcommand{\tii}{t_{2}}
\newcommand{\tiii}{t_{3}}
\newcommand{\tiv}{t_{4}}
\newcommand{\psla}{\hbox{\rlap/p}}
\newcommand{\qsla}{\hbox{\rlap/q}}
\newcommand{\nsla}{\hbox{\rlap/n}}
\newcommand{\lsla}{\hbox{\rlap/l}}
\newcommand{\msla}{\hbox{\rlap/m}}
\newcommand{\cnsla}{\hbox{\rlap/N}}
\newcommand{\clsla}{\hbox{\rlap/L}}
\newcommand{\cmsla}{\hbox{\rlap/M}}
\newcommand{\blmt}{\lrbr - 3\rrbr}
\newcommand{\blfo}{\lrbr 4 1\rrbr}
\newcommand{\bltp}{\lrbr 2 +\rrbr}
\newcommand{\clitwo}[1]{{\rm{Li}}_{2}\lpar{#1}\rpar}
\newcommand{\clitri}[1]{{\rm{Li}}_{3}\lpar{#1}\rpar}
\newcommand{\xt}{x_{\ft}}
\newcommand{\zt}{z_{\ft}}
\newcommand{\Ht}{h_{\ft}}
\newcommand{\xts}{x^2_{\ft}}
\newcommand{\zts}{z^2_{\ft}}
\newcommand{\Hts}{h^2_{\ft}}
\newcommand{\ztc}{z^3_{\ft}}
\newcommand{\Htc}{h^3_{\ft}}
\newcommand{\ztq}{z^4_{\ft}}
\newcommand{\Htq}{h^4_{\ft}}
\newcommand{\ztv}{z^5_{\ft}}
\newcommand{\Htv}{h^5_{\ft}}
\newcommand{\ztx}{z^6_{\ft}}
\newcommand{\Htx}{h^6_{\ft}}
\newcommand{\ztz}{z^7_{\ft}}
\newcommand{\Htz}{h^7_{\ft}}
\newcommand{\sht}{\sqrt{\Ht}}
\newcommand{\atan}[1]{{\rm{arctan}}\lpar{#1}\rpar}
\newcommand{\dbff}[3]{{\hat{B}}_{_{{#2}{#3}}}\lpar{#1}\rpar}
\newcommand{\ztbs}{{\bar{z}}^{2}_{\ft}}
\newcommand{\ztb}{{\bar{z}}_{\ft}}
\newcommand{\Htbs}{{\bar{h}}^{2}_{\ft}}
\newcommand{\Htb}{{\bar{h}}_{\ft}}
\newcommand{\Hztb}{{\bar{hz}}_{\ft}}
\newcommand{\Ln}[1]{{\rm{Ln}}\lpar{#1}\rpar}
\newcommand{\Lns}[1]{{\rm{Ln}}^2\lpar{#1}\rpar}
\newcommand{\wt}{w_{\ft}}
\newcommand{\wts}{w^2_{\ft}}
\newcommand{\wtb}{\overline{w}}
\newcommand{\fra}{\frac{1}{2}}
\newcommand{\frb}{\frac{1}{4}}
\newcommand{\frc}{\frac{3}{2}}
\newcommand{\frd}{\frac{3}{4}}
\newcommand{\fre}{\frac{9}{2}}
\newcommand{\frf}{\frac{9}{4}}
\newcommand{\frg}{\frac{5}{4}}
\newcommand{\frh}{\frac{5}{2}}
\newcommand{\fri}{\frac{1}{8}}
\newcommand{\frj}{\frac{7}{4}}
\newcommand{\frl}{\frac{7}{8}}
\newcommand{\Spzzh}{\hat{\Sigma}'_{_{\zb\zb}}}
\newcommand{\sqs}{\sqrt{s}}
\newcommand{\Rtg}{R_{_{3Q}}}
\newcommand{\Rtt}{R_{_{33}}}
\newcommand{\Rww}{R_{_{\wb\wb}}}
\newcommand{\ssA}{{\scriptscriptstyle{A}}}
\newcommand{\ssB}{{\scriptscriptstyle{B}}}
\newcommand{\ssC}{{\scriptscriptstyle{C}}}
\newcommand{\ssD}{{\scriptscriptstyle{D}}}
\newcommand{\ssE}{{\scriptscriptstyle{E}}}
\newcommand{\ssF}{{\scriptscriptstyle{F}}}
\newcommand{\ssG}{{\scriptscriptstyle{G}}}
\newcommand{\ssH}{{\scriptscriptstyle{H}}}
\newcommand{\ssI}{{\scriptscriptstyle{I}}}
\newcommand{\ssJ}{{\scriptscriptstyle{J}}}
\newcommand{\ssK}{{\scriptscriptstyle{K}}}
\newcommand{\ssL}{{\scriptscriptstyle{L}}}
\newcommand{\ssM}{{\scriptscriptstyle{M}}}
\newcommand{\ssN}{{\scriptscriptstyle{N}}}
\newcommand{\ssO}{{\scriptscriptstyle{O}}}
\newcommand{\ssP}{{\scriptscriptstyle{P}}}
\newcommand{\ssQ}{{\scriptscriptstyle{Q}}}
\newcommand{\ssR}{{\scriptscriptstyle{R}}}
\newcommand{\ssS}{{\scriptscriptstyle{S}}}
\newcommand{\ssT}{{\scriptscriptstyle{T}}}
\newcommand{\ssU}{{\scriptscriptstyle{U}}}
\newcommand{\ssV}{{\scriptscriptstyle{V}}}
\newcommand{\ssW}{{\scriptscriptstyle{W}}}
\newcommand{\ssX}{{\scriptscriptstyle{X}}}
\newcommand{\ssY}{{\scriptscriptstyle{Y}}}
\newcommand{\ssZ}{{\scriptscriptstyle{Z}}}
\newcommand{\ssWF}{\rm{\scriptscriptstyle{WF}}}
\newcommand{\OLA}{\rm{\scriptscriptstyle{OLA}}}
\newcommand{\QED}{\rm{\scriptscriptstyle{QED}}}
\newcommand{\QCD}{\rm{\scriptscriptstyle{QCD}}}
\newcommand{\EW}{\rm{\scriptscriptstyle{EW}}}
\newcommand{\UV}{\rm{\scriptscriptstyle{UV}}}
\newcommand{\IR}{\rm{\scriptscriptstyle{IR}}}
\newcommand{\OMS}{\rm{\scriptscriptstyle{OMS}}}
\newcommand{\GMS}{\rm{\scriptscriptstyle{GMS}}}
\newcommand{\sszero}{\rm{\scriptscriptstyle{0}}}
\newcommand{\DiagramFermionToBosonFullWithMomenta}[8][70]{
  \vcenter{\hbox{
  \SetScale{0.8}
  \begin{picture}(#1,50)(15,15)
    \put(27,22){$\nearrow$}      
    \put(27,54){$\searrow$}    
    \put(59,29){$\to$}    
    \ArrowLine(25,25)(50,50)      \Text(34,20)[lc]{#6} \Text(11,20)[lc]{#3}
    \ArrowLine(50,50)(25,75)      \Text(34,60)[lc]{#7} \Text(11,60)[lc]{#4}
    \Photon(50,50)(90,50){2}{8}   \Text(80,40)[lc]{#2} \Text(55,33)[ct]{#8}
    \Vertex(50,50){2,5}          \Text(60,48)[cb]{#5} 
    \Vertex(90,50){2}
  \end{picture}}}
  }
\newcommand{\DiagramFermionToBosonPropagator}[4][85]{
  \vcenter{\hbox{
  \SetScale{0.8}
  \begin{picture}(#1,50)(15,15)
    \ArrowLine(25,25)(50,50)
    \ArrowLine(50,50)(25,75)
    \Photon(50,50)(105,50){2}{8}   \Text(90,40)[lc]{#2}
    \Vertex(50,50){0.5}         \Text(80,48)[cb]{#3}
    \GCirc(82,50){8}{1}            \Text(55,48)[cb]{#4}
    \Vertex(105,50){2}
  \end{picture}}}
  }
\newcommand{\DiagramFermionToBosonEffective}[3][70]{
  \vcenter{\hbox{
  \SetScale{0.8}
  \begin{picture}(#1,50)(15,15)
    \ArrowLine(25,25)(50,50)
    \ArrowLine(50,50)(25,75)
    \Photon(50,50)(90,50){2}{8}   \Text(80,40)[lc]{#2}
    \BBoxc(50,50)(5,5)            \Text(55,48)[cb]{#3}
    \Vertex(90,50){2}
  \end{picture}}}
  }
\newcommand{\DiagramFermionToBosonFull}[3][70]{
  \vcenter{\hbox{
  \SetScale{0.8}
  \begin{picture}(#1,50)(15,15)
    \ArrowLine(25,25)(50,50)
    \ArrowLine(50,50)(25,75)
    \Photon(50,50)(90,50){2}{8}   \Text(80,40)[lc]{#2}
    \Vertex(50,50){2.5}          \Text(60,48)[cb]{#3}
    \Vertex(90,50){2}
  \end{picture}}}
  }
\newcommand{\expgw}{\frac{\gf\mws}{2\srt\,\pi^2}}
\newcommand{\expgz}{\frac{\gf\mzs}{2\srt\,\pi^2}}
\newcommand{\Spww}{\Sigma'_{_{\wb\wb}}}
\newcommand{\shf}{{\hat s}^4}
\newcommand{\acz}{\scff{0}}
\newcommand{\acoo}{\scff{11}}
\newcommand{\acod}{\scff{12}}
\newcommand{\acdo}{\scff{21}}
\newcommand{\acdd}{\scff{22}}
\newcommand{\acdt}{\scff{23}}
\newcommand{\acdq}{\scff{24}}
\newcommand{\acto}{\scff{31}}
\newcommand{\actd}{\scff{32}}
\newcommand{\actt}{\scff{33}}
\newcommand{\actq}{\scff{34}}
\newcommand{\actc}{\scff{35}}
\newcommand{\acts}{\scff{36}}
\newcommand{\acoA}{\scff{1A}}
\newcommand{\acdA}{\scff{2A}}
\newcommand{\acdB}{\scff{2B}}
\newcommand{\acdC}{\scff{2C}}
\newcommand{\acdD}{\scff{2D}}
\newcommand{\actA}{\scff{3A}}
\newcommand{\actB}{\scff{3B}}
\newcommand{\actC}{\scff{3C}}
\newcommand{\ada}{\sdff{0}}
\newcommand{\adb}{\sdff{11}}
\newcommand{\adc}{\sdff{12}}
\newcommand{\add}{\sdff{13}}
\newcommand{\ade}{\sdff{21}}
\newcommand{\adf}{\sdff{22}}
\newcommand{\adg}{\sdff{23}}
\newcommand{\adh}{\sdff{24}}
\newcommand{\adi}{\sdff{25}}
\newcommand{\adj}{\sdff{26}}
\newcommand{\adl}{\sdff{27}}
\newcommand{\adm}{\sdff{31}}
\newcommand{\adn}{\sdff{32}}
\newcommand{\ado}{\sdff{33}}
\newcommand{\adp}{\sdff{34}}
\newcommand{\adq}{\sdff{35}}
\newcommand{\adr}{\sdff{36}}
\newcommand{\ads}{\sdff{37}}
\newcommand{\adt}{\sdff{38}}
\newcommand{\adu}{\sdff{39}}
\newcommand{\adw}{\sdff{310}}
\newcommand{\adv}{\sdff{311}}
\newcommand{\ady}{\sdff{312}}
\newcommand{\adz}{\sdff{313}}
\newcommand{\admt}{\frac{\tman}{\sman}}
\newcommand{\admu}{\frac{\uman}{\sman}}
\newcommand{\frm}{\frac{3}{8}}
\newcommand{\frn}{\frac{5}{8}}
\newcommand{\fro}{\frac{15}{8}}
\newcommand{\frp}{\frac{3}{16}}
\newcommand{\frq}{\frac{5}{16}}
\newcommand{\frr}{\frac{1}{16}}
\newcommand{\frs}{\frac{7}{2}}
\newcommand{\frt}{\frac{7}{16}}
\newcommand{\fru}{\frac{1}{3}}
\newcommand{\frw}{\frac{2}{3}}
\newcommand{\frz}{\frac{4}{3}}
\newcommand{\fry}{\frac{13}{3}}
\newcommand{\fraa}{\frac{11}{4}}
\newcommand{\bee}{\beta_{e}}
\newcommand{\beW}{\beta_{_\wb}}
\newcommand{\beWs}{\beta^2_{_\wb}}
\newcommand{\etaW}{\eta_{_\wb}}
\newcommand{\toDdrh}{{\ds\frac{2}{{\hat{\varepsilon}}}}}
\newcommand{\bqas}{\begin{eqnarray*}}
\newcommand{\eqas}{\end{eqnarray*}}
\newcommand{\mhcub}{M^3_{_H}}
\newcommand{\adComA}{\sdff{A}}
\newcommand{\adComB}{\sdff{B}}
\newcommand{\adComC}{\sdff{C}}
\newcommand{\adComD}{\sdff{D}}
\newcommand{\adComE}{\sdff{E}}
\newcommand{\adComF}{\sdff{F}}
\newcommand{\adComG}{\sdff{G}}
\newcommand{\adComH}{\sdff{H}}
\newcommand{\adComI}{\sdff{I}}
\newcommand{\adComJ}{\sdff{J}}
\newcommand{\adComL}{\sdff{L}}
\newcommand{\adComM}{\sdff{M}}
\newcommand{\adComN}{\sdff{N}}
\newcommand{\adComO}{\sdff{O}}
\newcommand{\adComP}{\sdff{P}}
\newcommand{\adComQ}{\sdff{Q}}
\newcommand{\adComR}{\sdff{R}}
\newcommand{\adComS}{\sdff{S}}
\newcommand{\adComT}{\sdff{T}}
\newcommand{\adComU}{\sdff{U}}
\newcommand{\adComAc}{\sdff{A}^c}
\newcommand{\adComBc}{\sdff{B}^c}
\newcommand{\adComCc}{\sdff{C}^c}
\newcommand{\adComDc}{\sdff{D}^c}
\newcommand{\adComEc}{\sdff{E}^c}
\newcommand{\adComFc}{\sdff{F}^c}
\newcommand{\adComGc}{\sdff{G}^c}
\newcommand{\adComHc}{\sdff{H}^c}
\newcommand{\adComIc}{\sdff{I}^c}
\newcommand{\adComJc}{\sdff{J}^c}
\newcommand{\adComLc}{\sdff{L}^c}
\newcommand{\adComMc}{\sdff{M}^c}
\newcommand{\adComNc}{\sdff{N}^c}
\newcommand{\adComOc}{\sdff{O}^c}
\newcommand{\adComPc}{\sdff{P}^c}
\newcommand{\adComQc}{\sdff{Q}^c}
\newcommand{\adComRc}{\sdff{R}^c}
\newcommand{\adComSc}{\sdff{S}^c}
\newcommand{\adComTc}{\sdff{T}^c}
\newcommand{\adComUc}{\sdff{U}^c}
\newcommand{\adComAf}{\sdff{A}^f}
\newcommand{\adComBf}{\sdff{B}^f}
\newcommand{\adComCf}{\sdff{F}^f}
\newcommand{\adComDf}{\sdff{D}^f}
\newcommand{\adComEf}{\sdff{E}^f}
\newcommand{\adComFf}{\sdff{F}^f}
\newcommand{\adComGf}{\sdff{G}^f}
\newcommand{\adComHf}{\sdff{H}^f}
\newcommand{\adComIf}{\sdff{I}^f}
\newcommand{\adComJf}{\sdff{J}^f}
\newcommand{\adComLf}{\sdff{L}^f}
\newcommand{\adComMf}{\sdff{M}^f}
\newcommand{\adComNf}{\sdff{N}^f}
\newcommand{\adComOf}{\sdff{O}^f}
\newcommand{\adComPf}{\sdff{P}^f}
\newcommand{\adComQf}{\sdff{Q}^f}
\newcommand{\adComRf}{\sdff{R}^f}
\newcommand{\adComSf}{\sdff{S}^f}
\newcommand{\adComTf}{\sdff{T}^f}
\newcommand{\adComUf}{\sdff{U}^f}
\newcommand{\adComBfc}{\sdff{B}^{fc}} 
\newcommand{\adComCfco}{\sdff{C}^{fc1}}
\newcommand{\adComCfcd}{\sdff{C}^{fc2}} 
\newcommand{\adComCfct}{\sdff{C}^{fc3}} 
\newcommand{\adComDfc}{\sdff{D}^{fc}}
\newcommand{\adComEfc}{\sdff{E}^{fc}}
\newcommand{\adComFfc}{\sdff{F}^{fc}}
\newcommand{\adComGfc}{\sdff{G}^{fc}}
\newcommand{\adComHfc}{\sdff{H}^{fc}}
\newcommand{\adComLfc}{\sdff{L}^{fc}}
\newcommand{\afba}[1]{A^{#1}_{_{\rm FB}}}
\newcommand{\alra}[1]{A^{#1}_{_{\rm LR}}}
\newcommand{\adComAt}{\sdff{A}^t}
\newcommand{\adComBt}{\sdff{B}^t}
\newcommand{\adComCt}{\sdff{T}^t}
\newcommand{\adComDt}{\sdff{D}^t}
\newcommand{\adComEt}{\sdff{E}^t}
\newcommand{\adComFt}{\sdff{T}^t}
\newcommand{\adComGt}{\sdff{G}^t}
\newcommand{\adComHt}{\sdff{H}^t}
\newcommand{\adComIt}{\sdff{I}^t}
\newcommand{\adComJt}{\sdff{J}^t}
\newcommand{\adComLt}{\sdff{L}^t}
\newcommand{\adComMt}{\sdff{M}^t}
\newcommand{\adComNt}{\sdff{N}^t}
\newcommand{\adComOt}{\sdff{O}^t}
\newcommand{\adComPt}{\sdff{P}^t}
\newcommand{\adComQt}{\sdff{Q}^t}
\newcommand{\adComRt}{\sdff{R}^t}
\newcommand{\adComSt}{\sdff{S}^t}
\newcommand{\adComTt}{\sdff{T}^t}
\newcommand{\adComUt}{\sdff{U}^t}
\newcommand{\adComAtt}{\sdff{A}^{\tau}}
\newcommand{\adComBtt}{\sdff{B}^{\tau}}
\newcommand{\adComCtt}{\sdff{T}^{\tau}}
\newcommand{\adComDtt}{\sdff{D}^{\tau}}
\newcommand{\adComEtt}{\sdff{E}^{\tau}}
\newcommand{\adComFtt}{\sdff{T}^{\tau}}
\newcommand{\adComGtt}{\sdff{G}^{\tau}}
\newcommand{\adComHtt}{\sdff{H}^{\tau}}
\newcommand{\adComItt}{\sdff{I}^{\tau}}
\newcommand{\adComJtt}{\sdff{J}^{\tau}}
\newcommand{\adComLtt}{\sdff{L}^{\tau}}
\newcommand{\adComMtt}{\sdff{M}^{\tau}}
\newcommand{\adComNtt}{\sdff{N}^{\tau}}
\newcommand{\adComOtt}{\sdff{O}^{\tau}}
\newcommand{\adComPtt}{\sdff{P}^{\tau}}
\newcommand{\adComQtt}{\sdff{Q}^{\tau}}
\newcommand{\adComRtt}{\sdff{R}^{\tau}}
\newcommand{\adComStt}{\sdff{S}^{\tau}}
\newcommand{\adComTtt}{\sdff{T}^{\tau}}
\newcommand{\adComUtt}{\sdff{U}^{\tau}}
\newcommand{\etavz}[1]{\eta^{\zb #1}_{_V}}
\newcommand{\phanst}{$\hphantom{\sigma^{s+t}\ }$}
\newcommand{\phanat}{$\hphantom{A_{FB}^{s+t}\ }$}
\newcommand{\phanss}{$\hphantom{\sigma^{s}\ }$}
\newcommand{\phanas}{$\hphantom{A_{FB}^{s}\ }$} 
\newcommand{\pbb}{\,\mbox{\bf pb}}
\newcommand{\pe}{\,\%\:}
\newcommand{\pc}{\,\%}
\newcommand{\temiv}{10^{-4}}
\newcommand{\temv}{10^{-5}}
\newcommand{\temvi}{10^{-6}}
\newcommand{\di}[1]{d_{#1}}
\newcommand{\delip}[1]{\Delta_+\lpar{#1}\rpar}
\newcommand{\propbb}[5]{{{#1}\over {\lpar #2^2 + #3 - \ib\varepsilon\rpar
\lpar\lpar #4\rpar^2 + #5 -\ib\varepsilon\rpar}}}
\newcommand{\cfft}[5]{C_{#1}\lpar #2;#3,#4,#5\rpar}    
\newcommand{\ppl}[1]{p_{+{#1}}}
\newcommand{\pmi}[1]{p_{-{#1}}}
\newcommand{\bpox}{\beta^2_{\xi}}
\newcommand{\bffdiff}[5]{B_{\rm d}\lpar #1;#2,#3;#4,#5\rpar}             
\newcommand{\cffdiff}[7]{C_{\rm d}\lpar #1;#2,#3,#4;#5,#6,#7\rpar}    
\newcommand{\affdiff}[2]{A_{\rm d}\lpar #1;#2\rpar}             
\newcommand{\Dqf}{\Delta\qf}
\newcommand{\bposx}{\beta^4_{\xi}}
\newcommand{\svverti}[3]{f^{#1}_{#2}\lpar{#3}\rpar}
\newcommand{\Mods}{\mbox{$M^2_{12}$}}
\newcommand{\Mots}{\mbox{$M^2_{13}$}}
\newcommand{\Motq}{\mbox{$M^4_{13}$}}
\newcommand{\Mdts}{\mbox{$M^2_{23}$}}
\newcommand{\Mdos}{\mbox{$M^2_{21}$}}
\newcommand{\Mtds}{\mbox{$M^2_{32}$}}
\newcommand{\dffpt}[3]{D_{#1}\lpar #2,#3;}           
\newcommand{\quu}{Q_{uu}}
\newcommand{\qdd}{Q_{dd}}
\newcommand{\qud}{Q_{ud}}
\newcommand{\qdu}{Q_{du}}
\newcommand{\msPj}[6]{\Lambda^{#1#2#3}_{#4#5#6}}
\newcommand{\bdiff}[4]{B_{\rm d}\lpar #1,#2;#3,#4\rpar}             
\newcommand{\bdifff}[7]{B_{\rm d}\lpar #1;#2;#3;#4,#5;#6,#7\rpar}             
\newcommand{\adiff}[3]{A_{\rm d}\lpar #1;#2;#3\rpar}  
\newcommand{\aw}{a_{_\wb}}
\newcommand{\az}{a_{_\zb}}
\newcommand{\sct}[1]{sect.~\ref{#1}}
\newcommand{\dreim}[1]{\varepsilon^{\rm M}_{#1}}
\newcommand{\drem}{\varepsilon^{\rm M}}
\newcommand{\hcapV}[2]{{\hat{\cal F}}^{#2}_{_{#1}}}
\newcommand{\swww}{{\scriptscriptstyle \wb\wb\wb}}
\newcommand{\szhz}{{\scriptscriptstyle \zb\hb\zb}}
\newcommand{\shzh}{{\scriptscriptstyle \hb\zb\hb}}
\newcommand{\bwith}[3]{\beta^{#3}_{#1}\lpar #2\rpar}
\newcommand{\Shhh}{{\hat\Sigma}_{_{\hb\hb}}}
\newcommand{\Sphhh}{{\hat\Sigma}'_{_{\hb\hb}}}
\newcommand{\seWilc}[1]{w_{#1}}
\newcommand{\seWtilc}[2]{w_{#1}^{#2}}
\newcommand{\eilc}{\gamma}
\newcommand{\eilcs}{\gamma^2}
\newcommand{\eilcc}{\gamma^3}
\newcommand{\eilcb}{{\overline{\gamma}}}
\newcommand{\eilcbs}{{\overline{\gamma}^2}}
\newcommand{\Sttww}{\Sigma_{_{33;\wb\wb}}}
\newcommand{\bSttww}{{\overline\Sigma}_{_{33;\wb\wb}}}
\newcommand{\Pggtg}{\Pi_{\ph\ph;3Q}}
\def\rmL{{\rm L}}
\def\rmT{{\rm T}}
\def\rmM{{\rm M}}
\newcommand{\hsm}{\hspace{-0.4mm}}
\newcommand{\hsmm}{\hspace{-0.2mm}}
\def\negs{\hspace{-0.26in}}
\def\negss{\hspace{-0.18in}}
\def\negsss{\hspace{-0.09in}}
\def\scee{{\mbox{\lowercase{$\fep\fem$}}}}
\def\scffb{{\mbox{\lowercase{$\ff\barf$}}}}
\def\scqq{{\mbox{\lowercase{$\fq\barq$}}}}
\def\scln{{\mbox{\lowercase{$\fl\fnu$}}}}
\def\app#1#2 {{\it Acta. Phys. Pol.} {\bf#1},#2}
\def\cpc#1#2 {{\it Computer Phys. Comm.} {\bf#1},#2}
\def\np#1#2 {{\it Nucl. Phys.} {\bf#1},#2}
\def\pl#1#2 {{\it Phys. Lett.} {\bf#1},#2}
\def\prep#1#2 {{\it Phys. Rep.} {\bf#1},#2}
\def\prev#1#2 {{\it Phys. Rev.} {\bf#1},#2}
\def\prl#1#2 {{\it Phys. Rev. Lett.} {\bf#1},#2}
\def\zp#1#2 {{\it Zeit. Phys.} {\bf#1},#2}
\def\sptp#1#2 {{\it Suppl. Prog. Theor. Phys.} {\bf#1},#2}
\def\mpl#1#2 {{\it Modern Phys. Lett.} {\bf#1},#2}
\def\jetp#1#2 {{\it Sov. Phys. JETP} {\bf#1},#2}
\def\fpj#1#2 {{\it Fortschr. Phys.} {\bf#1},#2}
\def\afp#1#2 {{\it Acta.Phys. Polon.} {\bf#1},#2}
\def\err#1#2 {{\it Erratum} {\bf#1},#2}
\def\ijmp#1#2 {{\it Int. J. Mod. Phys} {\bf#1},#2}
\def\nc#1#2 {{\it Nuovo Cimento} {\bf#1},#2}
\def\ap#1#2 {{\it Ann. Phys.} {\bf#1},#2}
\def\cmp#1#2 {{\it Comm. Math. Phys.} {\bf#1},#2}
\def\el#1#2 {{\it Europhys. Lett.} {\bf#1},#2}
\def\hpa#1#2 {{\it Helv. Phys. Acta} {\bf#1},#2}
\def\yf#1#2 {{\it Yad. Fiz.} {\bf#1},#2}
\def\nim#1#2 {{\it Nucl. Instrum. Meth.} {\bf#1},#2}
\def\spz#1#2 {{\it Sov. Pisma Zhetf} {\bf#1},#2}
\def\jetpl#1#2 {{\it JETP Lett.} {\bf#1},#2}
\def\sjnp#1#2 {{\it Sov. J. Nucl. Phys.} {\bf#1},#2}
\def\ptp#1#2 {{\it Progr. Theor. Phys. (Kyoto)} {\bf#1},#2}
\def\rmp#1#2  {{\it Rev. Mod. Phys.} {\bf#1},#2}
\def\zhetf#1#2 {{\it ZhETF} {\bf#1},#2}
\def\prs#1#2 {{\it Proc. Roy. Soc.} {\bf#1},#2}
\def\phys#1#2 {{\it Physica} {\bf#1},#2}
\def\itetal{{\it et al.}}
\newcommand{\dalpha}{\Delta\alpha}
\newcommand{\drho}{\Delta\rho}
\newcommand{\drhov}{\delta\rho}
\newcommand{\dkapv}{\delta\kappa}
\newcommand{\drhovh}{\delta{\hat{\rho}}}
\newcommand{\drhovb}{\delta{\hat{\rho}}}
\newcommand{\epsb}{\bar\varepsilon}
\newcommand{\gll}{\Gamma_{\fl}}
\newcommand{\gqq}{\Gamma_{\fq}}
\newcommand{\Imsi}[1]{I^2_{#1}}
\newcommand{\reni}[1]{R_{#1}}
\newcommand{\renis}[1]{R^2_{#1}}
\newcommand{\sreni}[1]{\sqrt{R_{#1}}}
\newcommand{\dalphav}{\Delta\alpha^{(5)}(\mzs)}
\newcommand{\Rvaz}[1]{g^{#1}_{\sss{\zb}}}
\newcommand{\rvab}[1]{{\bar{g}}_{#1}}
\newcommand{\rvabs}[1]{{\bar{g}}^2_{#1}}
\newcommand{\rab }[1]{{\bar{a}}_{#1}}
\newcommand{\rvb }[1]{{\bar{v}}_{#1}}
\newcommand{\rabs}[1]{{\bar{a}}^2_{#1}}
\newcommand{\rva }[1]{g_{#1}}
\newcommand{\rvas}[1]{g^2_{#1}}
\newcommand{\Rva }[1]{G_{#1}}
\newcommand{\Rvac}[1]{G^{*}_{#1}}
\newcommand{\Rvah }[1]{{\hat{G}}_{#1}}
\newcommand{\Rvahc}[1]{{\hat{G}}^{*}_{#1}}
\newcommand{\Rvas}[1]{G^2_{#1}}
\newcommand{\rhobi} [1]{{\bar{\rho}}_{#1}}
\newcommand{\kappai}[1]{\kappa_{#1}}
\newcommand{\rhois}[1]{\rho^2_{#1}}
\newcommand{\rhohi} [1]{{\hat{\rho}}_{#1}}
\newcommand{\rhobpi}[1]{{\bar{\rho}}'_{#1}}
\newcommand{\bm}[1]{\mbox{\boldmath $#1$}}
\newcommand{\Kx}{{x}}
\newcommand{\KxT}{{x^{\ssT}}}
\newcommand{\Kxi}{{x}_{_K}}
\newcommand{\KxiT}{{x}^{\ssT}_{_K}}
\newcommand{\Kvec}[1]{{#1}}
\newcommand{\KvecT}[1]{{#1}^{\ssT}}
\newcommand{\Pmat}{\tilde{P}}
\newcommand{\Pmati}{\tilde{P}^{-1}}
\newcommand{\detg}[1]{\mbox{det}_{#1}}

%% file: 2f-Chapt-Title.tex
\title{\bf Two-Fermion Production in Electron-Positron Collisions\\ 
}

\author{
Michael Kobel$^{1,a}$ and Zbigniew W\c{a}s$^{2}$ \\[0.5cm]
  C.~Ainsley$^3$, 
  A.~Arbuzov$^{4,5}$,
   S.~Arcelli$^{6}$,
   D.~Bardin$^{7}$,
   I.~Boyko$^{7}$,
   D.~Bourilkov$^{8,b}$,
   P.~Christova$^{7,9}$,
   J.~Fujimoto$^{10}$, 
   M.~Gr\"unewald$^{11}$,  
   T.~Ishikawa$^{10}$, 
   M.~Jack$^{12}$, 
   S.~Jadach$^{13,2}$,
   L.~Kalinovskaya$^{7}$, 
   Y.~Kurihara$^{10}$,
   A.~Leike$^{14}$,
   R.~Mcpherson$^{15}$,  
   M-N.~Minard$^{16,c}$,
   G.~Montagna$^{17,18}$, 
   M.~Moretti$^{19,20}$, 
   T.~Munehisa$^{21}$,
   O.~Nicrosini$^{18,17}$, 
   A.~Olchevski$^{7,22,d}$,
   F.~Piccinini$^{18,17}$,   
   B.~Pietrzyk$^{16}$,
   W.~P\l{a}czek$^{23}$, 
   S.~Riemann$^{12}$, 
   T.~Riemann$^{12}$,
   G.~Taylor$^{24,e}$,
   Y.~Shimizu$^{10}$, 
   M.~Skrzypek$^{2}$,
   S.~Spagnolo$^{25}$, 
   B.F.L.~Ward$^{13,26}$}
\maketitle

\begin{itemize}
\item[$^{1}$] 
             Physikalisches Institut, Universit\"at Bonn, Germany 
\item[$^{2}$]
             Institute of Nuclear Physics, ul. Kawiory 26a, 30-055,
             Cracow, Poland
\item[$^{3}$]
             High Energy Physics Group, Cavendish Laboratory, University of Cambridge, UK
\item[$^{4}$]
             DFT, Universit\`a di Torino; 
             INFN, Sezione di Torino;
             via Giuria 1, I-10125 Torino, Italy
\item[$^{5}$] 
             BLTPh, Joint Institute for Nuclear Research, Dubna, 141980, Russia
\item[$^{6}$] 
             EPHEPG, Department of Physics and Astronomy, 
             University of Maryland, USA
\item[$^{7}$]
             LNP, JINR, RU-141980, Dubna, Russia 
\item[$^{8}$] 
             Institute for Particle Physics, ETH Zurich, Switzerland
\item[$^{9}$]
             Faculty of Physics, Bishop Preslavsky University,
             Shoumen, Bulgaria  
\item[$^{10}$]
             High Energy Accelerator Research Organization (KEK),
             Tsukuba, Ibaraki 305-0801, Japan
\item[$^{11}$]
             Humboldt University Berlin, Germany
\item[$^{12}$]
             DESY, D-15738 Zeuthen, Germany 
\item[$^{13}$]  
             Department of Physics and Astronomy
             The University of Tennessee, Knoxville, Tennessee 
             37996-1200, USA
\item[$^{14}$] 
             Sektion Physik, Ludwig-Maximilians-Universit\"at M\"unchen,
             Germany
\item[$^{15}$] 
             Department of Physics and Astronomy, TRIUMF, 
             University of Victoria, Canada
\item[$^{16}$]
             LAPP, IN2P3-CNRS, F-74941 Annecy-le-Vieux, France
\item[$^{17}$]
             Dipartimento di Fisica Nucleare e Teorica, 
                        Universit\`a di Pavia, Italy
\item[$^{18}$] 
             INFN, Sezione di Pavia, Italy
\item[$^{19}$] 
             Dipartimento di Fisica, Universit\`a di Ferrara, Italy
\item[$^{20}$] 
             INFN, Sezione di Ferrara, Italy
\item[$^{21}$]
             Yamanashi University, Kofu, Yamanashi 400-8510, Japan
\item[$^{22}$]
             EP Division, CERN, CH-1211 Geneva 23, Switzerland  
\item[$^{23}$] 
             Institute of Computer Science, Jagellonian University, 
             Cracow, Poland
\item[$^{24}$] 
             SCIPP Natural Sciences II, University of California, USA
\item[$^{25}$] 
             Rutherford Appleton Laboratory, UK
\item[$^{26}$] 
             SLAC, Stanford University, Stanford, California 94309, USA.
\item[$^{a}$]
              coordinator for OPAL   collab.,   
      $^{b}$  coordinator for L3     collab.,   
      $^{c}$  coordinator for ALEPH  collab.,   
\item[$^{d}$] 
              coordinator for DELPHI collab., 
      $^{e}$  experimental coordinator for $\nu \bar \nu \gamma$ channels.
\end{itemize}

\clearpage

\hbox{  }
\vspace{40mm}
\begin{center}
{\bf
        The LEP-2 Monte Carlo Workshop 1999/2000
}
\end{center}
\vspace{10mm}
\begin{center}
{\bf\LARGE
       Two-Fermion Production in Electron-Positron Collisions    
}
\end{center}
\vspace{10mm}
\begin{center}
{\bf
        Two-Fermion Working Group Report
}
\end{center}
\vspace{30mm}
\begin{center}
{\bf\LARGE
        Abstract
}
\end{center}
This report summarizes the results of the two-fermion working group
of the LEP2-MC workshop, held at CERN from 1999 to 2000.
Recent developments in the theoretical calculations of the
two fermion production process in the electron-positron
collision at LEP2 center of the mass energies are reported.
The Bhabha process and the production of muon, tau, neutrino
and quark pairs is covered.
On the basis of comparison of various calculations,
theoretical uncertainties are estimated and compared with those
needed for the final LEP2 data analysis.
The subjects for the further studies are identified.

\clearpage

%% file: 2f-Chapt-Part1.tex
\section{Introduction}
At LEP2 the two-fermion production has the highest cross section 
of all hard processes.
At LEP1 it serves as the unique reaction to study the
properties of the Z boson
and it had a very distinct two-body character,
as the photon initial state emission (ISR) was highly suppressed.
At LEP2, far above the Z pole, the ISR is strong and frequent,
the radiative tail of the Z develops to such an extent
that the ISR QED radiative
corrections are several times as large as the Born cross sections.
Another important fact is that one needs to account for
the production of secondary real fermion-anti-fermion pairs 
(usually light and soft)
due to the radiation of off-shell photons and Z bosons from the initial- 
or final-state.  
This makes the task of the ``signal definition'', that is what
we really mean by the two-fermion final state, rather nontrivial,
in other words there is the question
of the separation between the radiative corrections to
two-fermion production and the genuine four-fermion production.
This aspect of the two-fermion process was highlighted 
in the discussion of the LEP Electroweak Group 
and also in the presentations in the beginning
of the current workshop.
One aim of the workshop therefore was to come up with a 2-fermion
signal definition, which is applicable to all 2-fermion final states,
and suited equally well for theoretical predictions and experimental
measurements.

The other topics which emerged as important theoretical issues for
the work in two-fermion group were the question of the reliability of the existing
QED calculations, especially of the so called initial-final state interference,
and the question of the reliability of the pure electroweak corrections.
The outstanding performance of the machine gave LEP experiments
sizable samples of events with one and two explicitly (tagged) photons
which are very useful for searches of the phenomena beyond
the Standard Model (SM).
The evaluation of the rates of these events from QED was high on the agenda
of our work from the beginning.

The layout of our chapter reflects to a large extent the evolution of the
work of our two-fermion group.
In general we pursued two ways of collecting, evaluating and improving
theoretical calculations for two-fermion precess.
On one hand, we collected all available theoretical calculations,
as implemented in Monte Carlo (MC) and semi-analytical programs (codes)
and we applied them to get predictions for all
cross sections, asymmetries, etc. measured in LEP2 experiments,
which were identified in first place, with the help of our experimental
coordinators, one for each LEP collaboration.
All codes bear their own theoretical error specification and applicability range.
This what we call the ``wide-range comparisons'' of many codes 
for many observables has given
us confidence into individual error specifications, or has led to some
questions to be solved either within this workshop or beyond. 
On the other hand, the alternative path was followed of the so called ``tuned-comparisons''
or ``theme-comparisons'' which either concentrated on the more detailed comparisons
of 2-3 codes, usually concentrating just on one theoretical problem
and trying to reduce just one source of the theoretical errors, 
for example from QED effects.
The prominent part of the ``theme-comparisons'' was the study on the effects
of the secondary pair production process.

Having this in mind the outline of the report is not surprising.
In the first section we amass the rich list of processes and measurable
quantities for all 2-fermion channels, 
that is the Bhabha process, the quark-, $\mu$-, $\tau$-pair channels, 
without and with tagged photons
which are one pillar of the ``wide-range comparisons''.
Another pillar is the third section in which all theoretical calculations/codes
are collected -- each of them includes its individual total theoretical
error and range of applicability.
In the second section we present the harvest of the wide-range comparisons,
summarizing in a quantitative way results of them, channel per channel.

The ``theme-comparisons'' are located in the fourth section and their
subset related to secondary pair production was important enough
to be awarded the status of the separate fifth section.

In the last section we summarize all important results 
and list the problems in two-fermion production
which are still left out for further work.

\section{Experimental observables and theoretical precision requirements}
\label{list}

This section collects specifications
of the quantities measured in the two-fermion process at LEP2
which we call for short ``observables'',
and we also try for each listed observable to define the necessary
precision level of the theoretical prediction, keeping in mind
the total experimental error which will be achieved at the end of LEP2
operation, for data combined for all four LEP experiments.

The great diversity of these observables is
a distinct feature of the two-fermion process,
as compared to the $WW$ channel or QCD studies, see other sections.
It is partly due to the fact that various final states
like muon-pairs, tau-pairs, quark-pairs, neutrino-pair with gamma and
the Bhabha process have very different experimental characteristics,
different methods of measurement and each of them comes in
two versions, accepting Z radiative return or rejecting it.
Furthermore, the two-fermion process
cannot be experimentally completely disentangled
from the four-fermion process, multiplying again the possible option
for defining the {\em two-fermion observables}.
On top of that there are still (and will be) differences
between the ways the four experiments define and measure their
cross-sections, asymmetries and distributions.

In this section we make a sort of ``frontal attack'' on the problem
of defining what is measured as a two-fermion process at LEP2,
by doing the most complete list of two-fermion observables used in LEP2
data analysis.
The primary aim of this is
to help theorists to understand what is really measured in LEP2 and what
are the ultimate precision targets in the LEP2 data.
However, such an exhaustive  list can also be useful for experimental collaborations
when combining data from four LEP collaborations.

We never had a hope to have a complete theoretical prediction
and full discussion of theoretical errors for all this impressive
list of two-fermion observables.
It is not even necessary as some of them are quite similar,
and some of them are rather difficult to implement for the average
theorist.
In the process of scrutinizing various theoretical calculations
we use only part of these observables,
mostly of the ``simplified type''.
The simplified observables are also necessary because
semi-analytical programs can provide predictions only for them
and not for the realistic ones.
(MC event generators have no such limitations.)
Another role of simplified observables is that they are
prototypes of the observables used for combining data from the four LEP
collaborations.

For the so called ``tuned comparisons'' which were made for instance among
\KKMC\ and \zf\ or between BHWIDE and LABSMC the authors of these codes
have used their own, even more simple kinematic cuts,
tailored specifically to these tests. 
They are not discussed in this section.

The observables which {\em include} the Z resonance in the phase space,
that is Z-{\em inclusive}, 
and which {\em exclude} the Z resonance, that is Z{\em -exclusive},
we usually denote them using the short-hand notation ``inclusive observables''
and ``exclusive observables'' instead of the full ``Z-inclusive''
and ``Z-exclusive''.
We hope that this will not lead to confusion, and wherever necessary
we shall expand to the full terminology.

For the purpose of our main aim, that is of establishing
theoretical errors for the {\em typical two-fermion observables}
this section contains too much information.
We think, however, that it is a valuable asset of this report
and we decided to keep it to the full extent,
accepting that only some of them will be really used in the actual
{\em theoretical} studies.

\subsection{Precision requirements for theoretical predictions}

One of the most important ingredient of the observable definition is its precision
tag. Obviously the higher precision the more complicated the study of its theoretical
uncertainties will be. Also more of the details of experimental cuts will be needed
to estimate the theoretical systematic errors. 


The following rule of thumb 
with respect to the errors obtained for the data taken at
$\sqrt{s}=189$~GeV
was suggested for estimating the required precision
in cases where there is no better information available.

\begin{enumerate}
\item
  The experimental statistical error is decreased by factor of $\sqrt{12} \simeq 3.5$
  with respect  to present (summer 1999) one for the single collaboration. 
  We still expect statistics to
  grow by factor of 3 and combination of all 4 experiments makes the total statistics 
a factor of 12 bigger.
\item
  The experimental systematic error can be expected to go down by a factor of 2
  (may be 3) due to improvements and partial non-correlations among experiments.
\item
  The above estimates of the experimental statistical and systematic error should be
  added in quadrature.
\item 
  The required  precision should be $1/3$ of that to assure that theoretical 
  effects will not deteriorate experimental results (will not increase an error)
  by more than 10\% .
\end{enumerate}

The demand of a maximum of 10 \% increase of the final experimental overall error due 
to theoretical uncertainties is not an over-demand. It is equivalent to a decrease
in running time of experiments by 20 \% in cases when the statistical error dominates.
A similar 10\% deterioration  of an overall detection performance would occur
only after a 30 \% decrease in data taking for measurements where the statistical
and systematic errors are similar in size. 
Having this in mind and also an enormous effort of so many people over so 
many years, not mentioning the costs, may even lead to the conclusion that our
rule for precision requirements from theory is not strict enough.  

In general as inclusive (incl.) we understand the cross section for $\sqrt{s'/s}>0.1$
and as exclusive (excl.) for $\sqrt{s'/s}>0.85$. For Bhabha excl$_1$ denotes 
$|\cos\theta|<0.9$ and excl$_2$ denotes $|\cos\theta|<0.7$. For asymmetries
the (absolute) error tag can be obtained from the one of thecross section
multiplying it by 0.01$\cdot \sqrt{2}$ and dropping the \% symbol.

The following precision tags can be assumed if explicit 
numbers do not overrule them in the text:

\begin{enumerate}
\item cross sections for $q\bar q$ final states: incl. 0.11 \%, excl. 0.23 \%
\item cross sections  for $e^-e^+$ final states: excl$_1$ 0.13 \%, excl$_2$ 0.21 \%
\item cross sections and asymmetries for $\mu^-\mu^+$ final states: incl. 0.41 \%, excl. 0.53 \%
\item cross sections and asymmetries for $\tau^-\tau^+$ final states: incl. 0.44 \%, excl. 0.61 \%
\item searches background cross sections  from quarks and tagged hard photons 0.3 \% 
\item searches background cross sections  from leptons and tagged hard photons:\\
single photons 1.5 \%, multiple photons 5\% 
\item searches background  cross sections from neutrinos and tagged hard photons\\
single photons 0.5 \%, multiple photons 2\%

\end{enumerate} 
These numbers serve as a starting point for the definition of precision tags
required from complete theoretical calculations. For separate ingredients
such as interferences pair corrections etc, the physical precision tag must be even
more strict, to assure that their combination will not overcome the required
tag. Finally the tag for technical tests must be set another factor of few smaller.

\subsection{General comments on observable definition}

In the following chapters the observables will be defined. For more details see respectively:
ALEPH \cite{ALPUB183}, DELPHI \cite{DELPUB172} for all observables, for 
   (\ffbar($\gamma$)) processes L3 \cite{L3PUB136,L3PUB172,L3PUB189,L3RUNALF},
OPAL \cite{OPAL2F172,OPAL2F183,OPAL2F189};  for  ($\llbar+\gamma$)
\cite{L3GAML189,OPALGAML172}; and  for ($\vvbar+\gamma$)
\cite{L3GAMNU189,OPALGAMNU172,OPALGAMNU183}.

The main groups of
observables are formed by physics processes, they are later divided into
{\em realistic} and {\em idealized} ones corresponding to different stages of 
experimental analysis. Finally specific solutions adopted by experiments
are placed in the subsections. Some points which could be extracted from 
the observables definition like a glossary or the approach to the extra pair are
discussed at the beginning of the chapter to avoid unnecessary repetition.

The purpose of the list of observables is to review all necessary conditions
for calculation of theoretical predictions. More explanatory examples are
given later in the section.

\subsection{Notation}
In definition of observables we will use some short-hand notations to make
it easier. Let us illustrate just a few of them. 
\begin{itemize}
\item
$E_{cm}$: center of mass energy.
\item
$\theta_\gamma$: The polar angle of particle $\gamma$ with respect to the 
electron beam.
\item
$|\cos\theta_{f/\bar f}| < 0.9$: The polar angle of both particles $f$ and $\bar f$
must satisfy the cut.
\item
$E_\gamma$: The energy of particle $\gamma$.
\item
$x_\gamma=2E_\gamma/E_{cm}$: The energy fraction of particle $\gamma$.
\item
$N_{trk}$: Number of charged tracks in the event.
\item 
${\mathrm acol}(e^+ e^-)$: The collinearity angle between particles $e^+$ 
and $e^-$.
\item
$A_{FB}$: forward-backward asymmetry constructed on the basis  of final state
charged particles.
\item
$M_{prop-}(f\bar{f})$: invariant mass of the $s$-channel propagator, with
ISR/FSR interference subtracted.
\item
$M_{prop+}(f\bar{f})$: invariant mass of the $s$-channel propagator, with
ISR/FSR interference not subtracted. In this case, the propagator mass
is ambiguous for the interference contribution, so this part is actually 
evaluated using the $f\bar{f}$ invariant mass excluding radiative photons.
\item
$M_{inv}(f\bar{f})$: invariant mass of $f$ and $\bar{f}$ - all other 
particles such as collinear photons are excluded. For realistic observables,
this is determined from the reconstructed 4-momenta of the two particles.
\item
$M_{ang}(f\bar{f})$: invariant mass determined from the measured polar
angles of particles $f$ and $\bar{f}$ with respect to the electron beam.
These are taken from jet angles in case of hadrons.
This calculation is based on four-momentum conservation assuming that
only one radiative photon is present. If no photon is seen in the detector,
this radiative photon is assumed to go along the beam axis. If an energetic,
isolated photon is seen in the detector, then its reconstructed polar
angle is used.
\item
$M_{kine}(f\bar{f})$: invariant mass determined from a kinematic fit.
This uses 4-momentum conservation to improve the mass estimate based on the 
reconstructed 4-momenta of the $f$ and $\bar{f}$.
\item  
s'$_{L3}$ for $\mu^+\mu^-$ or $\tau^+\tau^-$ from observed ISR photon (E more 
than 10 GeV, $ |\cos\theta_\gamma|<0.985$, separation to nearest fermion
more than 10 degrees), {\em or} from fermion angles assuming photon
escapes along the beam pipe
\item  
s'$_{L3}$ for $q\bar q$:  
to reconstruct the effective centre-of-mass energy two different methods
are used. For the first one, all events are reclustered into two
jets using the JADE algorithm.
A single ISR photon is assumed to be emitted along the beam axis and to result in
a missing momentum vector. From the polar angles of the jets, $\theta_{1}$ and
$\theta_{2}$, the photon energy can be estimated.
The second method uses the clustered jets ($y_{\mathrm{cut}}=0.01$)
obtained using the JADE algorithm.
A kinematic fit is performed on the jets and the missing 
four-momentum vector using different hypotheses for the emitted ISR photons.
The missing energy is attributed to zero, one or two ISR photons.
From the differences given by the two methods, systematic errors
on the effective centre-of-mass energy reconstruction are calculated.
When an isolated energetic photon (energy larger than 10 GeV) is detected,
the energy and momentum of this photon are added to the undetected ISR photons.

\item 
s'$_{\rm OPAL}$ for \qqbar:  from a series of kinematic fits using
observed ISR photons ( $ |\cos\theta_\gamma|<0.985$, 
isolated with less than 1 GeV energy flow in a cone of half angle 0.2~rad
around them) plus hadronic jets. The fits allow zero, one or two additional
photons emitted close to the beam direction. s' is taken from the fit
with the lowest number of extra photons giving an acceptable $\chi^2$.


\end{itemize}

\subsection{Additional pair treatment}

\noindent{\bf ALEPH}\\
Apart from the cuts listed for realistic observables, no additional requirement is added to reject events 
with real or virtual pair emission, their contribution is taken in account for efficiency calculation. 
The main rejection to 4-fermions process is due to invariant mass 
cuts associated to the topological cut on the event shape.
For a two-fermion final state they are considered as backgrounds and their residual contribution estimated 
from Monte-Carlo four-fermion events for which  both 
fermion pair invariant mass \ffbar\ or $\fpfp$ are above 65~GeV. 
This excludes from the signal definition the 4-fermions arising from virtual 
Z or W contribution, but not the contribution from virtual photon.
The above requirement are valid for hadronic, muon pairs and tau pairs.
Note that for ALEPH the pair contribution is kept both in  realistic 
and idealized observables.

\noindent{\bf DELPHI}\\
In the case of idealized observables it is assumed that theoretical predictions should 
not include either real or virtual pair corrections. These are subtracted 
from the data with the help of the Monte Carlo.
The appropriate systematic error is to be discussed with the help of Monte
Carlo program or programs. Monte Carlo predictions for idealized 
observables with pair corrections excluded are required as well as for 
realistic observables with pair corrections included. 
In the case of realistic observables implicit cuts on additional 
fermion pairs are given, where appropriate, with the sufficient detail 
for DELPHI observables.
Exceptions are $\nu\bar \nu \gamma$ observables, where events are required to have no jets/leptons
($ch$) within the DELPHI acceptance, that is of energy $E_{ch}>0.5$ GeV and $|\cos\theta_{ch}|<0.97$.

\noindent{\bf L3}\\
Let us define first the cut on the secondary pairs for idealized observables:
\begin{itemize}
\item
 We assume that all the phase space for the secondary pairs is included
and integrated over, however contribution form resonant diagrams from 
4-fermion processes etc. must be subtracted from the data first by means
of Monte Carlo.
\end{itemize}
For the secondary pair cut to be used
for realistic  L3 observables our aim was to describe it in a form 
as brief as possible. In particular we accept ambiguities which may 
lead to differences much smaller than precision tag. 
The cut off on the secondary pair is performed in the following way:
\begin{itemize}
\item
  The invariant mass of the primary pair must be at least 60 GeV for $qq$, 
  75 GeV for $\mu\mu/\tau\tau$. The invariant mass is determined from collinearity
  of primary pair and other tracks under assumption that missing is only
  single photon (collinear to the beam). The primary pair means lepton
  same-flavor anti-lepton or two jets, of highest energies. 
\item
 cut on invariant mass of the secondary pair to be smaller than invariant mass
  of the primary pair is not explicitly required, but it is in practice 
  through the selection targets on the 2 highest energy jets to form the primary 
  pair, seen in the detector. 
\item
  The W and Z-pair production background is reduced by applying the following
  cuts.
  Semi-leptonic W-pair decays are rejected by requiring the transverse
  energy imbalance to be smaller than 0.3 $E_{\mathrm{vis}}$. The background from
  hadronic W(Z)-pair decays is reduced by rejecting events with at least 
  four jets each with energy larger than 15  GeV. The jets are obtained
  using the JADE algorithm with a fixed jet resolution parameter 
  $y_{\mathrm{cut}}=0.01$ .
  The remaining part of the 4-fermion processes due to W and Z contributions
  remain and must be subtracted later with the help of Monte Carlo of well
  established theoretical uncertainty.  


\item
  For Bhabha there is no explicit cut on primary mass, but implicit cut 
  on secondary pair is present through the collinearity cut. 
  The primary electrons are selected as the two highest in energy clusters 
  which are matched to charged tracks. Note that in this way we allow 
  the presence of photon(s) which is harder than one
  or both of the final state $e^{+}e^{-}$.
\item
  $M_{inv}(e^+e^-)$ ($L3$) and electron clusters definitions:  In L3 electrons
  are considered dressed, absorbing photons (or extra pairs) in a half-cone 
  2.5 deg, all these can be seen as detector coverage is larger than 
  used phase space.
\item
   our  definition is not practical  for primary pair being neutrinos.
  Then, instead we request veto cut, no visible charged energy deposits
  above 1 GeV and $\cos_{\theta_X} <0.97$.  
\end{itemize}

\noindent{\bf OPAL}\\
To define our strategy for additional pair treatment let us recall the main
points from the ref.~\cite{OPAL2F172}.\\
No additional cuts, apart those listed for the respective observables,
are applied to reject events with pair emission.
In   general, we compare  our   measurements with analytical predictions
including pair  emission. This means   that pair emission  via virtual
photons  from both the  initial and  final  state must be included  in
efficiency calculations, and be excluded from background estimates. In
order to perform the separation,  we ignore interference between  $s$-
and $t$-channel diagrams contributing to  the same four-fermion  final
state, and generate separate   Monte Carlo samples for   the different
diagrams for each final state.  For a two-fermion  final state \ffbar\ we
then  include as   signal   those four-fermion   events   arising from
$s$-channel  processes for which $m_{\ffbar}  > m_{\fpfp}$, $m_{\fpfp} <$
70~GeV   and $m_{inv}(\ffbar)/\sqrt{s} > 0.1$  
($m_{inv}(\ffbar)/\sqrt{s}    > 0.85$ in  the
non-radiative case).  This kinematic classification closely models the
desired classification of $\ffbar\fpfp$  in terms of intermediate bosons,
in that pairs  arising from virtual photons  are generally included as
signal whereas those arising from virtual Z bosons are not. All events
arising  from $s$-channel processes failing   the above cuts, together
with those arising from the  $t$-channel process (Zee) and  two-photon
processes are regarded as background. Four-fermion processes involving
WW or single W production  are also background in  all cases.  The
overall efficiency, $\epsilon$, is calculated as 
\begin{equation}
  \epsilon = \left(1 -
    \frac{\sigma_{\ffbar\fpfp}}{\sigma_{\mathrm{tot}}}\right)\epsilon_{\ffbar}
  + \frac{\sigma_{\ffbar\fpfp}}{\sigma_{\mathrm{tot}}}\epsilon_{\ffbar\fpfp}
\end{equation}
where $\epsilon_{\ffbar}$, $\epsilon_{\ffbar\fpfp}$ are the efficiencies 
derived from the two-fermion and four-fermion signal Monte Carlo events
respectively, $\sigma_{\ffbar\fpfp}$ is the generated four-fermion 
cross-section, and $\sigma_{\mathrm{tot}}$ is the total cross-section
from the analytical prediction (e.g. \zf\ )
including pair emission. Using this definition of efficiency,
effects of cuts on soft pair emission in the four-fermion generator
are correctly summed with vertex corrections involving virtual pairs.
The inclusion of the four-fermion part of the signal produces
negligible changes to the efficiencies for hadronic events and for
lepton pairs with $\sqrt{s'_{OPAL}/s}>0.85$. The efficiencies for lepton pairs
with $\sqrt{s'_{OPAL}/s}>0.10$ are decreased by about 0.5\%.
\\
The discussion in the above paragraph applies to hadronic, muon pair
and tau pair final states. In the case of electron pairs, the situation
is slightly different. In principle the $t$-channel process with a second
fermion pair arising from the conversion of a virtual photon emitted from 
an initial- or final-state electron should be included as signal. 
As this process is not included in any program we use for
comparison we simply ignore such events: they are not included as 
background as this would underestimate the cross-section.


\subsection{ Realistic $e^+e^-\to q\bar q(\gamma)$ observables }

\noindent{\bf ALEPH}

\begin{enumerate}\setcounter{enumi}{0}
\item 
   {\bf Inclusive Selection:}\labobs{Aleph1} 
   Event clustered into jets with JADE algorithm until 
   $(M_{jet-jet}/E_{cm})^2 > 0.008$. Low mass jets with high electromagnetic
   energy fraction are assumed to be radiative photons. Non-photon jets then
   cluster together until only two left, corresponding to the $q\bar q$
   system. Then 

  \begin{enumerate}
    \item \labobs{Aleph1a} $M_{inv}(q\bar q)>50$~GeV (excluding photon jets),
    \item \labobs{Aleph1b} $M_{ang}(q\bar q)>0.1 E_{cm}$. 
  \end{enumerate}
\item 
  {\bf Exclusive Selection:}\labobs{Aleph2} 
  Same as inclusive selection, then following extra cuts,

    \begin{enumerate} 
    \item \labobs{Aleph2a} $M_{inv}(q\bar q)>0.7 E_{cm}$ (excluding photon jets),
    \item \labobs{Aleph2b} $M_{ang}(q\bar q)>0.9 E_{cm}$,
    \item \labobs{Aleph2c} $|\cos\theta_{q/\bar q}|<0.95$, 
    \item \labobs{Aleph2d} Event $\hbox{thrust} > 0.85$.
  \end{enumerate}
\end{enumerate}

\vspace{5mm}
\noindent{\bf DELPHI}\\
Events were retained if they contained at least 7 charged tracks and
if the charged energy was greater than 15~\% of the collision energy. In
addition, the quantity $ E_{rad}=\sqrt{E_F^2+E_B^2} $, where $E_F$ and $E_B$
stand for the total energy seen in the Forward and Backward electromagnetic
calorimeters, was required to be less than 90 \% of the beam energy.

\begin{enumerate}\setcounter{enumi}{2} 
\item 
{\bf Inclusive Selection:}~\labobs{Delphi1}~$M_{\rm ang}(q\bar q)>75$~GeV
\item 
{\bf Exclusive Selection:}~\labobs{Delphi2}~$M_{\rm ang}(q\bar q)/\sqrt{s}>0.85$
\end{enumerate}

\noindent{\bf L3}\\
Events are selected by restricting the visible energy, 
$ E_{\mathrm{vis}} $, to $ 0.4 < E_{\mathrm{vis}}/\sqrt{s} < 2.0 $.
The longitudinal energy imbalance must satisfy
$|E_{\mathrm{long}}|/E_{\mathrm{vis}}<0.7$. 
These cuts account for a large reduction of the two-photon background.
In order to reject background originating from lepton pair events, more
than 18 calorimetric clusters with an energy larger than 300 MeV are 
requested.

\begin{enumerate}\setcounter{enumi}{4} 
\item 
{\bf Inclusive Selection 1:} \labobs{LT1}~$\sqrt{s^\prime_{\rm L3}/s}>0.10$
\item 
{\bf Inclusive Selection 2:} \labobs{LT2}~$\sqrt{s^\prime_{\rm L3}}>60$~GeV
\item 
{\bf Exclusive Selection:} \labobs{LT3}~$\sqrt{s^\prime_{\rm L3}/s}>0.85$
\end{enumerate}

\noindent{\bf OPAL}\\
Events are required to have at least 
7~electromagnetic clusters and at least 5~tracks. 
The total energy deposited in the 
electromagnetic calorimeter has to be  at least 14\% of the centre-of-mass energy. 
The energy balance $R_{bal}$ along the beam direction 
has to satisfy \mbox{$ R_{bal} \equiv \mid \Sigma (E_{clus} \cdot \cos \theta) \mid / \Sigma E_{clus} < 0.75$}, 
where the sum runs over all clusters in the electromagnetic calorimeter, 
$\theta$ is the polar angle, and $E_{clus}$ is the energy of  each cluster. 
Events selected as W-pair candidates according to the criteria of \cite{OPALWW183}
are rejected.

\begin{enumerate}\setcounter{enumi}{7} 
\item 
   {\bf Inclusive Selection:}~\labobs{OPAL1}~$\sqrt{s'_{OPAL}/s}>0.10$
\item 
   {\bf Exclusive Selection:}~\labobs{OPAL2}~$\sqrt{s'_{OPAL}/s}>0.85$
\end{enumerate}

\subsection{ Idealized $e^+e^-\to q\bar q(\gamma)$ observables }

\noindent{\bf ALEPH}

\begin{enumerate}\setcounter{enumi}{9} 
\item 
{\bf Inclusive Selection:}~\labobs{IAleph1}~$M_{prop+}(q\bar q)/E_{cm}>0.1$ . 

\item 
{\bf Exclusive Selection:}~\labobs{IAleph2}~$M_{prop+}(q\bar q)/E_{cm}>0.9$ 
and $|\cos\theta_q| < 0.95$. 
\end{enumerate}

\noindent{\bf DELPHI}

\begin{enumerate}\setcounter{enumi}{11} 
\item 
   {\bf Inclusive Selection:}~\labobs{IDelphi1}~$M_{prop+}(q\bar q)/E_{cm}>0.1$ and 
$|\cos\theta_q| < 1.0$.

\item 
   {\bf Exclusive Selection:}~\labobs{IDelphi2}~$M_{prop+}(q\bar q)/E_{cm}>0.85$ and 
$|\cos\theta_q| < 1.0$.
\end{enumerate}

\noindent{\bf L3}

\begin{enumerate}\setcounter{enumi}{13} 
\item 
   {\bf Inclusive Selection 1:}~\labobs{ILT1}~$M_{prop-}(q\bar q)/E_{cm}>0.10$ and 
   $|\cos\theta_q| < 1.0$.

\item 
{\bf Inclusive Selection 2:}~\labobs{ILT2}~$M_{prop-}(q\bar q)>60~GeV$ and 
$|\cos\theta_q| < 1.0$.

\item 
{\bf Exclusive Selection:}~\labobs{ILT3}~$M_{prop-}(q\bar q)/E_{cm}>0.85$ and 
$|\cos\theta_q| < 1.0$.
\end{enumerate}

\noindent{\bf OPAL}

\begin{enumerate}\setcounter{enumi}{16} 
\item 
{\bf Inclusive Selection:}~\labobs{IOpal1}~$M_{prop-}(q\bar q)/E_{cm}>0.10$ and 
$|\cos\theta_q| < 1.0$.

\item 
{\bf Exclusive Selection:}~\labobs{IOpal2}~$M_{prop-}(q\bar q)/E_{cm}>0.85$ and 
$|\cos\theta_q| < 1.0$.
\end{enumerate}

\subsection{ Realistic $e^+e^-\to e^+ e^-(\gamma)$ observables }

\noindent{\bf ALEPH}

\begin{enumerate}\setcounter{enumi}{18} 
\item 
{\bf Exclusive Selection 1:}~\labobs{Aleph3}~
$2\le N_{trk}\le 8$. Two of tracks must be identified as electrons, have 
$|\cos\theta| < 0.95$ and opposite charge. The scalar sum of 
their momenta should exceed $0.3 E_{cm}$. The sum of their energies
including photons within $20^\circ$ of each tracks should exceed 
$0.4 E_{cm}$. Then $M_{inv}(e^+ e^-)>80$~GeV (reconstructed from track
momenta), $M_{ang}(e^+ e^-)>0.9 E_{cm}$ (where radiative photons 
reconstructed as for $q\bar q$ selection), and $-0.9 < \cos\theta^*_{e^-} < 0.9$.
\item 
{\bf Exclusive Selection 2:}~\labobs{Aleph4}~
Same as exclusive selection 1, but $-0.9 < \cos\theta^*_{e^-} < 0.7$.
\end{enumerate}

\noindent{\bf DELPHI}

\begin{enumerate}\setcounter{enumi}{20} 
\item 
{\bf Exclusive Selection:}~\labobs{Delphi3}~
The electron and positron were required to be in the polar angle range
$44^\circ<\theta<136^\circ$ and the non-radiative events were selected
by asking the collinearity to be smaller than $20^\circ$.
\end{enumerate}

\noindent{\bf L3}\\
The electron and positron are required to be in the polar angle range
$44^\circ<\theta<136^\circ$ or $20^\circ<\theta<160^\circ$
and non-radiative events are selected by asking the collinearity to
be smaller than $20^\circ$ OR $M_{inv}(e^+ e^-)/E_{cm}>0.85$.

\begin{enumerate}\setcounter{enumi}{21} 
\item 
{\bf Exclusive Selection 1:}~\labobs{LT4}~$M_{inv}(e^+ e^-)/E_{cm}>0.85$ and 
$|\cos\theta_{{e^+}/{e^-}}| < 0.71934$.
\item 
{\bf Exclusive Selection 2:}~\labobs{LT5}~${\mathrm acol}(e^+ e^-) < 25 deg.$ and 
$|\cos\theta_{{e^+}/{e^-}}| < 0.71934$.
\item 
{\bf Exclusive Selection 3:}~\labobs{LT6}~${\mathrm acol}(e^+ e^-) < 25 deg.$ and 
$|\cos\theta_{{e^+}/{e^-}}| < 0.94$.
\item 
{\bf Inclusive Selection 1:}~\labobs{LT7}~${\mathrm acol}(e^+ e^-) < 120 deg.$ and 
$|\cos\theta_{{e^+}/{e^-}}| < 0.71934$.
\item 
{\bf Inclusive Selection 2:}~\labobs{LT8}~$M_{inv}(e^+ e^-)/E_{cm}>0.10$ and 
$|\cos\theta_{{e^+}/{e^-}}| < 0.94$.
\end{enumerate}

\noindent{\bf OPAL}\\
Events selected as electron pairs are required to have at least two and not more than eight clusters in the 
electromagnetic calorimeter, and not more than eight tracks in the central tracking chambers. At least two 
clusters must have an energy exceeding 20\% of the beam energy, and the total energy deposited in the 
electromagnetic calorimeter must be at least 50\% of the centre-of-mass energy. For the inclusive selection
and the exclusive selection 1, at least two of the three highest energy clusters must each have an associated 
central detector track. If all three clusters have an associated track, the two highest energy clusters are chosen 
to be the electron and positron. For the large acceptance exclusive selection 2,  no requirement is placed on 
the association of tracks to clusters, but the requirement on the total electromagnetic energy is increased to 
70\% of the centre-of-mass energy.

\begin{enumerate}\setcounter{enumi}{26} 
\item 
{\bf Inclusive Selection :}~\labobs{Opal3}~$\mathrm{acol}(e^+ e^-)<170^\circ$ and 
$|\cos\theta_{{e^+}/{e^-}}| < 0.9$.
\item 
{\bf Exclusive Selection 1:}~\labobs{Opal4}~$\mathrm{acol}(e^+ e^-) < 10^\circ$ and 
$|\cos\theta_{e^-}| < 0.7$.
\item 
{\bf Exclusive Selection 2:}~\labobs{Opal5}~$\mathrm{acol}(e^+ e^-) < 10^\circ$ and 
$|\cos\theta_{{e^+}/{e^-}}| < 0.96$.
\end{enumerate}

\subsection{ Idealized $e^+e^-\to e^+ e^-(\gamma)$ observables }

\noindent{\bf ALEPH}
\begin{enumerate}\setcounter{enumi}{29} 
\item 
   {\bf Exclusive Selection 1:}~\labobs{IAleph3}~$M_{inv}(e^+ e^-)/E_{cm}>0.9$ and 
   $-0.9 < \cos\theta^*_{e^-} < 0.9$. 
\item 
   {\bf Exclusive Selection 2:}~\labobs{IAleph4}~$M_{inv}(e^+ e^-)/E_{cm}>0.9$ and 
   $-0.9 < \cos\theta^*_{e^-} < 0.7$. 
\end{enumerate}

\noindent{\bf DELPHI}
\begin{enumerate}\setcounter{enumi}{31} 
\item 
{\bf Exclusive Selection:}~\labobs{IDelphi4}~$M_{inv}(e^+ e^-)/E_{cm}>0.85$ and 
$|\cos\theta_{{e^+}/{e^-}}| < 0.7$.
\end{enumerate}

\noindent{\bf L3}
\begin{enumerate}\setcounter{enumi}{32} 
\item 
   {\bf Exclusive Selection 1:}~\labobs{ILT4}~$M_{inv}(e^+ e^-)/E_{cm}>0.85$ and 
   $|\cos\theta_{{e^+}/{e^-}}| < 0.71934$.
\item 
   {\bf Exclusive Selection 2:}~\labobs{ILT5}~${\mathrm acol}(e^+ e^-) < 25 deg.$ and 
   $|\cos\theta_{{e^+}/{e^-}}| < 0.71934$.
\item 
   {\bf Exclusive Selection 3:}~\labobs{ILT6}~${\mathrm acol}(e^+ e^-) < 25 deg.$ and 
   $|\cos\theta_{{e^+}/{e^-}}| < 0.94$.
\item 
   {\bf Inclusive Selection 1:}~\labobs{ILT7}~${\mathrm acol}(e^+ e^-) < 120 deg.$ and 
$|\cos\theta_{{e^+}/{e^-}}| < 0.71934$.
   \item 
   {\bf Inclusive Selection 2:}~\labobs{ILT8}~$M_{inv}(e^+ e^-)/E_{cm}>0.10$ and 
   $|\cos\theta_{{e^+}/{e^-}}| < 0.94$.
\end{enumerate}

\noindent{\bf  OPAL}
\begin{enumerate}\setcounter{enumi}{37} 
\item 
   {\bf Inclusive Selection :}~\labobs{IOpal3}~$\mathrm{acol}(e^+ e^-)<170^\circ$ and 
   $|\cos\theta_{{e^+}/{e^-}}| < 0.9$.
\item 
   {\bf Exclusive Selection 1:}~\labobs{IOpal4}~$\mathrm{acol}(e^+ e^-) < 10^\circ$ and 
   $|\cos\theta_{e^-}| < 0.7$.
\item 
   {\bf Exclusive Selection 2:}~\labobs{IOpal5}~$\mathrm{acol}(e^+ e^-) < 10^\circ$ and 
   $|\cos\theta_{{e^+}/{e^-}}| < 0.96$.
\end{enumerate}

\subsection{ Realistic $e^+e^-\to \mu^+\mu^-(\gamma)$ observables}

\noindent{\bf ALEPH}

\begin{enumerate}\setcounter{enumi}{40} 
\item 
   {\bf Inclusive Selection:}~\labobs{Aleph5}~
   $2\le N_{trk}\le 8$. Two tracks must be identified as muons, have 
   $p > 6$~GeV, $|\cos\theta| < 0.95$ and opposite charge. The scalar sum of 
   their momenta should exceed 60~GeV. Photons are identified from jets clustered with 
   JADE 
   algorithm until $(M_{jet-jet}/E_{cm})^2 > 0.008$. 
   Then $M_{ang}(\mu^+\mu^-)/E_{cm}>0.10$ (where radiative photons 
   reconstructed as for $q\bar q$ selection).
\item 
   {\bf Exclusive Selection:}~\labobs{Aleph6}~
   Same as inclusive selection, then following extra cuts, 
   $M_{ang}(\mu^+\mu^-)$ $/E_{cm}>0.9 E_{cm}$ (where radiative photons 
   reconstructed as for $q\bar q$ selection) and $M_{inv}(\mu^+\mu^-)>0.74$ 
   (excluding photons).
\end{enumerate}

\noindent{\bf DELPHI}\\
An event was required to have two identified muons in the polar angle range
$20^\circ\le\theta_{\mu}\le160^\circ$ and the highest muon momentum of at 
least $30~GeV/\it{c}$.
$M_{inv}(\mu^+\mu^-)$ was calculated from a kinematic fit, 
where four different topologies were investigated for each event:
i) no photon radiated, ii) one photon radiated along the beam line, iii) one
seen and one unseen photon in any direction, iv) a single unseen photon in
any direction. The seen photon fit was performed if a neutral energy deposit
greater than $5~GeV$ was measured in the electromagnetic calorimeters.
\begin{enumerate}\setcounter{enumi}{42} 
\item 
   {\bf Inclusive Selection:}~\labobs{Delphi4}~
$M_{inv}(\mu^+\mu^-)>75~GeV$ 
\item 
   {\bf Exclusive Selection:}\labobs{Delphi5}~
   $M_{inv}(\mu^+\mu^-)/E_{cm}>0.85$  
\end{enumerate}

\noindent{\bf L3}\\
An event must have two identified muons in the polar angle range
$20^\circ\le\theta_{\mu}\le160^\circ$ and the highest muon momentum should
exceed 35 GeV.
The effective centre-of-mass energy for each event is determined
assuming the emission of a single ISR photon. In case the photon is found in the
detector ($|\cos\theta_\gamma| < 0.985$) it is required to have an energy, 
$E_{\gamma}$, larger than 15~GeV in
the electromagnetic calorimeter and an angular separation to the nearest muon of
more than 10 degrees. Otherwise the photon is assumed to be emitted along the
beam axis and its energy is calculated from the polar angles of the outgoing
muons. 

\begin{enumerate}\setcounter{enumi}{44} 
\item 
   {\bf Inclusive Selection:}~\labobs{LT9}~$M_{ang}(\mu^+\mu^-)>75~GeV$ 
\item 
   {\bf Exclusive Selection:}~\labobs{LT10}~$M_{ang}(\mu^+\mu^-)/E_{cm}>0.85$  
\end{enumerate}

\noindent{\bf OPAL}\\
$N_{trk}\ge 2$. A pair of tracks is taken as muon pair candidate,
if both tracks are identified as muons, have 
$p > 6$~GeV, $|\cos\theta| < 0.95$ and are separated by 320~mrad in azimuthal angle $\phi$. 
If more than one pair of tracks satisfies the above conditions, the pair with the largest
scalar momentum sum is chosen. The event is rejected if more than one other track
has a transverse momentum greater than 0.7~GeV.
Finally, $E_{vis}$, defined as the scalar sum of the two muon momenta plus the
energy of the highest energy cluster in the electromagnetic calorimeter
($|\cos\theta| < 0.985$), has to be larger than
$0.35\sqrt{s} + 0.5M^2_z/\sqrt{s}$. For photon to be used in definition of 
$M_{ang}$ it has to be separated at least by $200$~mrad from charged muons.
\begin{enumerate}\setcounter{enumi}{46} 
\item 
{\bf Inclusive Selection:}~\labobs{Opal6}~$M_{ang}(\mu^+\mu^-)/E_{cm}>0.10$\\
        and $M_{inv}(\mu^+\mu^-) > 70$~GeV, if 
        $0.35\sqrt{s} < E_{vis}- 0.5M^2_z/\sqrt{s} < 0.75\sqrt{s}$.
\item 
{\bf Exclusive Selection:}~\labobs{Opal7}~$M_{ang}(\mu^+\mu^-)/E_{cm}>0.85$\\
        and $M_{inv}(\mu^+\mu^-) > \sqrt{M^2_Z+0.1s}$. 
\end{enumerate}

\subsection{ Idealized $e^+e^-\to \mu^+\mu^-(\gamma)$  observables }

\noindent{\bf ALEPH}
\begin{enumerate}\setcounter{enumi}{48} 
\item 
   {\bf Inclusive Selection:}~\labobs{IAleph5}~$M_{inv}(\mu^+\mu^-)/E_{cm}>0.1$ and 
   $|\cos\theta_{\mu^-}| < 0.95$. 
\item 
   {\bf Exclusive Selection:}~\labobs{IAleph6}~$M_{inv}(\mu^+\mu^-)/E_{cm}>0.9$ and 
   $|\cos\theta_{\mu^-}| < 0.95$. 
\end{enumerate}

\noindent{\bf DELPHI}
\begin{enumerate}\setcounter{enumi}{50} 
\item 
   {\bf Inclusive Selection:}~\labobs{IDelphi5}~$M_{inv}(\mu^+\mu^-)>75~GeV$ and 
   $|\cos\theta_{\mu^-}| < 0.95$.
\item 
   {\bf Exclusive Selection:}~\labobs{IDelphi6}~$M_{inv}(\mu^+\mu^-)/E_{cm}>0.85$ and 
   $|\cos\theta_{\mu^-}| < 0.95$.
\end{enumerate}

\noindent{\bf L3}
\begin{enumerate}\setcounter{enumi}{52} 
\item 
   {\bf Inclusive Selection:}~\labobs{ILT9}~$M_{inv}(\mu^+\mu^-)>75~GeV$ and 
   $|\cos\theta_{\mu^-}| < 0.90$.
\item 
   {\bf Exclusive Selection 1:}~\labobs{ILT10}~$M_{inv}(\mu^+\mu^-)/E_{cm}>0.85$ and 
   $|\cos\theta_{\mu^-}| < 0.90$.
\item 
   {\bf Exclusive Selection 2:}~\labobs{ILT11}~$M_{inv}(\mu^+\mu^-)/E_{cm}>0.85$ and 
   $|\cos\theta_{\mu^-}| < 1.0$.
\end{enumerate}

\noindent{\bf OPAL}
\begin{enumerate}\setcounter{enumi}{55} 
\item 
   {\bf Inclusive Selection:}~\labobs{IOpal6}~$M_{prop-}(\mu^+\mu^-)/\sqrt{s}>0.10$ and 
   $|\cos\theta_{\mu^-}| < 0.95$.
\item 
   {\bf Inclusive Selection:}~\labobs{IOpal7}~$M_{prop-}(\mu^+\mu^-)/\sqrt{s}>0.10$ and 
   $|\cos\theta_{\mu^-}| < 1.00$.
\item 
   {\bf Exclusive Selection:}~\labobs{IOpal8}~$M_{prop-}(\mu^+\mu^-)/\sqrt{s}>0.85$ and 
   $|\cos\theta_{\mu^-}| < 0.95$.
\item 
   {\bf Exclusive Selection:}~\labobs{IOpal9}~$M_{prop-}(\mu^+\mu^-)/\sqrt{s}>0.85$ and 
   $|\cos\theta_{\mu^-}| < 1.00$.
\end{enumerate}

\subsection{ Realistic $e^+e^-\to \tau^+\tau^-(\gamma)$ observables }

\noindent{\bf ALEPH}
\begin{enumerate}\setcounter{enumi}{59} 
\item 
   {\bf Inclusive Selection:}\labobs{Aleph7}~ 
   Two identified $\tau$ candidates with $M_{inv}(\tau^+\tau^-)>25$~GeV 
   and ${\mathrm acol}(\tau^+\tau^-) < 250$~mrad in plane perpendicular to beam 
   axis. Then $M_{ang}(\tau^+\tau^-)>0.1 E_{cm}$ (where radiative photons 
   are reconstructed as for $q\bar q$ selection).
\item 
   {\bf Exclusive Selection:}~\labobs{Aleph8}~ 
   Same as inclusive selection, then following extra cut
   $M_{ang}(\tau^+\tau^-)>0.9 E_{cm}$ (where radiative photons 
   reconstructed as for $q\bar q$ selection) and $|\cos\theta_{\tau^-}| < 0.95$.
\end{enumerate}

\noindent{\bf DELPHI}\\
The leading track in each hemisphere was required to lie in
the polar angle range $|\cos\theta|<0.94$, and
the observed charged particle multiplicity was requested to be unity in one
hemisphere and no more than five in the other. 
At least one of the leading tracks was required to have momentum
greater than~$0.025\times\sqrt{s}$. 
Full description of the cuts can be found in~\cite{DELPUB172}.
The $M_{ang}(\tau^+\tau^-)$ was calculated from fermion directions estimated by 
leading tracks. 
\begin{enumerate}\setcounter{enumi}{61} 
\item 
   {\bf Inclusive Selection:}~\labobs{Delphi6}~
   $M_{ang}(\tau^+\tau^-)>75~GeV$ 
\item 
   {\bf Exclusive Selection:}~\labobs{Delphi7}~
   $M_{ang}(\tau^+\tau^-)/E_{cm}>0.85$ 
\end{enumerate}

\noindent{\bf L3}\\
Tau leptons are identified as narrow, low multiplicity jets, containing from one to five
charged particles. Tau jets are formed by matching the energy depositions in the
electromagnetic and hadron calorimeters with tracks in the central tracker and
the muon spectrometer.
Two tau jets of at least 3.5~GeV are required to lie
within the polar angular range $|\cos{\theta}|<0.92$.
To reject background from two photon processes the most energetic jet must have an
energy larger than 20 GeV.
The reconstruction of the effective centre-of-mass energy follows the procedure
described in the muon subsection,  using the polar angles of the two tau jets.
Full description of the cuts can be found in~\cite{L3PUB172}.
\begin{enumerate}\setcounter{enumi}{63} 
\item 
   {\bf Inclusive Selection:}~\labobs{LT11}~$M_{ang}(\tau^+\tau^-)>75~GeV$ 
\item 
   {\bf Exclusive Selection:}~\labobs{LT12}~$M_{ang}(\tau^+\tau^-)/E_{cm}>0.85$ 
\end{enumerate}

\noindent{\bf OPAL}\\
Tau-pair candidates  are events with exactly two charged, low multiplicity
 cones with $35^{\circ}$  half-angle.
\begin{itemize}
\item The Event is rejected if it was selected as a $\mu$ pair.
\item $\mathrm{N_{trk}} < 7$, where $\mathrm{N_{trk}}$ is the number
  of tracks in the central detector.
\item $\mathrm{N_{trk} + N_{clus}} < 16$, where $\mathrm{N_{clus}}$ is
  the number of ECAL clusters.
\item $|\cos\theta_{\tau^+\tau^-}| < 0.90$, where $|\cos\theta_{\tau}|$ is the 
  cosine of the respective cone axis.
\item $0.02 < \mathrm{R_{shw}} < 0.7$, where $\mathrm{R_{shw}}$ is the summed
  ECAL energy scaled by the centre of mass energy.
\item $ \mathrm{R_{trk}} < 0.8$, where $\mathrm{R_{trk}}$ is the scalar sum
  of track momenta scaled by the centre of mass energy.
\item  $\mathrm{R_{shw}} > 0.2$ or $\mathrm{R_{trk}} > 0.2$.
\item  $\mathrm {R_{shw} + R_{trk}} < 1.05 (1.10)$, for 
  $|\cos\theta| > 0.7$ $(< 0.7)$,where $|\cos\theta|$ is the average
  of the two tau cones.
\item $\mathrm{|\cos\theta_{p^{ECAL}_{missing}}}| < 0.99$, where
  $\mathrm{\cos\theta_{p^{ECAL}_{missing}}}$ is the cosine of the
  direction of missing momentum calculated using the ECAL.
\item $\mathrm{p_t^{ECAL} > 0.015\sqrt{s}}$  where $p_t^{ECAL}$ 
    is the sum of the transverse ECAL energy. 
\item Events are rejected if $\mathrm{0.9 < E/p < 1.1}$ in both cones.
\item  Using the values of $\theta_{\tau}$, 
        the expected energy of each lepton is calculated assuming that
        the final state consists only of two leptons plus a single
        unobserved photon along the beam direction. We then require
        that $0.02 < \sqrt{(X_{E1}^2 + X_{E2}^2)} < 0.8$,
        and $ \sqrt{(X_{P1}^2 + X_{P2}^2)} < 0.8$,
        where $X_{E1,E2}$ are the total electromagnetic calorimeter
        energies in each tau cone normalized to the expected value
        calculated above, and $X_{P1,P2}$ are the scalar sums of track
        momenta in the two tau cones, also normalized to the
        expected values. 
\item $\mathrm{\theta_{acollinearity} < 180^{\circ}
  -\mathrm{2tan^{-1}\left(\frac{2M_{Z}\sqrt{s}}{s - M_{Z}^{2}} \right) 
     + 10^{\circ}}}$.
This cut is placed $\mathrm{10^{\circ}}$ degrees above the expected
radiative return peak.
\item $\mathrm{\theta_{acoplanarity} < 30^{\circ}}$.
\end{itemize}
\begin{enumerate}\setcounter{enumi}{65} 
\item 
   {\bf Inclusive Selection:}~\labobs{Opal8}~$M_{ang}(\tau^+\tau^-)/\sqrt{s}>0.10$.
\item 
   {\bf Exclusive Selection:}~\labobs{Opal9}~$M_{ang}(\tau^+\tau^-)/\sqrt{s}>0.85$.
\end{enumerate}

\subsection{ Idealized $e^+e^-\to \tau^+\tau^-(\gamma)$ observables }

\noindent{\bf ALEPH}
\begin{enumerate}\setcounter{enumi}{67} 
\item 
   {\bf Inclusive Selection:}~\labobs{IAleph7}~$M_{inv}(\tau^+\tau^-)/E_{cm}>0.1$ and 
   $|\cos\theta_{\tau^-}| < 0.95$. 
\item 
   {\bf Exclusive Selection:}~\labobs{IAleph8}~$M_{inv}(\tau^+\tau^-)/E_{cm}>0.9$ and 
   $|\cos\theta_{\tau^-}| < 0.95$. 
\end{enumerate}

\noindent{\bf DELPHI}
\begin{enumerate}\setcounter{enumi}{69} 
\item 
   {\bf Inclusive Selection:}~\labobs{IDelphi7}~$M_{inv}(\tau^+\tau^-)>75~GeV$ and 
   $|\cos\theta_{\tau^-}| < 0.95$.
\item 
   {\bf Exclusive Selection:}\labobs{IDelphi8}~ $M_{inv}(\tau^+\tau^-)/E_{cm}>0.85$ and 
   $|\cos\theta_{\tau^-}| < 0.95$.
\end{enumerate}

\noindent{\bf L3}
\begin{enumerate}\setcounter{enumi}{71} 
   \item 
   {\bf Inclusive Selection:}~\labobs{ILT12}~$M_{inv}(\tau^+\tau^-)>75~GeV$ and 
   $|\cos\theta_{\tau^-}| < 0.92$.
\item 
   {\bf Exclusive Selection 1:}~\labobs{ILT13}~$M_{inv}(\tau^+\tau^-)/E_{cm}>0.85$ and 
   $|\cos\theta_{\tau^-}| < 0.92$.
\item 
   {\bf Exclusive Selection 2:}~\labobs{ILT14}~$M_{inv}(\tau^+\tau^-)/E_{cm}>0.85$ and 
   $|\cos\theta_{\tau^-}| < 1.0$.
\end{enumerate}

\noindent{\bf OPAL}
\begin{enumerate}\setcounter{enumi}{74} 
\item 
   {\bf Inclusive Selection 1:}~\labobs{IOpal10}~$M_{prop-}(\tau^+\tau^-)/\sqrt{s}>0.10$ and 
   $|\cos\theta_{\tau^-}| < 0.90$.
\item 
   {\bf Inclusive Selection 2:}~\labobs{IOpal11}~$M_{prop-}(\tau^+\tau^-)/\sqrt{s}>0.10$ and 
   $|\cos\theta_{\tau^-}| < 1.00$.
\item 
   {\bf Exclusive Selection:}~\labobs{IOpal12}~$M_{prop-}(\tau^+\tau^-)/\sqrt{s}>0.85$ and 
   $|\cos\theta_{\tau^-}| < 0.90$.
\item 
   {\bf Exclusive Selection:}~\labobs{IOpal13}~$M_{prop-}(\tau^+\tau^-)/\sqrt{s}>0.85$ and 
   $|\cos\theta_{\tau^-}| < 1.00$.
\end{enumerate}

\subsection{\qqbar $\gamma$'s observables} 

This section contains precision requirements for regions
of phase space selected in searches for anomalous neutral gauge
couplings (ZZ$\gamma$,Z$\gamma\gamma$), especially for the 
visible $\gamma+Z\to\gamma+\qqbar$ final state.
The prominent Standard Model process populating this region is
the radiative return to the Z. Numerical results are not provided, we expect
uncertainties for initial state bremsstrahlung contributions to be similar to the
one in case of leptons, however the additional, often dominant uncertainties
due to interplay of the final state QCD interaction and photon emission, 
are not discussed in our section.

\noindent{\bf L3}  
\begin{enumerate}\setcounter{enumi}{78} 
\item 
   {\bf Total Cross-Section}~\labobs{LT13}
   Hadronic events are selected by asking for more than 6 charged
   tracks and more than 11 calorimetric clusters. The transverse
   and longitudinal energy imbalances should be below 15 and
   20 \% respectively.
   A photon with $14^\circ < \theta_\gamma < 166^\circ$
   and energy $80~GeV < \sqrt{s-2 E_{\gamma} \sqrt{s}} < 110~GeV$
   is required.
   The precision tag is 0.3-0.4 \%.
\end{enumerate}

\noindent{\bf OPAL} 
\begin{enumerate}\setcounter{enumi}{79} 
\item 
   {\bf Total Cross-Section}~\labobs{Opal10}
   Events are selected with two jets and one visible photon, fulfilling
   a 3C kinematic fit, allowing an additional photon of at most
   5\% of the beam energy to be radiated along the beam pipe.
   The relevant phase space is defined by

   \begin{enumerate}
      \item $E_\gamma >$ 50~GeV
      \item $15^\circ < \theta_\gamma < 165^\circ$
      \item angle between photon and closest jet, $\alpha_{\gamma-jet}>30^\circ$.
   \end{enumerate}

\vskip 5pt
\noindent
   The required precision tag for the radiative return cross-section
   in this phase space for a LEP-combined analysis
   is 0.3\%. It assumes uncorrelated experimental systematics and
   is dominated by the experimental statistics.
   For fully correlated experimental systematics the precision tag is 0.4\%.
   Currently, the predictions of
   the \KKMC\ and PYTHIA generators in this phase space differ by 15\%.
\item 
   {\bf Differential Cross-Section}~\labobs{Opal11}
   Selected Events in the above phase space are subjected to a 
   maximum likelihood fit in the differential distributions of
   $E_{\gamma}, \cos\theta_\gamma$, and $\cos\theta^\star_{q}$,
   which is the angle of the final state quark-jets in the Z rest frame.
   Precision tag:
       The ratio between the rate of events within $|\cos\theta_\gamma|<0.7$
       and the boundary of the acceptance has to be known 
       within about 1\%. The discrepancy between \KKMC\ and PYTHIA in
       the region $|\cos\theta_\gamma|<0.7$
       is currently at level of 30\% (stat. significance 2$\sigma$).
\end{enumerate}


\subsection{\llbar$ \gamma$'s observables} 
This final state is primarily used to search for single (one photon) or pair (two photons)
production of excited leptons. In principle, the radiative return peak in $M_{\ell\ell}$
can be rejected in these searches, which is not possible in the anomalous
neutral coupling analyses (see above). In practice, however, experimental resolution
and further (final state) radiation reduce the power of such an anticut.
Therefore both topologies (including and omitting the anticut) need to be studied.

\noindent{\bf ALEPH}\\ 
Photons are required to have at least 15~GeV and $|\cos\theta_{\gamma}|<$0.95.
Leptons are also required to have $|\cos\theta_{l}|<$0.95, 
they are identified from oppositely charged 
jet clustered from charged tracks (at most 3) with at least one jet above 5~GeV. 
The angle between the 
2 jets being at least 30 degree and the photon candidate at least 10 degree 
from the charged jet.  
\begin{itemize}
\item 
         {\bf \llbar\ plus one photon} \\
         Two leptons of same type are required accompanied of at least one photon.
         The two most energetic leptons and the most energetic photon are considered.

\vskip 2pt

         \begin{enumerate}\setcounter{enumi}{81}
             \item{\bf       ee$\gamma$ }~\labobs{ALEPH-11}
               Two tracks are identified as electron and the sum of 
               charged energy greater than 0.6*$E_{cm}$.
               Expected precision tag 1.4\%
             \item{\bf   $\mu\mu\gamma$ }~\labobs{ALEPH-12}
               Two muon have to be identified and 
               charged energy greater than $0.6*E_{cm}$. 
               Precision tag 1.5\%
             \item{\bf $\tau\tau\gamma$ }~\labobs{ALEPH-13}
               At least a 3 prong tau jet candidate in the event is 
               required. Precision tag 1.7\%
            \end{enumerate}
\item 
            {\bf \llbar\ plus two photons} \\
            Two electron (or muon or tau)
            have to be identified and 2 photon jet identified 
            Isolation cut on the 4 angles between photon and lepton 
            required to be at least 20 degree . The 2 reconstructed 
            lepton-gamma masses have to agree within 5 GeV/c**2
            \begin{enumerate}\setcounter{enumi}{84} 
              \item{\bf        ee$\gamma\gamma$ }~\labobs{ALEPH-14}
                 Precision tag 3 \%.
              \item{\bf    $\mu\mu\gamma\gamma$ }~\labobs{ALEPH-15}
                 Precision tag 5 \%.
              \item{\bf  $\tau\tau\gamma\gamma$ }~\labobs{ALEPH-16}
                 Precision tag 8 \%.
            \end{enumerate}
\end{itemize}

\noindent{\bf DELPHI}\\ 
Events with a visible energy above $0.2\sqrt{s}$ in the region
$|cos\theta|<0.9397$ and not more than 6 charged tracks are selected.
Photons are required to have an energy above 5 GeV and
$|cos\theta|<0.9848$.
The total charged (neutral) energy in a $15^\circ$ cone around the photon
is required to be below 1 GeV (2 GeV). 
Charged particles are clustered into jets. Jets are required to
have $|cos\theta|<0.9063$ and the most energetic to have an energy
above 5 GeV.  
In three- or four-body topologies the energies are rescaled by imposing 
energy momentum conservation and using the measured angles. An improved 
energy resolution is obtained. Compatibility between the measured and rescaled 
energies is required.
\begin{itemize}
\item {\bf \llbar\ plus one photon}
   The energy of the photon is required to be above 10 GeV. 
   Events with one photon and one or two jets in the final state
   are considered. This accounts for situations in which one of the
   leptons was lost in the beam pipe or has very low momentum
   (close to the kinematic limit).
   The expected backgrounds and thus the further selections and the
   efficiencies are very dependent on the lepton flavour.

   \begin{enumerate}\setcounter{enumi}{87}
   \item{\bf ee$\gamma$ }~\labobs{DELPHI8}
       At least one jet identified as an electron and the other not
       identified as a muon, the most energetic
       jet with E$>10$ GeV. One photon with $|cos\theta|<0.7660$. 
       If only one jet is found the angle between the jet and the photon
       must be in the between $100^\circ$ and $179^\circ$. 
       Precision tag: 1.4\%
   \item{\bf   $\mu\mu\gamma$ }~\labobs{DELPHI9}
       At least one jet identified as a muon and no jets identified as 
       electrons, the most energetic jet with E$>$10 GeV.                    
       Precision tag: 1.6\%
   \item{\bf $\tau\tau\gamma$ }~\labobs{DELPHI10}
       The most energetic jet with E$>$10 GeV. One photon with $|cos\theta|<0.94$. 
       Tau events are selected by requiring a     difference between the measured 
       and the rescaled energy of the jets, expected due to the presence of 
       neutrinos. If only one jet is found the angle between the jet and the 
       photon must be between $100^\circ$ and $179^\circ$.
       Precision tag: 1.8\%
   \end{enumerate}
\item{\bf \llbar\ plus two photons}
   Two jets and two photons in the event. Compatibility between 
   two reconstructed $\ell\gamma$ masses is required. The relevant
   quantity is the minimum of the electron-photon invariant mass
   differences ($\Delta m _{\ell\gamma}$).
   The statistical error of the selection clearly dominates. 

   \begin{enumerate}\setcounter{enumi}{90} 
   \item{\bf ee$\gamma\gamma$ }~\labobs{DELPHI11}
       $\Delta m _{e\gamma} <$ 15 GeV/c$^2$.     
       Precision tag: 5\%  
   \item{\bf $\mu\mu\gamma\gamma$ }~\labobs{DELPHI12}
       $\Delta m _{\mu\gamma} <$ 10 GeV/c$^2$
       Precision tag: 6\%
   \item{\bf $\tau\tau\gamma\gamma$ }~\labobs{DELPHI13}
       $\Delta m _{\tau\gamma} <$ 10 GeV/c$^2$
       Precision tag: 6\%
   \end{enumerate}
\end{itemize}

\noindent{\bf L3}\\ 
Hadronic events are rejected by requiring less than 8 charged tracks.
Photons are required to have at least 15~GeV and $|\cos\theta_{\gamma}|<$0.97.
Leptons are required to have the same flavor and $|\cos\theta_{l}|<$0.94.
They are identified from oppositely charged tracks/low multiplicity jets(for tau).

\begin{itemize}
\item 
     {\bf \llbar\ plus one photon} \\
     At least one lepton accompanied by at most one hard photon.
     The second lepton may be in the beam pipe.
     The photon is required to have at least 20~GeV and $|\cos\theta_{\gamma}|<$0.75.
     One mass (l$\gamma$) should be greater than 70 GeV.

            \begin{enumerate}\setcounter{enumi}{93}
             \item{\bf       ee$\gamma$ }~\labobs{LT14}
               Expected precision tag 1.2\%
             \item{\bf   $\mu\mu\gamma$ }~\labobs{LT15}
               Precision tag 1.5\%
             \item{\bf $\tau\tau\gamma$ }~\labobs{LT16}
               Precision tag 1.8\%
            \end{enumerate}
\item 
            {\bf \llbar\ plus two photons} \\
            Two electron (or muon or tau) and 2 photons seen.
            The photons are required to have at least 15~GeV and
            one should be in $|\cos\theta_{\gamma}|<$0.75,
            the second in $|\cos\theta_{\gamma}|<$0.94.
            The difference of the 2 reconstructed lepton-gamma masses should be 
            below 10 GeV and their sum above 100 GeV.

            \begin{enumerate}\setcounter{enumi}{96} 
            \item{\bf        ee$\gamma\gamma$ }~\labobs{LT17}
                Precision tag 4 \%.
            \item{\bf    $\mu\mu\gamma\gamma$ }~\labobs{LT18}
                Precision tag 6 \%.
            \item{\bf  $\tau\tau\gamma\gamma$ }~\labobs{LT19}
                Precision tag 6 \%.
            \end{enumerate}
\end{itemize}

\noindent{\bf OPAL}\\ 
Photon candidates are required to have an energy exceeding 5\% of the beam energy,
and have to lie within $|\cos\theta_\gamma| < 0.95$.  Leptons are also identified 
within $|\cos\theta_\ell| < 0.95$ and are required to have $p_t > 1$~GeV. 
Photons, electrons, and muons are required to be isolated by at least 20 degree 
from the nearest  charged track of momentum larger than 1~GeV.
\begin{itemize}
\item 
            {\bf \llbar\ plus one photon} \\
            There must be at least two identified  leptons of the same type and at least one photon. 
            The two most energetic leptons and the most energetic
            photon are used in the analysis. Their energy sum is called $E_{vis}$ in the following.
            The precision tag assumes uncorrelated experimental 
            systematics. Both, uncorrelated systematics, and 
            expected statistical precision of the LEP combined result
             contribute to about equal parts to the precision tag.
            For correlated experimental  systematics the required
            theoretical precision is loosened by about 0.9\%.

            \begin{enumerate}\setcounter{enumi}{99} 
            \item{\bf ee$\gamma$ w/o rad return rejection}~\labobs{Opal12}
                          \begin{itemize}
                          \item $E_{vis}>1.6E_{\rm beam}$
                          \item $|\cos\theta_\gamma| < 0.70$
                          \item $|\cos\theta_e| < 0.70$ for at least one electron.  
                          \end{itemize}
                         Precision tag: 1.3\%
            \item{\bf ee$\gamma$ with rad return rejection}~\labobs{Opal13}
                          \begin{itemize}
                          \item $E_{vis}>1.6E_{\rm beam}$
                          \item $|\cos\theta_\gamma| < 0.70$
                          \item $|\cos\theta_e| < 0.70$ for at least one electron.
                          \item reject events with 85~GeV $< M_{ee} <$ 95~GeV  
                          \end{itemize}
                          Precision tag: 1.4\%
            \item{\bf $\mu\mu\gamma$ w/o rad return rejection}~\labobs{Opal14}
                          \begin{itemize}
                          \item $E_{vis}>1.6E_{\rm beam}$
                          \end{itemize}
                          Precision tag: 1.5\%
            \item{\bf $\mu\mu\gamma$ with rad return rejection}~\labobs{Opal15}
                          \begin{itemize}
                          \item $E_{vis}>1.6E_{\rm beam}$
                          \item reject events with 85~GeV $< M_{\mu\mu} <$ 95~GeV  
                          \end{itemize}
                          Precision tag: 1.7\%
            \item{\bf $\tau\tau\gamma$ w/o rad return rejection}~\labobs{Opal16}
                          \begin{itemize}
                          \item $0.8E_{\rm beam} < E_{vis}>1.9E_{\rm beam}$
                          \end{itemize}
                          Precision tag: 1.4\%
            \item{\bf $\tau\tau\gamma$ with rad return rejection}~\labobs{Opal17}
                          \begin{itemize}
                          \item $ 0.8E_{\rm beam} < E_{vis} < 1.9E_{\rm beam}$
                          \item reject events with 85~GeV $< M_{\tau\tau} <$ 95~GeV  
                          \end{itemize}
                          Precision tag: 1.5\%
            \end{enumerate}
\item 
             {\bf \llbar plus two photons} \\
             There must be at least two identified leptons of the same type and at least two photons.  
             The two most energetic leptons and the two most energetic
             photons are used in the analysis. Their energy sum is called $E_{vis}$ in the following.
             The precision tag is dominated by the statistical error
             of the selection.

             \begin{enumerate}\setcounter{enumi}{105} 
             \item{\bf ee$\gamma\gamma$ w/o rad return rejection}~\labobs{Opal18}

                          \begin{itemize}
                          \item $E_{vis}>1.6E_{\rm beam}$
                          \item minimum opening angle among all electron-photon combinations
                                   $|\cos\alpha^{e\gamma}_{\rm min}| < 0.90$.
                          \end{itemize}

                         Precision tag: 3\%
             \item{\bf ee$\gamma\gamma$ with rad return rejection}~\labobs{Opal19}

                          \begin{itemize}
                          \item $E_{vis}>1.6E_{\rm beam}$
                          \item minimum opening angle among all electron-photon combinations
                                   $|\cos\alpha^{e\gamma}_{\rm min}| < 0.90$.
                          \item reject events with 85~GeV $< M_{ee} <$ 95~GeV  
                          \end{itemize}

                          Precision tag: 3\%

             \item{\bf $\mu\mu\gamma\gamma$ w/o rad return rejection}~\labobs{Opal20}

                          \begin{itemize}
                          \item $E_{vis}>1.6E_{\rm beam}$
                          \end{itemize}
                          Precision tag: 5\%

             \item{\bf $\mu\mu\gamma\gamma$ with rad return rejection}~\labobs{Opal21}

                          \begin{itemize}
                          \item $E_{vis}>1.6E_{\rm beam}$
                          \item reject events with 85~GeV $< M_{\mu\mu} <$ 95~GeV  
                          \end{itemize}
                          Precision tag: 6\%

             \item{\bf $\tau\tau\gamma\gamma$ w/o rad return rejection}~\labobs{Opal22}

                          \begin{itemize}
                          \item $0.8E_{\rm beam} < E_{vis}>1.9E_{\rm beam}$
                          \end{itemize}
                          Precision tag: 5\%

             \item{\bf $\tau\tau\gamma\gamma$ with rad return rejection}~\labobs{Opal23}

                          \begin{itemize}
                          \item $ 0.8E_{\rm beam} < E_{vis} < 1.9E_{\rm beam}$
                          \item reject events with 85~GeV $< M_{\tau\tau} <$ 95~GeV  
                          \end{itemize}
                          Precision tag: 6\%

             \end{enumerate}
\end{itemize}

\subsection{$\nu \bar \nu \gamma$'s observables} 
The observables listed below do not include any cut to eliminate the regions of phase 
space
dominated by the radiative return to the $Z$. Such events are usually selected by a cut 
on the mass of the invisible system. 
For such an extended observable the required precision tag should simply be 
scaled according to the somewhat decreased statistics. 
In general, {\em fully exclusive}
predictions with help of the full multi-dimensional distributions of photon(s)
energies and directions
are needed with a precision comparable to that of the total cross section.

\noindent{\bf ALEPH}

\begin{enumerate}\setcounter{enumi}{111} 
\item~\labobs{Nu1}~
   Single photon and missing energy:
   Exactly one photon with $|\cos\theta_\gamma| < 0.95$ and $P_t > 0.0375 *  E_{cm}$.
   No other photon with $|\cos\theta_\gamma| < 0.9997$ and $E_\gamma > 1$~GeV.
\item~\labobs{Nu2}~
   Two or more photons and missing energy:
   Two or more photons, each  with
   $E_\gamma > 1$~GeV and $|\cos\theta_\gamma| < 0.95$.
   The transverse momentum of the multi-photon system must be such that
   $\Sigma P_t > 0.0375 * (E_{cm}-\Sigma E)$. 
   No other photon with $|\cos\theta_\gamma|$ between 0.95 and 0.9997, and $E_\gamma > 1$~GeV.
\end{enumerate}

\noindent{\bf DELPHI}

\begin{enumerate}\setcounter{enumi}{113} 
\item
   Single photon and missing energy:
   Require exactly one observed photon with energy requirement varying with polar angle

   \begin{enumerate}
   \item   ~\labobs{Nu11}~Very forward (or backward): 
              $3.8^o<\theta<6.5^o$, $x_\gamma>0.3$ 
              or $x_\gamma>(9.2-\theta)/9$ 

   \item   ~\labobs{Nu12}~Forward (or backward):   $12^o<\theta<32^o$, $x_\gamma>0.2$

   \item   ~\labobs{Nu13}~Barrel:     $45^o<\theta<90^o$, $x_\gamma>0.06$
   \end{enumerate}
   No additional photons with $E_\gamma>0.8$~GeV and $\theta>38$ mrad unless 
   they are within $3^o$, $15^o$ and $20^o$
   from the primary photon for the very forward, forward and barrel regions, respectively.
\item
   Two or more photons and missing energy\labobs{Nu14}~:
   Two or more photons with $E_\gamma > 0.05*E_{beam}$ and $|\cos\theta_\gamma| < 0.985$,
   at least one of which has $|\cos\theta_\gamma| < 0.906$.
   No additional photon with $E_\gamma > 0.02*E_{cm}$ and $|\cos\theta_\gamma|$
   between 0.985 and 0.9994.
\end{enumerate}

\noindent{\bf  L3}

\begin{enumerate}\setcounter{enumi}{115} 
\item 
   Single photon and missing energy:
   Exactly one photon with $|\cos\theta_\gamma| < 0.97$ and:

   \begin{enumerate}
   \item~\labobs{Nu3}~energy and $P_t > 5 GeV$ if $14^o<\theta$ /endcaps/
   \item~\labobs{Nu4g}~energy and $P_t > 5 GeV$ if $43^o<\theta$ /barrel/
   \end{enumerate}
   No other photon with $|\cos\theta_\gamma| < 0.9997$ and $E_\gamma > 10$~GeV.
\item 
   Two or more photons and missing energy:
    each photon with $|\cos\theta_\gamma| < 0.97$ and:

   \begin{enumerate}
   \item~\labobs{Nu5}~energy and $P_t > 5 GeV$ if $14^o<\theta$ (endcaps)
   \item~\labobs{Nu6}~energy and $P_t > 1 GeV$ if $43^o<\theta$ (barrel)
   \end{enumerate}
   The transverse momentum of the multi-photon system must be greater than
   5 GeV and the collinearity greater than 2.5 deg.
   No other photon with $|\cos\theta_\gamma| < 0.9997$ and $E_\gamma > 10$~GeV.
\end{enumerate}

\noindent{\bf  OPAL}

\begin{enumerate}\setcounter{enumi}{117} 
\item
   Single photon and missing energy\labobs{Nu7}~:
   The maximum energy photon must have $|\cos\theta_\gamma| < 0.9660$ and 
   $P_t > 0.05 * E_{beam}$.
   There may be at most ONE additional photon with 
   $E_\gamma > 0.3$~GeV and $|\cos\theta_\gamma| < 0.9848$.
\item
   Two or more photons and missing energy:
   Two or more photons each with 

   \begin{enumerate}
   \item
      \labobs{Nu8}~$E_\gamma > 0.05*E_{beam}$ and $|\cos\theta_\gamma| < 0.9660$,
   \item
      \labobs{Nu9}~$E_\gamma > 1.75$~GeV, $|\cos\theta_\gamma| < 0.8$
      and the sum of their $P_t > 0.05 *E_{beam}$
   \item
      \labobs{Nu10}~$E_\gamma > 1.75$~GeV, $|\cos\theta_\gamma| < 0.966$
      and the sum of their $P_t > 0.05 *E_{beam}$
   \end{enumerate}
\end{enumerate}

%% file: 2f-Chapt-Part1a.tex
\section{Discussion of numerical results for idealized and realistic observables}
\label{theory_err}

In this section we compare numerical results from all MC and semi-analytical
programs available in in our working group.

The aim of this exercise is two-fold:
\begin{itemize}
\item
   to check if these codes give meaningful predictions.
   Usually they do but it can be a nontrivial test,
   we also check in this way the applicability range of the codes.
\item
   to probe the error specifications declared by the authors of the codes
   in the section~\ref{sec:programs}
\end{itemize}

\noindent
In order to make the comparison meaningful we defined a common input.
In general we took physical parameters from the latest
publication of world averages in the PDG.
For the Higggs mass we took $M_H=120$ GeV.

For the QCD coupling the average $\alpha_S(M_Z)$ was used as an input.
The value of 
$\Delta\alpha^{(5)}_{had}(M_Z)=0.027782 \pm 0.000254$ is not present in PDG,
and this value was taken from ref.~\cite{Jegerlehner:1999hg}.
It contains  updates of the analysis  
of ref.~\cite{Eidelman:1995ny}.
For additional discussion on this subject see subsection~\ref{LowHad}.

\subsection{Hadronic low energy vacuum polarisation\label{LowHad}}
\begin{figure}[!ht]
\centering
\setlength{\unitlength}{0.1mm}
\begin{picture}(1600,800)
\put(275,   0){\makebox(0,0)[lb]{\epsfig{file=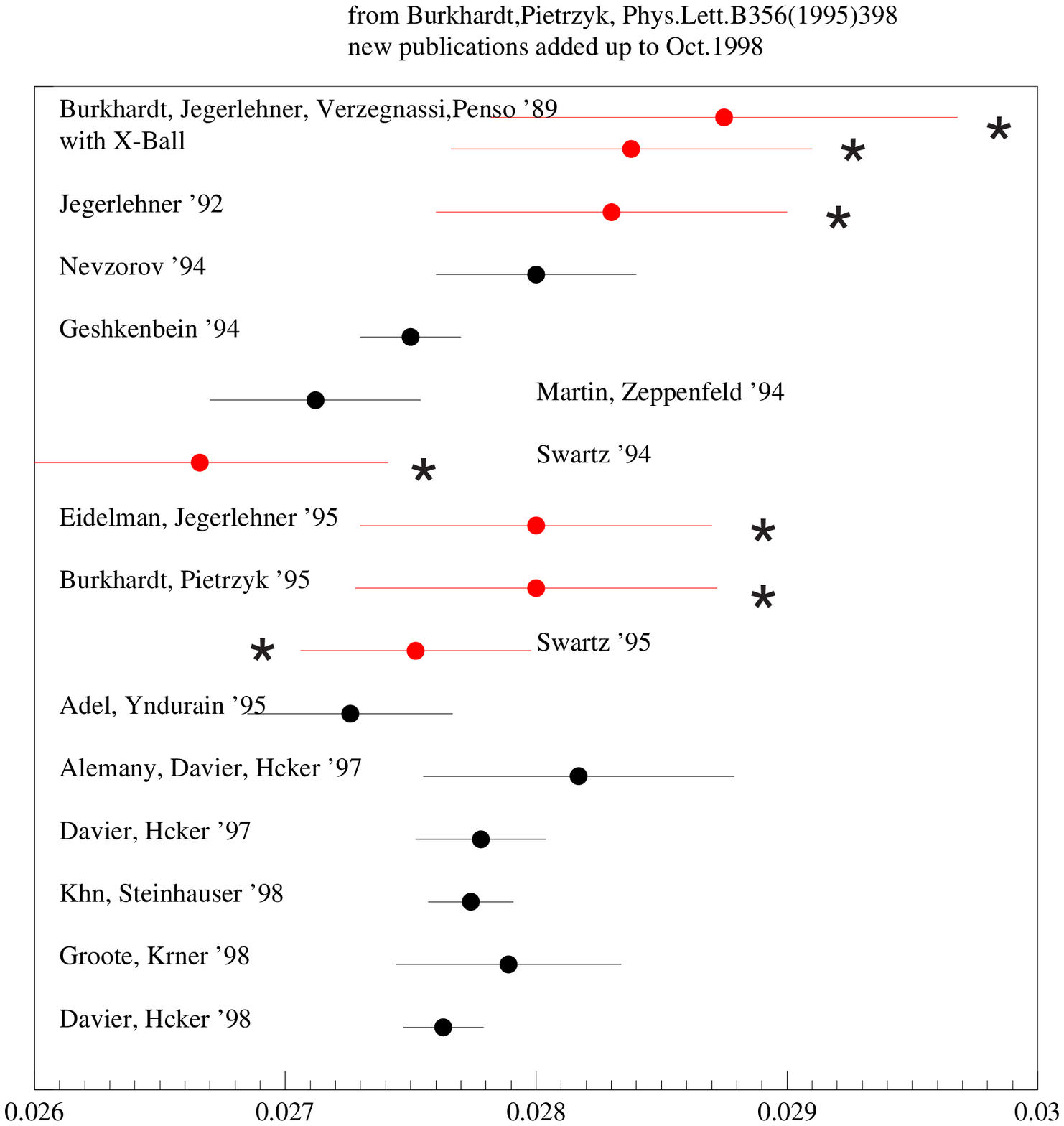,width=100mm,height=100mm}}}
\end{picture}
\caption{\small\sf Results of calculations of the hadronic contribution to 
the QED vacuum polarization. The results
marked with * are obtained using a dispersion integral of $R_{had}$
measured  experimentally. }
\label{alpha_figure}
\end{figure}

Vacuum polarization makes about 6\% correction to the value of $\alpha_{QED}$
at s = m$^2_Z$. The leptonic part of this contribution is
known with excellent precision \cite{Steinhauser:1998rq}. 
The quark part, however, is more difficult since
the quark masses are not unambiguously defined and there is no general 
agreement that the pertubartive QCD can be use for reliable calculations. 
Several reevaluations of the hadronic contribution to the QED
vacuum polarization have been performed to determine the effective QED
coupling $\alpha(M^2_Z)$ \cite{Burkhardt:1989ky, Eidelman:1995ny, VAP-HelmutBolek:1995,
VAP-Nevzorov:1994, VAP-Geshkenbein:1994, VAP-Zeppenfeld:1994,
VAP-Swartz:1995, VAP-AbYn:1995, VAP-Alemany:1998, VAP-Davier:1997, 
VAP-Kuhn:1998, VAP-Groote:1998, VAP-Davier:1998}. 
They are compared on Fig. \ref{alpha_figure}. 
The results
marked with * are obtained using a dispersion integral of $R_{had}$
$$R_{had} = \frac{\sigma(e^+e^- \rightarrow hadrons)}
                 {\sigma(e^+e^- \rightarrow \mu^+\mu^-)} $$
measured  experimentally. The uncertainties are dominated by the precision of 
measurement of $\sigma(e^+e^- \rightarrow hadrons)$ at the $e^+e^-$ 
centre-of-mass energies below 5 GeV. The values not marked with * are obtained
by using perturbative QCD in this region.

The values of hadronic contributions presented in Fig. \ref{alpha_figure} 
are in good 
agreement. The precision of calculations using 
a dispersion integral of $R_{had}$ measured  experimentally
is sufficient
for the precision required at LEP 2. The values of
the effective QED coupling $\alpha(s)$ 
at different center-of-mass are given in Table \ref{alpha_table}.

\begin{table}[h]
\begin{center}
\begin{tabular}{|c|c|}    \hline\hline
$\sqrt{s}$ (GeV) & $\alpha^{-1}$(s) \\ \hline \hline
80.364 & 129.08 $\pm$ 0.09   \\
91.1871 & 128.89 $\pm$ 0.09 \\
161 & 128.07 $\pm$ 0.09 \\
172 & 127.98 $\pm$ 0.09 \\
183 & 127.89 $\pm$ 0.09 \\
186 & 127.86 $\pm$ 0.09 \\
192 & 127.82 $\pm$ 0.09 \\
196 & 127.79 $\pm$ 0.09 \\
200 & 127.76 $\pm$ 0.09 \\
205 & 127.72 $\pm$ 0.09 \\
\hline \hline
\end{tabular}
\caption[]{\small $\alpha^{-1}$(s) using results of calculations of Ref.
\cite{VAP-HelmutBolek:1995}.}
\label{alpha_table}
\end{center}
\end{table}        

\begin{table}[!h]
\begin{center}
\epsfig{file=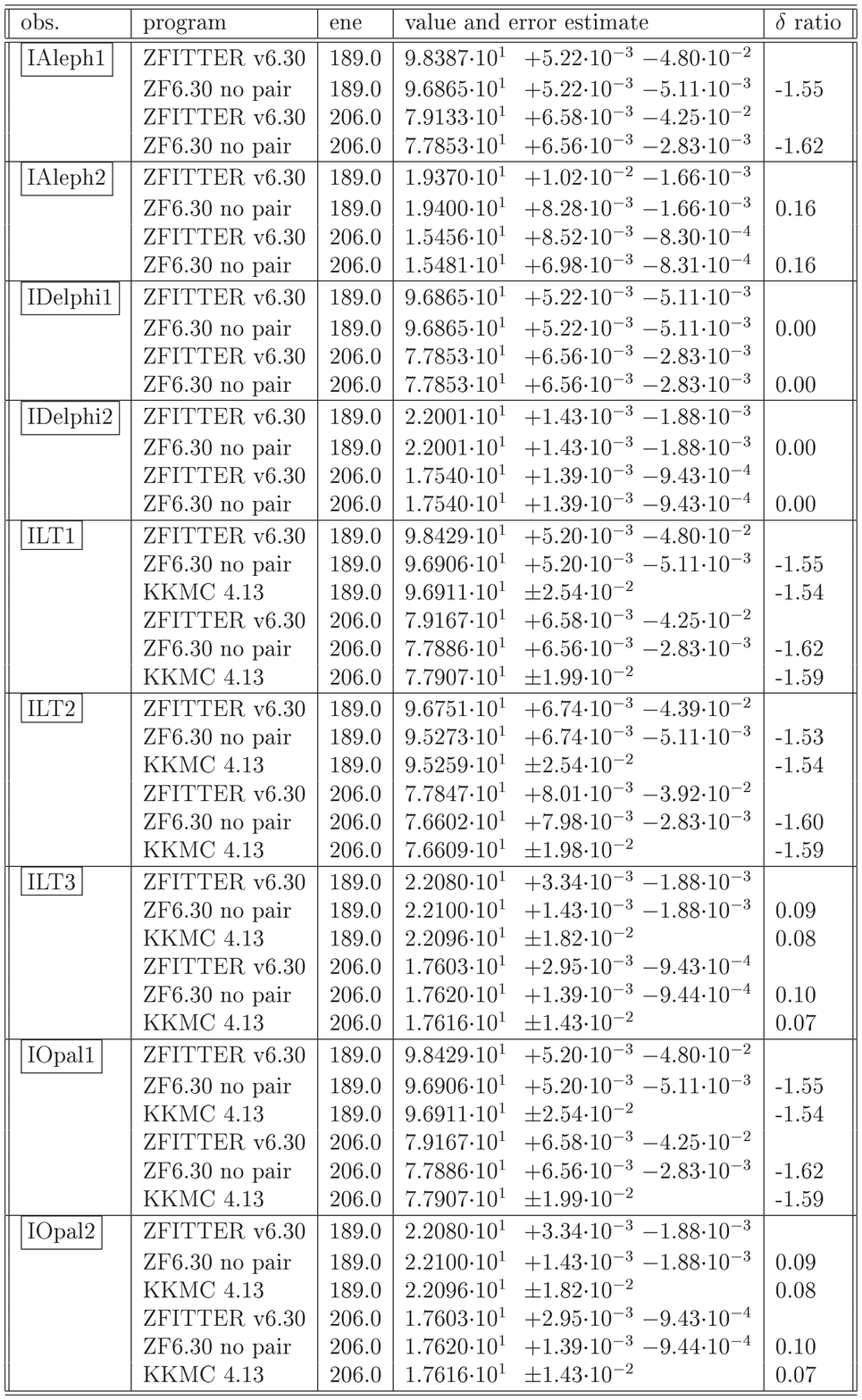,width=10cm,height=15cm,angle=0}
\caption{
Numerical predictions from theoretical calculations of the following            
idealized observables  in $q \bar q (\gamma )$ final states;  {\em cross sections}.   
Last field of the table shows relative
deviation (multiplied by 100) with respect to the first calculation.        
}
\label{tab:qqi4567890}
\end{center}
\end{table}

\subsection{Discussion of numerical results for $q \bar q (\gamma )$ observables}

There is good (0.1 \% or better) agreement between \KKMC\ and \zf\  
with pair corrections switched off, in all
available entries in table \ref{tab:qqi4567890}. For the observables where predictions from \KKMC\ are
missing it is due to experimental treatment of the interference correction,
which requires more runs of the Monte Carlo for every entry.
The level of agreement is consistent with the predictions of the systematic 
theoretical uncertainties from the program authors.
We can thus conclude that the systematic error from the QED/electroweak sector
for these processes is indeed smaller than  0.2 \%. If pair effects
and their uncetrainty is taken into account as discussed in section
\ref{pairconclusions} the total uncertainty from the QED/electroweak sector is 0.26 \%.

Some improvements in  \KKMC\ and \zf\  with respective
to published versions to obtain that level of agreements
were needed. Especially the question of choice of input parameters had 
to be revisited. See the sections \ref{KK-MC} and \ref{ZFITTER}
describing the  programs and section on comparisons \ref{TunedZFKK}
for details.
We can conclude that in all cases the overall QED uncertainty in quark channels
are below the experimental precision tag. The effect due to pairs
can not be neglected but even if it is included in a rather approximate way, this would be enough.
We do not expect the appropriate uncertainty to be sizable enough 
to affect the experimental studies in any case.

Finally let us point that we were not addressing any questions related to 
uncertainties of final state QCD interactions, see report of the
QCD working group in this report~\cite{Ballestrero:2000ur}. 
The QCD FSR corrections are implemented in \KKMC\ and \zf\  as an overall
$K$-factor taking into account all available higher orders, 
however, they do not take into account any kinematic cuts on real gluons.
That is why their predictions for realistic observables are of the partial use only.

\subsection{Discussion of numerical results for observables in $e^+ e^- (\gamma )$ final states}

\begin{table}[!h]
\begin{center}
\epsfig{file=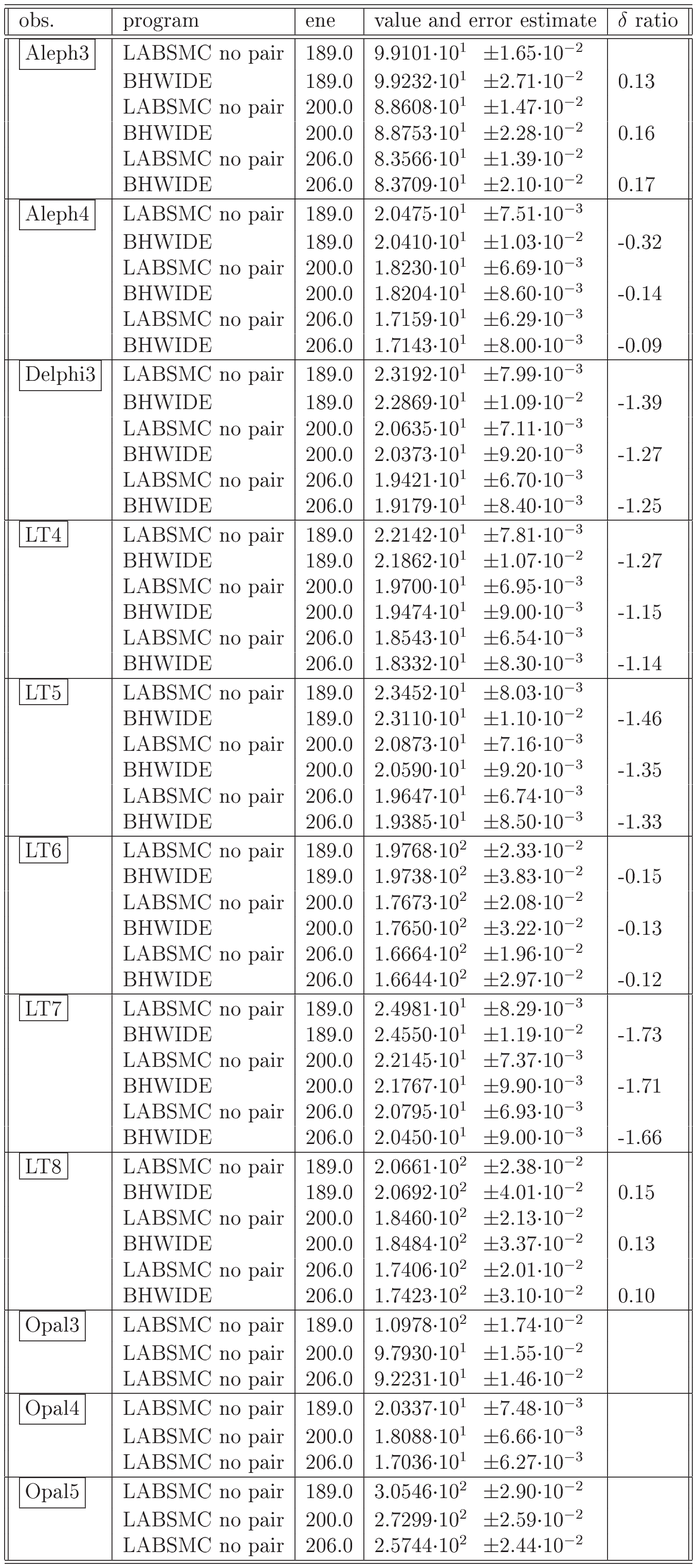,width=10cm,height=18cm,angle=0}
\caption{
Numerical predictions from theoretical calculations of the following            
realistic observables  in $e^+ e^- (\gamma )$ final states;  {\em cross sections}.   
Last field of the table shows relative
deviation (multiplied by 100) with respect to the first calculation.        
}
\label{tab:ee34567890}
\end{center}
\end{table}

\begin{table}[!h]
\begin{center}
\epsfig{file=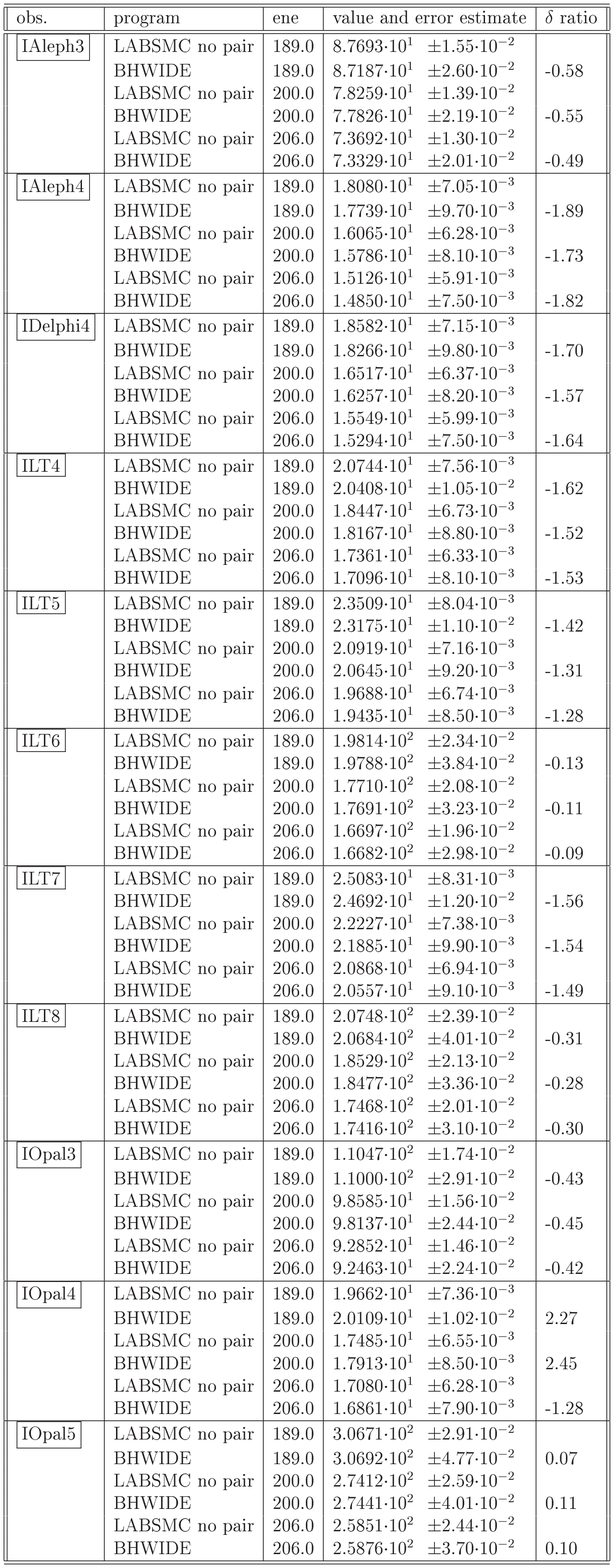,width=10cm,height=18cm,angle=0}
\caption{
Numerical predictions from theoretical calculations of the following            
idealized observables  in $e^+ e^-  (\gamma )$ final states;  {\em cross sections}.   
Last field of the table shows relative
deviation (multiplied by 100) with respect to the first calculation.        
}
\label{tab:eei4567890}
\end{center}
\end{table}
      \centerline{Tuned  BHWIDE LABSMC comparisons}
\vskip 1 mm

\noindent
Certain tuned comparisons of the BHWIDE and LABSMC  programs were
performed for the $e^+e^-(\gamma)$ observables. 
First, at the Born level an agreement was
found between the predictions of the two programs
(after adjusting the EW parameters) at the level of $0.1\%$ (stat. errors).
Then, cross-checks of the pure QED contribution to the Bhabha
process were performed, i.e. of only the $\gamma$-exchange contribution with the pure QED
corrections ( photonic corrections only -- no pairs). 
Here,  the two programs differ up to $0.4\%$ (the predictions of 
BHWIDE are in most cases lower than the ones of LABSMC). 
These differences can be explained by the different treatments
of higher order QED corrections in the two programs: the YFS exponentiation in
BHWIDE versus the structure function formalism in LABSMC. The biggest
differences are for the observables with a direct cut on the ``bare''
invariant mass $M_{inv}(e^+e^-)$. For such observables FSR plays an
important role and the differences in the higher order FSR implementation
in the two programs may be the reason for these discrepancies.
The agreement for other observables is at the level of $0.1\%$.

\centerline{Further comparisons}
\vskip 1 mm

All idealized observables are computed with LABSMC and BHWIDE,
for  realistic observables
only  LABSMC results are available for OPAL.
It would  be very desirable to compare the MC codes with
semi-analytic programs like TOPAZ0 that, unfortunately, did not contribute to the present Workshop.

At 189 GeV the agreement between BHWIDE and LABSMC is better than the following:
\begin{center}
\begin{tabular}{llll}
Type of observ.  &     Barrel       &    Endcaps  \\
 
Real. obs.       &        1.4\%     &     0.2\%   \\

Ideal. obs.      &     1.4-2.4\%    &     0.1-0.6\% \\
\end{tabular}
\end{center}

This is true for both cases: high-energy events and
observable where Z-return is included.
This does not contradicts  the BHWIDE estimate that the precision
in the barrel region is equal to or better than 1.5\%, and can be after tests
reduced to 0.5\%.
For the endcaps the precision is better or equal 0.5\%, which
is due to the dominant photon exchange in the t-channel 
when the forward region is included.

\noindent
The remaining differences between BHWIDE and LABSMC, 
as seen in tables \ref{tab:ee34567890},  \ref{tab:eei4567890}, for the full 
Bhabha process come from non-QED
contributions/corrections and are under investigation. At present
our estimate of systematic error must remain at 2 \% for the barrel region
and 0.5 \% for endcap.

The differences in case of observables with explicitly tagged photons,
($l\bar l \gamma$  final states) are, as expected bigger, but still
at the 3\% level.

For \citobs{DELPHI8} and \citobs{LT14} one of the electrons can be lost in the
beam pipe. BHWIDE doesn't describe such configurations (both $e^+$, $e^-$ are
required to be detected) - that is why BHWIDE does not provide any results
for these observables.
OPAL's realistic observables: 
\citobs{Opal3},  \citobs{Opal4},  \citobs{Opal5}, 
are not implemented with BHWIDE also.


\subsubsection{ Some comments on results for $e^+ e^- (\gamma )$ }

The issue of extrapolations is especially important in cases when 
theoretical uncertainties are not fully under control.

\noindent
In the case of L3 and OPAL selections, predictions for the realistic observables 
agree with the idealistic ones at 3\%.  No large
extrapolations are thus needed. This is due to 
use of collinearity cuts in both cases. The only
exception is the pair\citobs{LT4}-\citobs{ILT4}, where the difference is
around 6 \%. This is due to the use of an invariant mass cut,
where realistic observables sum electron energies with
the energies of photons close by.

In the cases of ALEPH and DELPHI the differences between
idealistic and realistic observables are larger (more than
20 \%). Here one of the, idealized observable -- realistic observable, pair
uses the invariant mass
cut and the other an collinearity cut (or mass from the 
angles). This leads to larger extrapolations.
The results presented here favor the use of collinearity
cuts in all cases (for both types of observables).

The precision tags set by experimental considerations are: 
0.21 \% barrel, 0.13 \% endcap.
The precision tag for the barrel is not met by the theoretical calculation.
A sizable factor of {\it ten} is still missing.
This will
reduce the sensitivity of searches for new phenomena like:
\begin{itemize}
\item
  contact interactions~\cite{Bourilkov:2000ap},
\item
  low scale gravity effects~\cite{Bourilkov:1999iz},
\item
  sneutrinos,
\item
  non-zero size of the electrons~\cite{Bourilkov:2000ap}.
\end{itemize}

The precision tag in the endcaps is closer to being met.
Even the precision of or below 0.5 \%, if confirmed,
will largely improve results like the measurement of
the running of the fine-structure constant, which are
limited by the theoretical uncertainty~\cite{L3RUNALF}.

As an example of the effects of theoretical uncertainties in the Bhabha channel
let us use as an example the  contact interactions. 
The  expected precision for the measurement in  the barrel detector (44-136 deg.) 
for four  LEP experimental combined,
and 600 pb$^{-1}$ at $\sim$ 200 GeV (average) energy
is about 0.45 \% (statistical) for the cross-section.  
The systematic error should be lower.
The  0.0026 statistical error is expected for $A_{fb}$, 
here systematic error can be a bit higher.

The numerical results for the limits on contact interactions are summarized in the table
\ref{contact_bhabha}:
%
\begin{table}[!ht]
\scriptsize
\centering
\caption{\small
 Limits for Contact Interaction models at 95 \% CL
 expected from combined data of large angle
 Bhabha scattering (barrel) at LEP2 /rough estimate/}
\begin{tabular} {||l|l|l|l||}
\hline\hline
 Theoretical error:        &  2 \%         &    1 \%       &    0.5 \%      \\
\hline\hline
   Contact Interaction     &   Sensitivity &  Sensitivity  &    Sensitivity \\
  Model                    &    [TeV]      &   [TeV]       &      [TeV]     \\
\hline\hline
   LL     &       7.6     &       9.3     &       10.6     
\\
   RR     &       7.5     &       9.2     &       10.4
\\
   LR     &       9.3     &      10.7     &       11.8
\\
   VV     &      16.1     &      19.5     &       22.0
\\
   AA     &      12.1     &      12.1     &       12.2
\\
\hline\hline
\end{tabular}
 \label{contact_bhabha}
\end{table}
\pagebreak

\centerline{\bf Running of $\alpha_{QED}$}
In the following L3 paper \cite{L3RUNALF}, the cross section for Bhabha scattering from 
20 to 36 deg. at 189 GeV was measured to be:\\
$\sigma = 145.6 \pm 0.9 (stat.) \pm 0.8 (sys.) \pm 2.2 (theory)$ pb.\\
This can be converted to a measurement of the running of the
fine-structure constant between the $Q^2$ of the luminosity monitor
and the endcap calorimeter:\\
${1\over \alpha}(-12.25 GeV^2) - {1 \over \alpha}(-3434 GeV^2)  =  
3.80 \pm 0.61 (exp.) \pm 1.14 (theory)$\\
or total error on ${1 \over \alpha}$ of 1.29 dominated by the theory 
uncertainty of BHWIDE taken to be 1.5 \%. The running is established at 3 sigma level.
A LEP combined measurement can reduce the statistical error by factor more
than 3 and the systematic error by factor of about two.

Clearly the theory uncertainty is the key for improvement. As a result of the
MC workshop the theory uncertainty seems  to be reduced to 0.5\%, but not yet 
matching the
experimental precision. This gives a promise to measure the running of
$\alpha$ in this range with a precision on ${1 \over \alpha}$  of 0.3 (4 times more
precise than now).

\begin{table}[!p]
\begin{center}
\setlength{\unitlength}{1cm}
\begin{picture}(16,20)
\put( 0,10.1){\makebox(0,0)[lb]{\epsfig{file=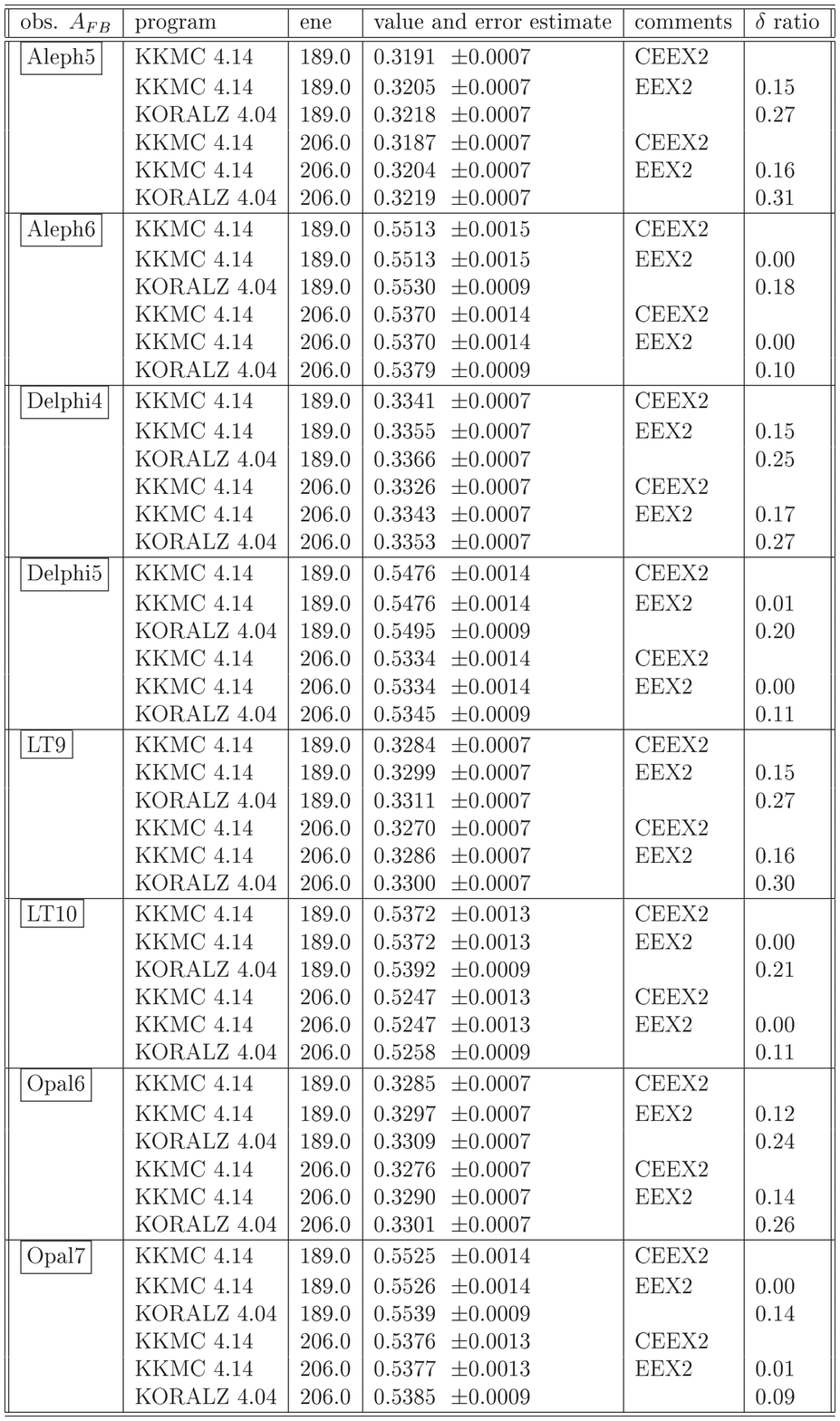,width=10cm,height=16cm,angle=90}}}
\put( 0, 0  ){\makebox(0,0)[lb]{\epsfig{file=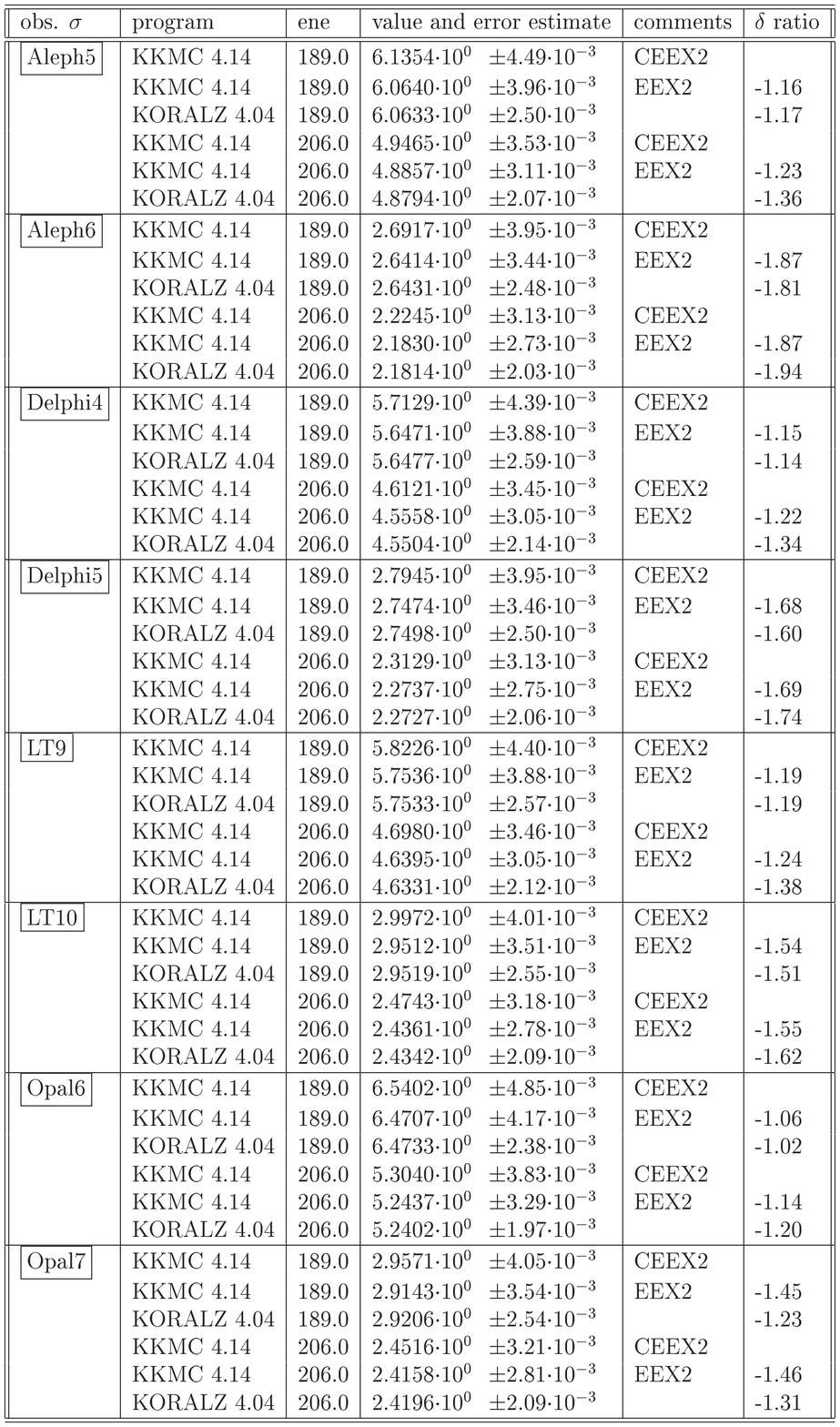,width=10cm,height=16cm,angle=90}}}
\end{picture}
\caption{
Numerical predictions from theoretical calculations of the following            
realistic observables in $\mu^+ \mu^- (\gamma )$ final states;
{\em cross sections and symmetries}.
Last field of the table shows relative
deviation (multiplied by 100) with respect to the first calculation.        
}
\label{tab:mumu34567890}
\end{center}
\end{table}

 
\begin{table}[!p]
\begin{center}
\setlength{\unitlength}{1cm}
\begin{picture}(16,20)
\put( 0,10.1){\makebox(0,0)[lb]{\epsfig{file=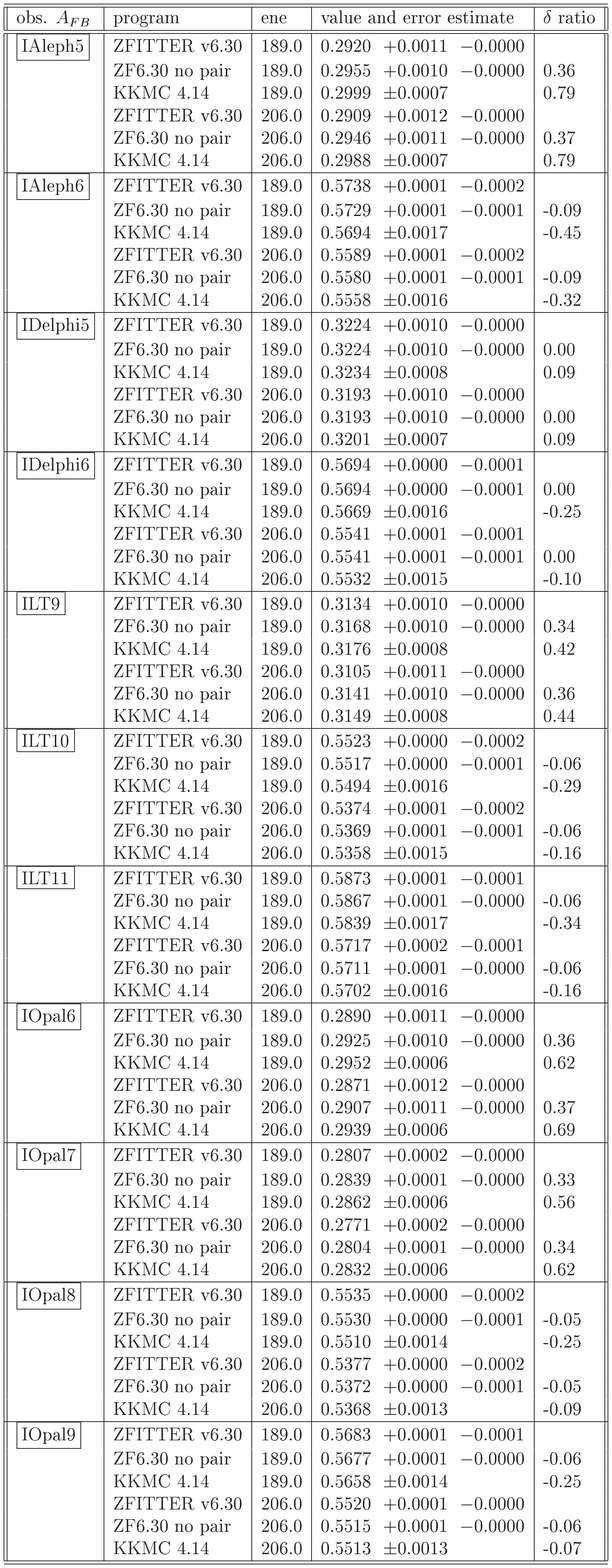,width=10cm,height=16cm,angle=90}}}
\put( 0, 0  ){\makebox(0,0)[lb]{\epsfig{file=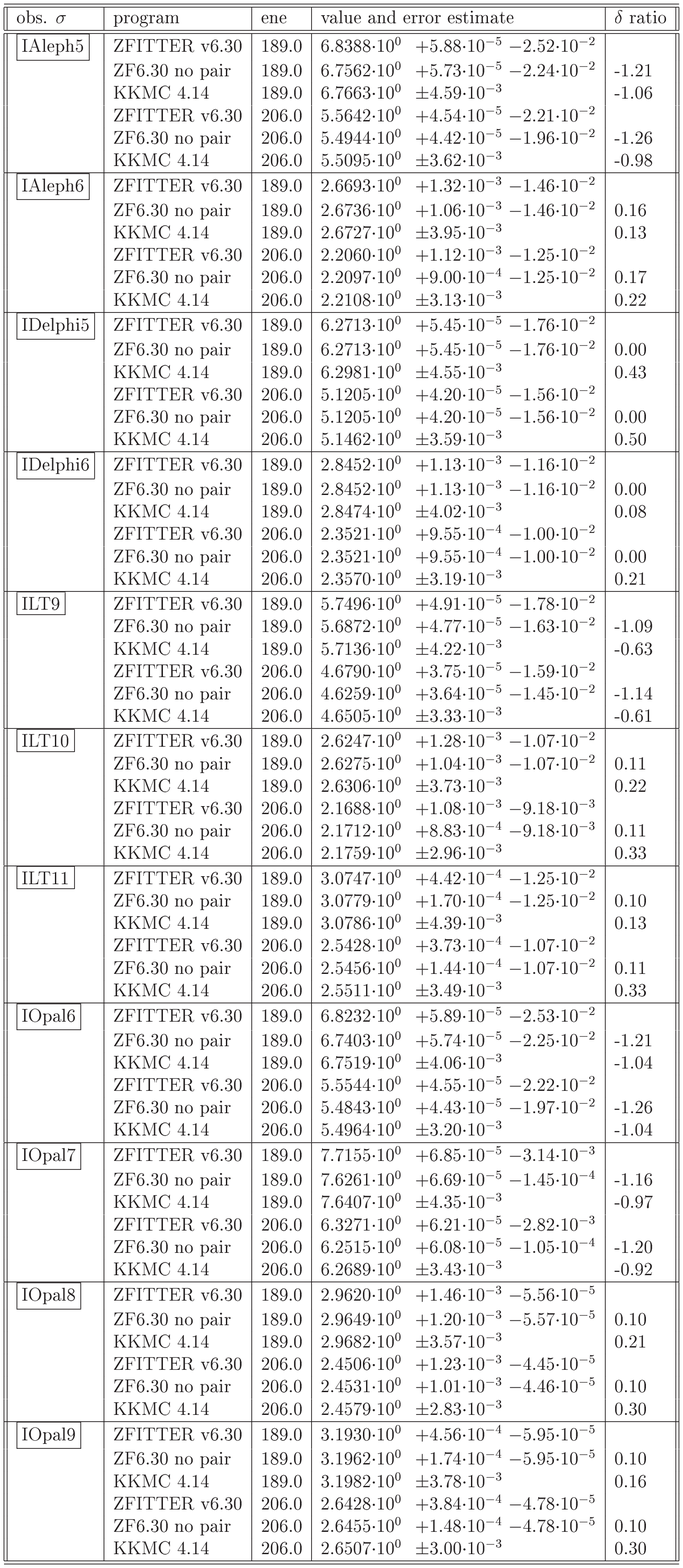,width=10cm,height=16cm,angle=90}}}
\end{picture}
\caption{
Numerical predictions from theoretical calculations of the following            
idealized observables in $\mu^+ \mu^- (\gamma )$ final states;
{\em  cross sections and asymmetries}. 
Last field of the table shows relative
deviation (multiplied by 100) with respect to the first calculation.        
}
\label{tab:mumui67890}
\end{center}
\end{table}



\begin{table}[!p]
\begin{center}
\setlength{\unitlength}{1cm}
\begin{picture}(16,20)
\put( 0,10.1){\makebox(0,0)[lb]{\epsfig{file=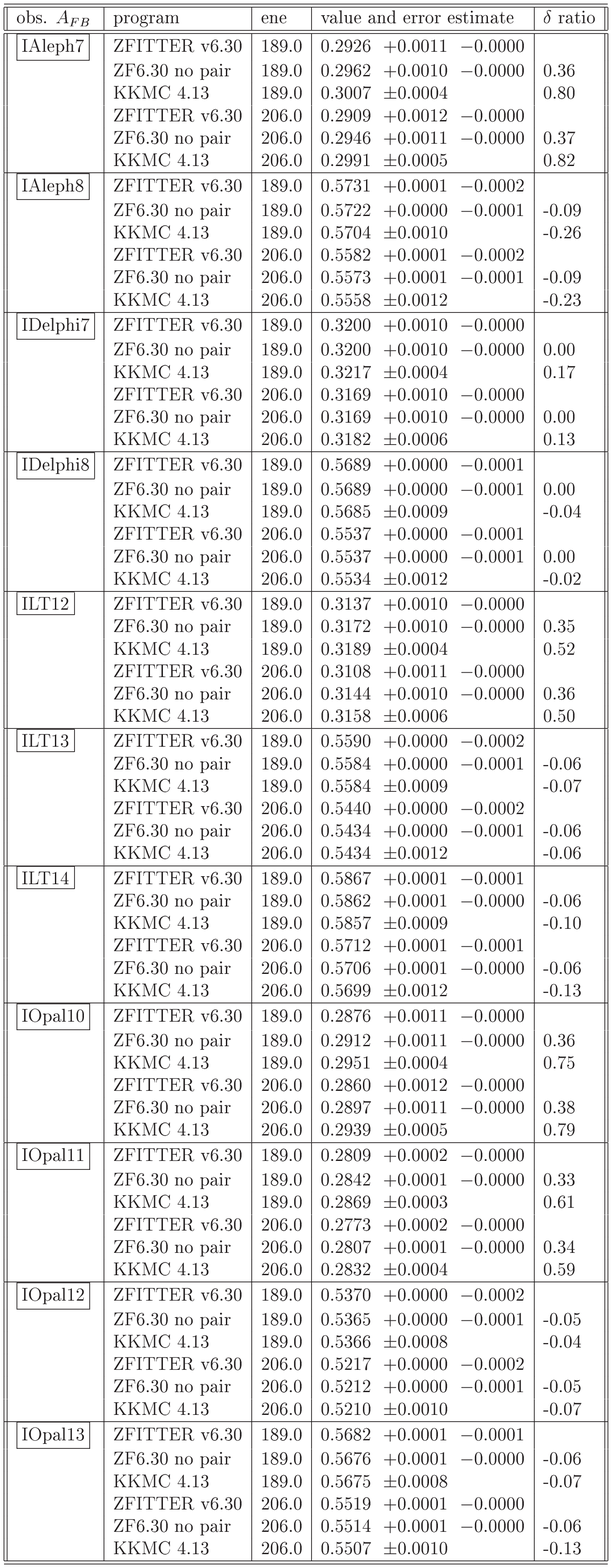,width=10cm,height=16cm,angle=90}}}
\put( 0, 0  ){\makebox(0,0)[lb]{\epsfig{file=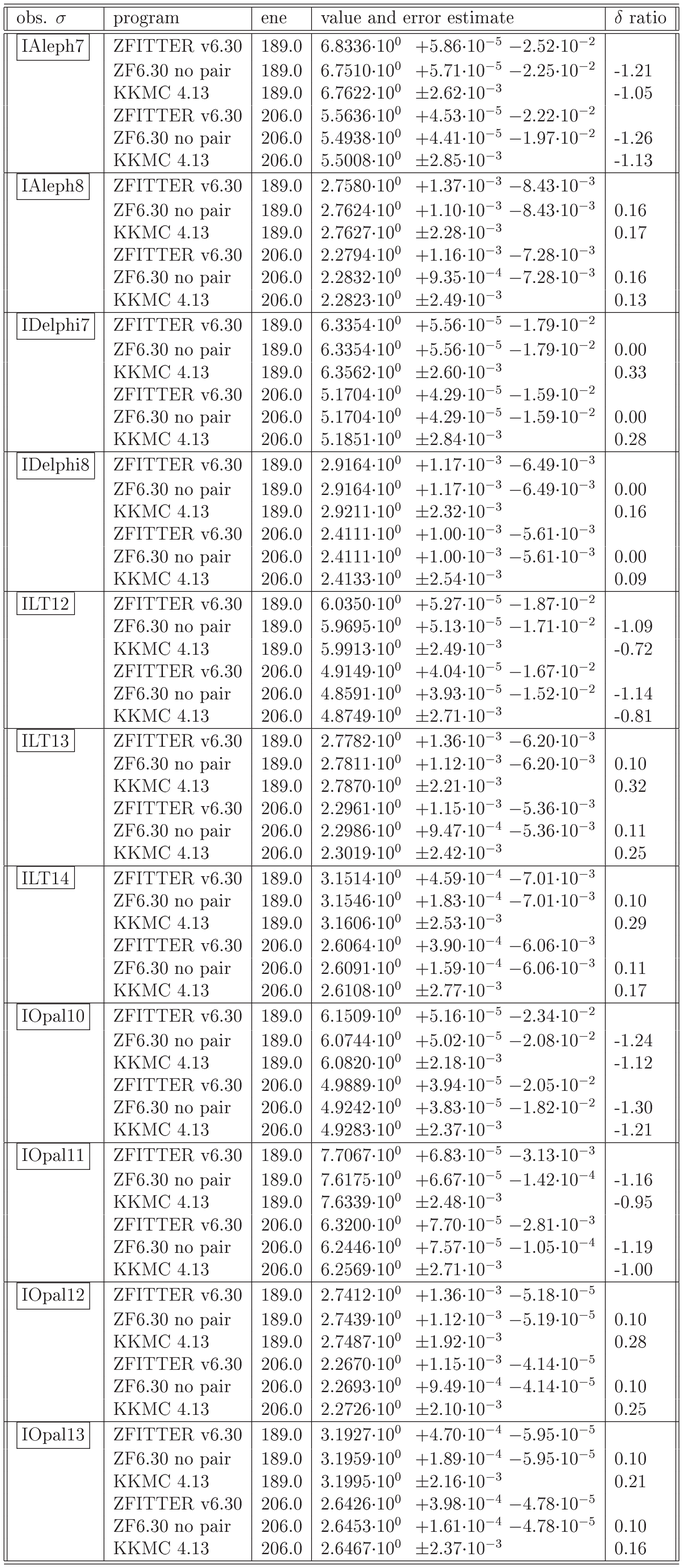,width=10cm,height=16cm,angle=90}}}
\end{picture}
\caption{
Numerical predictions from theoretical calculations of the following            
idealized observables in $\tau^+ \tau^- (\gamma )$ final states;
{\em  cross sections and asymmetries}. 
Last field of the table shows relative
deviation (multiplied by 100) with respect to the first calculation.        
}
\label{tab:tatai67890}
\end{center}
\end{table}

\subsection{Discussion of numerical results for observables in 
            $\mu^+\mu^-(\gamma)$ and $\tau^+ \tau^- (\gamma )$ final states}

This comparison is essential for the workshop because it tests theoretical 
uncertainties on different strategies of moving from raw data, 
realistic observables, to hard physics parameters with the  intermediate step 
of idealized observables. That is the strategy actually in use by all 
experiments. The size of effects such as QED interferences in context of truly 
complicated analyses including sophisticated cuts should be documented. 
The results can serve as a benchmark for old simulations also. 

The group of realistic observables in tab. \ref{tab:mumu34567890} includes 
comparisons of the old KORALZ Monte Carlo with \KKMC. As one can see the 
differences were 
in all cases due to the interference correction
and new method of exponentiation%
\footnote{Note that in the case of the observables with the tagged photons 
(see table \ref{tab:llg67890}) the pattern of differences is more complicated 
and \KKMC\ CEEX2 results do not coincide with KORALZ.}, 
other effects such as different way of implementing electroweak corrections
were not important. Differences between results from\KKMC\ option EEX2 and KORALZ
were always below 0.25 \%, which is no surprise as exponentiation in KORALZ is quite similar to the option
EEX2 for \KKMC. The CEEX2 option of \KKMC\ includes better scheme of 
exponentiation and in particular effects due to interference.

The comparisons for $\tau$ and $\mu$ leptons idealized observables
(tables \ref{tab:mumui67890},\ref{tab:tatai67890} include calculations 
performed with the help of \KKMC\ and \zf\  programs. 
Almost everywhere the differences between \KKMC\ and \zf\  predictions 
are  below 0.4\%, which is the precision tag of experiments (See also sections 
\ref{TunedZFKK} and \ref{ZFKKifi}). The exception are the cases with the 
intermediate cut on $M_{inv}$. Here the differences are slightly larger, 
up to 0.6 \%, but also acceptable. 
The effect due to pairs can not be neglected but even if included in a rather 
approximate way would be enough. The interference effect and the effect of 
CEEX exponentiation combined with respect to EEX are sizable also and should be taken 
into account. The three classes of effects as can be seen from the tables
are respectively up to  1.3  and 1.6 \% thus respectively 3 to 4 times the 
experimental precision tag.

We can conclude that in all cases the overall theoretical  uncertainty in  $\mu$ and 
$\tau$ channels are 0.4 \%, just
below (or very close to) the experimental precision tag. 
Therefore, we do not expect the overall QED  uncertainty in these channels 
to be sizable enough to affect the experimental studies in any case now. 
This comfortable situation for the experimental analyses is the result of 
better understanding reached in comparisons which have been performed 
recently, in particular, in the framework of this workshop.

One should note that, for example, for the dimuon channel where expected 
ultimate experimental error on the cross section measurement from four LEP 
experiments is about 1.2\% the decrease of theory uncertainty from 1\% to the 
present 0.4--0.5\% is roughly equivalent to the additional year of LEP running.

In some of the analyses performed with the two-fermion LEP data (like the 
fits for searches extra dimension gravity) the sensitivity depends on 
the differential 
distribution over the fermion production angle. The question of theoretical 
uncertainties in such differential distributions was, at least approximately, 
addressed by comparing forward-backward asymmetry values with different 
angular cuts. The agreement between different calculations in these 
quantities satisfies the experimental requirements also. 

\begin{center}
{\bf  Realistic $\tau$ observables:}
\end{center}
Numerical results for $\tau$-lepton  
realistic observables are not collected. 
This is not only  due to complexity, but also  due to relative similarity 
to the easier $\mu$-lepton case. The remaining, rather historical issue is
the  spin implementation in KORALZ. It can be of some concern
for those analysis which rely on simulation with that program.
The eventual cross check of this aspect is rather easy; 
it should follow exactly the 
same procedure as presented in table~\ref{tab:mumu34567890}.

\begin{table}[!p]
\begin{center}
\setlength{\unitlength}{1cm}
\begin{picture}(16,20)
\put( 0,10.1){\makebox(0,0)[lb]{\epsfig{file=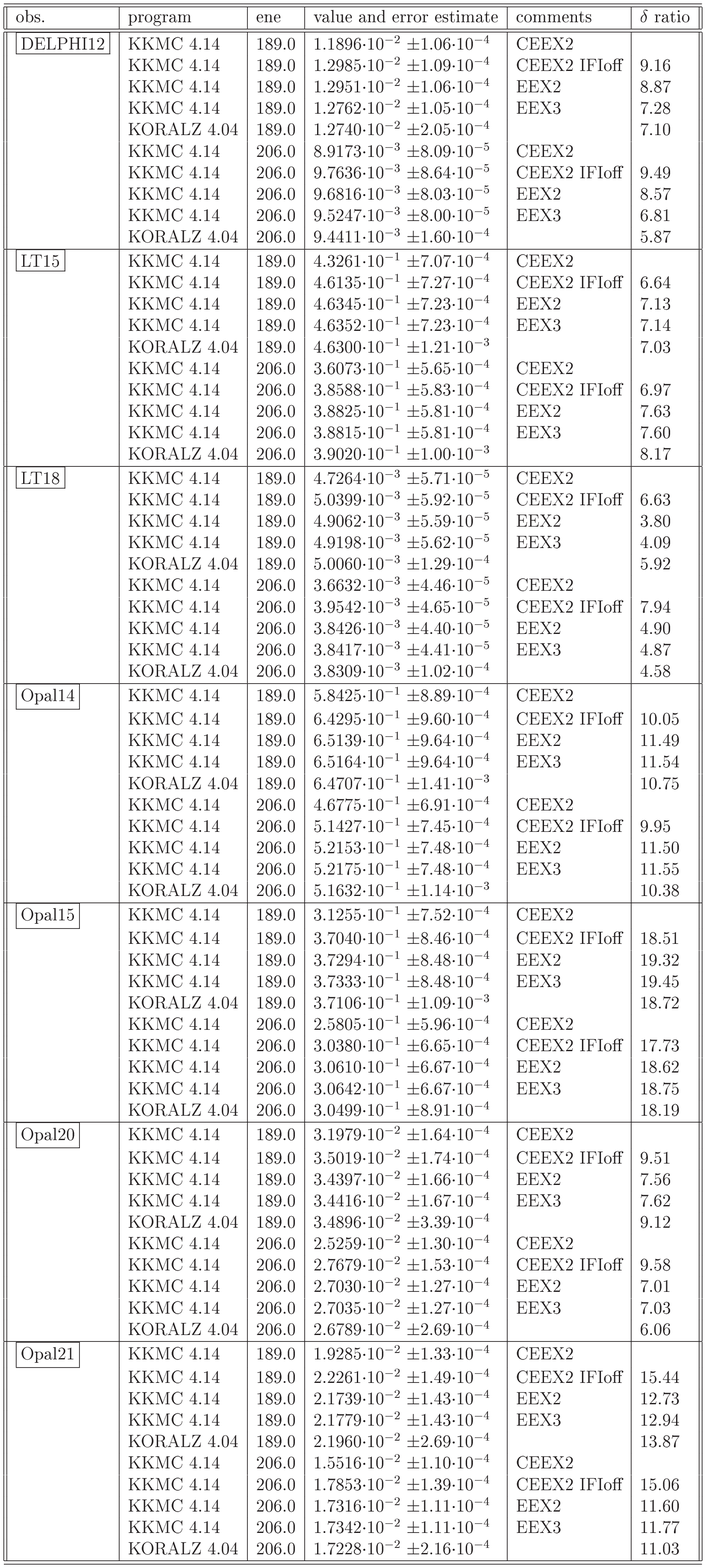,width=10cm,height=16cm,angle=90}}}
\put( 0, 0  ){\makebox(0,0)[lb]{\epsfig{file=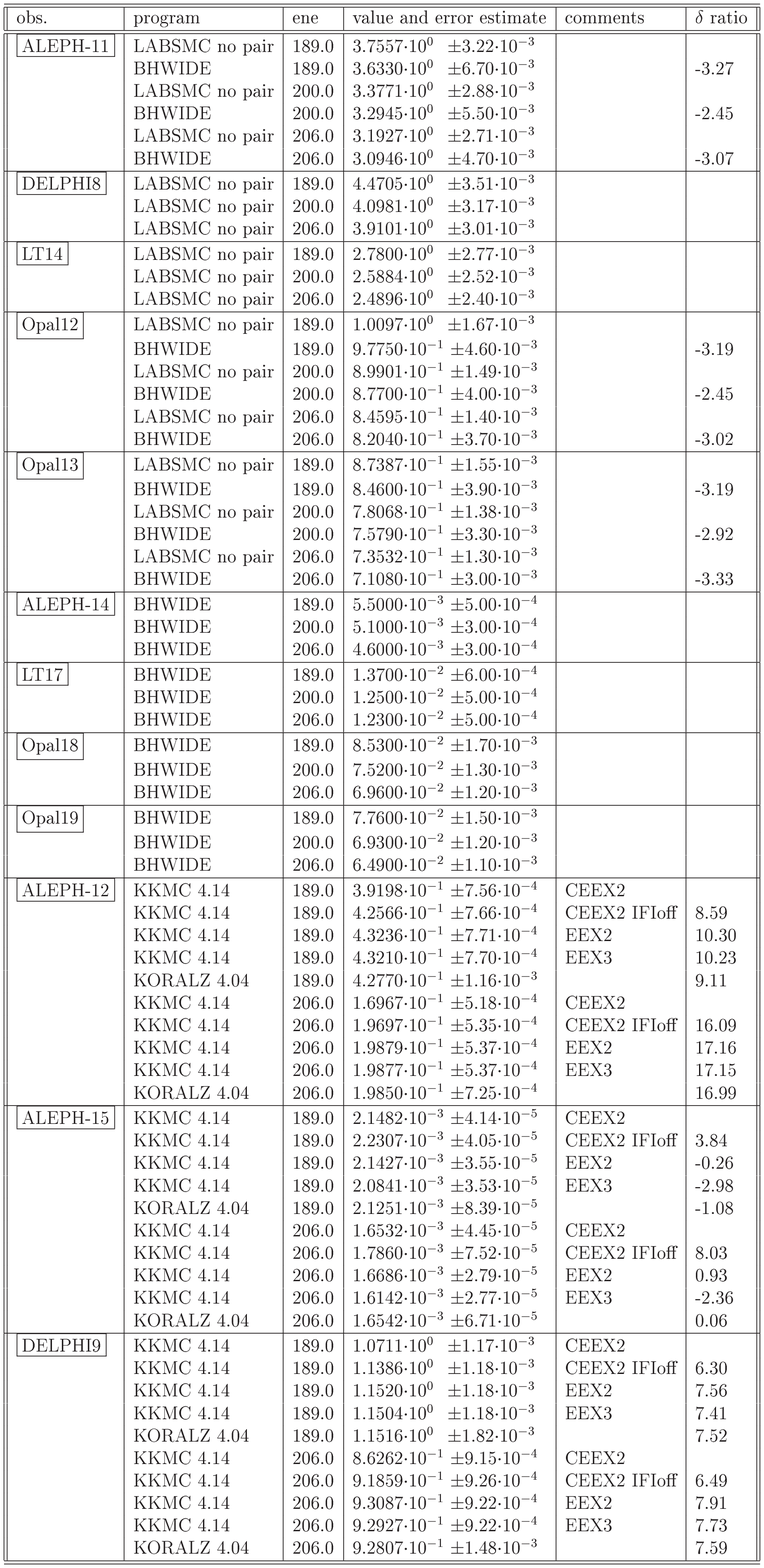,width=10cm,height=16cm,angle=90}}}
\end{picture}
\caption{
Numerical predictions from theoretical calculations of the following            
realistic observables in $l^+ l^- \gamma $ final states;
{\em  cross sections}. 
Last field of the table shows relative
deviation (multiplied by 100) with respect to the first calculation.        
}
\label{tab:llg67890}
\end{center}
\end{table}

\subsection{ Discussion of numerical results for  $l^+ l^- \gamma $ observables }
\label{Discllg}
One can see (Tab.~\ref{tab:llg67890}) that in all cases KORALZ and \KKMC\ 
(options CEEX2 IFIoff, EEX2 EEX3) give  results which are similar within requested
precision tags. Different choices of the exponentiation
etc., lead to effects at the level of 1 to 2 \%. As the case of the \KKMC\ and the matrix element
CEEX2 is expected to be the best, and the typical numerical size of the pair effects does not exceed
the precision tag, we can conclude that all effects, except those of the 
interference correction (difference between \KKMC\ results, option CEEX2 and CEEX2 IFIoff),
are well under sufficient control since a rather long time for this group of observables.
The size of the interference correction
is however sizable,  as expected, depending on selection it  can vary 
from 0 to nearly 20 \% and definitely must be taken into account in comparison of data 
with  theoretical predictions.
Comparisons with other possible calculations of interference corrections,
like in single photon mode of KORALZ,
or as in ref.~\cite{Fujimoto:1993ym} for two hard photons,
were not performed for the observables as in table \ref{tab:llg67890}.

The predictions of KORALZ due to the older exponentiation used (which is similar to
EEX2)
should coincide with \KKMC\ EEX2 results. One can see that it is not always the case.
This may indicate e.g.  deficiencies of the way how electroweak corrections 
are implemented in KORALZ.
The method designed for LEP1 works still quite good, but in configurations with massive bremsstrahlung,
such as radiative return to $Z$, limits become visible.
These differences are important to evaluate other places where similar systematic error
may play a role, that is $\nu \bar \nu \gamma$ and $\tau^+ \tau^- (\gamma) $ final states
in case of observables including radiative return to $Z$.   

For the observables from $e^+ e^- \gamma $ the differences between the two available codes
were at 3 \% maximum. For the moment this can be used as the estimate of theoretical
uncertainty. It is consistent with the 2 \% estimation from LABSMC. 




\begin{table}[!p]
\begin{center}
\setlength{\unitlength}{1cm}
\begin{picture}(16,20)
\put( 0,10.1){\makebox(0,0)[lb]{\epsfig{file=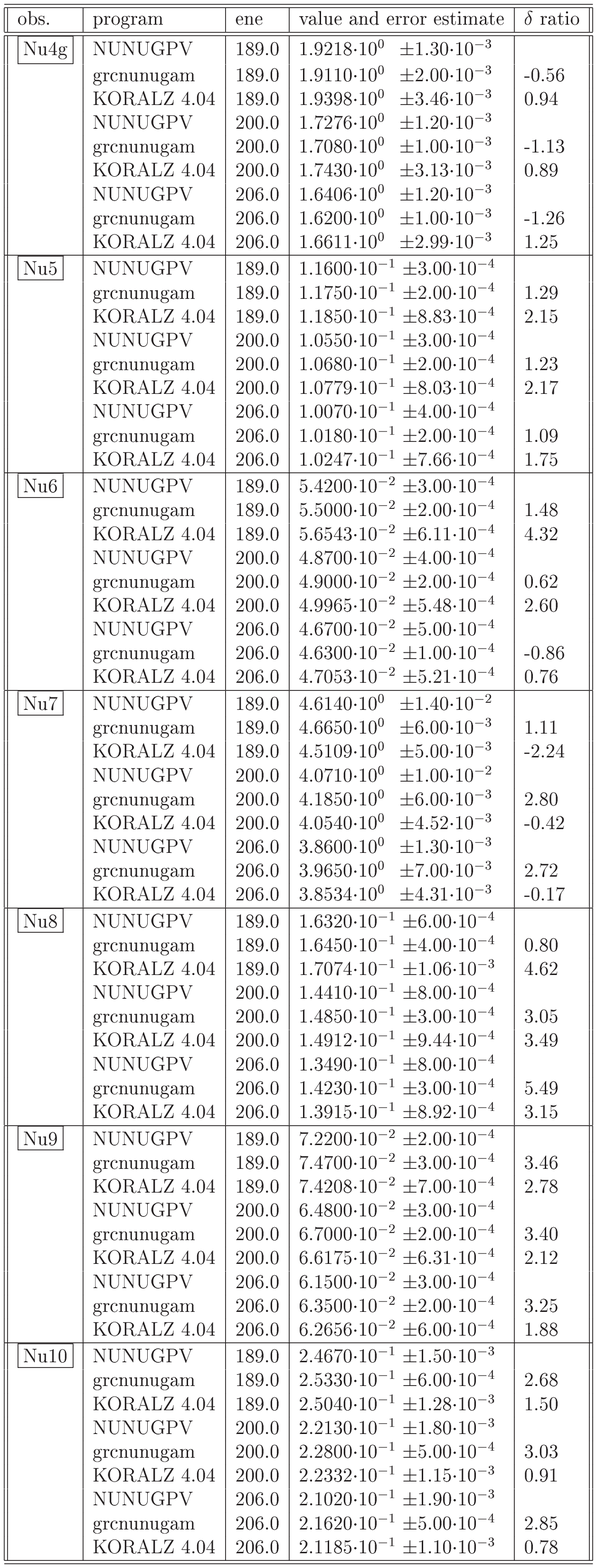,width=10cm,height=16cm,angle=90}}}
\put( 0, 0  ){\makebox(0,0)[lb]{\epsfig{file=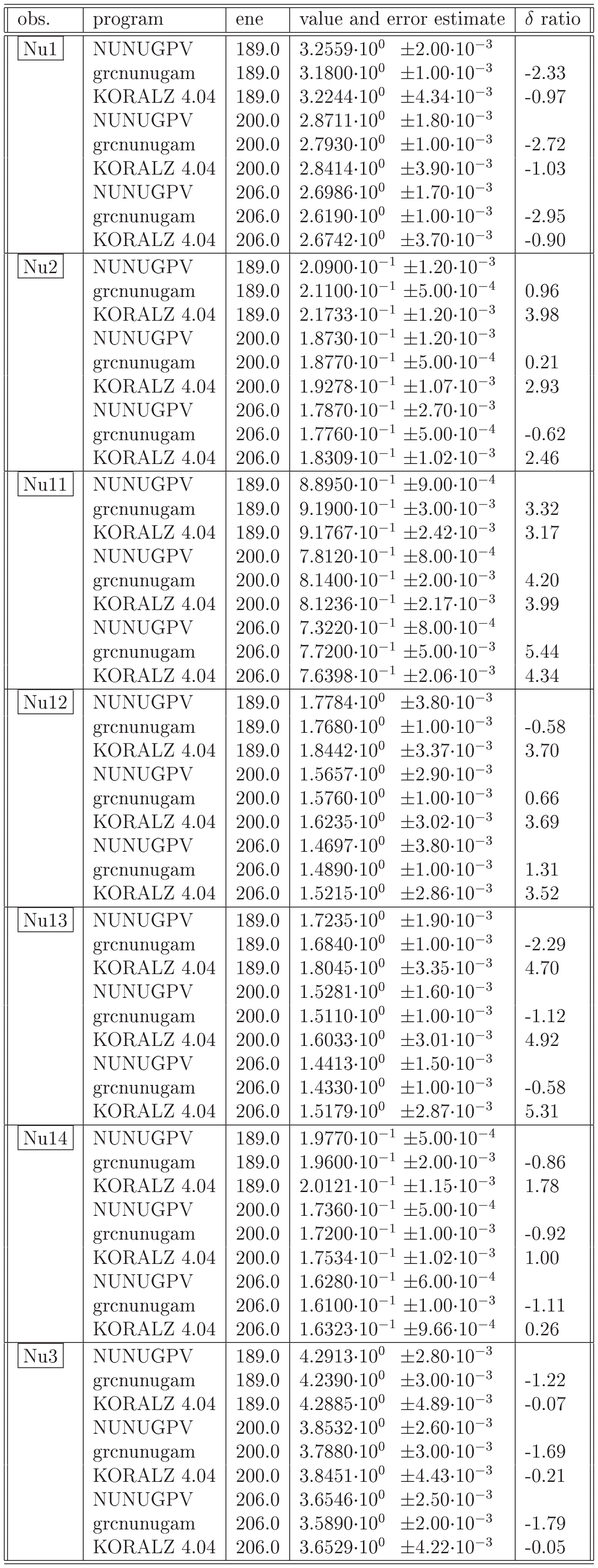,width=10cm,height=16cm,angle=90}}}
\end{picture}
\caption{
Numerical predictions from theoretical calculations of the following            
realistic observables in $\nu \bar \nu \gamma $ final states.                   
Last field of the table shows relative
deviation (multiplied by 100) with respect to the first calculation.        
}
\label{tab:nunug67890}
\end{center}
\end{table}

\subsection{ Discussion of numerical results for  $\nu \bar \nu \gamma $ observables }
Events where one or more photons are accompanied by missing
energy are the characteristic signature of many new physics 
processes. For example, in the framework of both the MSSM 
and GMSB models of supersymmetry, neutralino pair production 
can give rise to events where one or two photons are accompanied by
missing energy. This final state may also be produced in
theories where quantum gravity is propagating in extra spatial
dimensions. In such theories, gravitons may be produced copiously 
in association with a photon. The graviton subsequently escapes
detection giving rise to the photon and missing energy signature. 
Such events can also be used to study the trilinear WW$\gamma$
vertex and thereby to search for anomalous couplings.
The standard model background for such searches comes from
two processes: radiative returns to the Z resonance, with the
Z decaying to neutrinos, and $t$-channel W exchange with the photon(s)
radiated from the beam electrons or the W.

The large integrated luminosity provide by LEP at high 
energies allows such searches to be performed with high precision.  
At the end of the LEP2 running period each experiment will have 
around 1800 single photon and missing energy events and close
to 90 events with two photons and missing energy. When the data
from the four experiments are combined the single photon cross
section will be measured with a statistical precision of around
1.2\%. The combined systematic uncertainties from the photon 
selection efficiency and luminosity measurement is expected to be
around 0.5\%. Clearly, to have a negligible contribution to the 
overall cross section measurement, and hence to the search for new
physics, the theoretical uncertainty on the SM background  
prediction must also be at the 0.5\% level. 
Much less precision is 
required for the two photon and missing energy channel, where the
combined statistical uncertainty at the end of LEP2 will be around 5\%. 
The precision required of the theoretical estimate of the SM background
in this case is only 2\%. For this final state the other sources of 
experimental systematic uncertainty are negligible.

The level of precision which is now being achieved by LEP is impressive.
The initial estimate of the total integrated luminosity of LEP 2 was 
only half what was finally achieved.
Furthermore, techniques for combining the data from the photon and missing
energy searches of all four experiments have been developed in the framework
of the LEP SUSY Working Group.
That is why, the required precision is so much higher than the 
2\% level which was though to be sufficient until now.

There were three independent Monte Carlo programs available for comparison 
of numerical results.
The main sources of differences between the results of these calculations 
were expected to arise from the following effects:
\begin{enumerate}
\item
   Although the cuts are (supposed to be)
   the same for all programs, the input parameters were not set to the same values, 
   we are not performing "tuned comparisons"; this means in particular that we have to
   expect discrepancies of about 2\% due to the different
   renormalization schemes implemented, as was for instance shown by
   the Japanese group in~\cite{Kurihara:1999vc}.
\item
   The QED corrections arising from missing non-log terms are expected to lead to
   a theoretical uncertainty of about 1-2\%. 
\end{enumerate}   
Taken these two effects into account, the size of the observed
discrepancies%
\footnote{ The observable~\citobs{Nu4g} was an exception,
until the cut on the energy of the trigger photon was increased from 
previous 1 GeV to present 5 GeV. 
This may be good starting point for further investigation.}
is essentially what was expected.

\begin{figure}[!ht]
\centering
\setlength{\unitlength}{0.1mm}
\begin{picture}(1600,800)
\put( 375,750){\makebox(0,0)[b]{\large \citobs{Nu1}}}
\put(1225,750){\makebox(0,0)[b]{\large \citobs{Nu2}}}
\put(-20,   0){\makebox(0,0)[lb]{\epsfig{file=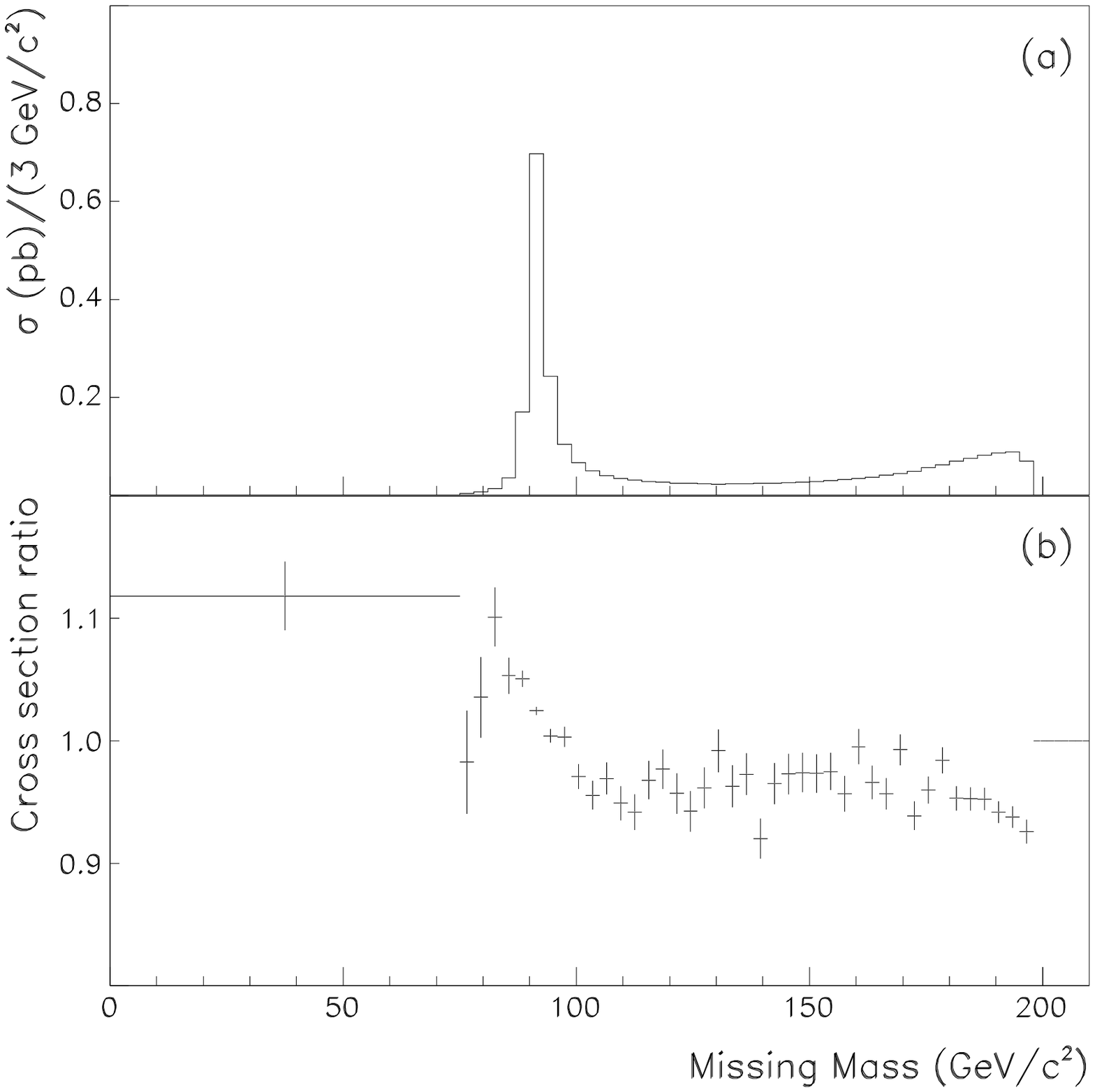,width=80mm,height=80mm}}}
\put(800,   0){\makebox(0,0)[lb]{\epsfig{file=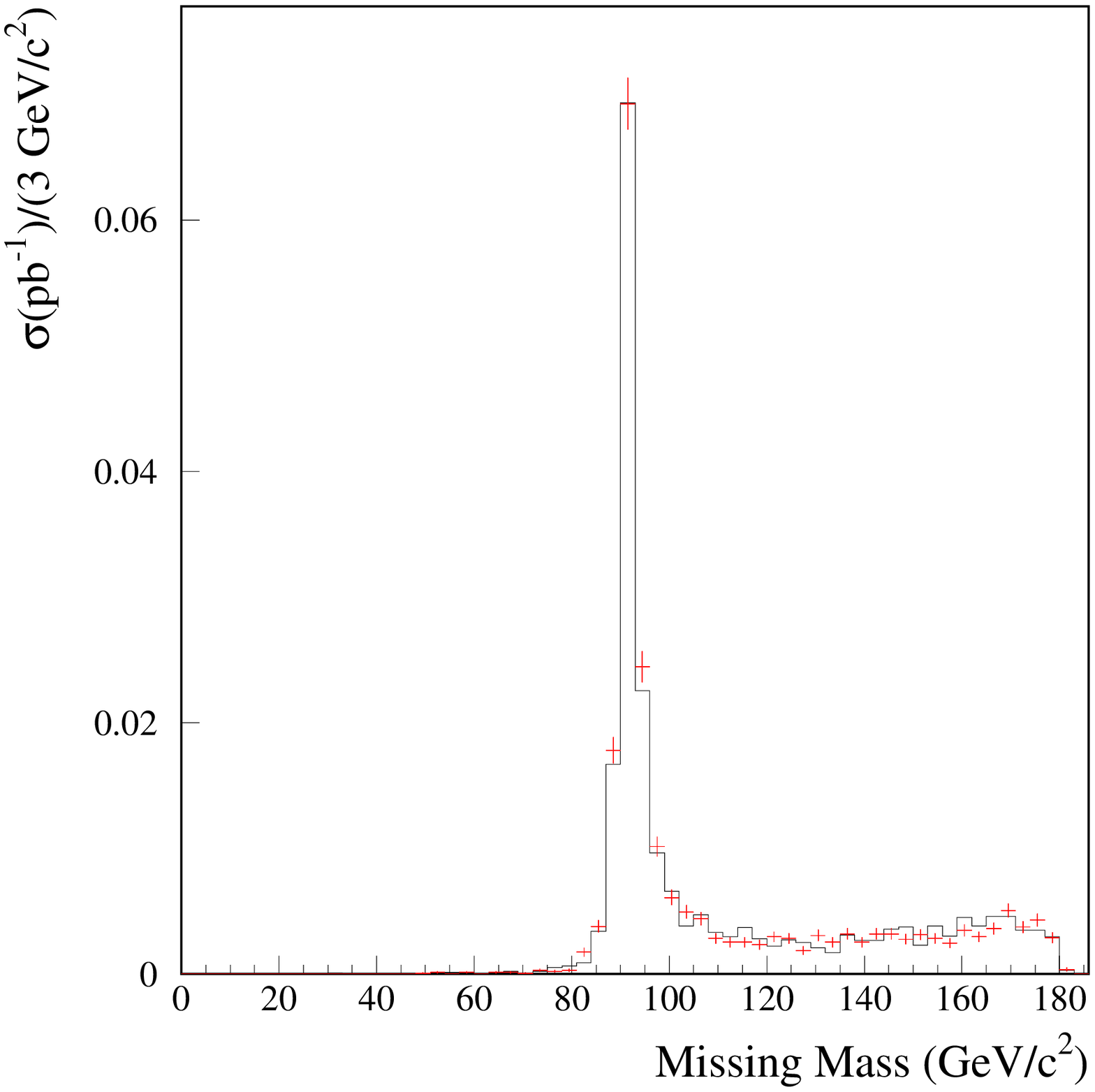,width=80mm,height=80mm}}}
\end{picture}
\caption[]{\small\sf {\bf Left side:} 
 Missing mass distribution from the NUNUGPV Monte Carlo:
   part (a);  and ratio of KORALZ to NUNUGPV predictions part (b). Plot was made
   for  Centre-of-mass energy of 206 GeV
   and selection cuts as defined for observable  \citobs{Nu1}. 
\\
{\bf Rigrt side:}
 Missing mass distribution from the nunugpv Monte Carlo
  (histogram) and KORALZ (error bars) for Center of mass energy of 189 GeV
  and selection cuts as defined for observable  \citobs{Nu2}.
}
\label{NuFig}
\end{figure}

The comparison of observable between the different Monte Carlo programs may
be summarized as follows:\\
$\bullet$ In the worst case the difference between 
the programs is at the level of 4-5\%,\\
$\bullet$
Moreover, when the event selections are particularly clear/simple: and there are no
sharp cuts (no selection of narrow bands in angular dependences, as in \citobs{Nu13} 
or cuts on soft photons \citobs{Nu13})
the level of agreement is better. This could arise as a result of
systematic differences between the codes simply via different implementations
of {\it very complicated} cuts.
More likely, this could be explained by the different way how
hard matrix elements and/or soft photons are treated in some corners of the 
phase space.
\\
$\bullet$
This explanation seems to be supported by the following two plots \ref{NuFig}, 
representing the missing mass spectrum 
for one and two photon events compared between KORALZ and
NUNUGPV with cuts as for observables~\citobs{Nu1} and~\citobs{Nu2}.
The KORALZ predictions tend to be higher than NUNUGPV 
for the part of the spectrum of missing mass smaller or comparable
to $Z$ and lower for events of large missing
mass (which have relatively soft photons).

\noindent
$\bullet$
On the other hand the implementation of  $W$ contribution of 
$e^+e^- \to \nu_e \bar \nu_e \gamma$ channel in KORALZ is affected by 
approximation. 
\vskip 1 mm

\subsubsection{ Conclusions for  $\nu \bar \nu \gamma $}

In the case of relatively {\it simple} observables 
(no selection of narrow bands in angular dependences or cuts
on soft photons)
agreement was found at the 
level 2-3\% for both the  single- and double- photon observables
otherwise differences
of around 3-5\% are observed. 
Single- and double-tagged photon observables provide rather  similar pattern
of agreements and differences.
It is possible that the contributions of electroweak box diagrams and pair corrections,
which have not yet been fully studied, may introduce  theoretical 
uncertainty of around 2-3\%. This should not affect our
estimate of the final theoretical uncertainty of around 4\% for 
{\it simple} observables 
and 5\% otherwise.

As an example of what this level of theoretical uncertainty may mean in practice 
let us use the search for TeV scale quantum gravity propagating in two extra dimensions. 
The combined LEP limit on the mass scale associated with this new physics
at the end of LEP2 would be (assuming no hint of a signal)

1.23   TeV   for 0.4\%  theoretical syst
 
1.21   TeV   for 1.2\%  theoretical syst

1.12   TeV   for 5.0\%  theoretical syst

\noindent
For this topology the cross section for the new physics varies as
$(1/M)^4$.
The current (preliminary) limit from the ALEPH data taken up till now is 1.10 TeV.
From the point of view of this analysis, until further theoretical developments
will become available with the systematic error reduced 
below 4-5\%, there is little point in analyzing the final years data nor, 
indeed in combining the results of the four LEP experiments.

%% file: 2f-Chapt-Part2.tex
\section{Monte Carlo and Semi-analytical codes and their own error specifications}
\label{sec:programs}

In this section we present the Monte Carlo and the semi-analytical codes
used in the work of our working group.
The last subsection in the description of each code represents
the theoretical error specification of each calculation,
as seen by the authors of the codes.
They are the starting point for the comparisons in our working group. 
Throughout the comparisons of the codes and discussion among the authors
of the codes we could verify these statements, improve the understanding
of the problems and add more value to them.
But first of all we need to define our starting point.
And this is to be done here in this section.

\subsection{Presentation of the program BHWIDE}
\label{BHWIDE}
\begin{tabbing}
1) \underline{Authors:}\hspace{4cm} \= {\bf 
S. Jadach, W. P\l{}aczek and B.F.L. Ward
} \\[.3cm]
2) \underline{Program:}    \> {\bf
BHWIDE v.1.10, December 1998
} \\[.3cm]
3)\underline{ Can be obtained from}  \> 
{\tt http://enigma.phys.utk.edu/pub/BHWIDE/} 
\\[.3cm]
4) Reference to main description \>
\cite{bhwide:1997} 
\\[.3cm]
5) Reference to example \>
\cite{lepybk:1996jn}
\\[.3cm]
6) advertisement
\\[.3cm]
\end{tabbing}
  
In this subsection, we  briefly describe our Monte Carlo (MC) event generator 
for large angle Bhabha (LABH) scattering 
called {\tt BHWIDE} and discuss some important cross-checks of the program.

{\tt BHWIDE} is based on the YFS exclusive exponentiation 
procedure~\cite{Yennie:1961ad},
where all the IR singularities are summed-up to infinite order
and canceled out properly in the so-called YFS form factor. 
The remaining non-IR residuals, $\bar{\beta}_n^{(l)}$, corresponding to 
the emission of $n$-real photons, are calculated perturbatively up to a given
order $l$, where $l\geq n$, and $(l-n)$ is a number of loops in 
the $\bar{\beta}_n^{(l)}$ calculation. 
In {\tt BHWIDE} an arbitrary number $n$ of real photons with non-zero $p_T$
are generated according to the YFS MC method of Ref.~\cite{bhlumi2:1992}.   
The non-IR residuals $\bar{\beta}_n^{(l)}$ are calculated up to 
${\cal O}(\alpha)$, i.e. $\bar{\beta}_0^{(1)}$ and $\bar{\beta}_1^{(1)}$
corresponding to zero-real (one-loop) and one-real (zero-loop) photons,
respectively, are included.
In $\bar{\beta}_0^{(1)}$ we implemented two libraries of the ${\cal O}(\alpha)$
virtual EW corrections: 
{\bf (1)} the older one of Refs.~\cite{Bohm:1988fg,Berends:1988jm}, which is
not up to date but can be useful for some tests/cross-checks, and 
{\bf (2)} the more recent one of Ref.~\cite{Beenakker:1991mb}. 
When the genuine weak corrections are switched off
(or numerically negligible) they are equivalent. 
In $\bar{\beta}_0^{(1)}$ we implemented two independent matrix elements
for single-hard-photon radiation: 
{\bf (1)} our calculation~\cite{bhwide:1997} 
in terms of helicity amplitudes, and 
{\bf (2)} the formula of 
CALKUL~\cite{calkul:1982} for the squared matrix element. 
We have checked that the above two representations agree numerically
up to at least 6 digits on an event-by-event basis.

The MC algorithm of {\tt BHWIDE} is based on the algorithm of the program 
{\tt BHLUMI} for small angle Bhabha scattering~\cite{bhlumi2:1992}, however 
with some important extensions: 
{\bf (1)} QED interferences between the electron and positron lines 
(``up-down'' interferences) had to be reintroduced as they are important
in LABH;
{\bf (2)} the full YFS form factor for the $2\rightarrow 2$ process, including
all $s$-, $t$- and $u$-channels, was implemented~\cite{bhwide:1997};
{\bf (3)} the exact ${\cal O}(\alpha)$ matrix element for the full Bhabha 
process was included.
The multi-photon radiation is generated at the low-level MC stage as
for the $t$-channel process, while the $s$-channel as well as all
interferences are reintroduced through appropriate MC weights.
This means that the program is more efficient when the $t$-channel
contribution is dominant, as e.g. at LEP2 energies; however, it proved
to work well also near the $Z$ resonance.  

The program is written in {\tt FORTRAN77} and is particularly suited
for use under the {\tt Unix} operating system\footnote{%
    However, it can be used, in principle, on any operating system
    with a {\tt FORTRAN77} compiler.} %
for which a special directory structure has been created with useful
{\tt Makefile}'s for easy compiling and linking. 
The program runs in three stages: (1) initialization -- where all
input parameters are read and transmitted to the program as well
as all necessary initializations are performed, (2) event generation
-- here a single event is generated, and (3) finalization -- final
bookkeeping for a generated event statistics is done and some useful
information is provided (printed-out).
There are two main modes of event generation: one can generate either 
variable weight events (useful for various tests) or constant (=1) 
weight events (useful for apparatus MC simulations). 
Various input parameter options, to be set 
by the user, allow to choose between different contributions/corrections
to the cross section, such as weak corrections (two libraries),
vacuum polarization (three parametrizations), etc. Other input 
parameters allow to specify the necessary ingredients for the
cross section calculation and the event generation, such as the CMS
energy, physical parameters (masses, widths, etc.), phase space cuts, etc.    
For each generated event, four-momenta of the final state electron,
positron and all radiative photons are provided. In the 
variable-weight-event mode they are supplemented with the main 
(best) event weight as well a vector of weights corresponding to various
models/approximations. In the finalization stage, the total cross 
section corresponding to the generated event sample is calculated
and provided (printed-out) together with some other useful information.

\subsection{Error specifications of BHWIDE}

So far, several tests/cross-checks of the program have been
performed, see e.g. Ref.~\cite{Barcelona:1998wp}.
First comparisons with other MC programs for LABH were done
during the LEP2 Workshop in 1995~\cite{lepybk:1996jn}.
They showed a general agreement of {\tt BHWIDE} with most
of those programs within $2\%$ at LEP2 energies.
At that time such a level of precision was expected to be sufficient
for LEP2. Discrepancies between various calculations can be
explained by the fact that most of the programs were designed
for LEP1 where the $Z$ $s$-channel contribution was dominant,
while at the LEP2 energy range the $t$-channel $\gamma$ exchange
dominates. Thus, the physical features of the Bhabha process
at LEP1 and LEP2 are very different.  
Recently, a more detailed study of the theoretical precision
of {\tt BHWIDE} has been carried out~\cite{bhwide:1999ws}.
Comparisons have been made with the MC programs: 
{\tt OLDBIS}~\cite{bhlumi2:1992}
(a modernized version of the program {\tt OLDBAB}~\cite{oldbab:1983})
and {\tt BHLUMI}~\cite{bhlumi4:1996}, and the semi-analytic code 
{\tt ALIBABA}~\cite{Beenakker:1991mb}. Tests were done at ${\cal O}(\alpha)$
and with higher order corrections for various cuts -- by starting
from the pure $t$-channel $\gamma$-exchange and switching on
gradually other contributions/corrections.
This study shows that at ${\cal O}(\alpha)$ {\tt BHWIDE} agrees
with {\tt OLDBIS} within $0.1\%$ for the pure QED process, 
while {\tt ALIBABA} differs by up to $0.3\%$. 
When higher order corrections are included,
{\tt BHWIDE} is generally within $1\%$ of {\tt BHLUMI}
($0.5\%$ in the forward region: $\cos\theta_e>0.7$)
for the pure $t$-channel $\gamma$-exchange process
and within $1.2\%$ of {\tt ALIBABA} for all kinds of contributions/corrections.
From these test we have estimated the overall theoretical
precision of {\tt BHWIDE} at $1.5\%$ for the LEP2 energy range.
We expect that by making some improvements of the program
(e.g. modifying the ``reduction procedures'' for the matrix element 
calculations, including ${\cal O}(\alpha^2)$ LL corrections)
and performing some new cross-checks (e.g. with the program
LABSMC~\cite{Arbuzov:1999db}) we can reduce this precision to $\sim 0.5\%$.
Further improvements of the theoretical precision can be made, in our
opinion, with the help of the {\tt KK} MC program~\cite{kkcpc:1999}
after implementing in it the $e^+e^-$ channel.

\newpage
\subsection{Presentation of the program KORALZ}
\label{KORALZ}
\begin{tabbing}
1) \underline{Author:}\hspace{4cm} \= {\bf 
S. Jadach, B.F.L. Ward and Z. W\c as
} \\[.3cm]
2) \underline{Program:}    \> {\bf
KORALZ v.4.04, 
} \\[.3cm]
3)\underline{ Can be obtained from}  \> 
Library of Computer Physics Communication, \\
\> or from
the author ({\tt z.was$\at$cern.ch}) upon request
\\[.3cm]
4) \underline{ Reference to main description} \>
\cite{Jadach:1999tr} and references therein.
\\[.3cm]
5) \underline{Reference to example, use:}\>
\cite{Jachol,AFBnowa}\\[.3cm]
\end{tabbing}

Initially the {\tt KORALZ} event generator was 
written~\cite{koralzearly}
to simulate $\tau$-pair production and decay for LEP1 physics at the first
order of QED bremsstrahlung without any exponentiation. 
Only longitudinal $\tau$ spin effects were included. 
Later~\cite{koralz:1991,Jadach:1994yv}, longitudinal beam
 polarization was included and  higher
order QED effects were incorporated using powerful exponentiation techniques
\cite{yfs2:1990,yfs3:1992}
of initial state bremsstrahlung first, but later 
of final state bremsstrahlung as well. The interference of initial
and final state bremsstrahlung was always neglected, except as a parallel mode
of operation at the single bremsstrahlung level. This assumption was good at the 
peak of the $Z$ resonance, due to suppression of the correction due to 
$Z$ life-time, and the estimation of the error based on the single 
bremsstrahlung calculations was sufficient \cite{KANCEL,AFBnowa}. 
At that time a quite complete system of tests and cross-checks was developed
for the effects due to corrections of initial state bremsstrahlung 
using dedicated methods based on comparison of
semi-analytical results and the Monte Carlo \cite{AFBstara}
using importance sampling. Tests at the technical precision of $10^{-4}$ and better
could be obtained. In general the total precision of 0.2 \% was achievable 
for quite a range spectrum of observables for $\mu^+\mu^-$ or $\tau^+\tau^-$
final states. The $\nu\bar \nu \gamma$ final states were also introduced
\cite{nunu} using the assumption that the $t$-channel $W$-exchange forms
a contribution which is rather small so that, in particular, the 
$W-W-\gamma$ interaction can be completely neglected.

The use of KORALZ at LEP2 energies impedes substantial improvement of precision.
For $\mu^+\mu^-$ or $\tau^+\tau^-$ it is mainly due to the lack of interference
effects in exponentiation. For $\nu\bar \nu \gamma$ final states it is
due to the approximate treatment of $t$-channel $W$-exchange. 
Even though the $W-W-\gamma$ interaction was included, 
the method explained in \cite{Jachol}
is not enough for the single  photon observables
and any precision meant to be better than 1-2 \%. For double photon
observables precision decreases even further as the matrix element 
for the two photon configuration is approximate to the pragmatic
order $\alpha^2$ only.

Electroweak corrections are implemented in KORALZ using the reduced Born
method. This means that the effects of electroweak corrections 
beyond the crude Born level are implemented at  the leading-log level only.
Recently, the final version of KORALZ was published and documented
\cite{Jadach:1999tr}. That version uses DIZET  version 6.05, however,
for the sake of tests versions with the up to date DIZET library 
may be maintained.

Use of the program is expected to be gradually replaced by
\KKMC Monte Carlo which already at present is  superior in all
applications (for the time being except $\nu \bar \nu \gamma$ final states) 
and at all energy ranges as far as precision is concerned.
Some tools for studying anomalous effects in $\tau\tau\gamma$ final states
\cite{Paul:1999tt}, $\nu\bar\nu\gamma$ \cite{Jacholkowska:1999ei}, and leptoquarks
\cite{leptoki} are available at present for KORALZ only.
A number of flags, to be set by the user, allow the user to switch between
different options and perform specific comparisons and investigations,
e.g. for calculation of the program physical precision.

The program uses the following libraries:
YFS 3.4 \cite{yfs3:1992} for multiple photon bremsstrahlung, 
TAUOLA \cite{tauola:1993}
for $\tau$-lepton decay and PHOTOS \cite{photos:1994} for radiative corrections
in $\tau$-lepton decays.

\subsection{Error specifications of KORALZ}
The main purpose of the program was to serve the Monte Carlo simulation for
LEP1 observables. The program was adapted to become useful at LEP 2 energies,
but it was known that the backbone of its construction is not best suited for that 
purpose. Also, as the new program \KKMC was developed in parallel. The effort 
to push the limits of the KORALZ program precision were not exploited.

Let us recall the main points and present crude estimates of the related 
systematic errors.

$\bullet$ 
The electroweak section of the program is functionally equivalent to the 
one used in \KKMC and based now on the DIZET part of \zf\ . The related 
contribution to the systematic error can be thus taken as 0.15 \%.

$\bullet$ 
There is no pair correction included, as the size of pair effects 
is typically of order of 1.5 \%; the appropriate contribution has to be calculated 
independently with the help of a semianalytical program or other means. 
In case of non-idealized observables, this leads to an uncertainty which we 
can estimate as  0.4 \%  (0.2 \% for idealized ones). 

$\bullet$
   The similar situation holds for the QED initial-final state interference which is
not included in the program also and affects observables for all final states 
including charged fermions. At LEP 2 the interference effects are 
at 1-2 \% level for photon non-tagging observables. For non-neutrino final states and
observables where one or more photons are tagged the uncertainty is bigger (5 to 20 \%)
and  \KKMC\ should be used.

$\bullet$
The matrix element is limited to pragmatic second-order. The related uncertainty 
is about 0.1 \% for observables where photons are not tagged, about 0.2 \% for single
photon tagged observables, but can be  more for observables where more
than one photon are tagged. 

$\bullet$
In KORALZ exponentiation is based on the relatively old algorithm 
\cite{yfs3:1992} (with some later improvements but of incomplete tests only)
and 1 (0.2)  \% uncertainty for observables including (not including) radiative
return to Z should be added due to that point.
This is especially important for $\nu \bar \nu \gamma$ observables.

$\bullet$
 For $\nu\bar\nu \gamma$ final states some rather simple approximations are used
in implementation of the contribution of $t$-channel $W$-exchange and the $W-W-\gamma$ 
coupling. It was shown in \cite{Jachol} that the corresponding uncertainty is
not exceeding 1 or 2 \% for observables including single tagged photons.
For double tagged photons we expect the related contribution to uncertainty to be 
of order of 3-10 \% depending on the average requested $p_T$ of the second hardest
photon.

The final numbers for uncertainties for observables can be obtained as the sum 
in quadrature of the above uncertainties. It will be calculated at the 
end of the workshop as the individual contributions  can still change thanks 
to the comparisons, in particular with \KKMC.

\newpage
\subsection{Presentation of the program \KKMC\ }
\label{KK-MC}
\begin{tabbing}
1) \underline{Author:}\hspace{4cm} \= {\bf 
S. Jadach, B.F.L. Ward and Z. W\c as
} \\[.3cm]
2) \underline{Program:}    \> {\bf \KKMC\  v.4.13 and v.4.14
} \\[.3cm]
3)\underline{ Can be obtained from}  \> 
Library of Computer Physics Communication, 
\\ \>
or from http://home.cern.ch/jadach,
\\[.3cm]
4) Reference to main description \>
\cite{kkcpc:1999}
\\[.3cm]
5) Reference to example, use: \>
\cite{ceex2:2000}
\\[.3cm]
\end{tabbing}

\KKMC\  is the Monte Carlo event generator providing weighted and
constant weight events for $e^+e^-\to f+\bar{f}+n\gamma,
f=\mu,\tau,d,u,s,c,b$ within the complete phase space.  Technical
description and users guide of the version 4.13 can be found in
ref.~\cite{kkcpc:1999} while physics content and numerical results are
contained in ref.~\cite{ceex2:2000}.  The current version with minor
improvements which was used during this workshop is 4.14.  It will be
publicly available at the time of publishing this report.  In the
following we describe the main features of the program and we discuss in a
detail the critical issue of the overall technical and physical
precision of the program, stressing that, although it can be viewed
from outside as a monolithic single code, in reality almost every vital
aspect/component of its total precision is relies on the comparison
with another independent code, quite often with several other ones.
Since this aspect was highlighted in the discussion during the workshop,
we elaborate on this at some length.

\subsubsection{ QED in \KKMC\  }
The QED part the program does not rely for the photon emission, on
the structure functions (SF) or the parton shower (PS) model but
rather on the new Coherent Exclusive Exponentiation
(CEEX)~\cite{ceex1:1999,ceex2:2000} which is an extension of the
Yennie-Frautschi-Suura (YFS) exponentiation~\cite{Yennie:1961ad}.  This
older Exclusive Exponentiation (EEX)~\cite{yfs2:1990}, more closely
related to the original YFS formulation, the same as in KORALZ, is kept as an option
in \KKMC\ , for tests of precision and for the purpose of the
backward compatibility.  The CEEX matrix element in \KKMC\  is
entirely based on spin amplitudes, which helps to treat exactly spin
effects and to include the QED initial-final state interference.  CEEX is
based entirely on Feynman diagram calculations and the present version
includes the complete \Order{\alpha^2} for ISR and almost complete
\Order{\alpha^2} for FSR%
\footnote{ For FSR the 2-$\gamma$ and 1-$\gamma$ real matrix element
  are exact, while the 1-loop corrections to the $1-\gamma$ real matrix
  element is still in the LL approximation. This in principle should be
  good enough, at the precision level of $\sim 0.1\%$.}.
It is important
to realize that the ISR calculation in \KKMC\  is the first
\Order{\alpha^2} independent calculation since the work by Burghers,
Berends and Van Neerven (BBVN)~\cite{Berends:1988ab}%
\footnote{ BBVN calculated \Order{\alpha^2} ISR also directly from Feynman
  rules.  The resulting inclusive/integrated distributions they have
  cross-checked with the renormalization group techniques, down to the
  second order next-to-leading logarithmic (NLL) term.  }.
On the contrary, semi-analytical programs like \zf\ , TOPAZ0 \cite{topaz0} 
or ${\cal
KK}$sem rely on the SF's (called also radiator functions) which are
derived from BBVN, as far as the \Order{\alpha^2} sub-leading terms are
concerned.  For the real photon emissions CEEX employs the
Weyl-spinor methods of Kleiss and
Stirling~\cite{kleiss-stirling:1985}.  The 2-loop virtual corrections
are derived from ref.~\cite{Barbieri:1972} and one-loop corrections to
single photon emission are from
refs.~\cite{igarashi:1987,Berends:1988ab} and were also cross checked
independently by our collaborators~\cite{yost-privcom}.

\subsubsection{  Electroweak corrections}
The complete \Order{\alpha} electroweak corrections with higher order
extensions are included with help of the DIZET
library~\cite{Bardin:1989tq}, the same version as that used in \zf\ 
6.30~\cite{Bardin:1999yd-orig}.  The complex electroweak form-factors (EWFFs),
dependent on $s$ and $t$ variables, are calculated by DIZET and used
in the construction of the CEEX matrix element.  In order to speed up
calculations they are stored in the look-up tables (using a finite grid
in the $s$ and $t$ variables) and interpolated.  The basic uncertainty of
EW corrections in \KKMC\  is therefore the same as that of
DIZET/\zf\  (but this is not true of the QED corrections).  
We have good reasons to
believe that our CEEX matrix element offers a better way of combining EW
corrections with QED corrections than that used in the semianalytical codes like
\zf\ , basically because in \KKMC\  it is done at the amplitude level,
using Feynman diagrams instead of the SF's.
The QCD FSR corrections are taken also
from DIZET, keeping properly track of their $s$-dependence (through
look-up tables and interpolation).

\subsubsection{ Spin effects}
Complete spin effects are included for the decaying $\tau$-pairs and
for beam polarizations in an exact way, valid from the $\tau$ threshold up to
multi-TeV linear collider energies.  Due to the use of the improved
Kleiss-Stirling spinor technique, the appropriate Wigner rotation of
the spin amplitudes is done in the rest frame of the outgoing fermions
and of the beam electrons~\cite{Jadach:1998-235}.
For $\tau$ channel the program implements
spin-sensitive $\tau$-decays using TAUOLA \cite{tauola:1993} for
$\tau$-lepton decay and PHOTOS \cite{photos:1994} for radiative
corrections in $\tau$-lepton decays.

\subsubsection{  Virtual pairs in \KKMC }
\label{subsec:kkmc-virt}
The effect of virtual initial and final state pairs is optionally  added to the $F_1$ electric 
form-factor, see Feynman diagram of fig.~\ref{fig:pair-bouble},
using an old well known formula~\cite{Burgers:1985qg}

\begin{eqnarray}
F_1^{pair}(s)&=& \sum\limits_f \bigg\{ -{1\over36} L_f^3
                 +{19\over72}L_f^2 +\big({1\over18}\pi^2-{265\over216}\big)L_f +C_F
\bigg\},
\\
C_F&=& \left\{ \begin{array}{ll} {383\over108}
-{11\over6}{\pi^2\over6},                                      & m_f =   m_F, \\
-{1\over3}\zeta(3)+{3355\over1296} -{19\over18}{\pi^2\over6},  & m_f \gg m_F.  
\end{array} \right.
\\ 
L_f&=& \log{s\over m_f^2},
\end{eqnarray}
with $m_f$ denoting the mass of virtual fermion in the fermion loop 
and  $m_F$ mass of the fermion flowing through the vertex (typically an electron).
The two cases correspond to correction due to identical and heavy 
fermion in the virtual loop.


Technically, virtual pairs in the initial and final state are added in \KKMC\ as  alternative weights:
        {\tt  WtList(213) } represents the case with Virtual Pairs and IFI on,
        {\tt   WtList(263)} represents the case with Virtual Pairs and IFI off.
Masses $m_f$ are taken 0.2GeV for $f=d,u,s$ and PDG values for the rest.
Changing $m_f$ of light quarks by factor two induces 
only $\delta\sigma_{virt}/\sigma=0.04\%$!

This option should be used in conjunction with adding the signal contribution of real pairs
via a full 4-fermion Monte Carlo generator, like KORALW.
%
Concerning the proper cancellation  of the virtual pairs mass-logs from
\KKMC\ and from KORALW, there should be no technical (precision) problems,
especially for the precision level 0.1\% required for LEP2.
The main complication will be a proper matching of these mass-logs in the presence
of the QED bremsstrahlung.
Here, the loading-logarithmic approximation and renormalization group 
will be used as a guide, as usual.
For the moment we use effective-quark masses instead of dispersion relations,
because we do not see clear indication that it is really necessary to use the latter method
at the 0.1\% level. However, if it turns out to be necessary, it is possible
to introduce $R_{had}(s)$ in both \KKMC\ and KORALW.

Summarizing, this new feature will allow the use of \KKMC\ together with the KORALW Monte
Carlo, according to the scheme already suggested in
\cite{koralw:1998}, to produce predictions for the observables with the
real/virtual pair contribution.

\subsubsection{ Recent improvements not yet documented elsewhere updates}
In version 4.14 the QCD FSR corrections to the final states of quarks
were cross checked and some necessary modifications were introduced.

The $F_1$ form-factor -- the virtual correction factor corresponding
to initial and final state emission of non-singlet
and singlet pairs was introduced, see above.

Note also that \KKMC\  is expected to take over all functionality of
the {\tt KORALZ} event generator.  The most important feature of 
{\tt KORALZ} which is still missing in \KKMC\  is
the neutrino channel.

\subsection{Error specifications of \KKMC}
 
\subsubsection{ Technical precision}
The overall technical precision due to phase space integration
is estimated to be 0.02\% in terms of the typical total cross section
with Z-exclusive or Z-inclusive cuts.  The basic test of the
normalization of the phase space integration is the following: we do
not cut on photon transverse momenta, but only on the total photon
energy through $s'=M^2_{inv}(f\bar{f})>s'_{\min}$ and downgrade the
ISR or ISR+FSR matrix element without ISR$\otimes$FSR interference to
most the simple CEEX \Order{\alpha^0} case, that is the product of the
real photon soft-factor times the YFS/Sudakov form-factor and $\sigma_{\rm
Born}(s')$, with $s'$ shifted due to ISR.  For this simplified QED
model we integrate analytically over the phase space, keeping for ISR
the terms of \Order{\alpha,L\alpha,L\alpha^2,L^2\alpha^2,L^3\alpha^3},
that is enough terms to reach 0.01\% precision even for Z-inclusive
cuts, and for FSR we limit ourselves to
\Order{\alpha,L\alpha,L^2\alpha^2}, also enough for this precision
tag%
\footnote{We see that for ISR and $\sigma_{\rm Born}(s')=const$
  switching off the \Order{L\alpha^2,L^3\alpha^3}
  terms changes results only by 0.01\%.}.  
Within such a simplified QED model we compare a very high
statistics MC run ($\sim 10^9$ events) with the analytical formula and
we get agreement, see ref.~\cite{kkcpc:1999}, better than 0.02\%.  The
possible loophole in this estimate of precision is that it may break
down when we cut the transverse momenta of the real photons, or switch to
a more sophisticated QED model.  The second is very unlikely as the
phase space and the actual SM model matrix element are separated into
completely separate modules in the program.  The question of the cut
transverse momenta of the real photons requires further discussion.
Here, it has to be stressed that in our MC the so-called big-logarithm
\begin{equation}
  L= \ln\Big( {s\over m_f^2}\Big) - 1
\end{equation}
is the  {\em result of the phase space integration} and if this
integration were not correct then we would witness the breakdown of the
infrared (IR) cancellation and the fermion mass cancellation for FSR.
We do not see anything like that at the 0.02\% precision level.  In
addition there is a wealth of comparison with many {\em independent
codes} of the phase space integration for $n_\gamma=1,2,3$ real
photons, with and without cuts on photon $p_T$.  It should be
remembered that the multi-photon phase space integration module/code in
\KKMC\  is unchanged since last 10 years.  For ISR it is based
on YFS2 algorithm of ref.~\cite{yfs2:1990} and for FSR on YFS3
algorithm of ref.~\cite{yfs3:1992}, these modules/codes were part of
the KORALZ~\cite{Jadach:1994yv} multi-photon MC from the very
beginning, already at the time of the LEP1 1989
workshop~\cite{Z-physics-at-lep-1:89}, and they were continuously
tested since then.  The phase space integration for $n_\gamma=1$ was
tested very early by the authors of YFS2/YFS3 against the older MC
programs MUSTRAAL~\cite{mustraal} and KORALB~\cite{koralb:1985} and with
analytical calculations, at the precision level $<0.1\%$, with and
without cuts on photon $p_T$.  The phase space integration for
$n_\gamma=2,3$ with cuts on photon $p_T$ was tested very many times
over the years by the authors of the YFS2/YFS3/KORALZ and
independently by all four LEP collaborations, using other
integration programs like COMPHEP, GRACE and other ones, in the context of the
search of the anomalous $2\gamma$ and $3\gamma$ events.  Another
important series of tests was done in ref.~\cite{nunu} for ISR
$n_\gamma=1,2$ photons (with cuts sensitive to $p_T$ of photons),
comparing KORALZ/YFS2 with the MC of ref.~\cite{Miquel:1990eg} for
the $\nu\bar{\nu}\gamma(\gamma)$ final states.  
Typically, these tests, in which QED matrix element was programmed in several
independent ways, showed agreement at the level of 10\% for the cross
section for $n_\gamma=2$ which was of order 0.1\% of the Born, or
0.2-0.5\% for $n_\gamma=1$ which was of order 1\% of the Born, so they
never invalidated our present technical precision of 0.02\% in terms
of Born cross section (or total cross section in terms of Z-inclusive cut).

We conclude therefore that the technical precision of \KKMC\ 
due to phase space integration is 0.02\% of the integrated cross
section, for any cuts on photon energies Z-inclusive and Z-exclusive,
stronger than%
\footnote{It downgrades to 0.5\% for $M_{inv}(\mu\bar{\mu}) \leq
   2m_\mu$, i.e. full phase space.}
$M_{inv}(f\bar{f})>0.1\sqrt{s}$ and any mild cut on the
transverse photon energies due to any typical realistic experimental
cuts.  For the cross sections with a single photon tagged it is about
0.2-0.5\% and with two photon tagged it is $\sim 10\%$ of the
corresponding integrated cross section.  These conclusions are based on
the comparisons with at least six other independent codes.

\subsubsection{ Physical precision of pure QED ISR and FSR}
In the following we shall discuss mainly the physical precision of ${\cal
KK}$MC, that is the magnitude of the missing higher orders in the
QED/SM matrix element implemented in \KKMC\ .  This will also
include the technical precision of the matrix element implementation
not related to phase space integration discussed previously.

As we already mentioned, in \KKMC\  we have also the older EEX-type
matrix element, similar to the one of KORALZ/YFS2 and BHLUMI.  Its
crucial role in establishing physical precision is that of ``second
line of defense'' because it has its own estimate of the physical and
technical precisions (unrelated to phase space integration) which are
factor 2 worse than for CEEX, but a very solid and independent one.  
The basic test of the EEX matrix element is based again on the comparison
with the analytical integration over the photon phase space, this time
within the \Order{\alpha,L\alpha,L^2\alpha^2} only, but with the
additional bonus that the analytical integration is exact in the soft
limit.  Furthermore, the EEX matrix element is split into about six
pieces, so called $\bar{\beta}$-functions and each of them is
cross-checked separately.  The comparison is done for ISR and FSR
separately, taking $\sigma_{\rm Born}(s')=const$ in addition to the
normal one with Z resonance.  Since some of $\bar{\beta}$-functions
like $\bar{\beta}_{1,2}$ are concentrated in the region of the phase
space with $n_\gamma=1,2$ real hard photons, their separate tests
provide an independent non-trivial cross-check of the phase space
integration.  The above detailed tests lead for $\sigma_{\rm
Born}(s')=const$ to differences between MC and analytical results
$<0.1\%$, vanishing to zero for strong cuts on total photon energy.
This is our basic estimate of the technical precision of the
implementation of the EEX (unrelated to phase space integration).

There is an eternal ongoing discussion how to estimate the physical
precision.  Our approach is the conservative one, just take the
difference of the two consecutive perturbative calculations at hand%
\footnote{ One possible pitfall with the above rule is that the
  difference between the two consecutive perturbative calculations may
  be accidentally zero for a given value of the cuts, one should
  therefore vary the values of the cuts before drawing conclusions.}.
In order to be not over-conservative we usually take half of such a
difference, which means that we assume that the convergence of the
perturbative expansion is like $(1/2)^n$ at least, which is not a bad
assumption for QED where $2L_e\alpha/\pi\sim 0.07$ and $1/L=0.05$.

In the case of EEX we check the differences of EEX3-EEX2 and
EEX2-EEX1, where EEX1= \Order{\alpha,L\alpha}$_{\rm EEX}$,
EEX2= \Order{\alpha,L\alpha,L^2\alpha^2}$_{\rm EEX}$ and
EEX3= \Order{\alpha,L\alpha,L^2\alpha^2,L^3\alpha^3}$_{\rm EEX}$.  We
find $(1/2)$(EEX2-EEX1)$\sim 0.1\%$ for Z-exclusive cuts and $\sim
0.5\%$ for Z-inclusive cuts, and this we take as a physical precision
of the EEX2 and EEX3 QED matrix element (no ISR$\otimes$FSR interf.).
The difference $(1/2)$(EEX3-EEX2) is generally negligible $<0.1\%$ for
any cuts.

Having fortified our position on the physical precision of EEX, how do
we proceed to determine physical precision of CEEX matrix element?
We can compare with EEX2 or EEX3 and in this way we get a handle on
the \Order{L\alpha} ISR which is missing in EEX2 and
\Order{L^3\alpha^3} missing in CEEX (which is negligible, however).
The other possibility is to look into differences of
CEEX2= \Order{\alpha,L\alpha,L^2\alpha^2,L\alpha^2}$_{\rm CEEX}$ and
CEEX1= \Order{\alpha,L\alpha}$_{\rm CEEX}$.  We did both and we treat
the latter difference (1/2)(CEEX2$-$CEEX1) as our basic source of the
physical precision and the former CEEX2$-$EEX3 as an additional
cross-check.  In ref.~\cite{ceex2:2000} we have found
(1/2)(CEEX2$-$CEEX1) to be for both Z-exclusive and Z-inclusive
observables below 0.2\%.  The difference CEEX2$-$EEX3 is rather large,
up to 0.8\% for Z-inclusive cross section which suggests that the
proper inclusion of the \Order{L^1\alpha^2} ISR is important and we
need in fact the third independent calculation with the complete
\Order{L^1\alpha^2} ISR.  This however is available since long, from
BBVN~\cite{Berends:1988ab}.  In ref.~\cite{ceex2:2000} we compared
cross section and charge asymmetries from \KKMC\  with
semianalytical calculation based on ISR SF's based on
BBVN~\cite{Berends:1988ab}, with added complete \Order{L^3\alpha^3}
ISR and YFS exponentiation, essentially with the JSW formula of
ref.~\cite{jadach:1991}, upgraded with the corresponding FSR SF (in
the case FSR is switched on).  The above analytical formula is
implemented in the ${\cal KK}$sem code which is part of the ${\cal
KK}$MC package.  The results of the comparison of the ${\cal KK}$sem
code and \KKMC\  fully confirms our estimate of 0.2\% in the
cross section and in charge asymmetry, for Z-exclusive and Z-inclusive
cuts, excluding still ISR$\otimes$FSR from consideration.

We may summarize once again how solid is the \Order{L^1\alpha^2}
ISR: The two-loop \Order{L^1\alpha^2} component was already
triple-cross-checked at the time of BBVN~\cite{Berends:1988ab} work,
the two-loop \Order{L^1\alpha^2} component comes from at least
two independent sources~\cite{igarashi:1987,Berends:1988ab} and was
recently recalculated independently once again%
\footnote{ We thank Scott Yost for this valuable cross-check.},
while the two real photon emission exact massive matrix element was doubly
cross-checked with two independent codes.  

On top of that comes the cross check with \zf\  presented in this
report, which from the point of view of QED ISR and FSR (no
ISR$\otimes$FSR) is in the same class as BBVN, ${\cal KK}$sem while
\KKMC\  is rather independent because of the independent full
phase space evaluation, and the independent one-loop-one-real and
two-real-photon matrix elements.

Summarizing, the physical precision of 0.2\% in total cross section
and charge asymmetry due to QED ISR and FSR is estimated in a rather
solid and conservative way, using many independent codes/calculations,
with the triple cross-check being rather the rule than the exception.

\subsubsection{ Physical precision of QED ISR$\otimes$FSR}
The QED ISR$\otimes$FSR is characterized in the separate
section~\ref{subsec:IFI}
of this report so here only mention that
the effect of the QED ISR$\otimes$FSR is included in the exponentiated
form in our program with help of the new coherent exponentiation technique
based entirely on spin amplitudes.

The ISR$\otimes$FSR result of \KKMC\  were
debugged/tested first of all by comparing it
with the results of \Order{\alpha^1} KORALZ without exponentiation,
see ref.~\cite{ceex2:2000} where
we have found typical agreement $<0.2\%$ for both
Z-exclusive and Z-inclusive cuts.
The biggest discrepancy in ref.~\cite{ceex2:2000} was noticed to be
0.4\% for the charge asymmetry for a Z-inclusive cut and for the cross section
for certain values (far from experimental ones) for the Z-exclusive cut,
see also the section on ISR$\otimes$FSR in this report,
where we add more comparisons with \zf\  code.
Summarizing, the  inclusion of the ISR$\otimes$FSR 
does not worsen our total theoretical
error of $0.2\%$ estimate for the Z-exclusive cuts, 
while it makes it go to $0.4\%$ level for Z-inclusive cuts.
The new comparisons with \zf\  on ISR$\otimes$FSR  presented in
section \ref{ZFKKifi} in this report are consistent with the above estimate.

Note also that the most complete summary/discussion on 
the subject ISR$\otimes$FSR can
be found in the presentation of S.J. at June 1999 meeting of LEPEWG
(see transparencies on http://home.cern.ch/jadach).

\subsubsection{ Physical precision of electroweak corrections}
The uncertainty due to pure electroweak corrections is the same as of
DIZET, and can be determined for instance by playing with the user options
of DIZET, which are available for the user of \KKMC\ .  
We would like to stress, however, that some
physical/technical uncertainties in \zf\  are really related to the
way the EW corrections in \zf\  are combined with the QED part.  In
general, the way it is done in \KKMC\  is simpler and these
uncertainties are therefore reduced.

\subsubsection{ Tagged photons}
Precision is not less than  1 \% for observables
with a single photon tagged and  3 \% for observables with double
photon tagged.

\newpage

\subsection{Presentation of the program LABSMC}
\label{LABSMC}

\begin{tabbing}
1) \underline{Author:}\hspace{4cm} \= {\bf 
A.B. Arbuzov
} \\[.3cm]
2) \underline{Program:}    \> {\bf
LABSMC v.2.05, 5 May 2000
} \\[.3cm]
3)\underline{ Can be obtained from}  \> 
the author ({\tt arbuzov$\at$to.infn.it}) upon request 
\\[.3cm]
4) Reference to main description \>
\cite{Arbuzov:1999db}
\\[.3cm]
5) Reference to example \>
\cite{Arbuzov:1999ud}
\\[.3cm]
6) advertisement
\\[.3cm]
\end{tabbing}


Initially the semi--inclusive {\tt LABSMC} event generator was 
created~\cite{Arbuzov:1999db}
to simulate large--angle Bhabha scattering at energies
of about a few GeV's at electron positron colliders like
VEPP--2M and DA$\Phi$NE. The code included the
Born level matrix element, the complete set of ${\mathcal O}(\alpha)$ 
QED RC, and the higher order leading logarithmic RC by means
of the electron structure functions. The relevant set of formulae
can be found in Ref.~\cite{Arbuzov:1997pj}. The generation of events is
performed using an original algorithm, which combines advantages of
semi--analytical programs and Monte Carlo generators.

The structure of our event generator was described in detail in 
paper~\cite{Arbuzov:1999db}. The extension for LEP2 energies is done
by introducing electroweak (EW) contributions, such as $Z$-exchange, 
into the matrix elements. The third~\cite{Skrzypek:1992vk} and 
fourth~\cite{Arbuzov:1999cq} order leading logarithmic
photonic corrections were also included in the new version.
The version of the program under consideration is suited for
large--angle scattering. The small--angle version, which incorporates
some additional second--order corrections~\cite{Arbuzov:1997qd}, will be
described elsewhere. 

Starting from the ${\mathcal O}(\alpha^2)$ order
the emission of photons is treated semi--inclusively by means of structure
functions. Such photons are treated as effective particles, which
go at zero angles in respect to the relevant charged particles.
The conservation of 4-momenta is fulfilled for each generated event.
This feature of the program does not allow to generate realistic events
with two photons at large angles. 

The code contains:
\begin{itemize}

\item[$\bullet$]
the tree level electroweak Born cross section;

\item[$\bullet$]
the complete set of ${\mathcal O}(\alpha)$ QED radiative corrections (RC);

\item[$\bullet$]
vacuum polarization corrections by leptons, hadrons~\cite{Eidelman:1995ny}, 
and $W$-bosons;

\item[$\bullet$]
one--loop electroweak RC  according to 
Ref.~\cite{Bardin:1999yd}
by means of DIZET~\cite{Bardin:1989tq} package;

\item[$\bullet$]
higher order leading log photonic corrections by means of electron 
structure functions~\cite{Kuraev:1985hb,Skrzypek:1992vk,Arbuzov:1999cq};

\item[$\bullet$]
matrix element for radiative Bhabha scattering with both $\gamma$- 
and $Z$-exchange~\cite{Bohm:1988fg,Berends:1988jm}, 
vacuum polarization RC, and optionally ISR leading log RC
(with exponentiation according to Ref.~\cite{Kuraev:1985hb});

\item[$\bullet$] 
pair corrections in the ${\mathcal O}(\alpha^2L^2)$ leading log 
approximation~\cite{Arbuzov:1996vi,Arbuzov:1997vj}, including the 
two--photon (multi-peripheral) mechanism of pair production.

\end{itemize}

A number of flags, to be set by user, allows to switch between
different options and perform specific comparisons and investigations.
In particular one can switch to generation of only radiative events
with visible photons. That allows to avoid technical problems due
to low statistics in this case.

The inclusion of the third and fourth order LLA photonic corrections
allows not to use exponentiation. A simple estimate~\cite{Arbuzov:1999cq}
shows that the difference between the two treatments 
at LEP2 is negligible, while the exponentiation requires a specific event
generation procedure. 

{\tt LABSMC} is a {\tt FORTRAN} program. 
It works as follows. First, the code makes initialization and
reads flags and parameters from a list provided by user.
Then it performs an integration (in semi--analytical branch)
and generates events. The 4-momenta of generated particles
are to be analyzed or recorded in a user subroutine. 
A certain control of technical precision is
provided by comparison of the results from semi--analytical
and Monte Carlo branches. Note, that for a case of complicated cuts,
which can not be done in the semi--analytical branch,
one has to increase the number of generated events to reach
the ordered precision.

The accuracy of the code is defined by two main points:
technical precision (numerical precision in integrations,
errors due to limited statistics, possible bugs {\it etc.})
and the theoretical uncertainty.
Of course, one has than choose the proper, corresponding to his
concrete problem, set of flags and parameters.

The technical precision has to be checked and improved, if required,
by detailed tests and comparisons with results of other codes.
The theoretical uncertainty is defined by:
absence of complete set of ${\mathcal O}(\alpha^2L)$ corrections
(for photonic and pair corrections),
an uncertainty in definition of vacuum polarization, 
approximate description of hadronic pair production.
There was observed a discrepancy in the treatments of electroweak RC 
in \zf\  and ALIBABA~\cite{Beenakker:1991mb}.
The theoretical uncertainty of the code in description
of large--angle Bhabha scattering at LEP2 is
estimated now to be of about 0.3\%.
The corresponding uncertainty for radiative Bhabha scattering
with a visible photons is about 2\%.
But for the latter, we have an additional theoretical systematic 
uncertainty (about 1\%),
coming from non--standard radiative corrections~\cite{Arbuzov:1998ax}.

\subsubsection{Note about pair corrections in LABSMC}
\label{pairLAB}
In LABSMC there are included contributions due to pair 
production according to Ref.~\cite{Arbuzov:1996vi,Arbuzov:1997vj}.
The secondary hadronic pairs are estimated within the leading 
log approximation.

The double resonant (ZZ) contribution, in which both the primary and 
secondary pair are produced via virtual $Z$-bosons, 
is not taken into account. This contribution will be subtracted
from the experimental data by means of some Monte Carlo event generator.

The impact of the multi-peripheral (two--photon) mechanism of pair production
and the one of the singlet pairs can be analyzed by means of the program.
But the default option is to drop these contributions as in the
event generator as well as in the experimental data.

The corrections in per-mil are given in table \ref{LABS_sp} of section \ref{LABSMC_pair}.
The quantities there  were calculated in respect to the cross sections,
where all other types of RC have been already applied.
There is a simple dependence of the size of corrections
on the applied cuts. The most strong cuts on real emission are there, 
the most large (and negative) effect is coming out. 
The largest corrections are found for some idealized observables, where
also the final state corrections do give a lot.

As concerning the two--photon mechanism, there are visible contributions
only for a few event selections (see table \ref{LABS_MP} of section \ref{LABSMC_pair}). 
In the rest of ES the multi-peripheral reaction is cut away by 
the corresponding sets of conditions.
The only large correction to \citobs{IOpal3} is because of wide range of allowed
collinearity and a very low energy threshold for electrons (1 GeV).

The accuracy on the above numbers for pair corrections
can be estimated to be about 20\%,
which is mainly coming from the uncertainty in the  description of
secondary hadronic pairs.

\subsection{Error specifications of LABSMC}

The theoretical uncertainty of {\tt LABSMC} 
is estimated by the analysis of the following sources of errors. 

\begin{itemize}

\item{}
A considerably large amount of about 0.10\%
is coming from the hadronic contribution into vacuum polarization.

\item{}
Unknown ${\mathcal O}(\alpha^2L)$ photonic and pair corrections 
can give as large as 0.20\%. 
Note, that for small--angle Bhabha at LEP1 we had the corresponding
contribution of the order 0.15\%~\cite{Arbuzov:1997qd},
and so we can estimate the uncertainty, taking into account that the large
log $L$ in the large--angle kinematics is greater .

\item{}
The approximate treatment of hadronic pair corrections typically 
contributes by not more than 0.05\%, 
depending on the concrete event selection.  
For observables \citobs{IAleph3}, \citobs{IAleph4}, and \citobs{ILT4} 
we have more: about 0.1\%.

\item{}
Photonic corrections in high orders
${\mathcal O}(\alpha^3L^2,\alpha^5L^5,\ldots)$ are not calculated 
in the code, but they are really small (0.02\%).

\item{}
Uncertainties coming from the treatment 
electroweak constants and loop corrections
can give up to 0.2\%
for the case barrel angular acceptance.
For the case with endcaps we have lower contribution from $Z$-exchange,
and the error is less than 0.1\%. 

\end{itemize}

Taking into account the limited technical precision, we derive the
resulting uncertainty of the code for description 
of large--angle Bhabha scattering at LEP2 to be of the order 0.3\%.
As concerning radiative Bhabha with a photon tagged at large angles, 
the uncertainty is defined by missing ${\mathcal O}(\alpha)$ 
corrections. It can be estimated to be of about 2\%.

\newpage
\subsection{Presentation of the program  {\bf \tt grc$\nu\nu\gamma$}}

\vskip 1.5cm   
\begin{enumerate}
\item Authors: Y. Kurihara, J. Fujimoto, T. Ishikawa, Y. Shimizu, T. Munehisa
\item Program: {\bf \tt grc$\nu\nu\gamma$} v.1.0, 1999.08.20
\item Can be obtained from:  {\tt http://www-sc.kek.jp/minami/}
\item Reference to main description: hep-ph/9908422 to be appeared in CPC.
\item Reference to example discussion of the prediction and its systematic
 uncertainty: hep-ph/9908422 to be appeared in CPC.

\vskip 1cm   
{\bf \tt grc$\nu\nu\gamma$} is an event generator which combines
the exact matrix elements for
$e^+e^- \rightarrow \nu \bar{\nu} \gamma (\gamma)$,
produced by the {\tt GRACE} system\cite{GRACE}, with
{\tt QEDPS}\cite{QEDPS} for ISR.
The advantages of these packages are:
\vskip 1cm
\begin{itemize}
\item The exact matrix elements up to the double-photon emission,
including the $\nu_e$ process, are used.
Double-photon emission is practically sufficient for experimental
analysis.
\item {\tt QEDPS} keeps the complete kinematics for the emitted
photons and virtual electrons before collisions. It allows a more
flexible treatment of the ISR effects in avoiding the double-counting.
\item 
Besides the above-mentioned pure QED corrections,
{\bf \tt grc$\nu\nu\gamma$} equips with
another class of the electroweak higher order corrections.
There is a switch to choose it from the following three schemes;
1) the running coupling constant scheme: 
the coupling constant of
the $fermion$-$fermion$-$Z$ vertex, $g_{ffZ}$,
is determined by the evolution from zero momentum transfer
to the mass squared of the $\nu\bar\nu$ system, $q_Z^2$, which differs
from one event to another. It varies according to the
renormalization group equation (RGE).
2) $G_\mu$ scheme\cite{gmu}:
It is such that the weak couplings are determined
through the weak-mixing angle, $\sin{\theta_W}$, which is given by
\begin{eqnarray}
{\rm sin}^2 \theta_W=\frac{\pi \alpha(q^2)}{\sqrt{2} G_\mu M_W^2}
\frac{1}{1-\Delta r}, \nonumber
\end{eqnarray}
where $M_W$ being the $W$-boson mass and $G_\mu$ the muon decay
constant. 
3) on-shell scheme: the weak couplings are simply fixed by $M_W$ and
$M_Z$ though the on-shell relation,
$\sin^2{\theta}_W=1-{M_W^2 \over M_Z^2}$, where $M_Z$ is the mass of
the $Z$-boson.

\item
For the $\nu_\mu$  case
the total cross sections and the hard-photon distributions of
{\bf \tt grc$\nu\nu\gamma$} are compared with those by
the $\cal{O}$($\alpha$) calculations\cite{Berends:1988zz,1-loop,IN}, 
{\tt KORALZ}\cite{Jadach:1999tr} and {\tt NUNUGPV}\cite{mmnp:1999}.
The theoretical error uncertainty for the ISR corrections is under control at the
1\% level. The systematics of the $G_{\mu}$ scheme,
coming from the double energy scales($M_Z$,$\sqrt{s}$) 
involved in the reaction, is estimated to be around 1\%. 
The energy spectrum of the hard-photons is in a reasonable agreement
with {\tt KORALZ} and {\tt NUNUGPV} up to the double-photon emission.
\item
Concerning $\nu_e$ a similar comparison with {\tt NUNUGPV} has been
done, though in this case some programs were lacking complete ${\cal O}(\alpha)$.

\item
In the package
the anomalous coupling of the $W$-$W$-$\gamma$ vertex is implemented.
The program includes only those terms which conserve $C$ and $P$
invariance, derived from the following effective
Lagrangian\cite{hagiwara:1987}:
\begin{eqnarray*}
L_{eff} =-ie[(1+\Delta g_{1\gamma})(W^{\dag}_{\mu \nu} W^{\mu} -
                  W^{\dag \mu} W_{\mu \nu}) A^\nu
+(1+\Delta \kappa_\gamma) W^{\dag}_{\mu}W_{\nu}A^{\mu \nu} \\
+\frac{\lambda_\gamma}{M_W^2}W^{\dag}_{\lambda \mu}
           W^{\mu}_{\nu}A^{\lambda\nu}],
\end{eqnarray*}
where $W_{\mu\nu}=\partial_{\mu}W_{\nu}-\partial_{\nu}W_{\mu}$,
$A_{\mu\nu}=\partial_{\mu}A_{\nu}-\partial_{\nu}A_{\mu}$.
Here $\Delta g_{1\gamma}$,
$\Delta \kappa_\gamma$ and $\lambda_\gamma$ stand for the anomalous
coupling parameters which vanish in the standard model.
\end{itemize}
\end{enumerate}

\subsection{Error specifications of {\bf \tt grc$\nu\nu\gamma$}}

(1) lack of constant term

One of the intrinsic limitation of the parton shower method 
is lack of constant terms. It is known that the leading logarithmic
solution of the $DGLAP$ equation can reproduce the exact perturbative 
calculations at LL order except constant terms. For the simple $e^+ e^-$
annihilation processes, this effect, so called $K-factor$, 
is known to be 0.6
cannot be better than this accuracy. Though there is no exact estimation 
of the $K-factor$ for the neutrino pair-production with hard photon(s), 
we can expect the $K-factor$ for these processes is at the same order 
as the simple $e^+ e^-$ annihilation processes. Then we assign 
the systematic error of 0.6

(2) internal consistency

        In the $grc\nu \nu \gamma$, the hard photon is treated using
exact matrix elements and the soft photon(s) are treated using QEDPS.
It is not necessarily that a definition of the hard photon is the
same as those of visible photon given by the experimental requirements.
The final result must be independent of the dividing point between
hard and soft photons. we checked the stability of the cross sections
when the dividing points are varied within a reasonable range.
If the experimental requirement is so tight, for example, no additional
photons with small energy in very forward region is required,
the final result is sensitive for the definition of the soft photon.
We assign this dividing-point dependence as a systematic error.
This error is much depend on the experimental cuts.

(3) multi photon limitation

        In the $grc\nu \nu \gamma$, up to two visible photon can be treated.
For the experimental requirement as 'two or more photons ..', we give
the results with only two visible photons. The probability to observe
third photons is negligible small in general. We estimate the error
of this limitation is less than 1

\newpage
\subsection{Presentation of the program NUNUGPV}
\label{NUNUGPVdescr}
\vskip 10pt\noindent
{\large \underline{Authors}: 
\hskip 4.1cm {\bf G.~Montagna, M.~Moretti, O.~Nicrosini and F. Piccinini}}
\\
\\
\noindent
{\large \underline{Program}: \hskip 4cm 
{\bf NUNUGPV  v.2.0, July 1998}}
\\
\\
\noindent
{\large \underline{Can be obtained from}:} \hskip 1.8cm
     {\tt  http://www.pv.infn.it/~nicrosi/programs/nunugpv/ }
\vskip 5pt\noindent
The code {\tt NUNUGPV}~\cite{mmnp:1999,mnp:1996} has been developed to simulate events for the 
signatures single- and multi-photon final states plus missing energy 
in the Standard Model at LEP and beyond.  
\\
\vskip 8pt
\leftline{{\it Matrix elements}}
\noindent
In the program the exact matrix elements are implemented for the reactions
$$e^+ e^- \to \nu_i \; {\bar \nu}_i \; n\gamma\; ,$$ 
with $i = e, \mu, \tau$ and $n = 1,2,3$. \\
The matrix element for single-photon production has been computed by means of 
helicity amplitude techniques~\cite{gp:1984}, while the amplitudes for 
multi-photon final states are calculated using the numerical algorithm 
{\tt ALPHA}~\cite{cm:1995} for the automatic evaluation of tree-level scattering 
amplitudes. The contribution of the anomalous couplings $\Delta k_\gamma$ and
$\lambda_\gamma$ to the $WW\gamma$ vertex is included analytically in 
the matrix element for $e^+ e^- \to \nu_e \; {\bar \nu}_e \; \gamma$. 
Trilinear and quadrilinear anomalous gauge couplings
   for the processes with more than one photon in the final state 
   have been recently implemented.
As an option the program contains also the contribution of a massive neutrino 
with standard couplings to the $Z$ boson.
\\
\vskip 8pt
\leftline{{\it Radiative corrections}}
\noindent
The phenomenologically relevant Leading Log (LL) QED radiative corrections, 
due to initial state radiation (ISR), 
are implemented via the Structure Function (SF) formalism. 
Due to the presence of a visible photon in the kernel cross section, 
the inclusion of ISR requires particular care. In order to remove 
the effects of multiple counting due to the 
overlap of the phase spaces of pre-emission photons (described by the 
SF's) and kernel photons (described by the matrix element), 
the $p_t/p_L$ effects are included in the SF's according to ref.~\cite{mnp:1996}.
The generation of the angular variables at the level of the ISR gives the 
possibility of rejecting in the event sample those pre-emission 
photons above the minimum detection angle and threshold energy, 
thus avoiding ``overlapping effects''.  
According to such a procedure, the cross section with 
higher-order QED 
corrections can be calculated as follows (for the 
data sample of at least one photon)
\begin{eqnarray}
\sigma^{1\gamma (\gamma)} &=& \int d x_1 d x_2 
d c_{\gamma}^{(1)} d c_{\gamma}^{(2)} \, 
\tilde{D} (x_1, c_{\gamma}^{(1)}; s) \tilde{D} 
(x_2, c_{\gamma}^{(2)}; s)  
\Theta(cuts) \nonumber \\
&& 
\quad \quad \qquad \quad \quad 
\qquad \times \left( d\sigma^{1\gamma} + d\sigma^{2\gamma} 
+ d\sigma^{3\gamma} + ... \right) ,
\label{eq:sfpt1}
\end{eqnarray}
where $c_{\gamma} = \cos \vartheta_\gamma$ and $\tilde{D} (x, c_{\gamma}; s)$ 
is 
a proper combination
of the collinear SF $D(x, s)$ with an 
angular factor inspired by the leading behavior 
$1 / (p \cdot k)$~\cite{mnp:1996} of the pre-emission photons. 
According to eq.~(\ref{eq:sfpt1}), 
an ``equivalent'' photon is generated for each colliding lepton
and accepted as a higher-order ISR contribution  if:
\begin{itemize}
\item the energy of the equivalent photon is below the
 threshold for the
observed photon $E_{\gamma, min}$, for arbitrary angles; 
\item or the angle of the  equivalent photon is outside the angular 
acceptance for
the observed photons, for arbitrary energies. 
\end{itemize} 
Within the angular acceptance of the detected photon(s), 
the cross section is evaluated by summing the exact matrix 
elements for 
the processes $e^+ e^- \to \nu \bar\nu n \gamma$, $n=1,2,3$ 
($d\sigma^{1\gamma}, d\sigma^{2\gamma}, d\sigma^{3\gamma}$). 
\vskip 8pt
\noindent
By means of the above sketched formulation, the signatures that can be handled 
by the program are:
\begin{itemize}
\item exactly one(two) visible photon(s) plus undetected radiation;
\item at least one(two) visible photon(s) plus undetected radiation;
\item exactly three visible photons with QED corrections in the collinear 
      approximation. 
\end{itemize}
Some improvements for Linear Collider energies have been recently 
introduced. Predictions for the single-photon 
signature are possible for polarized electron/positron beams.
Simulation of beamsstrahlung can be performed by means 
of the {\tt circe} library~\cite{to:1997}.  \\
\noindent
Both integration and unweighted event generation modes are available.
More details on technical and theoretical features can be found in 
ref.~\cite{mmnp:1999}.
Concerning the LEP2 energy regime, 
the present theoretical accuracy of {\tt NUNUGPV} is at the per cent level, 
as due to missing ${\cal O}(\alpha)$ electroweak corrections.

\subsection{Error specifications of NUNUGPV}

As discussed in Section~\ref{NUNUGPVdescr} 
and in the relevant literature there quoted, the
main ingredients NUNUGPV is based upon are

\begin{itemize}

\item exact matrix elements for the kernel reaction $e^+ e^- \to \nu_i 
\bar\nu_i n\gamma$, with $i=e,\mu,\tau$ and $n=1,2,3$, computed either
analytically ($n=1$) or numerically ($n=2,3$); 

\item convolution of the kernel cross section by means of $p_t$-dependent
structure functions, in order to take into account the huge effect of 
initial-state radiation, while avoiding double counting  in the presence of 
tagged photons. 

\end{itemize}

The main source of theoretical error is missing non-log $O(\alpha)$ electroweak
corrections, which can be estimated to be of the order of $1 - 2$~\%. 
Pushing the theoretical accuracy at the 0.1~\% level would require supplementing
the present formulation by a full $O(\alpha)$ calculation, at present not
available.

\newpage
\subsection{Presentation  of the program \zf\  with electroweak library DIZET}
\label{ZFITTER}
\begin{tabbing}
\underline{Authors:}\hspace{4cm} \= {\bf 
D.~Bardin, 
P.~Christova, 
M.~Jack, 
L.~Kalinovskaya,} 
\\
\> 
{\bf
A.~Olchevski, 
S.~Riemann, 
T.~Riemann
} 
\\[.3cm]
\underline{Program:}    \> 
{\bf ZFITTER v.6.21 (26 July 1999)
} \\[.3cm]
\underline{Can be obtained from:}  
\> 
 \tt{http://www.ifh.de/$\sim$riemann/Zfitter/zf.html}
\\
\> 
 {\tt /afs/cern.ch/user/b/bardindy/public/ZF6\_21}
\\[.3cm]
\underline{Reference to main description:} \>
 \cite{Bardin:1999yd-orig} 
\\[.3cm]
\underline{References to examples:} \>
\cite{Bohm:1989pb,Bardin:1989qr,Boudjema:1996qg,Bardin:1997xq,Bardin:1998nm%
,Bardin:1999gt,Christova:1999gh,Jack:1999af,Christova:2000zu}
\\[.3cm]
\underline{Program development:}  \>
The package is permanently updated, user requests are welcome;
\\ \> last update is v.6.30 (xx March 2000)
\\[.3cm]
\end{tabbing}

\zf\ is a Fortran program, based on a semi-analytical
approach to fermion pair production in \ee\ annihilation at a wide
range of centre-of-mass energies, including LEP1 and LEP2 energies.
The main body of the program relies on the analytical results presented in 
\cite{Bardin:1991fu,Bardin:1991de,Christova:1999cc}
for the QED part and in 
\cite{Bardin:1980fe,Bardin:1982sv,Akhundov:1986fc,Bardin:1986fi,%
Bardin:1989di,Bardin:1989tq}
for the electroweak physics part.
Some of the formulae used may be found only in 
 \cite{Bardin:1999yd-orig}. 

\zf\ version v.6.21 was the last one intended for the use at LEP1 
energies. The description of this subsection is mostly limited to this version.
During the 1999--2000 LEP2 Workshop there was a development which is briefly
summarized in subsection~\ref{zf_development}.

The calculation of realistic observables with potential account of
complete ${\cal O}(\alpha)$ QED and electroweak corrections plus soft photon 
exponentiation plus some higher order contributions 
is made possible with several calculational chains:
\begin{itemize}
\item   Born cross-sections;
\item   a fast option: cut on $s'$ or combined cuts on collinearity $\xi$ 
        and minimal energy $E_{\min}$ of the fermions
        for $\sigma_{\sss{T,FB}}$;
\item   cut on $s'$ (or on $\xi, E_{\min}$)
        for $d\sigma / d\cos\vartheta$;  for $\sigma_{\sss{T,FB}}$
        additional cut on the production angle of 
        {\em anti-fermions} ($\cos\vartheta$).
\end{itemize}
The scattering angle of {\em fermions}   remains unrestricted if the other
cut(s) do not impose an implicit restriction.

Numerical integrations are at most one-dimensional and performed with the 
Simpson method
\cite{Silin:19xy,Sedykh:19xy}.
This makes the code so fast and guarantees any practically needed numerical 
precision.

\zf\ calculates:
\begin{itemize}
\item
$\Delta r$ -- the Standard Model corrections to $G_{\mu}$;
\item $M_W$ -- the $W$ boson mass from $M_Z, M_H$, and fermion masses, and
  $\Delta r$; 
\item
$\Gamma_{Z,W} = \sum_f \Gamma_f$ --  total and partial $Z$ and $W$ boson
decay widths; 
\item
${d\sigma}/{d\cos\vartheta}$ -- differential cross-sections; 
\item
$\sigma_T$ -- total cross-sections;
\item
$A_{FB}$ -- forward-backward asymmetries;
\item
$A_{LR}$ -- left-right asymmetries;
\item
$A_{pol}, A_{FB}^{pol}$ -- final state polarization effects for $\tau$
leptons;  
\end{itemize}
Various interfaces allow fits to the experimental data to be performed with 
different sets of free 
parameters.
There are two options to parameterize the $Z$ boson propagator 
\cite{Bardin:1988xt}
(see also
\cite{Riemann:1997tj1,Bohm:2000jw} and the many references therein).

\zf\ uses pieces of code from other authors
(\cite{Silin:19xy,Sedykh:19xy,Matsuura:1987,Jegerlehner:1995ZZ,Degrassi:1996ZZ,%
Arbuzov:1999uq1,Kniehl:1990yc1}).
We find it important to mention explicitly that the programming of 
\zf\ accumulates the efforts of many theoreticians, whose work went into 
the code either as default programming or as options to be chosen by  
many flags.
A hopefully complete list (derived from \cite{Bardin:1999yd-orig}) comprises 
quite a few references for photonic radiative corrections
\cite{%
Bonneau:1971mk,%
Greco:1975rm,Greco:1975ke,Greco:1980,%
Kuraev:1985hb,Kniehl:1988id,Berends:1988ab,Bilenkii:1989zg,%
Bardin:1989qr,Beenakker:1989km,%
Montagna:1989at,Montagna:1991ku,Jadach:1992aa,Skrzypek:1992vk,%
Montagna:1997jv,Arbuzov:1999uq}
and radiative corrections contributing to the effective Born cross section 
%
\cite{%
Kallen:1955ks,Berman:1958,Kinoshita:1959ru,Kallen:1968,%
Dine:1979qh,Chetyrkin:1979bj,Celmaster:1980xr,%
Mann:1984dv,vanderBij:1984bw,vanderBij:1984aj,Djouadi:1987gn,%
vanderBij:1987hy,Djouadi:1988di,%
Bardin:1989aa,%
Gorishnii:1991hw,Kataev:1992dg,Arbuzov:1992pr,Fleischer:1992fq,%
Barbieri:1993ra,%
Avdeev:1994db,Chetyrkin:1994js3,Degrassi:1996mg,%
Degrassi:1997ps,%
Steinhauser:1998rq,vanRitbergen:1998hn,vanRitbergen:1998yd,Harlander:1998zb,%
Degrassi:1999jd}.
This is a feature of \zf\ which makes it very flexible for applications, but 
also for comparisons with other codes and checks of technical precisions in
program development phases. 
For a systematic presentation of the interplay of the many radiative 
corrections treated we refer also to \cite{Bardin:1999yd-orig}
and to  \cite{Bardin:1999ak}.

\zf\ is used optionally by other packages, among them are  
SMATASY \cite{Leike:1991pq,Riemann:1992gv,Kirsch:1995cf1}, 
ZEFIT,  \cite{Riemann:1997aa}.
Its electroweak library DIZET is used in
KORALZ, \cite{Jadach:1999tr},  
\KKMC,  \cite{kkcpc:1999}, 
BHAGENE, \cite{Field:1996dk}, and other programs like 
HECTOR, \cite{Arbuzov:1995id} for the study of $ep$ scattering.

\bigskip

\underline{QED initial--final interference:} 

The exponentiation~\cite{ABAprep} of initial--final interference (IFI) 
photonic corrections is implemented in \zf\ . The exponentiation
is done according the procedure developed in Ref.~\cite{Kuraev:1985hb}.
The base for the construction
is the general Yennie--Frautschi--Suura theorem~\cite{Yennie:1961ad}. 
The resulting formulae are close to the ones of Ref.~\cite{Greco:1990gk},
but the special treatment around the $Z$-peak is not
included in the program now. In the code the IFI option is governed 
by the flag {\tt INTF}. The effect of the IFI exponentiation was found to 
be important, especially for forward--backward asymmetry.

\bigskip

The Fortran package DIZET is part of the \zf\ distribution.
\\
It can be used in a stand-alone mode and is regularly used by other 
programs. 

On default, DIZET allows the following calculations:
\begin{itemize}
\item
by call of subroutine DIZET: $W$ mass and width, $Z$ and $W$ widths;
\item
by call of subroutine ROKANC: four weak NC form factors, running 
electromagnetic 
 and strong couplings needed for the composition of effective NC Born 
cross sections for the production of massless fermions (however, the mass
of the top quark appearing in the virtual state of the one-loop diagrams
for the process $e^+ e^- \to b \bar{b}$ is not ignored); 
\item
by call of subroutine RHOCC: the corresponding form factors and running 
strong coupling for the composition of effective CC Born cross sections. 
\end{itemize}
If needed, the form factors may be made to contain the contributions from 
$WW$ and $ZZ$ box diagrams thus ensuring (over a larger energy range than 
LEP~1) the correct kinematic behavior and gauge invariance.

\subsubsection{Pair corrections in \zf\ }
\label{pairZF}
One of particular contributions to the process of electron--positron
annihilation is the radiation of secondary pairs. 
In comparison with 
the photon radiation, it is relatively small, because it appears
only starting from the ${\mathcal O}(\alpha^2)$ order. Nevertheless,
the total effect of pair production could reach dozen per-mil and
should be taken into account in the data analysis. 
The secondary pair can be produced via a virtual photon or $Z$-boson.
The latter case is supposed to be subtracted from the experimental
date by means of some Monte Carlo event generator. 
(The $Z$ boson mediated
secondary pair production was also studied with GENTLE/4fan v.2.11,
see the description of results in subsection~\ref{subsec:gentle}.)
\\
\vskip 8pt
\leftline{\bf Lowest order pair corrections}
\noindent
The complete second order calculation for $e^+e^-$ and $\mu^+\mu^-$
initial state pairs was performed in Ref.~\cite{Berends:1988ab}. 
The contribution of hadronic and leptonic pairs (excluding electrons) 
was considered in paper~\cite{Kniehl:1988id}. 

The effect of secondary pair production in the final state was
calculated in Ref.~\cite{Hoang:1995ht}. It is worth to mention, that
the final state pair correction should be realized in a multiplicative way:
\begin{eqnarray} \label{mult}
\sigma = \sigma_{\mathrm{Born}}(1+\delta_\gamma)(1+\delta_{\mathrm{FSP}}),
\end{eqnarray}
where $\delta_\gamma$ stands for the initial state (IS) photonic correction,
and $\delta_{\mathrm{FSP}}$ give the final state (FS) pair one. 
At LEP2 energies,
when the radiative return to the $Z$-peak is allowed, we have very large
values of $\delta_\gamma$, and the multiplicative treatment provides
a correct counting of the simultaneous emission of IS photons 
and FS pairs. A cut on the invariant mass of the FS secondary pair
is allowed by setting parameter {\tt PCUT}.  

The pair contribution to the corrected cross section is presented
as the integral of the Born cross section
with the so--called pair radiator:
\begin{eqnarray} \label{integ}
\dd\sigma^{\mathrm{pair}} = \int\limits^1_{z_{\mathrm{min}}}
\dd z\; \tilde\sigma(zs) H(z)
= \sigma(s) (H_{\Delta}+H_{\mathrm{FSP}}) 
+ \int\limits^{1-\Delta}_{z_{\mathrm{min}}}
\dd z \tilde\sigma(zs) H_{\Theta}(z).
\end{eqnarray}
Here $H_{\Delta}$ represents the impact of virtual and soft pairs;
$H_{\mathrm{FSP}}$ stands for the final state pairs.
$\Delta$ is a {\em soft-hard separator} $(\Delta \ll 1)$, numerical
results should not depend on its value.

The singlet channel contribution and the interference of the
singlet and non--singlet channels  
are taken from Ref.~\cite{Berends:1988ab}. They can be called 
from \zf\  optionally (according to the {\tt IPSC} flag value).

A simple estimate of the interference between the ISR and FSR pairs
can be done: we can take the initial--final photon interference
multiplied by the conversion factor $(\alpha/(3\pi))\ln(s/m_e^2)$.
The smallness of the photonic interference and the additional
factor provide us the possibility to neglect the initial--final
pair interference completely.
\\
\vskip 8pt
\leftline{\bf Pair production in higher orders} 
\noindent
It was observed that the $O(\alpha^2)$ approximation is not enough
to provide the desirable precision. Really the interplay of the
initial state photon and pair radiation is very important.
So, one should consider higher orders. The first exponentiated
formula for pair production was suggested in Ref.~\cite{Kuraev:1985hb}.
The process of one pair production was supplied by emission
of arbitrary number of soft photons. This formula gives a
good approximation for leading logarithmic corrections 
close to the $Z$-peak. But it does not include the important
next--to--leading terms, and even the known third order leading
logs are not reproduced completely.

In Ref.~\cite{Jadach:1992aa} a
phenomenological formula for simultaneous exponentiation of
photonic and pair radiation was proposed. The correspondence
of the exponentiated formula to the perturbative results was
shown there for the case of real hard radiation. 
Nevertheless, the structure 
of the radiator function, suggested in~\cite{Jadach:1992aa}, does not
allow to check the correspondence for soft and virtual part
of the corrections analytically.

An alternative treatment of the higher order corrections due
to pair production was suggested in Ref.~\cite{Arbuzov:1999uq}.
In order to account the most
important part of the sub--leading corrections we consider the
convolution of the ${\mathcal O}(\alpha^2)$  
pair radiator with
the ordinary ${\mathcal O}(\alpha)$ photonic radiator, proportional
to the $P^{(1)}$ splitting function. In this way we receive 
the main part of the ${\mathcal O}(\alpha^3)$ leading logs,
proportional to $P^{(2)}$, and the sub--leading terms enhanced 
by $\ln(1-z)/(1-z)$, like $L^2\ln(1-z)/(1-z)$ and $L\ln^2(1-z)/(1-z)$.
Note that the convolution as well as exponentiation can not
give the correct complete sub--leading formula.
In fact the convolution 
gives a part of sub-leading terms coming from the kinematics, where
both the pair and the photon are emitted collinearly, while there
are other sources for the corrections, like, for instance, 
emission of a collinear pair and a large--angle photon.
But we suppose, that the main terms with enhancements are reproduced
correctly, that follows from the general experience in leading
log calculations. Note that the same background is under 
the exponentiation of such terms. For the case of pure
photonic radiation this was checked by direct perturbative calculations.

We checked that for real hard emission there is a agreement
between the most important terms in the third order contribution to 
$H_{\Theta}(z)$ and
the corresponding terms in expansion of the exponentiated formula
from Ref.~\cite{Jadach:1992aa}. Such a correspondence between the 
exponentiation and convolution procedures is well known also 
in the case of pure photonic radiation.

In the same way we derived the expressions both for leptonic and hadronic 
pairs. In contrast with Ref.~\cite{Jadach:1992aa} we extended the hadronic
pair contribution to the third order by means of convolution which takes into account the dynamical interplay between pairs and 
photons, when they are emitted at the same point, 
rather than by a static coefficient. 

The leading logs, which were not
reproduced by the convolution were supplied from Ref.~\cite{Arbuzov:1999uq}.
We estimated also at the fourth order contribution
by means of the leading logs (non--singlet channel only):
\begin{eqnarray} \label{quadr}
\dd\sigma_{e}^{(4)} = \int\dd z \tilde\sigma(zs) 
\biggl(\frac{\alpha}{2\pi}(L_e-1)\biggr)^4
\biggl[ 
\frac{1}{12}P^{(3)}(z)
+ \frac{11}{216}P^{(2)}(z) 
+ \frac{1}{108}P^{(1)}(z) \biggr].
\end{eqnarray}
In the ${\mathcal O}(\alpha^4)$ 
we keep only the leading logarithmic formula~(\ref{quadr}) 
for non--singlet electron pairs.

A good numerical agreement was observed in the treatment of higher order
leptonic ISR pairs by means of the convolution~\cite{Arbuzov:1999uq} and 
exponentiation~\cite{Jadach:1992aa} (see Table~2 in~\cite{Arbuzov:1999uq}).
The exponentiated treatment is implemented in \zf\  also (called by
setting {\tt ISPP=4}). The agreement with the exponentiated representation
from Ref.~\cite{Kuraev:1985hb} is not so good at LEP2 energies.
\\
\vskip 8pt
\leftline{\bf Numerical illustrations}
\noindent
In Table~1 we present the results for different contributions.
The value of correction due to pairs is defined in respect to
the cross section for annihilation into hadrons with pure
photonic corrections taken into account. The cut--off on both
pair and photonic corrections is equal: $z_{\mathrm{min}}=0.01 $
and $0.7225$, $s'>z_{\mathrm{min}}\cdot s$. 
In the FSR column we show the sum of leptonic and hadronic 
final state pair corrections ({\tt PCUT}=0.99).
In the last column the sum of ISR and FSR
pairs is given without the contribution of singlet pairs. 
Centre--of--mass energy is 200~GeV.

\begin{table}[ht]
\begin{centering}
\caption{Different contributions to $\delta$.}
\begin{tabular}[]{|c|c|c|c|c|c|c|c|} \hline
 & \multicolumn{5}{c|}{ISR pairs} & FSR pairs & sum  \\ \hline
 & $e$(NS) & $e$(NS+sing.) & $\mu$ & $\tau$ & hadr. &  & \\ \hline
 & \multicolumn{7}{c|}{$z_{\mathrm{min}}=0.01$} \\ \hline
${\mathcal O}(\alpha^2)$ & 6.41 & 42.00 & 1.99 & 0.67
 & 5.49 & 0.06 & 14.62 \\ 
${\mathcal O}(\alpha^3)$ & 7.28 & 42.86 & 2.19 & 0.72
 & 6.09 & 0.06 & 16.34 \\ 
${\mathcal O}(\alpha^4)$ & 7.24 & 42.82 & 2.19 & 0.72 
 & 6.09 & 0.06 & 16.30 \\ \hline
 & \multicolumn{7}{c|}{$z_{\mathrm{min}}=0.7225$} \\ \hline
${\mathcal O}(\alpha^2)$ & $-$0.38 & $-$0.40 & $-$0.11 & $-$0.03
 & $-$0.28 & $-$0.29 & $-$1.08 \\ 
${\mathcal O}(\alpha^3)$ & $-$0.56 & $-$0.59 & $-$0.17 & $-$0.05
 & $-$0.21 & $-$0.30 & $-$1.28 \\ 
${\mathcal O}(\alpha^4)$ & $-$0.53 & $-$0.56 & $-$0.17 & $-$0.05
 & $-$0.21 & $-$0.30 & $-$1.25 \\ \hline
\end{tabular}
\end{centering}
\end{table}

As could be seen from the Table, the contribution of singlet
pair production becomes important only for small values of 
$z_{\mathrm{min}}$. In data analysis at LEP, such events 
are supposed to be extracted from the data 
together with the two--photon process
$e^+e^-\to e^+e^-+{\mathrm{hadrons}}$. We emphasize, that
the procedure should be accurate and well understood, because 
in fact the events
with singlet pairs and multiperiferical production have
quite different signatures in the detector.
At LEP2 energies the contribution of singlet pairs becomes
really important, if the returning to the $Z$-peak is allowed
(for $z_{\mathrm{min}}\la 0.25$).

To estimate the uncertainty of our results we look at the 
relative size of different contributions and at the comparison
with the exponentiated formulae. The main source of the
uncertainty is the approximate treatment of the hadronic pairs.
Another indefiniteness is coming from the sub--sub--leading terms
of the third order, which can be received neither by
convolution nor by exponentiation, and from the 
fourth order correction. Our rough estimate for the theoretical
uncertainty due to pair production in description of 
electron--positron annihilation is 0.02 \% 
for without returning to the $Z$-peak. 
For the returning to the peak at LEP2 we estimate the uncertainty
to be at the level of 0.1\%.

\subsubsection{\zf\  development after v.6.21\label{zf_development} }

There was a certain development of \zf\  after version 6.21.
On 13 December 1999 we released \zf\ v.6.23
with an improved treatment of the second order corrections to angular 
distributions and $A_{FB}$.
The implementation relies on work done by A.B. Arbuzov and will be
described in an extended version of Ref.~\cite{Arbuzov:1999uq}.

For this workshop we have created \zf\ v.6.30 , which should be 
the last version for LEP2. It contains several new important user options. 

A new option governed by a new flag {\tt FUNA} is implemented, with:\\
{\tt FUNA}=0 -- old treatment,\\
{\tt FUNA}=1 -- new treatment.\\  
This is a new treatment of the second order ISR QED corrections,
in the presence of angular acceptance cuts {\tt ANG0}, {\tt ANG1},
based on a new calculation by A. Arbuzov (to appear as hep-ph report).
It is compatible with the use of {\tt ICUT=1,2,3}. 

The meaning of flag {\tt INTF} is extended in order to accommodate the new 
implementation of an exponentiation of IFI QED corrections, also
realized by A. Arbuzov (also to appear as hep-ph report):\\
{\tt INTF}=0,1 -- old options,     \\
{\tt INTF}=2 --  exponentiated IFI.

\vskip 2pt

Further, final state pair production corrections are implemented (A. Arbuzov).
The option is governed by a new flag:     \\
{\tt FSPP}=0 -- without FSR pairs,        \\
{\tt FSPP}=1 -- with FSR pairs, additive, \\
{\tt FSPP}=2 -- with FSR pairs, multiplicative. \\
For the FSPP corrections, the cut on the invariant mass of the secondary pair
is accessible.
In order to accommodate this cut value, the variable SIPP of the \\
{\tt SUBROUTINE ZUCUTS(INDF,ICUT,ACOL,EMIN,S$\_$PR,ANG0,ANG1,SIPP)}\\
is now used.                                                          
Therefore, the meaning of the variable SIPP has been changed.         
It has nothing to do with cutting of ISPP; there is no possibility to cut secondary 
pairs for ISPP, where the primary pair invariant mass cut should be equal to 
{\tt S$\_$PR}.

Finally, the new value of flag {\tt IPTO}=--1 allows to calculate pure 
virtual pair contributions separately.

\zf\  v.6.30 should be used together with DIZET v.6.23.
Two bugs are fixed in DIZET v.6.23.
A bug in the calculation of $\Gamma_{\sss{W}}$ is fixed (resulting in a 0.3\%
shift), and another one in the calculation of running $\alpha_{em}$
(of no numerical importance).

Further, an option to fit $V_{tb}$ is implemented to DIZET
(D.~Bardin, L.~Kalinovskaya, A.~Olshevsky, March 2000).
For this, a main program (interface) {\tt zwidthtb6\_30.f} 
has to be used together with a standaside DIZET version 6.30;
the argument list of that DIZET is changed to accommodate this possibility.

Some more small changes were implemented during this workshop in the result 
of tuned comparison with \KKMC\ . The range of variation of two flags
was extended.

New value {\tt WEAK=2} allows to switch off some tiny second order EWRC which
do not propagate via DIZET and therefore can't be taken into account by 
the other codes which use only DIZET.
It was proved that the numerical influence of these terms at LEP2 energies is 
one order of magnitude less than the typical precision tag.

New value {\tt CONV=-1} accomodates the choice $\alpha_{em}(0)$ for the $\gamma$
exchange amplitude allowing the calculation of the ``pure'' Born observables,
that was used for cross-checks of ISR QED convolution.

The interested reader may find further details on recent program
developments at \\
{\tt /afs/cern.ch/user/b/bardindy/public/ZF6\_30/}\\
and at \\
{\tt http://www.ifh.de/$\sim$riemann/Zfitter/}.

\bigskip

The most important conclusion which emerged from the tuned comparison with \KKMC\ 
is that at LEP2 energies it is not possible anymore to relay on a simplified treatment
of EW boxes realized in {\tt ZUTHSM} branch of \zf\ . EW boxes should be 
considered as a part of EW form factors and due to their angular dependence the only
way is to access them via {\tt ZUATSM} branch of \zf\  which was already
accessible in v.6.21 and was not specially updated during this workshop.
However, one should emphasize that the way of using of \zf\  at LEP1 fails
at LEP2 energies completely. In particular, one should use {\tt CONV=2} option 
allowing for {\em running} of EW form factors under the ISR convolution integral.
For more detail see section \ref{TunedZFKK}.
 
\bigskip

\underline{Statement on the precision and the systematic errors:} 

See the other parts of this report, especially sections 
\ref{theory_err}, section \ref{sec:dedicated}, and 
section \ref{conclusion}. 

\bigskip

\underline{Statement on limitations}: 

\zf\ should not be used for precision calculations of Bhabha cross sections.
The corresponding QED corrections have to be recalculated.
The effective Born approximation for Bhabha scattering is fixed to LEP~1 
kinematics.
It is relatively easy to improve the latter, but this has to be done yet.

\zf\ should be used below $t\bar{t}$ threshold. The implementation of the
channel $e^{+}e^{-}\to t\bar{t}$ is underway.

\subsection{Error specifications of \zf\ }


One should distinguish two main classes of sourses of theoretical errors.
First are, so-called {\em parametric uncertainties, PU's}, which are 
trivial: propagation of uncertainties of INPUT parameters results 
for an uncertainties of the predictions.
Here we present a study of PU's done with \zf\ .

It seems reasonable to assume that the only PU's which are worth studying,
are those due to uncertainties in:
\begin{itemize}
\item the running QED coupling $\alpha(s)$, due to errors in 
$\Delta\alpha^{5}_{had}$ for which we use
\begin{equation}
\Delta\alpha^{5}_{had}=0.027782\pm 0.000254;
\end{equation}
\item the pole masses of $b$ and $c$ quarks, for which we adopt
\begin{equation}
\mb = 4.70\pm 0.15\mbox{GeV}, \qquad\mb = 1.50\pm 0.25\mbox{GeV},
\end{equation}
\item the pole masses of the top quark, for which we take, PDG'98 value
\begin{equation}
\mt = 173.8\pm 5.2\mbox{GeV}, 
\end{equation}
\item Higgs boson mass, deserving more explanations.
\end{itemize}
Conventionally, we use in this report $\mh=120$ GeV as a preferred value.
For its lower limit it is reasonably to take $\mh\approx100$ GeV as the
present lower limit established from direct searches at LEP1. 
For upper limit we scanned the interval $\mh=125-200$ GeV, because
the total (hadronic) cross section shows up a non-monotonic behavior 
as a function of Higgs mass with a maximum at some value from this
interval. Parametric uncertainties due to $\mh$ variation are non-symmetric,
since the value $\mh=120$ GeV is chosen as the preferred value.

For the idealized quark observables we found the following largest 
variations (in per mil),
when we varied five above mentioned input parameters within
indicated limits: 
\[
\begin{array}{|l|l|}
\Delta\alpha^{5}_{had}\quad\mbox{only}\qquad&\qquad\pm 0.05 \\ 
\Delta\alpha^{5}_{had},\mb,\mc \quad \mbox{simultaneously}
\qquad&\qquad\pm 0.07\;\mbox{with negligible contribution from}\;\mc\\
\mt\quad\mbox{only}\qquad&\qquad\pm 1 \\
\mh\quad\mbox{only}\qquad&\qquad +0.50\div -0.85 \\
\end{array}
\]

One sees that parametric uncertainties
due to $\Delta\alpha^{5}_{had},\mb,\mc,\mh,\mt$, 
do not exceed 1 per mil, therefore the measurement of the total 
hadronic cross section at LEP2 will not contribute to further improvement 
of top mass and of the upper limit for the Higgs boson mass.

Similar study for muonic idealized observables is summarized in two following
tables.\\
Total cross-section (in per mil):
\[
\begin{array}{|l|l|}
\Delta\alpha^{5}_{had},\mb,\mc \quad \mbox{simultaneously}
\qquad&\qquad\pm 0.4\;\mbox{with negligible contribution from}\;\mc\\
\mt\quad\mbox{only}\qquad&\qquad\pm 0.45 \\
\mh\quad\mbox{only}\qquad&\qquad +0.40\div -0.60 \\
\end{array}
\]
Forward--backward asymmetry (in absolute units $10^{-3}$): 
\[
\begin{array}{|l|l|}
\Delta\alpha^{5}_{had},\mb,\mc \quad \mbox{simultaneously}
\qquad&\qquad\pm 0.1\;\mbox{with negligible contribution from}\;\mc\\
\mt\quad\mbox{only}\qquad&\qquad\pm 0.15 \\
\mh\quad\mbox{only}\qquad&\qquad +0.15\div -0.21 \\
\end{array}
\]
For taus, one should expect similar estimates.

Given precision tag of LEP2 measurements, one shouldn't expect that they
will add any improvement to our knowledge of input parameters.   




\newpage
\subsection{Program GENTLE: tool for the 2-fermion physics}
\label{GENTLE1}
\begin{tabbing}
1) \underline{Author:}\hspace{2.5cm} \= {\bf 
 Dmitri Bardin,    
 Jochen Biebel,    
 Michail Bilenky, 
 Dietrich Lehner,} 
\\ \> {\bf Arnd Leike,
 Alexander Olshevsky,
 Tord Riemann }
\\[.3cm]
2) \underline{Program:}  \> {\bf GENTLE/4fan v.2.11,  June 2000} 
\\[.3cm]
3)\underline{ Can be obtained from}  
\> {\tt /afs/cern.ch/user/b/bardindy/public/Gentle2\_11/}
\\
\> {\tt http://www.ifh.de/$\sim$riemann/doc/Gentle/gentle.html}

\end{tabbing}

In this section we describe a new version  
of the code {(\tt GENTLE/4fan v.2.11)}, where
several features to extract effects of  pair production in $2f$ processes 
have been added.
This version is an update of the original {\tt GENTLE/4fan v.2.00}
published in 
ref.\cite{Bardin:1996g} (see also \cite{Grunewald:2000ju}).
The results presented in this section use intensively the approach of
ref. \cite{Bardin:1995a}.
It was extended for a calculation of low-invariant-mass 
fermionic pairs of the NC24 family. We remind first of all Feynman diagrams 
describing this family. The NC24 process is a $4f$ process
\begin{equation}
e^{+}e^{-}\to f_{1}\bar{f}_{1}f_{2}\bar{f}_{2}
\label{4f_process}
\end{equation}
where $f_{1}\neq f_{2}\neq e$.
There are eight diagrams of conversion type, or NC08 sub-set (Fig. \ref{eightone}):
\begin{figure}[h]
\vspace*{-6mm}
\begin{eqnarray*}
\begin{array}{cccc}
\quad\;\vcenter{\hbox{
  \begin{picture}(150,100)(0,0)
    \ArrowLine(50,75)(25,80)        \Text(10,80)[lc]{$e^{+}$}
    \ArrowLine(50,25)(50,75)        \Text(40,50)[lc]{$ e$}
    \ArrowLine(25,20)(50,25)        \Text(10,20)[lc]{$e^{-}$}
    \Photon(50,75)(80,80){2}{7}     \Text(65,72.5)[ct]{$\gamma,Z$}
    \Photon(50,25)(80,20){2}{7}     \Text(65,27.5)[cb]{$\gamma,Z$}
    \ArrowLine(105,65)(80,80)       \Text(120,65)[rc]{$\bar{f}_{1}$}
    \ArrowLine(80,80)(105,95)       \Text(120,95)[rc]{$f_{1}$}
    \ArrowLine(105,5)(80,20)        \Text(120,5)[rc]{$\bar{f}_{2}$}
    \ArrowLine(80,20)(105,35)       \Text(120,35)[rc]{$f_{2}$}
    \Vertex(50,75){2.5}
    \Vertex(50,25){2.5}
    \Vertex(80,80){2.5}
    \Vertex(80,20){2.5}
  \end{picture}}}
&\quad+\qquad
\vcenter{\hbox{
  \begin{picture}(150,100)(0,0)
    \ArrowLine(50,75)(25,80)        \Text(10,80)[lc]{$e^{+}$}
    \ArrowLine(50,25)(50,75)        \Text(40,50)[lc]{$ e$}
    \ArrowLine(25,20)(50,25)        \Text(10,20)[lc]{$e^{-}$}
    \Photon(50,75)(80,80){2}{7}     \Text(65,72.5)[ct]{$\gamma,Z$}
    \Photon(50,25)(80,20){2}{7}     \Text(65,27.5)[cb]{$\gamma,Z$}
    \ArrowLine(105,65)(80,80)       \Text(120,65)[rc]{$\bar{f}_{2}$}
    \ArrowLine(80,80)(105,95)       \Text(120,95)[rc]{$f_{2}$}
    \ArrowLine(105,5)(80,20)        \Text(120,5)[rc]{$\bar{f}_{1}$}
    \ArrowLine(80,20)(105,35)       \Text(120,35)[rc]{$f_{1}$}
    \Vertex(50,75){2.5}
    \Vertex(50,25){2.5}
    \Vertex(80,80){2.5}
    \Vertex(80,20){2.5}
  \end{picture}}}
&\quad\mbox{\large{\bf IPPS=1}}
\end{array}
\end{eqnarray*}
\vspace*{-5mm}
\caption[]{The NC08 sub-family of diagrams.\label{eightone}}
\vspace*{-1mm}
\end{figure}
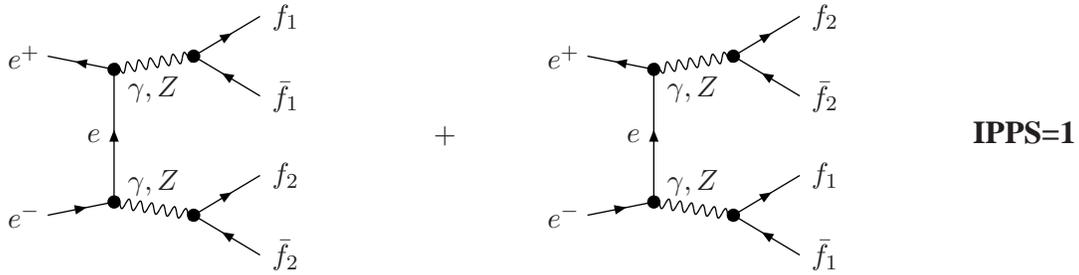

\noindent     
Next, there are eight pair-production-type diagrams(Fig. \ref{eighttwo}):

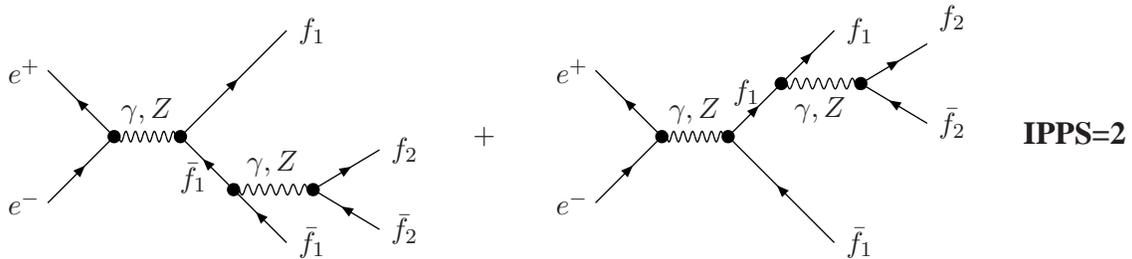
\begin{figure}[h]
\label{fig:pair-bouble}
\vspace*{-8mm}
\begin{eqnarray*}
\begin{array}{ccc}
\vcenter{\hbox{
  \begin{picture}(165,100)(0,0)
    \ArrowLine(25,25)(50,50)        \Text(10,25)[lc]{$e^{-}$}
    \ArrowLine(50,50)(25,75)        \Text(10,75)[lc]{$e^{+}$}
    \Photon(50,50)(75,50){2}{7}     \Text(62.5,55)[cb]{$\gamma,Z$}
    \ArrowLine(95,30)(75,50)        \Text(75,40)[lt]{$\bar{f}_{1}$}
    \ArrowLine(115,10)(95,30)       \Text(130,10)[rc]{$\bar{f}_{1}$}
    \ArrowLine(75,50)(115,90)       \Text(130,90)[rc]{$f_{1}$}
    \Photon(95,30)(125,30){2}{7}    \Text(110,35)[cb]{$\gamma,Z$}
    \ArrowLine(150,15)(125,30)      \Text(165,15)[rc]{$\bar{f}_{2}$}
    \ArrowLine(125,30)(150,45)      \Text(165,45)[rc]{$f_{2}$}
    \Vertex(50,50){2.5}
    \Vertex(75,50){2.5}
    \Vertex(95,30){2.5}
    \Vertex(125,30){2.5}
  \end{picture}}}
&\quad+&
\vcenter{\hbox{
  \begin{picture}(165,100)(0,0)
    \ArrowLine(25,25)(50,50)        \Text(10,25)[lc]{$e^{-}$}
    \ArrowLine(50,50)(25,75)        \Text(10,75)[lc]{$e^{+}$}
    \Photon(50,50)(75,50){2}{7}     \Text(62.5,55)[bc]{$\gamma,Z$}
    \ArrowLine(115,10)(75,50)       \Text(130,10)[rc]{$\bar{f}_{1}$}
    \ArrowLine(75,50)(95,70)        \Text(77,62)[lb]{$f_{1}$}
    \ArrowLine(95,70)(115,90)       \Text(130,90)[rc]{$f_{1}$}
    \Photon(95,70)(125,70){2}{7}    \Text(110,65)[ct]{$\gamma,Z$}
    \ArrowLine(150,55)(125,70)      \Text(165,55)[rc]{$\bar{f}_{2}$}
    \ArrowLine(125,70)(150,85)      \Text(165,95)[rc]{$f_{2}$}
    \Vertex(50,50){2.5}
    \Vertex(75,50){2.5}
    \Vertex(95,70){2.5}
    \Vertex(125,70){2.5}
  \end{picture}}}
\qquad\mbox{\large{\bf IPPS=2}}
\end{array}
\end{eqnarray*}
\vspace*{-5mm}
\caption[]{Second eight diagrams belonging to the NC24 process.\label{eighttwo}}
\vspace*{-1mm}
\end{figure}

\noindent And finally eight diagrams obtained by interchanging 
$f_{1}\leftrightarrow f_{2}$ (Fig. \ref{eighttre}):

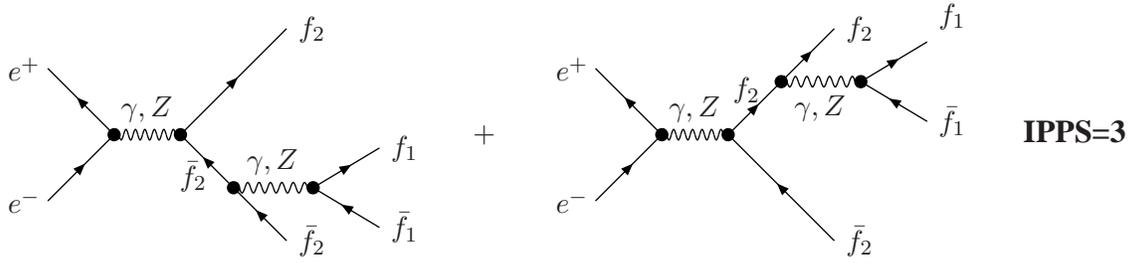
\begin{figure}[h]
\vspace*{-8mm}
\begin{eqnarray*}
\begin{array}{ccc}
\vcenter{\hbox{
  \begin{picture}(165,100)(0,0)
    \ArrowLine(25,25)(50,50)        \Text(10,25)[lc]{$e^{-}$}
    \ArrowLine(50,50)(25,75)        \Text(10,75)[lc]{$e^{+}$}
    \Photon(50,50)(75,50){2}{7}     \Text(62.5,55)[cb]{$\gamma,Z$}
    \ArrowLine(95,30)(75,50)        \Text(75,40)[lt]{$\bar{f}_{2}$}
    \ArrowLine(115,10)(95,30)       \Text(130,10)[rc]{$\bar{f}_{2}$}
    \ArrowLine(75,50)(115,90)       \Text(130,90)[rc]{$f_{2}$}
    \Photon(95,30)(125,30){2}{7}    \Text(110,35)[cb]{$\gamma,Z$}
    \ArrowLine(150,15)(125,30)      \Text(165,15)[rc]{$\bar{f}_{1}$}
    \ArrowLine(125,30)(150,45)      \Text(165,45)[rc]{$f_{1}$}
    \Vertex(50,50){2.5}
    \Vertex(75,50){2.5}
    \Vertex(95,30){2.5}
    \Vertex(125,30){2.5}
  \end{picture}}}
&\quad+&
\vcenter{\hbox{
  \begin{picture}(165,100)(0,0)
    \ArrowLine(25,25)(50,50)        \Text(10,25)[lc]{$e^{-}$}
    \ArrowLine(50,50)(25,75)        \Text(10,75)[lc]{$e^{+}$}
    \Photon(50,50)(75,50){2}{7}     \Text(62.5,55)[bc]{$\gamma,Z$}
    \ArrowLine(115,10)(75,50)       \Text(130,10)[rc]{$\bar{f}_{2}$}
    \ArrowLine(75,50)(95,70)        \Text(77,62)[lb]{$f_{2}$}
    \ArrowLine(95,70)(115,90)       \Text(130,90)[rc]{$f_{2}$}
    \Photon(95,70)(125,70){2}{7}    \Text(110,65)[ct]{$\gamma,Z$}
    \ArrowLine(150,55)(125,70)      \Text(165,55)[rc]{$\bar{f}_{1}$}
    \ArrowLine(125,70)(150,85)      \Text(165,95)[rc]{$f_{1}$}
    \Vertex(50,50){2.5}
    \Vertex(75,50){2.5}
    \Vertex(95,70){2.5}
    \Vertex(125,70){2.5}
  \end{picture}}}
\qquad\mbox{\large{\bf IPPS=3}}
\end{array}
\end{eqnarray*}
\vspace*{-5mm}
\caption[]{Third eight diagrams belonging to the NC24 process.\label{eighttre}}
\vspace*{-1mm}
\end{figure}

\noindent
There is one more diagram with the Higgs boson exchange which is
termed the {\em Higgs signal} or {\em Higgsstrahlung} contribution
which is not taken into account in this study.

\subsubsection*{Terminology, notation}

These 24 diagrams may be considered as a $4f$ background for a $2f$ process.
Their contribution to the $2f$ {\em signal} could be naturally defined 
by imposing cuts on the four fermion state. Events surviving cuts mimic 
the $2f$ process. 

To go further on, we have to provide several {\em definitions}.

\underline{The Born approximation for $2f$ process} is defined as 
\begin{equation}
\mbox{ISR~convolution}\,\{e^{+}e^{-}\to f_1\bar{f}_1\}\;,
\label{2f_signal}
\end{equation}
i.e. an ISR convolution of a $2f$ process with
$f_1\bar{f}_1$ being termed as a the "primary pair".

Relative contribution of $4f$ background processes, 
figs~\ref{eightone}--\ref{eighttwo},
may be conveniently described in terms of \underline{correction due to pair 
production (PP)}, which is defined by the ratio
\begin{equation}
\delta_{pairs}=
\frac{\mbox{ISR~convolution}\,\{e^{+}e^{-}\to f_1\bar{f}_1 f_2\bar{f}_2\}}
     {\mbox{ISR~convolution}\,\{e^{+}e^{-}\to f_1\bar{f}_1\}}\;.
\label{delta_pairs}
\end{equation}
In two last equations ``ISR~convolution'' stands for a rather standard
approach 
\begin{equation}
\sigma(s) = \int\,d x H(x,s){\hat\sigma}\bigl[( 1-x)s\bigr],
\label{fluxtotal}
\end{equation}
where $H(x,s)$ is a flux function and ${\hat\sigma}[( 1-x)s]$
is a kernel ($4f$ or $2f$) cross-section.

As far as $f_1 \neq f_2$ we have no questions which pair should be considered 
to be a ``primary'' one and which one --- a ``secondary''. 
We may distinguish them by imposing 
different cuts $R_{cut}$ and $P_{cut}$ on "primary pair", $f_1\bar{f}_1$ and 
"secondary pair" $f_2\bar{f}_2$\footnote{
Some question arises what to do if $f_1=f_2$, say $\mu$.
One may argue that one may distinguish them by requiring that one pair has 
large invariant mass and another one small.
Due to different cuts imposed the effects of Fermi statistics should be 
negligible.
Furthermore, {\tt GENTLE/4fan} allows symmetric treatment of two pairs.
Therefore, at least when all 24 diagrams are included everything should 
be correct (modulo above mentioned interferences contributions)
if one treats two muon pairs as two pairs of different particles. 
Moreover, $\mu\mu$ is only one of eight 4f-channels in $e^{+}e^{-}\to\mu\mu$. 
A similar problem occurs in the consideration of the total hadronic
cross-section, where five 4f-channels out of total 40 channels contain identical
particles.}. 

The \underline{invariant mass cuts} are defined as
\begin{eqnarray}
R_{cut}&=&\frac{M^2_{f_1\bar{f}_1}}{s}\geq 0.01\mbox{~inclusive},\;
                                         0.7225\mbox{~exclusive},
\\ \nonumber
P_{cut}&=&\frac{M^2_{f_2\bar{f}_2}}{s}\leq 10^{-4},10^{-3},10^{-2},10^{-1},1
                                      \mbox{~all values}.
\label{inv_mass_cuts}
\end{eqnarray}
From Eqn.(\ref{inv_mass_cuts}) one sees that ``primary'' pair is demanded 
to have large invariant mass, while ``secondary'' --- small. 
We also present cut values which 
were used in this study. For $R_{cut}$ we used two standard LEP2 values:
$0.01$ (inclusive selection) and $0.7225$ (exclusive selection), 
while for $P_{cut}$ we studied all allowed range ranging from very tight cuts, 
$10^{-4}$, to a no cut situation, $P_{cut}=1$.

We studied two \underline{processes} with primary muon and hadron (quark) pairs:
\begin{eqnarray}
e^{+}e^{-}&\to&\mu^{+}\mu^{-}, \quad\; \mbox{primary~muons},
\\ \nonumber
e^{+}e^{-}&\to&\mbox{hadrons},\;\,\mbox{primary~quarks}.
\end{eqnarray}

Treatment of the secondary pairs deserves special discussion.
We may describe them using {\em fermionic} language similar for description of
both ``primary'' and ``secondary'' pairs, i.e. sum up over all fermion species:
\begin{equation}
e^{+}e^{-},\mu^{+}\mu^{-},\tau^{+}\tau^{-},\mbox{hadrons}=u,d,c,s,b\mbox{-pairs}.
\end{equation}
(NB: Neutrino secondary pairs are presently NOT included; they should and 
will be!)

This approach suffices, however, a serious drawback. 
As for primary pairs is concerned
fermionic language may be used without questions since pairs is requested 
to be hard.
Even for inclusive selection $M_{f_1\bar{f}_1}\geq 0.2 E_{beam} \geq 38$ GeV.
On the contrary, secondary pairs are integrated from the production threshold, 
$2 m_{f}$,
up to some typically large cut value $0.1-1$. Therefore, we unavoidably cross 
the region 
of low lying resonances where a description in terms of quarks fails completely.
Fortunately, an adequate language for the description of low-invariant-mass 
hadronic 
pairs using a parameterization for the experimentally measured ratio 
$R=\sigma(e^{+}e^{-}\to\mbox{hadrons})/\sigma(e^{+}e^{-}\to\mbox{muons})$
is elaborated in the literature
very well, see e.g.~\cite{Hoang:1995ht}.


\subsubsection*{Virtual pairs}
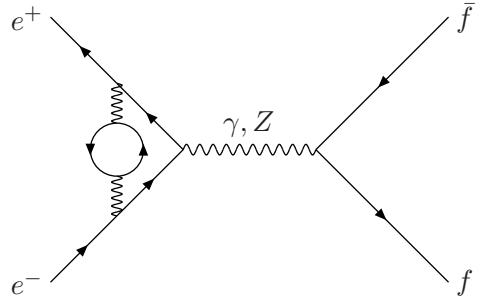
\begin{figure}[h]
\vspace{-4mm}
\begin{eqnarray*}
  \vcenter{\hbox{
  \begin{picture}(160,100)(0,0)
  \ArrowLine(150,100)(100,50)
  \ArrowLine(100,50)(150,0)
  \Photon(50,50)(100,50){2}{10}
  \ArrowLine(0,0)(25,25)
  \ArrowLine(25,25)(50,50)
  \ArrowLine(50,50)(25,75)
  \ArrowLine(25,75)(0,100)
  \Photon(25,75)(25,60){2}{5}
  \Photon(25,25)(25,40){2}{5}
  \ArrowArc(25,50)(10,90,270)
  \ArrowArc(25,50)(10,270,90)
  \Text(75,55)[bc]{$\gamma,Z$}
  \Text(-15,0)[lc]{$e^{-}$}
  \Text(-15,100)[lc]{$e^{+}$}
  \Text(160,0)[rc]{$f$}
  \Text(160,100)[rc]{$\bar{f}$}
  \end{picture}}}
\end{eqnarray*}
\vspace{-4mm}
\caption[]{A typical example of virtual pair correction.}
\label{fig:virtp}
\end{figure}

Virtual pairs have to be also added. There are ISR virtual pairs, 
see fig.\ref{fig:virtp}, FSR virtual pairs 
and initial--final interference (IFIPP) virtual pairs.
The latter are non-leading (see~Ref.~\cite{Arbuzov:1997vj})
and not included in this study.
\vspace*{-5mm}

\subsubsection*{Feynman Diagrams (FD) and their selection with {\bf IPPS, IGONLY}
 flags}

In order to study relative contribution of various Feynman diagrams we 
implemented 
in the code a ``user options'' {\tt IPPS} and {\tt IGONLY} which allows 
to select sub-groups of diagrams:

\noindent
{\bf IPPS=1} only ISPP is taken into account, see fig.\ref{eightone};\\
{\bf IPPS=2} only FSPP with the ordinary meaning of the "secondary pair",
             fig.\ref{eighttwo} is included;\\
{\bf IPPS=3} only FSPP of fig.\ref{eighttre} is accounted for;
{\bf IPPS=4} all final pairs, both fig.\ref{eighttwo} and fig.\ref{eighttre} 
together with interferences among them are included; \\
{\bf IPPS=5 -- IPPS=1$\oplus$ IPPS=2}; \\
{\bf IPPS=6 -- IPPS=1$\oplus$ IPPS=4}; \\
{\bf IPPS=7} only real IFIPP is considered as a separate contribution; \\
{\bf IPPS=8} all three above sets of 24 diagrams are included; \\
{\bf IGONLY=1} only $\gamma$ exchanges everywhere; \\
{\bf IGONLY=2} the ordinary secondary pair is produced via $\gamma$ exchange; \\
{\bf IGONLY=3} all $\gamma$ and $Z$ exchanges are allowed.

\newpage

\subsection{Program GRC4f: tool for the 2-fermion physics}

\label{GRC4f1}

\begin{tabbing}
1) \underline{Author:}\hspace{3.3cm} \= {\bf  J. Fujimoto et al.

} \\[.3cm]
2) \underline{Program:}    \> {\bf  GRC4f v 2.1.39,
{\tt http://is2.kek.jp/ftp/kek/minami/grc4f/}
} 
\end{tabbing}

A more complete references to  the GRC4f program can be found in the 
4-fermion chapter of this report\cite{Grunewald:2000ju}. Here we address only the features relevant 
for the generation of a 4-fermion signal and background sample for 2-fermion analyses,
i.e. splitting the 4-fermion events in 
a sample representing pair emission corrections to 2 fermions, and a true 4-fermion
background sample.
 
The GRC4f Monte Carlo package allows the generation of 4-fermion events
using Born-level matrix elements (ME), convoluted with ISR photon radiation.
It is possible to select the desired set of Feynman diagrams for each final state
\ffo\fft\ by the user. The generation of a signal sample for 2-fermion pair corrections can be done
in two ways:

A) Generating a "signal diagram sample" using only  those diagrams, which are  
considered as signal in the respective definition, and applying 
the $s^\prime$ (and  mass cuts) of the signal definition, to obtain a "4f signal sample".
In addition a second sample with all  non-signal diagrams
(e.g. MP and ISS) is created,  which forms the "4f background sample"
together with those events in the signal diagram sample 
which fail the $s^\prime$ or  mass cuts.
In this method the (generally small) interferences between signal and background
diagrams are neglected in the background subtraction. 

B) The "4f signal sample" is obtained from a set of Feynman diagrams, which 
is larger than the set of signal diagrams. For  each MC event  a weight $w$
is calculated with the help of the REW99 library~\cite{verzocchi:1999} ,  
which  is given by the squared ratio of the matrix 
          elements (ME) summed over all  signal diagrams, divided 
by the sum over all (signal+background) diagrams in the sample.
\begin{equation}
          w_{\rm signal}={                  | \sum ME_{\rm signal} |^2
                    \over
                           | \sum ME_{\rm signal} + \sum ME _{\rm background}|^2  }
\end{equation}
 where $s^\prime$ or  mass cuts can  be included in the signal weight,
by setting it to zero, if it fails the respective cut. Using the weight 
 $w_{\rm background} = 1 -  w_{\rm signal}$ one obtains a 4f background sample that
accounts for all interference effects between signal and background. 

\newpage
\subsection{Program KORALW: tool for the 2-fermion physics}
\label{KORALW1}

\begin{tabbing}
1) \underline{Author:}\hspace{4cm} \= {\bf S. Jadach, W. P\l{}aczek,
   M. Skrzypek, B.F.L. Ward and Z. W\c{a}s}
 \\[.3cm]
2) \underline{Program:}    \> {\bf KORALW 1.42.3}
 \\[.3cm]
3) \underline{Available at:} \>
   http://hpjmiady.ifj.edu.pl/programs/programs.html 
 \\[.3cm]
4) \underline{Main references:} \> 
\cite{Jadach:1998gi}
\\ \>
\cite{koralw:1995a}  
\\ \>
\cite{koralw:1995b}
%
 \end{tabbing}

 KORALW allows generation of 4-fermion events.
It is described in more detail in the 4-fermion chapter of this report
\cite{Grunewald:2000ju}.
Here, only the features relevant to 2-fermion pair corrections will be addressed.

It is possible to select in KORALW the desired set of Feynman diagrams 
for each final state \ffo\fft\ by the user. 
For the moment 
various approximations of the matrix element have been
introduced in KoralW for the $\mu\bar\mu\tau\bar\tau$ and 
partly for $\mu\bar\mu e\bar e$
channels only.
These approximations can be activated with the dip-switch {\tt ISWITCH} 
in the routine {\tt amp4f} in the file {\tt ampli4f.grc.all/amp4f.f}
The available settings are:
\begin{itemize}
\item
0: CC03 (old option for $WW$ final states), 
\item
1: all graphs,
\item
2: ISNS$_{\gamma+Z}$,
\item
3: FSNS$_{\gamma+Z}$ $\tau\bar\tau$ pair to $e\bar e\to \mu\bar \mu$,
\item
4: ISNS$_\gamma$ +FSNS$_{\gamma}$ $\tau\bar\tau$ pair to 
                                   $e\bar e\to \mu\bar\mu$.
\item
5: ISNS$_\gamma$ $\tau\bar\tau$ pair.
\item
6: FSNS$_{\gamma+Z}$ $\mu\bar\mu$ pair to $e\bar e\to\tau\bar\tau$,
\item
negative value: matrix element is 
calculated for all values of {\tt ISWITCH} 
that are declared (set to 1) in {\tt ISW4f} data statement. The appropriate
weights are in this case available as {\tt wtset(40+i)} with i=1,...,6
as above (note that these weights will be 
modified along with the principal weight by Coulomb correction and naive
QCD, whenever applicable). 
\end{itemize}
For other channels the approximations of matrix element 
can be introduced in a similar
manner in the file {\tt ampli4f.grc.all/grc4f$\_$init/selgrf.f}
Some demo programs are available in the {\tt demo.pairs} directory, see
{\tt README} file for more information.

Note that the above described extensions of {\tt ISWITCH} and {\tt
  demo.pairs} directory are {\em not} included in the distribution version
  1.42.3 but will be provided as a separate file at the same {\tt http}
  location or can be requested from the authors.

%% file: 2f-Chapt-Part4.tex
\section{Physics issues and dedicated studies on theoretical errors
\label{sec:dedicated}}

In the present section we concentrate on two related subjects:
the so called ``tuned comparisons'' of the codes done by the authors
of the given codes which are critical for the data interpretation
at LEP2 energies.
In the same time in these comparisons there is some leading
``physics precision theme'', in other words
they have in mind to clarify a certain aspect of the theoretical errors,
like for instance the question of the IRS$\otimes$FSR interference,
or secondary pair corrections.
The collections of these studies partly represents
what we really wanted to be discussed and partly
represent the  availability of the volunteers who had time and interest to provide them.

In this section we gather all studies of the above kind
except for material on the secondary pair contributions,
to which we dedicate the next two sections,
although they represent the same class of the workshop activity.

In the section on tuned comparisons of \KKMC\ and \zf\ 
some numerical results on the importance of the electroweak boxes
is included.
We regret that it was not possible to include a more complete
numerical study of the electroweak boxes.
In order to compensate for that at least partly, we start
the present section with a small section explaining what these EW boxes are
and what are their properties.

Since most of the studies in the present section
concentrate on QED effects, the second
small subsection is devoted to methods of QED calculations
and then we proceed to two sections
which present the tuned comparisons of \KKMC\ and \zf\ 
and a dedicated study on IRS$\otimes$FSR interference,
also prepared by the \KKMC\ and \zf\  teams.

\subsection{Electroweak boxes}

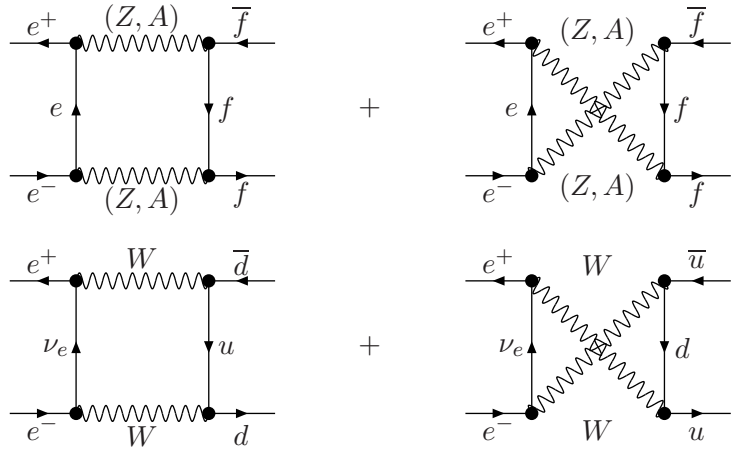
\begin{figure}[!h]
\vspace{-10mm}
\[
\ba{ccc}
\begin{picture}(100,70)(0,33)
  \ArrowLine(0,10)(25,10)
  \ArrowLine(25,10)(25,60)
  \ArrowLine(25,60)(0,60)
  \Photon(25,10)(75,10){3}{10}
  \Photon(25,60)(75,60){3}{10}
  \Vertex(25,10){2.5}
  \Vertex(25,60){2.5}
  \Vertex(75,10){2.5}
  \Vertex(75,60){2.5}
  \ArrowLine(100,60)(75,60)
  \ArrowLine(75,60)(75,10)
  \ArrowLine(75,10)(100,10)
  \Text(12.5,72)[tc]{$\fep$}
  \Text(50,75)[tc]{$(\zb,\ab)$}
  \Text(87.5,74)[tc]{$\fbf$}
  \Text(15,35)[lc]{$\fe$}
  \Text(85,35)[rc]{$f$}
  \Text(12.5,0)[cb]{$\fem$}
  \Text(50,-5)[bc]{$(\zb,\ab)$}
  \Text(87.5,-2)[cb]{$f$}
\end{picture}
\qquad
&+&
\qquad
\begin{picture}(100,70)(0,33)
  \ArrowLine(0,10)(25,10)
  \ArrowLine(25,10)(25,60)
  \ArrowLine(25,60)(0,60)
  \Photon(25,10)(75,60){3}{14}
  \Photon(25,60)(75,10){3}{14}
  \Vertex(25,10){2.5}
  \Vertex(25,60){2.5}
  \Vertex(75,10){2.5}
  \Vertex(75,60){2.5}
  \ArrowLine(100,60)(75,60)
  \ArrowLine(75,60)(75,10)
  \ArrowLine(75,10)(100,10)
  \Text(12.5,72)[tc]{$\fep$}
  \Text(50,70)[tc]{$(\zb,\ab)$}
  \Text(87.5,74)[tc]{$\fbf$}
  \Text(15,35)[lc]{$\fe$}
  \Text(85,35)[rc]{$f$}
  \Text(12.5,0)[cb]{$\fem$}
  \Text(50,0)[bc]{$(\zb,\ab)$}
  \Text(87.5,-2)[cb]{$f$}
\end{picture}
\ea
\]
\vspace*{1mm}
\[
\ba{ccc}
\begin{picture}(100,70)(0,33)
  \ArrowLine(0,10)(25,10)
  \ArrowLine(25,10)(25,60)
  \ArrowLine(25,60)(0,60)
  \Photon(25,10)(75,10){3}{10}
  \Photon(25,60)(75,60){3}{10}
  \Vertex(25,10){2.5}
  \Vertex(25,60){2.5}
  \Vertex(75,10){2.5}
  \Vertex(75,60){2.5}
  \ArrowLine(100,60)(75,60)
  \ArrowLine(75,60)(75,10)
  \ArrowLine(75,10)(100,10)
  \Text(12.5,72)[tc]{$\fep$}
  \Text(50,72.5)[tc]{$\wb$}
  \Text(87.5,72)[tc]{$\fbd$}
  \Text(12.5,35)[lc]{$\fnue$}
  \Text(85,35)[rc]{$\fu$}
  \Text(12.5,0)[cb]{$\fem$}
  \Text(50,-2.5)[bc]{$\wb$}
  \Text(87.5,-2)[cb]{$\fd$}
\end{picture}
\qquad
&+&
\qquad
\begin{picture}(100,70)(0,33)
  \ArrowLine(0,10)(25,10)
  \ArrowLine(25,10)(25,60)
  \ArrowLine(25,60)(0,60)
  \Photon(25,10)(75,60){3}{14}
  \Photon(25,60)(75,10){3}{14}
  \Vertex(25,10){2.5}
  \Vertex(25,60){2.5}
  \Vertex(75,10){2.5}
  \Vertex(75,60){2.5}
  \ArrowLine(100,60)(75,60)
  \ArrowLine(75,60)(75,10)
  \ArrowLine(75,10)(100,10)
  \Text(12.5,72)[tc]{$\fep$}
  \Text(50,70)[tc]{$\wb$}
  \Text(87.5,72)[tc]{$\fbu$}
  \Text(12.5,35)[lc]{$\fnue$}
  \Text(85,35)[rc]{$\fd$}
  \Text(12.5,0)[cb]{$\fem$}
  \Text(50,0)[bc]{$\wb$}
  \Text(87.5,0)[cb]{$\fu$}
\end{picture}
\ea
\]
\vspace*{10mm}
\caption[]{Full collection of QED and EW boxes.\label{boxes}}
\end{figure}

The one-loop {\em the non-QED} or purely {\em weak}
corrections may be represented as the sum of
{\em dressed} $\ph$ and $\zb$ exchange amplitudes plus the contribution
from {\em weak box} diagrams, i.e. $\zb\zb$ and $\wb\wb$ boxes, see Fig.~\ref{boxes}.
The $\zb\zb$ boxes are separately gauge-invariant.

If external fermion masses are neglected, then the complete one-loop amplitude 
(OLA) can be described by only four scalar functions and by the running
electromagnetic constant $\alpha^{\fer}(\sman)$.
Using notation of refs.~\cite{Bardin:1999ak,Bardin:1999yd-orig,Bardin:1999ak}
one may representing the dressed amplitude
in terms of four scalar form factors, $\vverti{}{ij}{\sman,\tman}$:
\bqa
\amp{\zb+\ab}{\OLA}&=&
\frac{\ecs\tcie\tcif}{4\siws\cows}\chizb(\sman)
\biggl\{
\gadu{\mu}\gdp\otimes\gadu{\mu}\gdp\vverti{}{\ssL\ssL}{\sman,\tman}
-4|\qe|\siws\gadu{\mu}\otimes\gadu{\mu}\gdp\vverti{}{\ssQ\ssL}{\sman,\tman}
\nl &&
-4|\qf|\siws\gadu{\mu}\gdp\otimes\gadu{\mu}\vverti{}{\ssL\ssQ}{\sman,\tman}
+16|\qe\qf|\siwf\gadu{\mu}\otimes\gadu{\mu}\vverti{}{\ssQ\ssQ}{\sman,\tman}
\biggr\};
\label{OLA_1}
\eqa
where the $\chizb(\sman)$ denotes the $Z$ boson propagator
\bq
\chizb(\sman)=\frac{1}{\sman-\mzs+\ib\sman\gz/\mz}\;.
\eq
The $\tman$-dependence is due to the weak boxes.
On top of the $\amp{\zb+\ab}{\OLA}$ there is                 
the corrected $\ph$-exchange amplitude, which contains, by construction, only 
the QED running coupling $\alpha^{\fer}(\sman)$:
\bq
\amp{\ab}{\OLA}=\frac{4\pi\alpha^{\fer}(\sman)}{\sman}
\gadu{\mu}\otimes\gadu{\mu}\,.
\label{OLA_3}
\eq

The above electroweak boxes are numerically negligible below
the WW threshold.
At very high energies, $\sim 1$ TeV they are known to be numerically
very large, as it was discussed recently in several papers
\cite{Beenakker:2000kb,Kuhn:2000hx,Ciafaloni:1999ub,Melles:2000ed}
in the context of the (im)possible exponentiation of the electroweak
corrections (so called Sudakov double logarithms)
in the non-abelian theories with the spontaneous symmetry breaking.
They are therefore part of rather interesting physical phenomenon.
At LEP2 the EW boxes are just rising from nothing to a few percents level,
see later this section.

EW box corrections are well known and they are theoretically under good control.
The only possible issue is the technical precision of their implementation
in the MC and other codes.
It would be therefore good to make additional tests of the existing codes
in this direction.

\subsection{Selected aspects of QED calculations}
{\em Structure function approach:}\\
The basic principles and also details of the structure function approach 
used in the presented programs are described in:
\cite{mnp:1996} for NUNUGPV,  and in 
refs.~\cite{Arbuzov:1997pj,Kuraev:1985hb,Skrzypek:1992vk,Arbuzov:1999cq}
for LABSMC.
The main goal of these approaches is to use the exact matrix element 
for the given process and one (two ...) extra photons and appropriate
phase space whenever they are available and combine them into single
prediction using LL structure function approach for fixing normalizations. 

\noindent{\em Parton shower approach:}\\
The QED radiative correction in the leading-log (LL) approximation
can also be obtained using the Monte Carlo method instead of the analytic
formulae of the structure function.
The details of this
method, {\tt QEDPS}, can be found in Ref.\cite{QEDPS}. 
Here we recall that the
algorithm can maintain the exact kinematics during the evolution
of an electron. This specific feature of the {\tt QEDPS} allows us to 
apply the {\tt QEDPS} to radiative processes with avoiding a double-counting
problem. If one needs to know the precise distributions of the hard photon(s)
associated with some kernel process such as neutrino pair-production,
one has to use the exact matrix-elements including hard photon(s)
with the soft-photon correction.
Since the {\tt QEDPS} can provide a complete kinematical information
about the emitted photons and the virtual electrons, it is easy
to separate
the soft photons from the parton shower not to go into the visible region.
In addition to this simple phase-space separation,
the ordering of the electron virtuality
is also required. During the evolution of an electron the virtuality
is monotonically increasing, which is realized naturally in the
{\tt QEDPS}
algorithm. A further condition must be imposed on the virtuality of
the electron in the matrix-elements after emitting the photon:
It  should be greater than the virtuality of the electron in
the last stage of {\tt QEDPS}.
These careful treatment to avoid the double-counting problem allows us
precise predictions of radiative photons.

\noindent{\em Exponentiation:}\\
The exponentiation of QED and its realization in the form of the Monte Carlo
is explained already in detail in  literature, see chapters \ref{KK-MC},
\ref{KORALZ} and \ref{BHWIDE} for references; in the following let us 
concentrate on relatively novel, and essential for establishing 
the precision required by experiments, subject of initial- final-state 
interference.

\subsection{QED ISR$\otimes$FSR interference in cross section and charge asymmetry}
\label{subsec:IFI}
Authors: \KKMC\ and \zf\  teams.

We start this section with characterizing the ISR$\otimes$FSR interference (IFI)
and listing/characterizing the relevant literature and existing tools/codes for calculating IFI.
The principal two subsections contain comparisons of \KKMC\  and \zf\ 
for the muon channel with ISR+FSR, with and without ISR$\otimes$FSR,
for the total cross section and charge asymmetry.
Finally we discuss the uncertainty of ISR$\otimes$FSR, as compared to LEP2 precision targets.

\subsubsection{Overview of properties of the ISR$\otimes$FSR interference}
At LEP2 the QED ISR$\otimes$FSR interference (IFI) is an order of magnitude
bigger than at LEP1 because it is not suppressed any more
by the factor $\Gamma_Z/M_Z$.
On the other hand the experimental errors are bigger, 
so its importance has to be measured
in terms of the target precision requirements 
defined in section 2.1.
All main characteristics if IFI can be understood looking at the leading term
in its the first order expression
\begin{equation}
  \label{eq:ifi-1-st-order}
  \delta^{\rm IFI}(\cos\theta) =
  4 Q_e Q_f{\alpha\over\pi}\; \ln {E^\gamma_{\max} \over E_{beam}}
                              \ln { 1-\cos\theta \over 1+ \cos\theta }
\end{equation}
(which is also the $\theta$ dependent part of the YFS/Sudakov form-factor).
The above factor multiplies the Born differential cross section.
We see immediately that:
\begin{itemize}
\item
  IFI is growing for stronger cuts on maximum photon energy $E^\gamma_{\max}\to 0$.
\item
  IFI always contributes to $A_{FB}$, however, not necessarily to the total
  cross section, unless the Born differential cross section is asymmetric itself.
  This is true at LEP2, where all muon and quark asymmetries are large.
\item
  IFI is proportional to the charge of the final fermion $Q_f$, and consequently
  it is smaller for quarks than for muons. In addition, for quarks the contributions
  from different channels tend to cancel each other.
\item
  It does not contain logs of fermion masses.
\end{itemize}
\begin{table}[!ht]
\centering
\setlength{\unitlength}{0.1mm}
\begin{picture}(1200,1200)
\put(-20,   0){\makebox(0,0)[lb]{\epsfig{file=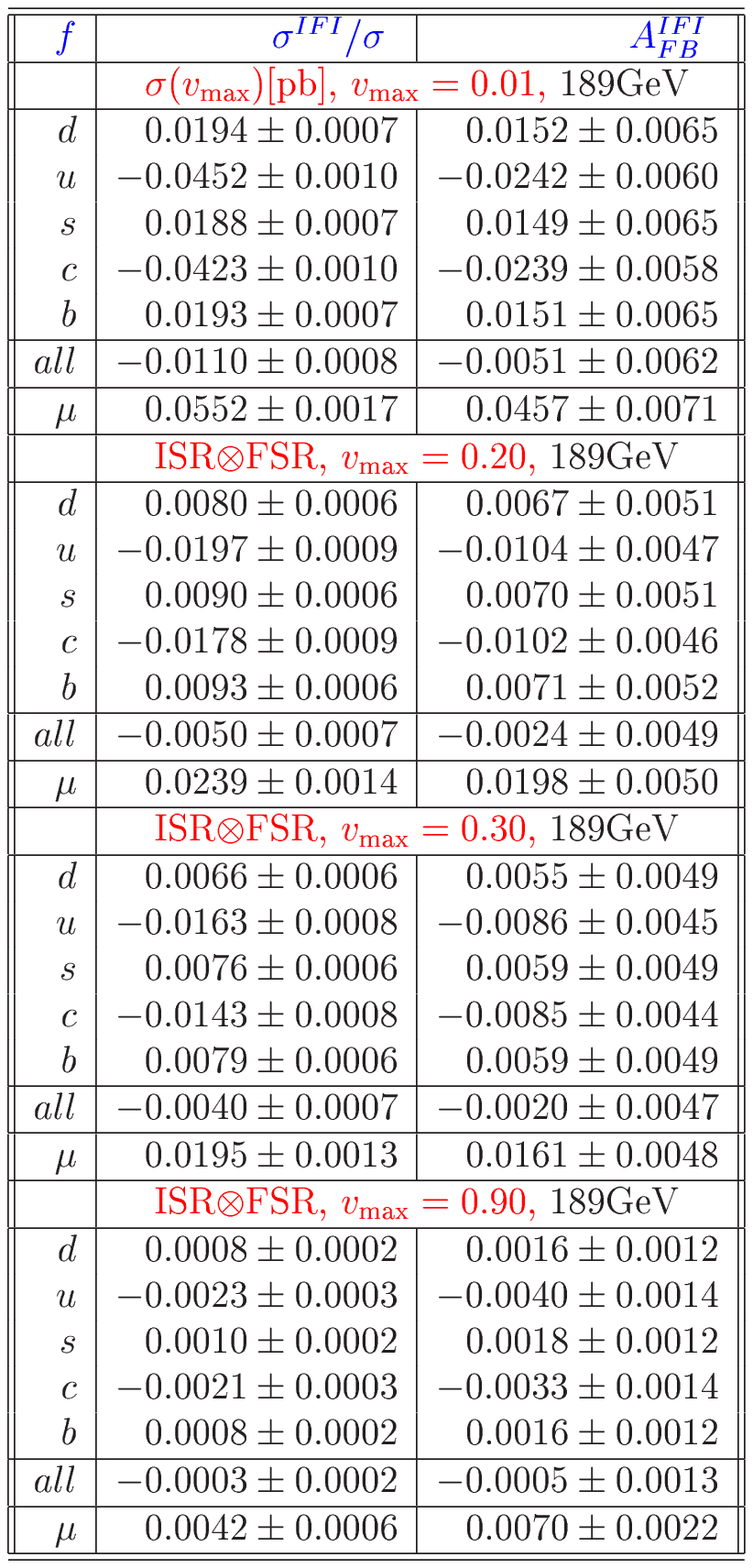,width=60mm,height=120mm}}}
\put(600,   0){\makebox(0,0)[lb]{\epsfig{file=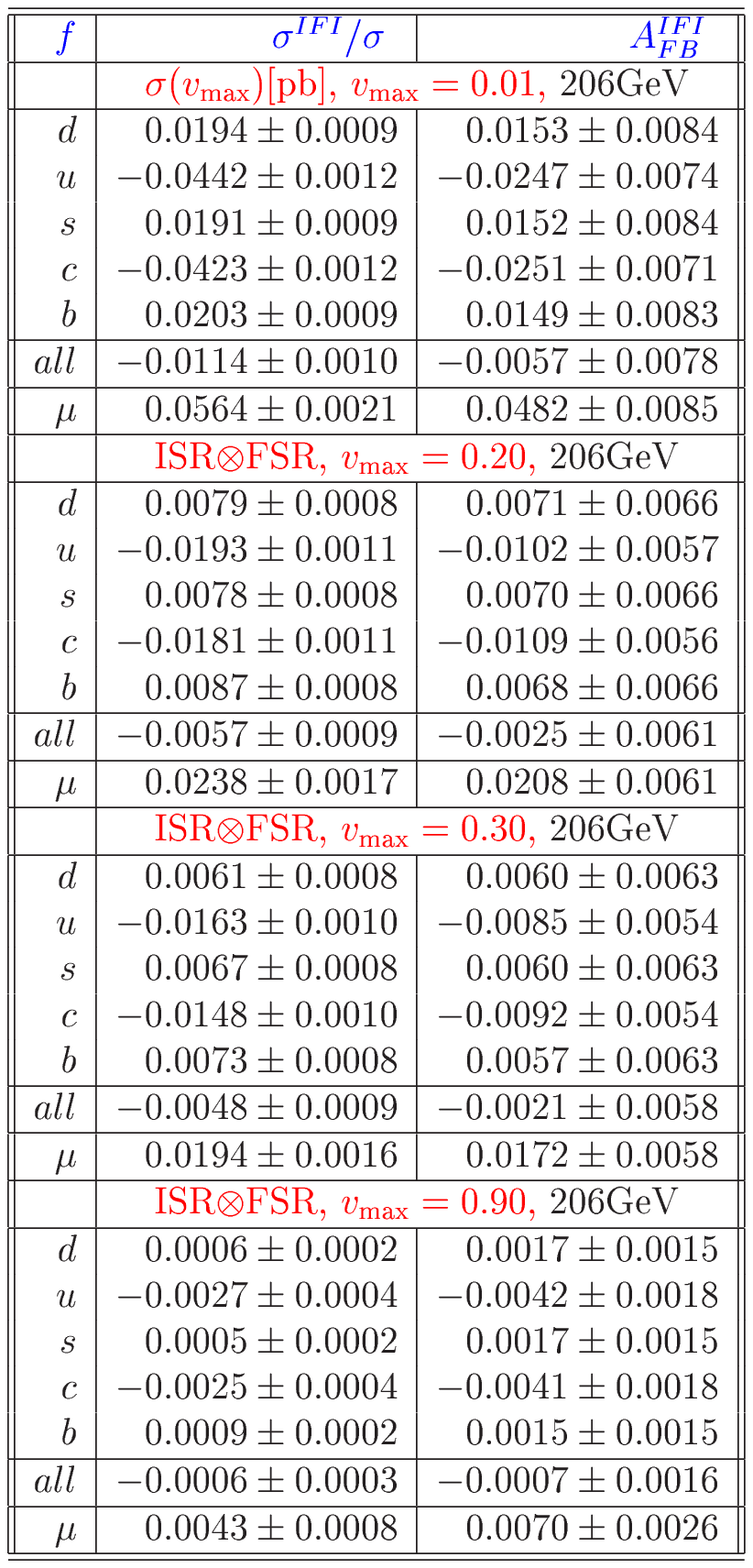,width=60mm,height=120mm}}}
\end{picture}
\caption{\small\sf 
 The quantitative illustration of the main properties of IFI for $q$- and $\mu$-pairs
 at two LEP energies.
}
\label{tab:ifi-illustr}
\end{table}
The above facts are illustrated in a more quantitative form 
in Tab.~\ref{tab:ifi-illustr}, where we show the values of the IFI
contributions to the $\mu-$ and $q$-pair channels at two LEP energies.
They are calculated with \KKMC\  for unrestricted $\cos\theta$ and a simple cut on the
fermion pair invariant mass%
\footnote{
  For realistic cuts the IFI contributions will be slightly smaller, by a factor $\sim 0.8$}
$M_{f\bar{f}}=\sqrt{s'}$.
As we see, the IFI contributions to $A_{FB}$ are the smallest for $d$,
twice bigger for $u$ with an alternating sign, and are the largest for $\mu$.
They increase for stronger cut on photon energy.
Since $A_{FB}\sim 0.6$ for all quarks and $\mu$, 
consequently the magnitude of the IFI
contribution of $\sigma$ shows the same pattern.
For the typical Z-exclusive $v_{\max}=1-s'_{min}=0.2$,
looking more closely into numbers, we find for the cross section
for muon pairs that the IFI contribution is about 
2.4\% of $\sigma^{\mu}$, that is 6 times bigger than the precision tag 0.4\%
of section 2.1.
For quarks it is 0.5\% of $\sigma^{h}$,
that is twice bigger than the precision tag of 0.2\% listed in section 2.1.
For the Z-inclusive $v_{\max}=0.9$
we have the IFI of 0.4\% of $\sigma^{\mu}$ versus the 0.4\% precision tag of section 2.1
and 0.03\% of $\sigma^{h}$ versus 0.2\% precision tag of section 2.1.
There is therefore no doubt that IFI is important for LEP data analysis.

\subsubsection{Exponentiation of IFI}
In the pure \Order{\alpha^1} calculation,
see eq.~(\ref{eq:ifi-1-st-order}), it is well known since long times that the
ISR$\otimes$FSR in the integrated cross section and in the charge
asymmetry goes to infinity for the strong cuts on photon energy $E^\gamma_{\max}\to 0$,
clearly an unphysical result.  
Furthermore, the angular distribution
close $\cos\theta=\pm 1$ gets singular behavior of the kind
$\ln((1-\cos\theta)/(1+\cos\theta)$.  
It is also well known since a long time~\cite{Yennie:1961ad,Greco:1976wq} 
that summing up properly the soft photon contributions cures both of these problems.  
This can be schematically demonstrated as
\begin{equation}
  \label{eq:ifi-exp}
  1+\delta^{\rm IFI}(\cos\theta) \to e^{\delta^{\rm IFI}(\cos\theta)}
\end{equation}
What kind of practical consequence we may expect?  
For the typical experimental Z-exclusive cut $s'>0.80s$ and
$|\cos\theta<0.95|$ the effect of the exponentiation will be rather small.  
Most probably it is equally or more important to convolute properly the
ISR$\otimes$FSR with the \Order{L^2\alpha^2} ISR.

If Z radiative return is included in the phase space then the situation is more delicate.
One hard photon is necessarily emitted and from the real-photon \Order{\alpha^1}
matrix element we know only that the ISR$\otimes$FSR is suppressed close to and
across the Z-peak in $s'$ distribution.
Exponentiation in this case means adding into the game a second and more
real photons and the \Order{\alpha^2} virtual corrections.
These additional \Order{\alpha^2} and higher corrections are not exactly known/available,
and in practice we can only add them in the soft photon approximation.
This is probably good enough for the LEP2 precision tag.
Such a scenario is already realized in the \KKMC\ , see below.

\subsubsection{Older works on IFI}
In the older literature a rather complete treatment of IFI can be
found in Ref.~\cite{Greco:1980}, where it is discussed in the soft-photon
approximation (no very hard photons), in exponentiated form%
\footnote{ 
  The authors of this paper point out the Yennie-Frautschi-Suura\cite{Yennie:1961ad}
  work as a prototype for IFI exponentiation.},
including Z-resonance and Z-radiative return (not too far from Z).
Later works, at the beginning of LEP1 era,
see Refs.~\cite{Lynn:1985rk,Kuhn:1988nh,AFBstara,Bardin:1991fu,Bardin:1991de},
see also LEP1 proceedings~\cite{Z-physics-at-lep-1:89}
have concentrated mainly on adding hard photons in the game 
and removing certain approximations in the virtual corrections.

\subsubsection{KORALZ Monte Carlo for IFI}
The KORALZ~\cite{Jadach:1994yv} Monte Carlo offers the most solid
benchmark for the \Order{\alpha^1} IFI without exponentiation.
The \Order{\alpha^1} part/option of KORALZ
is an improved version of the program of ref.~\cite{mustraal}
(exact $\gamma$-$Z$ boxes are added).
It was well tested to a precision  $<0.1\%$ against analytical calculations
in ref.~\cite{AFBstara}, also far away from Z-resonance.
It was also compared with the calculations of~ref.~\cite{Bardin:1991de}.
KORALZ was already used in the first
experimental studies of IFI at LEP1, see refs.~\cite{holt:1996,holt:1997}.

\subsubsection{IFI from \KKMC\ }
The IFI is now implemented in the exponentiated form
in the new MC event generator \KKMC\ ~\cite{kkcpc:1999}, see sect.~\ref{KK-MC}.
From the IFI point of view \KKMC\  represents the complete \Order{\alpha^1}
in exponentiated form (Coherent Exclusive Exponentiation),
however with some important extensions: 
(a) it convolutes IFI with the second order ISR (and FSR) and 
(b) it does have for IFI the exact second order
2-$\gamma$ matrix element.
It misses second order exact virtual corrections relevant for IFI (double boxes),
but not completely, they are included in the soft photon approximation.

The IFI numerical results from \KKMC\  were already
debugged/tested in ref.~\cite{ceex2:2000}
by comparing them with the results of \Order{\alpha^1} 
KORALZ without exponentiation (see above for more details).
It was found that the
IFI correction to the total cross section and charge asymmetry
from \KKMC\  and \Order{\alpha^1} KORALZ is about 2\% and agrees
to within $<0.2\%$ for the common examples of Z-exclusive cuts, even
without a cut on $\cos\theta$.  
One step further was also made in ref.~\cite{ceex2:2000}:
the \KKMC\  results without the ISR$\otimes$FSR were combined with
the ISR$\otimes$FSR of KORALZ \Order{\alpha^1} ISR%
\footnote{ This method was described in
 refs.~\cite{holt:1996,holt:1997} using KORALZ \Order{\alpha^1} and KORALZ/YFS3,
 and used to estimate higher orders to IFI at Z peak.}.
This kind of ``hybrid'' \KKMC\ +ISR$\otimes$FSR$_{1-st.ord.}$ result 
was compared with the exponentiated IFI of standard CEEX over the
wide range of photon energy cuts, for the total cross-section and charge
asymmetry.
Typical agreement of $<0.2\%$ was found for both Z-exclusive and Z-inclusive cuts.  
The biggest discrepancy was noticed to be
0.4\% for the charge asymmetry for a Z-inclusive cut and for the cross section
for certain values (far from the experimental ones) for the Z-exclusive cut.

In ref.~\cite{ceex2:2000} a preliminary 
comparison was also made for ISR$\otimes$FSR between \KKMC\  and \zf\  6.11.
Similar patterns of agreements and disagreements were found.
This is not surprising, as \zf\  is also combining the
ISR$\otimes$FSR$_{1-st.ord.}$ without exponentiation with the rest of
the calculation.
The authors of ref.~\cite{ceex2:2000} conclude that there is definitely
room of improvements of our understanding of ISR$\otimes$FSR,
especially for the Z-inclusive acceptance, but there is no emergency
situation%
\footnote{
  The more complete summary/discussion on these tests can
  be also found in the presentation of S.J. at the June 1999 meeting of LEPEWG,
  see transparencies on http://home.cern.ch/jadach}.

\begin{figure}[!ht]
\centering
\setlength{\unitlength}{0.1mm}
\begin{picture}(1600,800)
\put( 650,700){\makebox(0,0)[b]{\large (a)}}
\put(1500,700){\makebox(0,0)[b]{\large (b)}}
\put(-20,   0){\makebox(0,0)[lb]{\epsfig{file=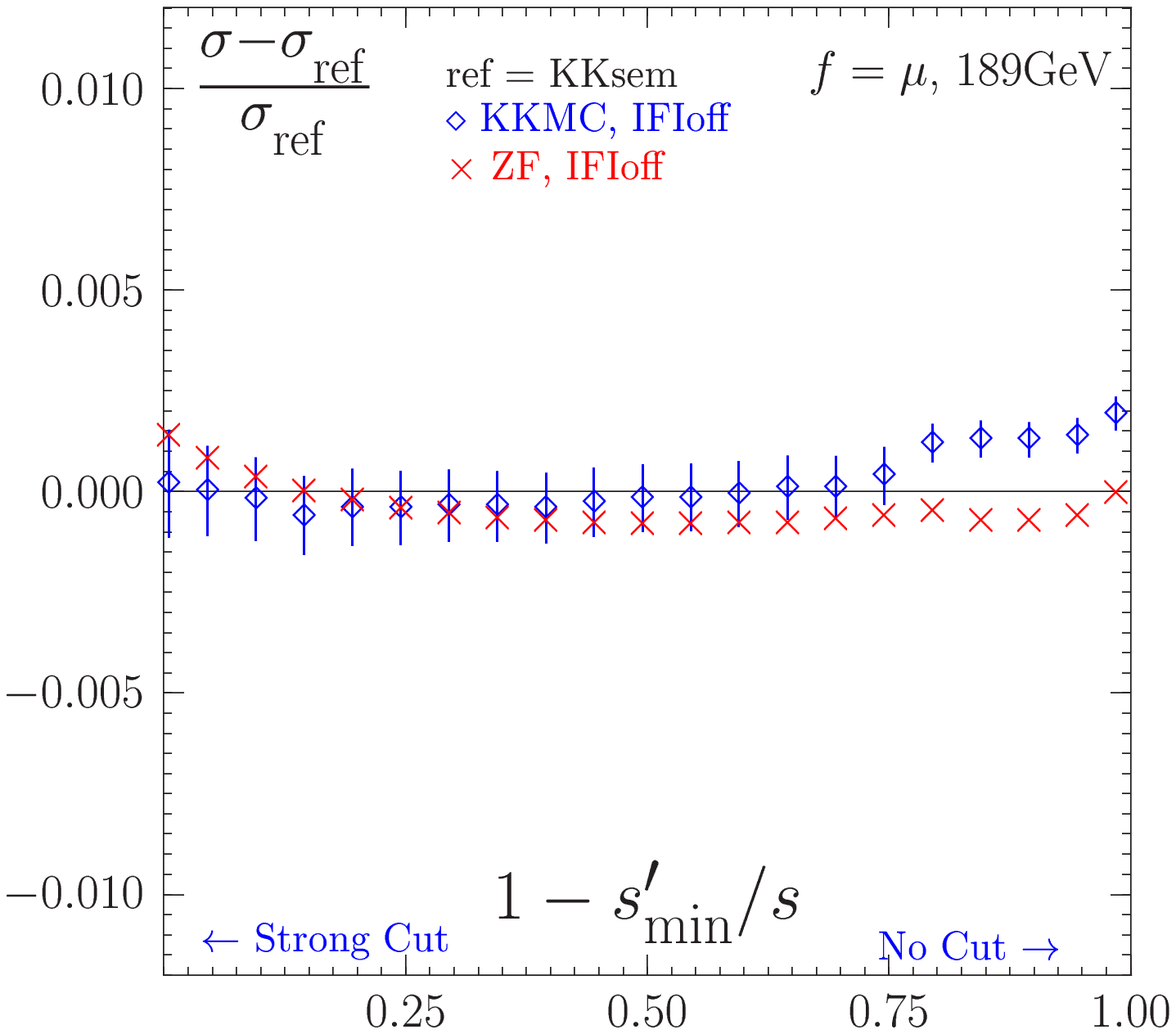,width=80mm,height=80mm}}}
\put(800,   0){\makebox(0,0)[lb]{\epsfig{file=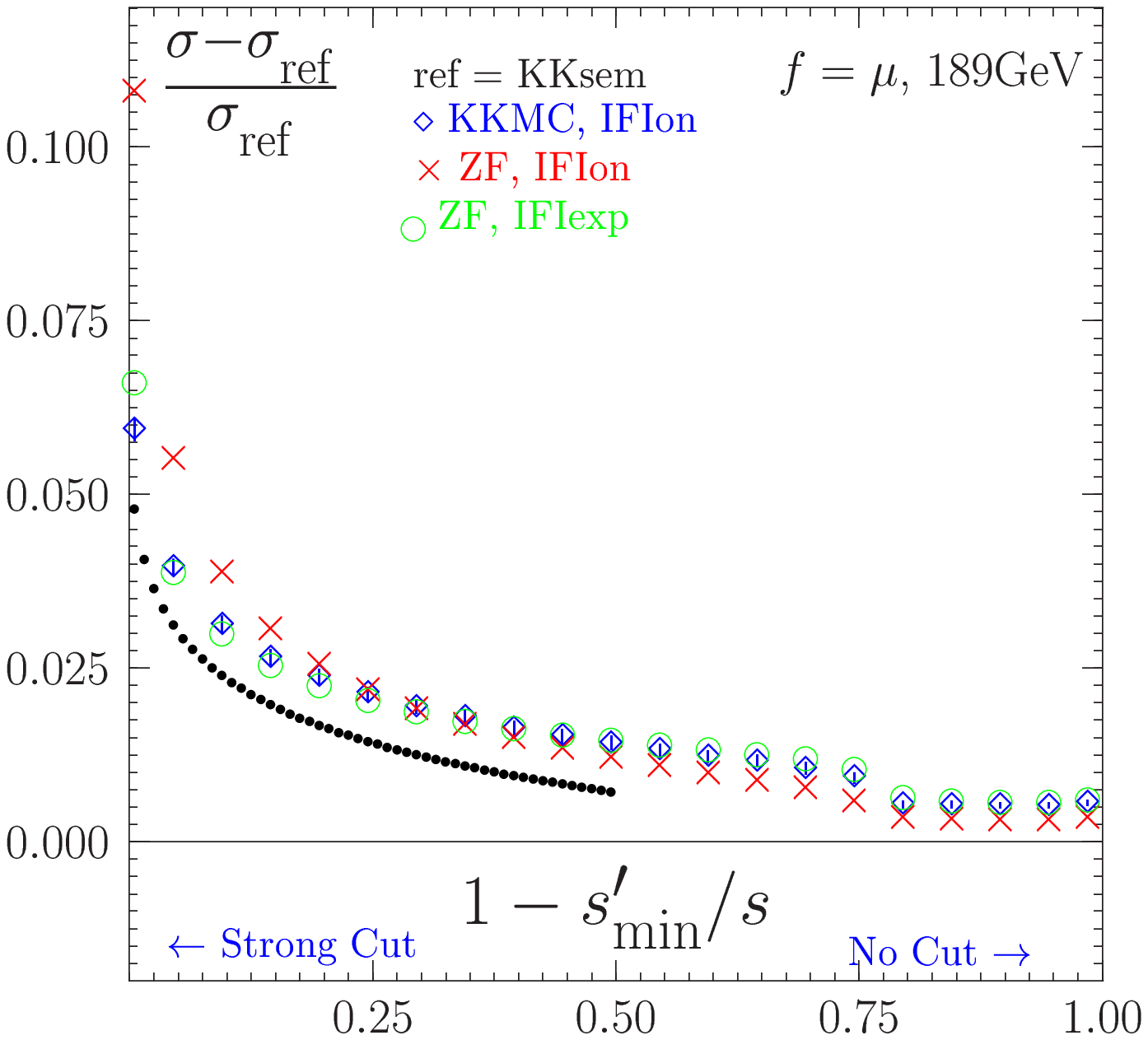,width=80mm,height=80mm}}}
\end{picture}
\vspace*{-5mm}
\caption{\small\sf 
 The comparison of \KKMC\ , ${\cal KK}$sem and \zf\ for cross section at 189GeV.
 The IFI is on/off for \KKMC\  and \zf\  and off for reference ${\cal KK}$sem.
 Black dots represent eq.~(\ref{eq:sigifi}).
}
\label{fig:ifi-warm-up}
\end{figure}

\begin{figure}[!ht]
\centering
\setlength{\unitlength}{0.1mm}
\begin{picture}(1600,800)
\put( 650,700){\makebox(0,0)[b]{\large (a)}}
\put(1500,700){\makebox(0,0)[b]{\large (b)}}
\put(-20,   0){\makebox(0,0)[lb]{\epsfig{file=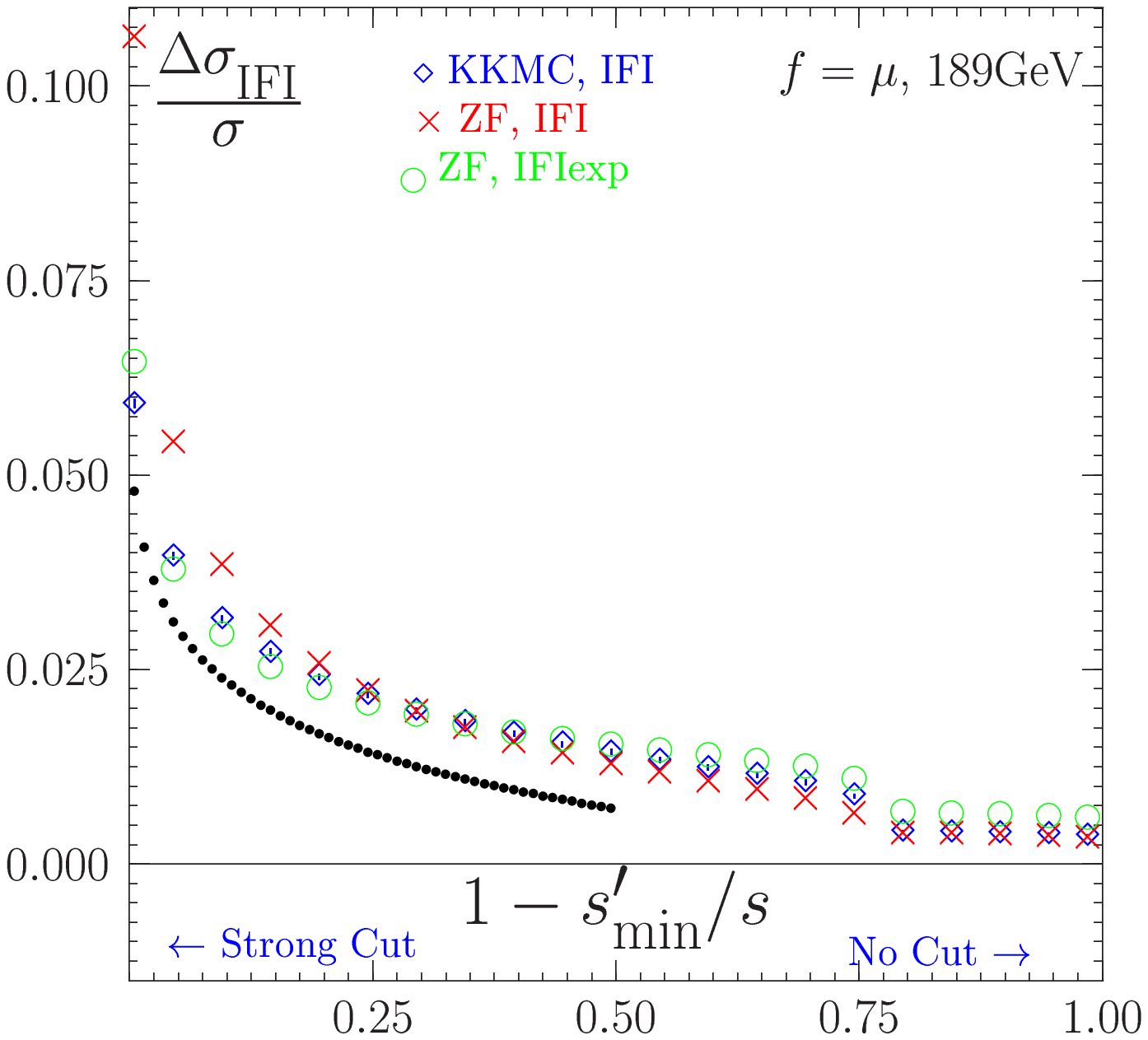,width=80mm,height=80mm}}}
\put(800,   0){\makebox(0,0)[lb]{\epsfig{file=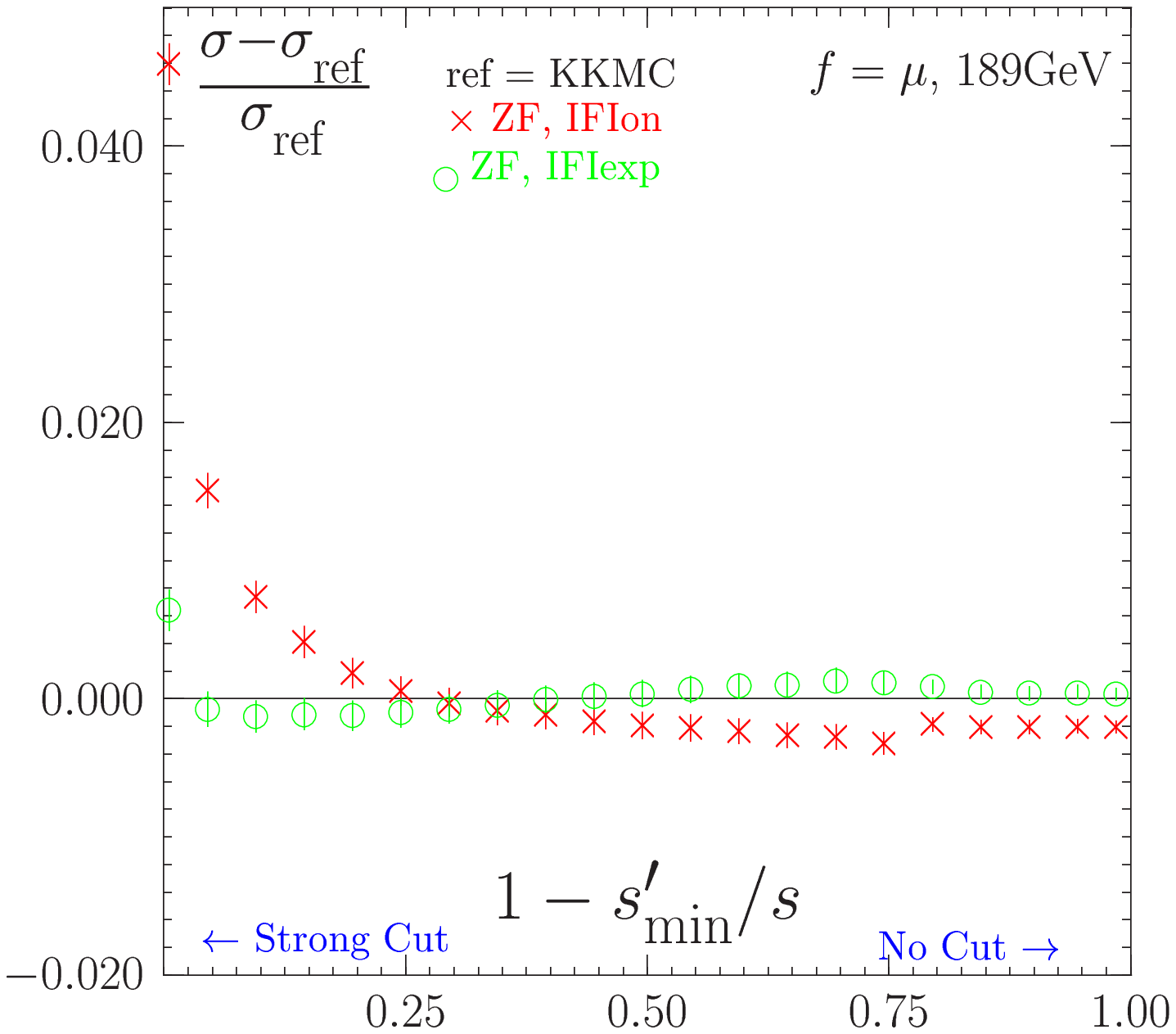,width=80mm,height=80mm}}}
\end{picture}
\vspace*{-5mm}
\caption{\small\sf 
 The comparison of \KKMC\  and \zf\  for cross section at 189GeV.
 The IFI is on/off for \KKMC\  and \zf\ .
 Black dots represent eq.~(\ref{eq:sigifi}).
}
\label{fig:ifi-first}
\vspace*{-2mm}
\end{figure}

\subsubsection{Exponentiated IFI from \zf\ }
The recent version of \zf\  includes ISR$\otimes$FSR
exponentiated according to Greco et.al.~\cite{Greco:1976wq}.
We call it in short ZFexp.
This option will be available in the future edition of \zf\ .
The first version which we tried in this comparison featured some numerical
problems but after extensive tests it now agrees rather well with 
\KKMC\ .
Note that if  the ISR$\otimes$FSR is correctly implemented
in both program then we expect the agreement of order 0.1\%
for any Z-exclusive cuts, 
in particular the difference between them should not increase for a strong cut.

\subsubsection{Semianalytical estimate of soft limit}
Before we come to numerical comparisons let us present a simple
semi-analytical estimate of the IFI contributions to cross-sections
and charge asymmetry in the soft fimit, in the case
of the exponentiation of IFI.
The purpose is two-fold: 
(a) such exprressions is useful in quick testing
more complicated programs like \KKMC\  and \zf\ ,
(b) they give non-trivial insight into higher orders.
The IFI correction to total cross section is
\begin{equation}
\label{eq:sigifi}
\delta_{IFI}(v_{\max})=
{ \sigma_{exp} - {\sigma_{exp}}^{\rm No IFI} \over {\sigma_{exp}}^{\rm No IFI}}
= 1 -2 A_{FB} \kappa\ln{v_{\max}}
    +\kappa^2\ln^2{v_{\max}} \Big( {1\over 2} +{\pi^2 \over 6}\Big)+const,
\end{equation}
where $\kappa= 4{\alpha\over \pi} Q_eQ_f $, $A_{FB}$ is Born asymmetry,
and $v_{\max}=1-s'_{\min}/s\simeq E^\gamma_{\max}/E_{beam}$
limits the maximum energy of all soft photons.
The constant part is related to non-IR parts of QED boxes.
The absence of big mass-logs along with $\ln{v_{\max}}$
in this formula is not an accident,
this is the rigorous result of the proper exponentiation of IFI to infinite order.
This gives argument for the lack of big enhancemet factors like
$ \ln E^\gamma_{\max}/E_{beam}$ in the IFI corrections.
Without such an enhancement factors IFI at higher orders will be always small,
for instance at \Order{\alpha^2} 
it is of order $\kappa {\alpha\over\pi}\ln{s\over m_e^2}$, that is $\sim 0.05\%$.
Similarly, one may estimate the IFI correction to $A_{FB}$:
\begin{equation}
\label{eq:afbifi}
\delta A_{FB}^{\rm IFI}(v_1) = 
- \kappa \ln{v_{\max}} \left( 2\ln(2) +{3\over 4} + 2 A_{FB} \right) 
  +{\cal O}(\kappa^2 \ln^2{v_{\max}}) + const.
\end{equation}
The precision of the above two formulas is $1\%$.
It is enough to test the correctness of the soft limit.
Later on in the relevant figures results  of the above formula are represented as
additional curve of black dots.

\subsubsection{Comparisons of \KKMC\  and \zf\  including IFI}
\label{ZFKKifi}
The material of present section, on tuned comparisons of \KKMC\  
and \zf\ with IFI switched on, will be continued later in the 
section~\ref{TunedZFKK}, for the case when IFI is switched off, and can be
thus regarded as its extension.
%
%
All presented numerical results will be for the muon channel
with the cut on the effective mass of muon pair $M_{f\bar{f}}=\sqrt{s'}$,
and no restriction on $\cos\theta$. 
The scattering angle $\theta$ is defined%
\footnote{ This angle definition makes little sense for Z radiative return at LEP2 energies
  where muon pair is very strongly boosted, but we keep it for historical reasons.}
as an angle of $\mu^-$ with respect $e^-$.
We shall discuss results for the total cross section first and
for the charge asymmetry later on%
\footnote{ However, we should always keep in mind that IFI contributes
  primarily to $A_{FB}$ and secondarily to $\sigma$, as already explained.}.

\subsubsection{IFI in the cross section}
As a warm-up exercise we present in fig.~\ref{fig:ifi-warm-up}
the comparison of \KKMC\  and \zf\  for IFI switched off and on.
Making the comparison for IFI switched off makes sense because in Sect.~\ref{TunedZFKK}
the distribution of $M_{f\bar{f}}=\sqrt{s'}$ was not affected by FSR,
and now it is.
As we see in fig.~\ref{fig:ifi-warm-up}(a), in the case of no IFI we
recover the same level of agreement among \KKMC\ , ${\cal KK}$sem and \zf\ 
at the level of 0.2\% as before.
Encouraged by this, we switch on IFI and find in fig.~\ref{fig:ifi-warm-up}(a)
that \KKMC\  and two  versions of \zf\  with and without exponentiation.
The later one \zf\  without exponentiation let us call ZFstd.
As we see, all curves 
departs for ${\cal KK}$sem (which has no IFI)
by $\sim 2\%$ for Z-exclusive cuts and $\sim 0.4\%$ for Z-inclusive,
and they agree fairly well to within $\sim 0.2-0.3\%$.
In the soft limit 
(the first point in the curves is for $s'_{\max}=0.99s$).
ZFstd diverges by 4\% from the \KKMC\ and ZFexp.

In the next Fig.~\ref{fig:ifi-first}(a) we look closer into 
the IFI effect in \KKMC\  and \zf\ ,
i.e. into the difference due to switching on IFI in each program (version).
In the Fig.~\ref{fig:ifi-first}(a) we view the same results
plotted as the differences  \zf\  $-$ \KKMC\ .
As we see the difference ZFstd$-{\cal KK}$MC
are within 0.4\% for a wide range
of the cuts, including typical Z-exclusive and Z-inclusive cuts,
while the difference ZFexp$-{\cal KK}$MC is twice smaller,
about 0.2\% only, again for a wide range of the cuts.
In the figures we also show (black dots) the analytical estimate of the IFI
exponentiated distribution. 
The estimate should be valid to within 1\% and its main aim is to test
the soft photon limit.
As we see the soft limit is correctly reproduced
for both \KKMC\  and ZFexp, and their difference at $s'_{\max}=0.99s$ is also below 1\%.

\begin{figure}[!ht]
\centering
\setlength{\unitlength}{0.1mm}
\begin{picture}(1400,1400)
\put( 900,1250){\makebox(0,0)[b]{\large (a)}}
\put( 550, 600){\makebox(0,0)[b]{\large (b)}}
\put(1300, 600){\makebox(0,0)[b]{\large (c)}}
\put(330, 700){\makebox(0,0)[lb]{\epsfig{file=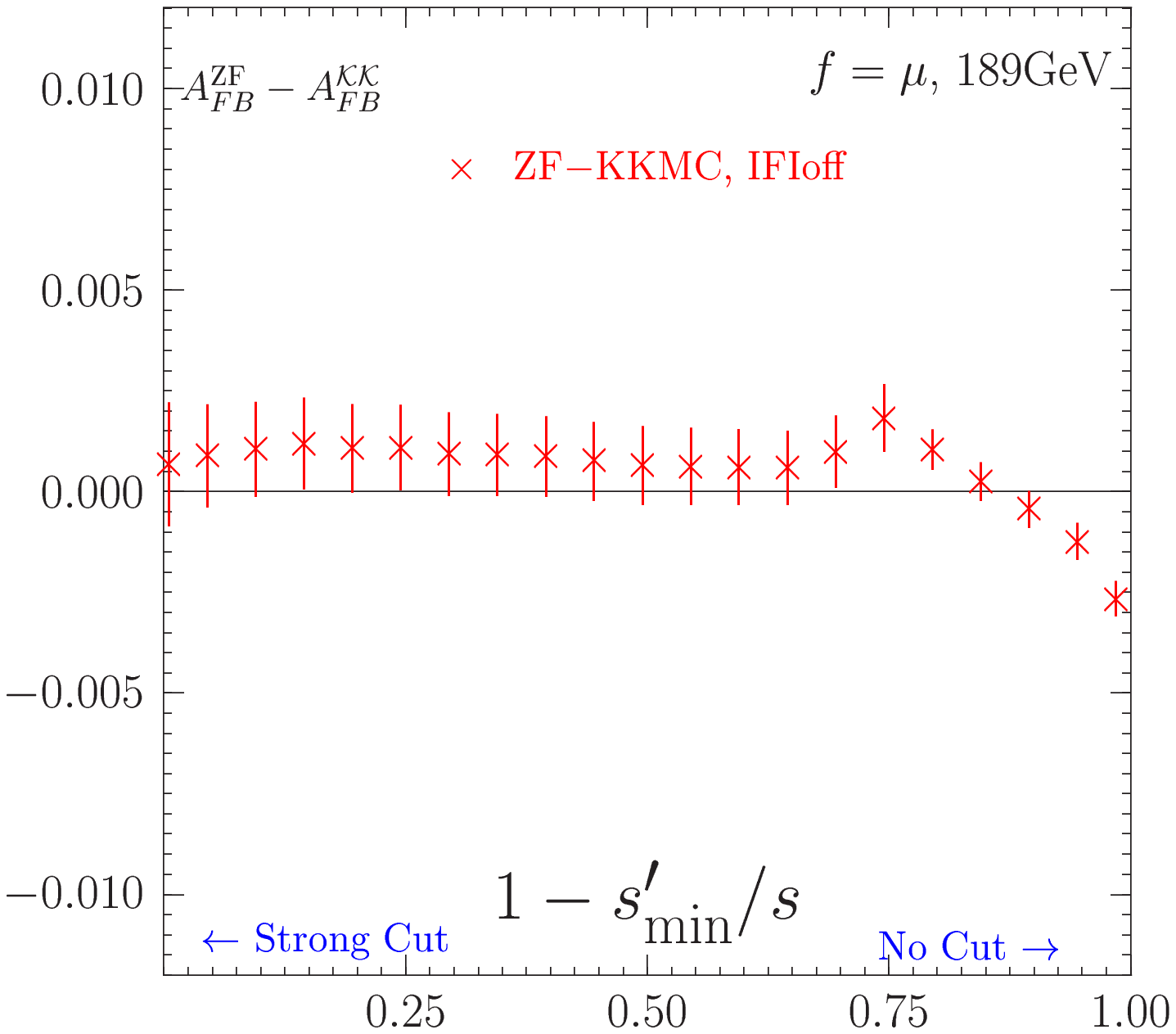,width=70mm,height=70mm}}}
\put(-20,   0){\makebox(0,0)[lb]{\epsfig{file=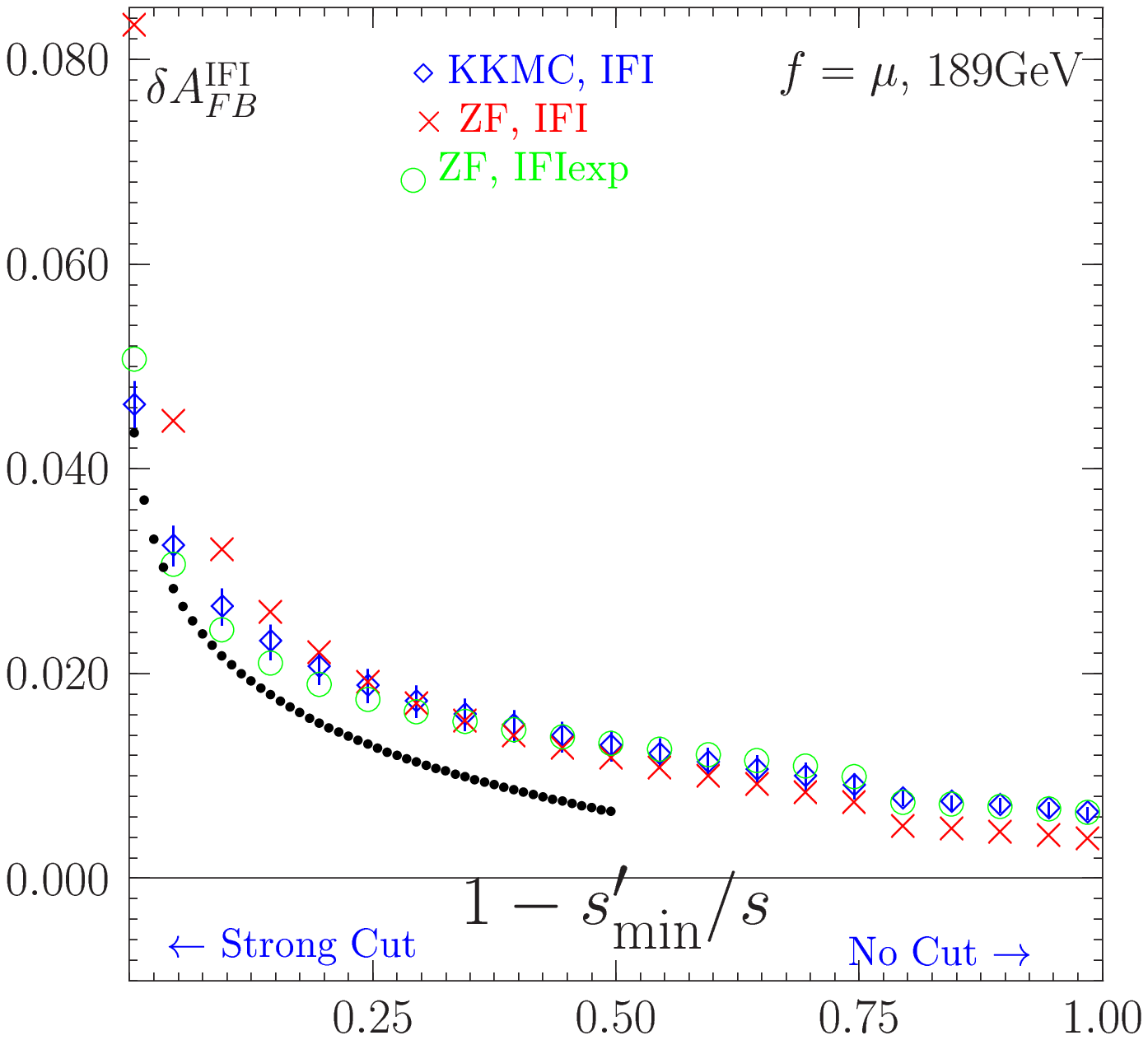,width=70mm,height=70mm}}}
\put(700,   0){\makebox(0,0)[lb]{\epsfig{file=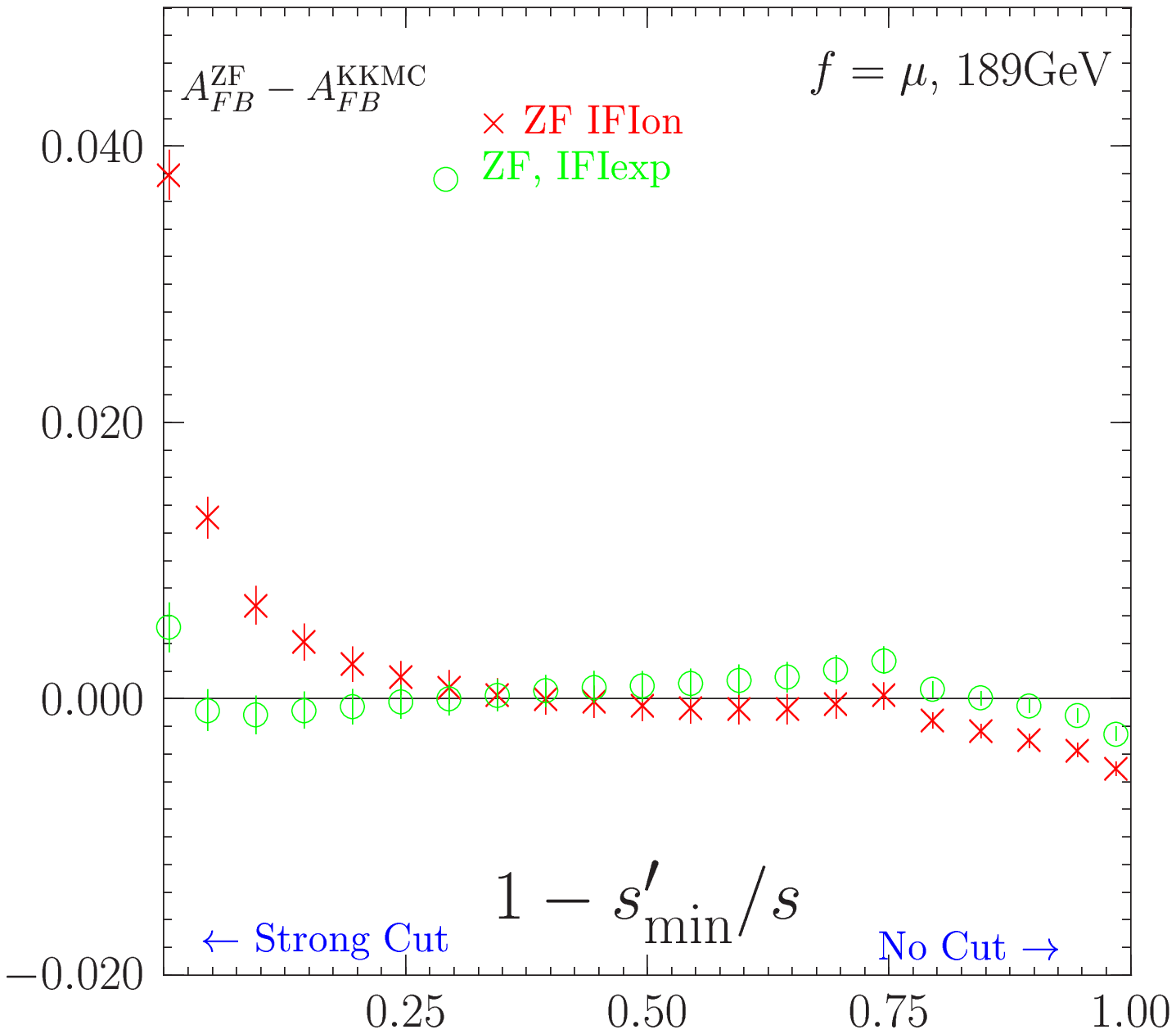,width=70mm,height=70mm}}}
\end{picture}
\caption{\small\sf 
 The comparison of \KKMC\  and \zf\  for $A_{FB}$ at 189GeV.
 Black dots represent eq.~(\ref{eq:afbifi}).
}
\label{fig:ifi-second}
\end{figure}

\subsubsection{IFI in the charge asymmetry}
In Fig.~\ref{fig:ifi-second}(a) we show the comparison of \KKMC\  and \zf\ 
for the charge asymmetry, in the case of IFI switched off.
The agreement is within $0.25\%$, and it should be $<0.20\%$
in view of the fact that ${\cal KK}$sem and \KKMC\  agree%
\footnote{
  We could not include ${\cal KK}$sem in the present comparison for $A_{FB}$
  because the agreement ${\cal KK}$sem - \KKMC\  was obtained
  for the $\theta$ definition in the Z rest frame\cite{ceex2:2000}.
  Unfortunately \zf\  cannot use such an angle,
  and we are forced to the CMS definition of $\theta$.}
for $A_{FB}$ to within $0.1\%$, see ref.~\cite{ceex2:2000}.
The quality of the test is also limited by MC statistics.

In the next plot of Fig.~\ref{fig:ifi-second}(b)
we look into the IFI effect in the \KKMC\  and in \zf\ ,
that is into the difference due to switching on IFI in each program (version).
In the Fig.~\ref{fig:ifi-second}(c) we view the same results
plotted as the difference \zf\   $-$  \KKMC\ .
As we see the the differences ZFstd$-{\cal KK}$MC
is within 0.4\% for a wide range
of the cuts, including typical Z-exclusive and Z-inclusive cuts.
The differences ZFexp$-{\cal KK}$MC is smaller, about 0.25\%.
This agreement is within the required precision 
tag of $0.4\%$-$0.5\%$ for $A_{FB}$ in section 2.1.
As for cross sections, we have also included in these plots 
the analytical estimate of IFI contribution to asymmetry
in the soft photon approximation. 
Results of \KKMC\  agree well with the analytical estimate
in the soft limit.
And what is also important the exponentiated version of \zf\ 
is much closer to \KKMC\  than the older one.

\begin{table}[!ht]
\centering
\setlength{\unitlength}{0.1mm}
\begin{picture}(1600,1000)
\put(0,   0){\makebox(0,0)[lb]{\epsfig{file=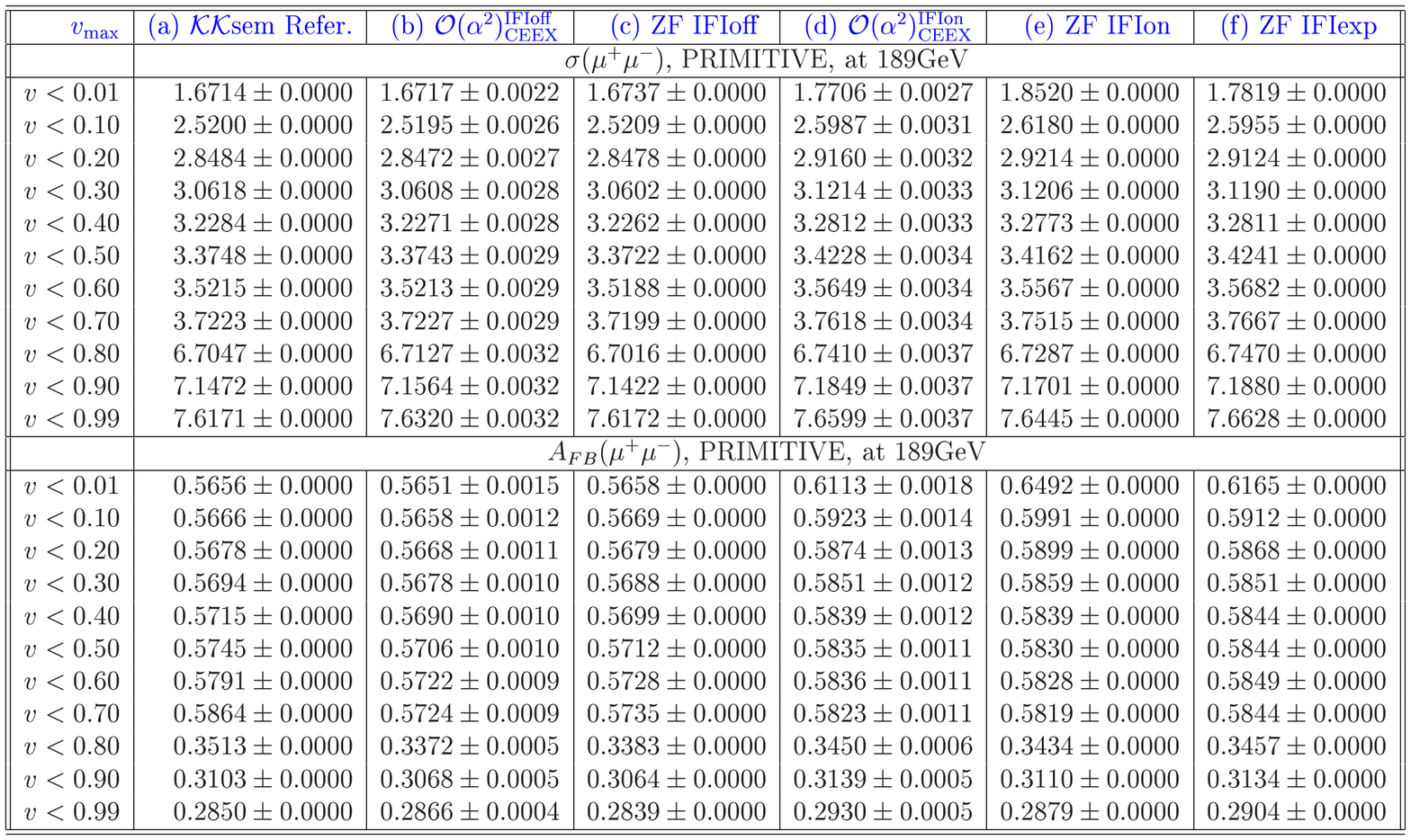,width=160mm,height=100mm}}}
\end{picture}
\caption{\small\sf 
  Cross sections and asymmetries from 
  \KKMC\ , ${\cal KK}$sem and \zf\  at 189GeV.
  The QED ISR$\otimes$FSR interference is switched on/off.
  No cut on $\cos\theta$. We define $v=1-s'/s$.
}
\label{tab:ifi-digital}
\end{table}

Finally, a subset of results which were presented visually
in Figs.~\ref{fig:ifi-first} and~\ref{fig:ifi-second}
we include also in Table~\ref{tab:ifi-digital} in a digital form,
as a reference benchmark for further studies.

\subsubsection{Conclusion on the uncertainty of IFI}

We have done the same at 206GeV and all results are practically the same,
because IFI depends on CMS energy only very weekly (logarithmically at most).

Summarizing, for the $\mu$-pair total cross section,
the uncertainty of the IFI is $\sim0.2\%$
well within the precision target $\delta\sigma^{\mu}/\sigma^{\mu}= 0.4\%$
of section 2.1 for both Z-inclusive and Z-exclusive cuts.
For the $\mu$-pair charge asymmetry the uncertainty of the IFI is $<0.3\%$,
also within the precision target  $\delta A^{\mu}_{FB}\sim 0.4-0.5\%$
of section 2.1 for both Z-inclusive and Z-exclusive cuts.

As seen in Table~\ref{tab:ifi-illustr}, the IFI corrections is factor 4-5 smaller
in the hadronic $\sigma^{h}$ than for $\sigma^{\mu}$,
so by scaling down the $\sim0.2\%$ uncertainty of the $\sigma^{\mu}$, 
we get something like $\sim0.05\%$, 
well below the precision target $\delta\sigma^{h}/\sigma^{h}= 0.1-0.2\%$ 
of section 2.1 for both Z-inclusive and Z-exclusive cuts.


\subsection{Tuned comparison of \zf\  6.30 and  \KKMC\  4.14}
\label{TunedZFKK}
Authors: \KKMC\ and \zf\  teams.

The main aim of this section is to compare {\cal KK}MC and \zf\ 
for hadronic total cross sections. 
In order to speed up calculations and make it easier to tune both programs,
we have switched off the ISR$\otimes$FSR interference.
We included the muon channel in all tests, just as a reference calculation.

Both of the programs \zf\ ~\cite{Bardin:1999yd-orig} and
\KKMC\ ~\cite{kkcpc:1999}
use the same library of electroweak form-factors (EWFF)
DIZET~\cite{Bardin:1989tq}.
The advantage of that is we can by comparing these programs
check very well the technical precision 
of the implementation of EW corrections
and the interplay of the EW and QED corrections. 
However, for these comparisons
we can draw little knowledge on the uncertainties of the pure EW corrections 
in DIZET.
For this one may consult the section on \zf\ in this report.
In the process of comparing \KKMC\  and \zf\  we have found out that
the simple semianalytical program ${\cal{KK}}$sem is very useful,
because it agrees always with \KKMC\  but is, of course, much faster.
The implementation of EW corrections in ${\cal{KK}}$sem is very similar 
to that in \KKMC\ , that it uses the same look-up tables of
$s$- and $t$-dependent EWFFs%
\footnote{ However, the effective Born distribution $d\sigma/d\cos\theta$ 
  is programmed in ${\cal{KK}}$sem independently and slightly differently,
  using a subprogram from KORALZ and not the Kleiss-Stirling spinors of 
  \KKMC\ .}.
We can use ${\cal{KK}}$sem also because in this section we restrict ourselves
to the simplest possible cut $v_{\max}=1-s'_{\min}/s$
on the invariant mass of the fermion pair,
or the propagator-mass, no cut on $\cos\theta$.

\begin{figure}[!ht]
\centering
\setlength{\unitlength}{0.1mm}
\begin{picture}(1600,1150)
\put(  100, 600){\makebox(0,0)[lb]{\epsfig{file=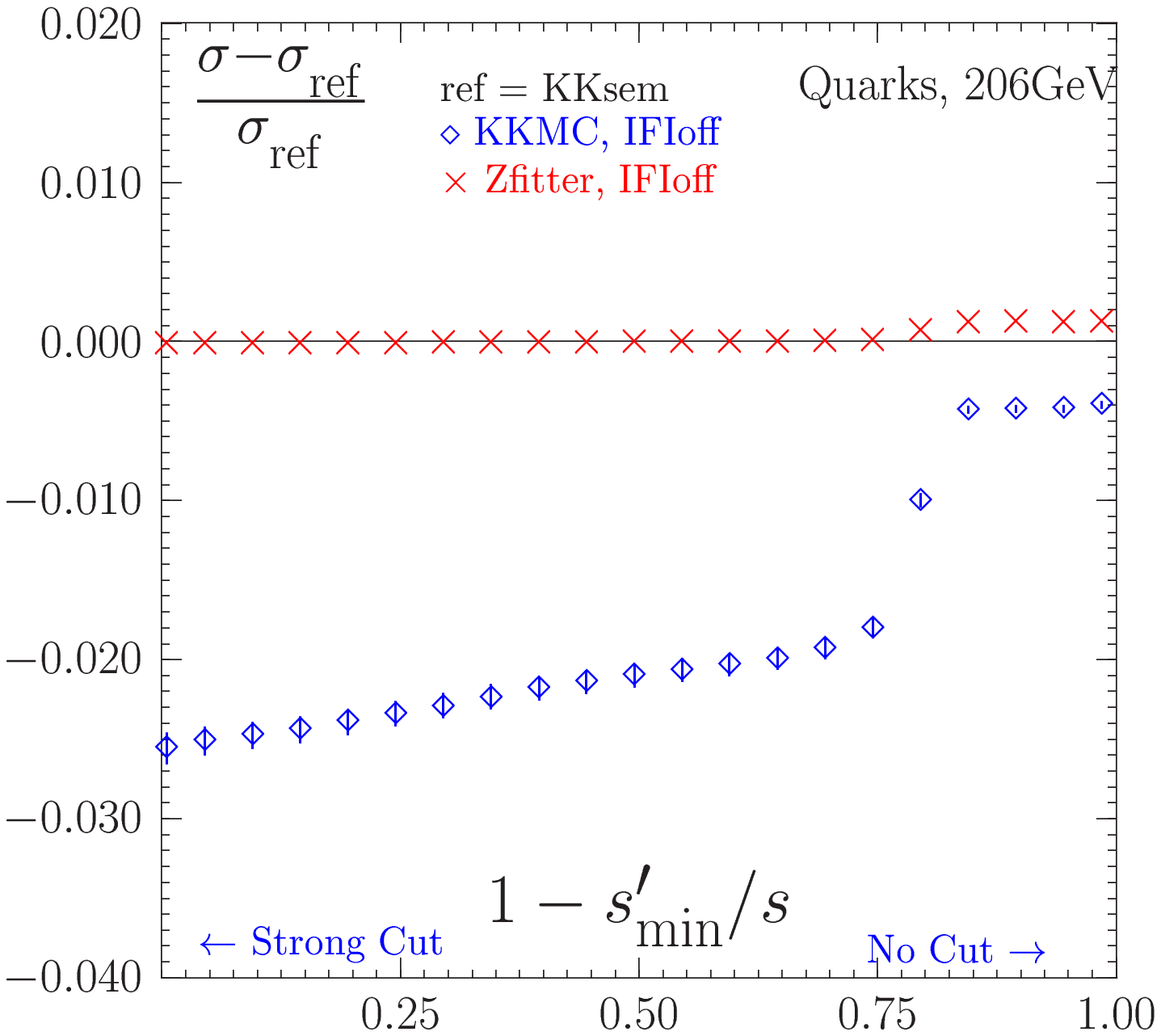,width=55mm,height=55mm}}}
\put(  920, 600){\makebox(0,0)[lb]{\epsfig{file=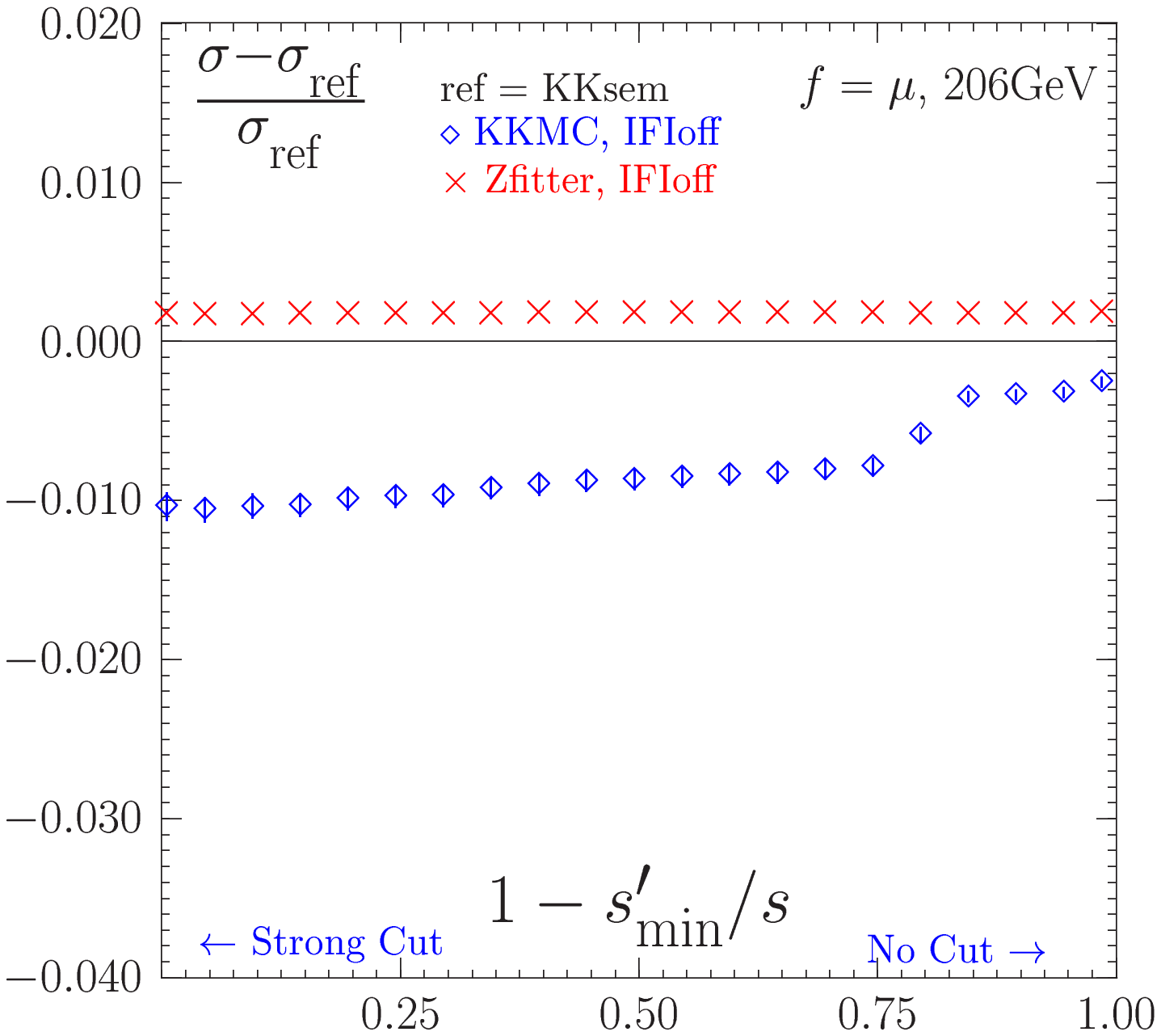,width=55mm,height=55mm}}}
\put(  -30,   0){\makebox(0,0)[lb]{\epsfig{file=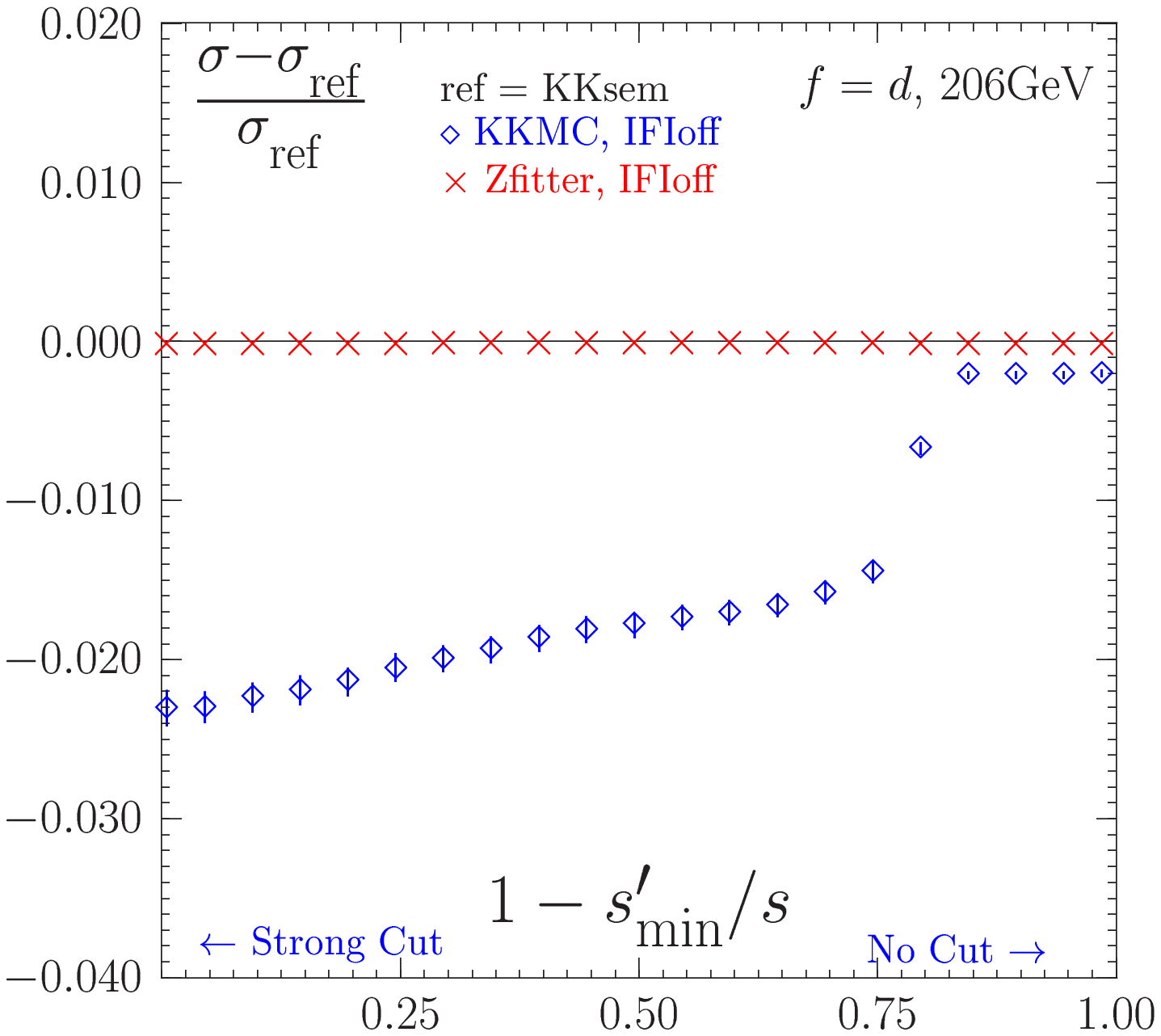,width=55mm,height=55mm}}}
\put(  520,   0){\makebox(0,0)[lb]{\epsfig{file=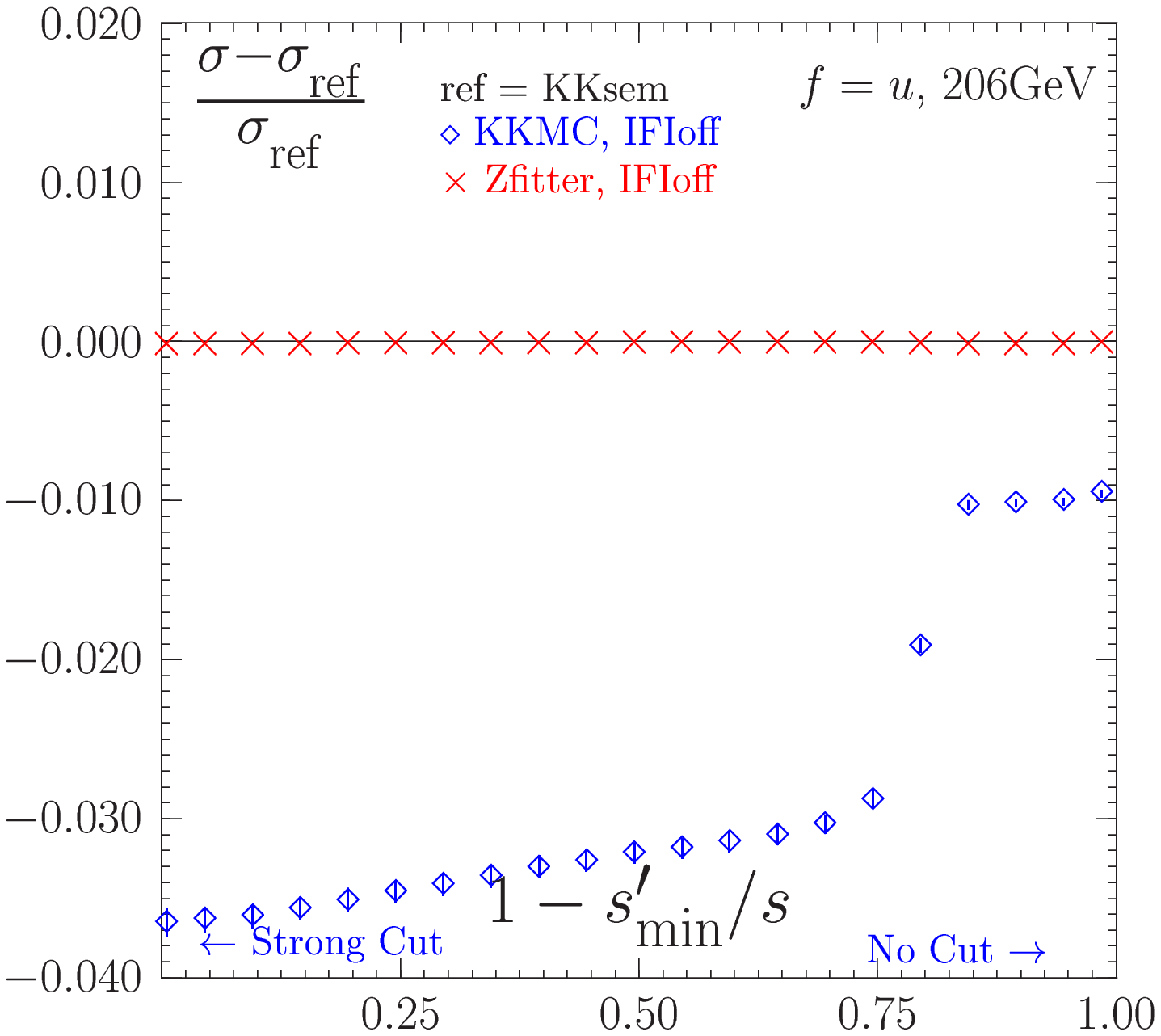,width=55mm,height=55mm}}}
\put( 1070,   0){\makebox(0,0)[lb]{\epsfig{file=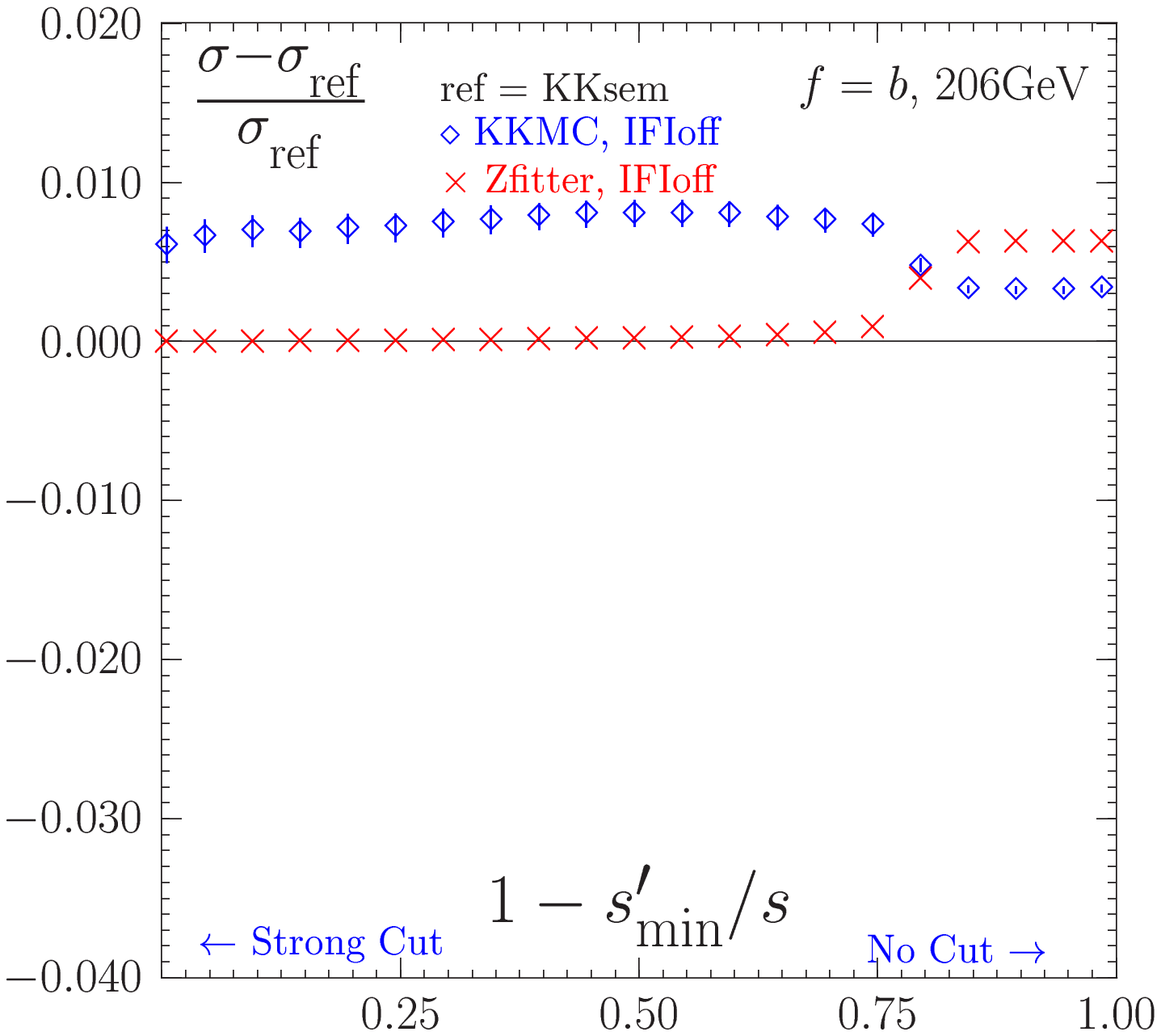,width=55mm,height=55mm}}}
\end{picture}
\caption{\small\sf 
 Electroweak boxes are OFF for \zf\  and ${\cal{KK}}$sem and ON for \KKMC\ .
 Cross section from \zf\ ,  \KKMC\  and ${\cal{KK}}$sem at 206GeV for quarks 
 and the muon.
 The ISR$\otimes$FSR is off.
 Results are plotted as a function of the cut of the propagator 
 mass $M_{f\bar{f}}=\sqrt{s'}$ with respect to ${\cal{KK}}$sem.
 No cut on $\cos\theta$.
 The QED and QCD FSR corrections are included.
\label{fig:zf-kk-isr206ibox}
}
\end{figure}

\subsubsection{The importance of the EW boxes and of running couplings}
Let us begin with emphasizing the fact that
the character of the electroweak corrections at LEP2 energies changes
dramatically with the onset of the so called ``EW-boxes'', that is to say box diagrams
with the exchange of the $W$ and $Z$ bosons, see Fig.~\ref{boxes}.
These genuinely quantum-mechanical contributions, which were negligible on Z resonances,
are above 2\% in the hadronic cross section at the highest LEP2 energies!
That is far bigger than the combined LEP2 experimental error,
almost as big as typical QED effects.
This point is illustrated by Fig.~\ref{fig:zf-kk-isr206ibox}
where we plot the cross section from ${\cal{KK}}$sem and \zf\  with EW boxes
switched off and from \KKMC\  in which EW boxes are switched on%
\footnote{We could of course switch on EW boxes in \zf\  and switch them off in 
  \KKMC\  -- the plot is just byproduct of one of our several tests}
(IBOXF=1 in DIZET).
As we see, at 206GeV, for the typical Z-exclusive cut $v_{\max}\sim 0.2$ ,
the EW boxes are the biggest for the $u$-quark, almost 4\%,
and after averaging over the five quarks%
\footnote{ The discrepancy for $b$-quark and Z-inclusive cut $v_{\max}\sim 0.9$
  is most probably due to simplified implementation of QCD FSR in ${\cal{KK}}$sem}
they contribute slightly above 2\%.
For the muon it is about 1\%.
We have also checked that at 189GeV
the contribution of the EW boxes is factor 2 smaller, both for quarks and muons.
Most probably the effect of EW boxes is slightly smaller for 
cross sections with the cut on $\cos\theta$.

We would like therefore to stress that the proper implementation of the EW boxes 
is of paramount
importance for the interpretation of the hadronic total cross section and
its energy dependence at the 
LEP2 energies.

\begin{figure}[!th]
\centering
\setlength{\unitlength}{0.1mm}
\begin{picture}(1600,1400)
\put(  -20, 600){\makebox(0,0)[lb]{\epsfig{file=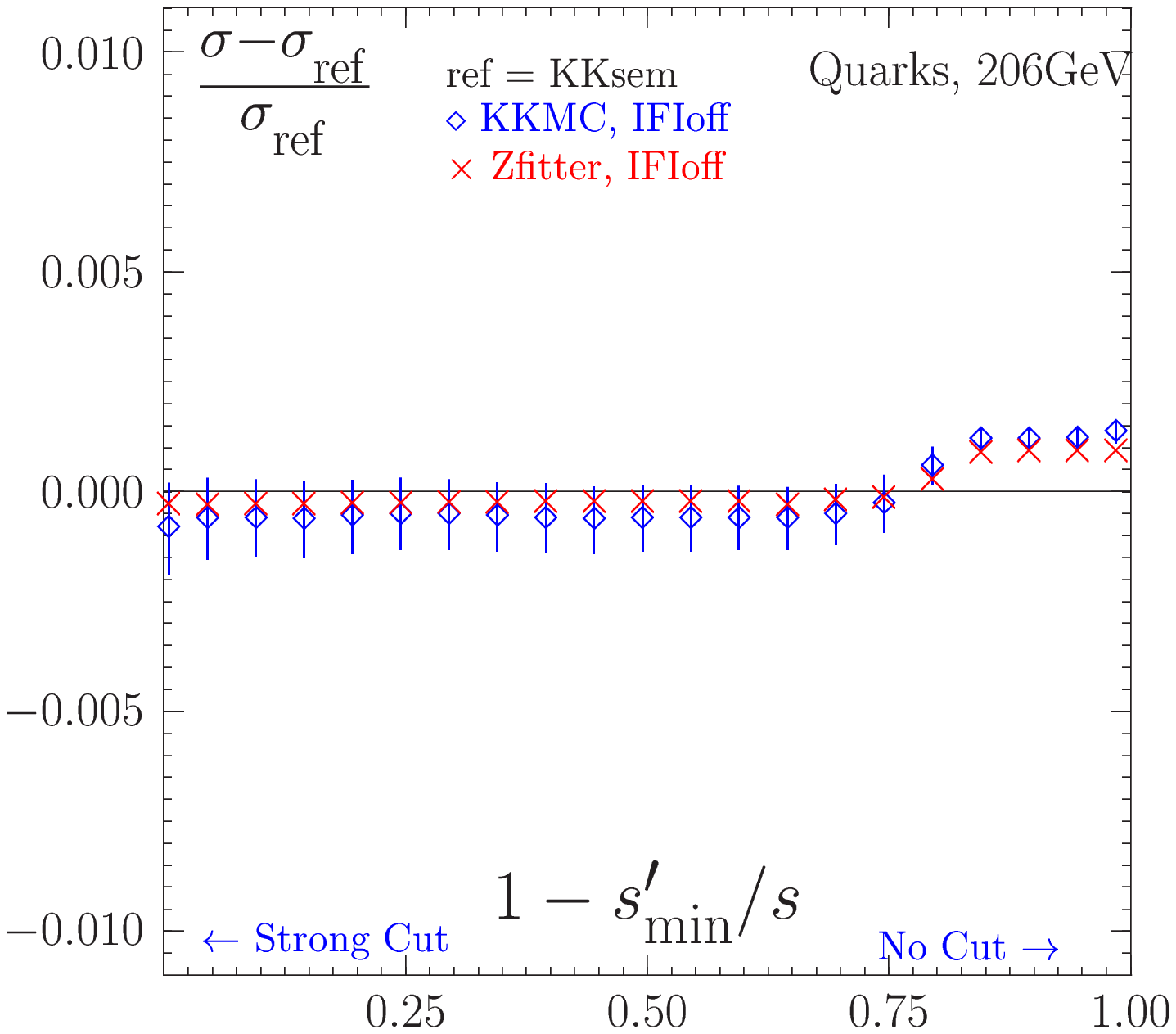,width=80mm,height=80mm}}}
\put(  800, 600){\makebox(0,0)[lb]{\epsfig{file=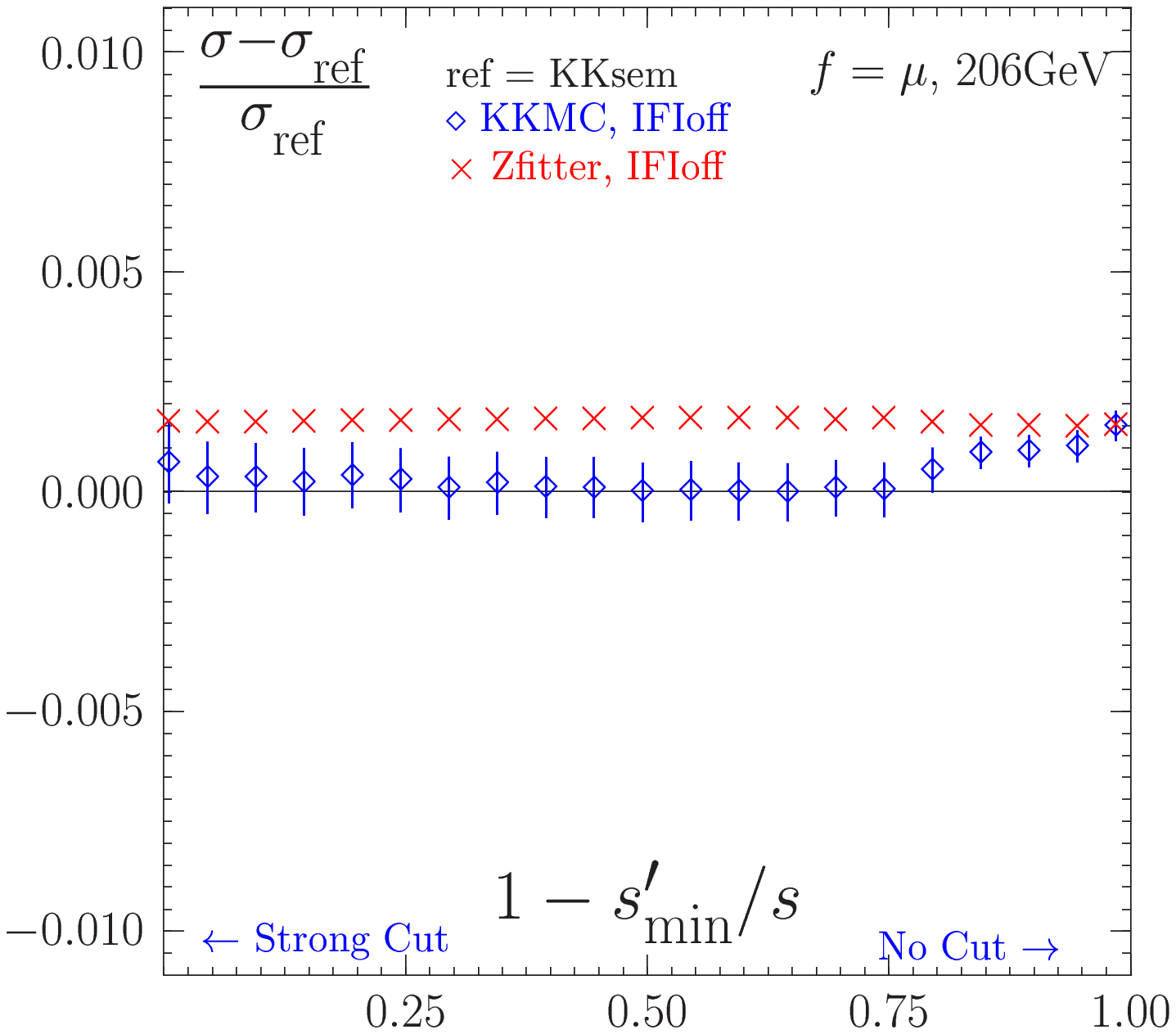,width=80mm,height=80mm}}}
\put(  -30,   0){\makebox(0,0)[lb]{\epsfig{file=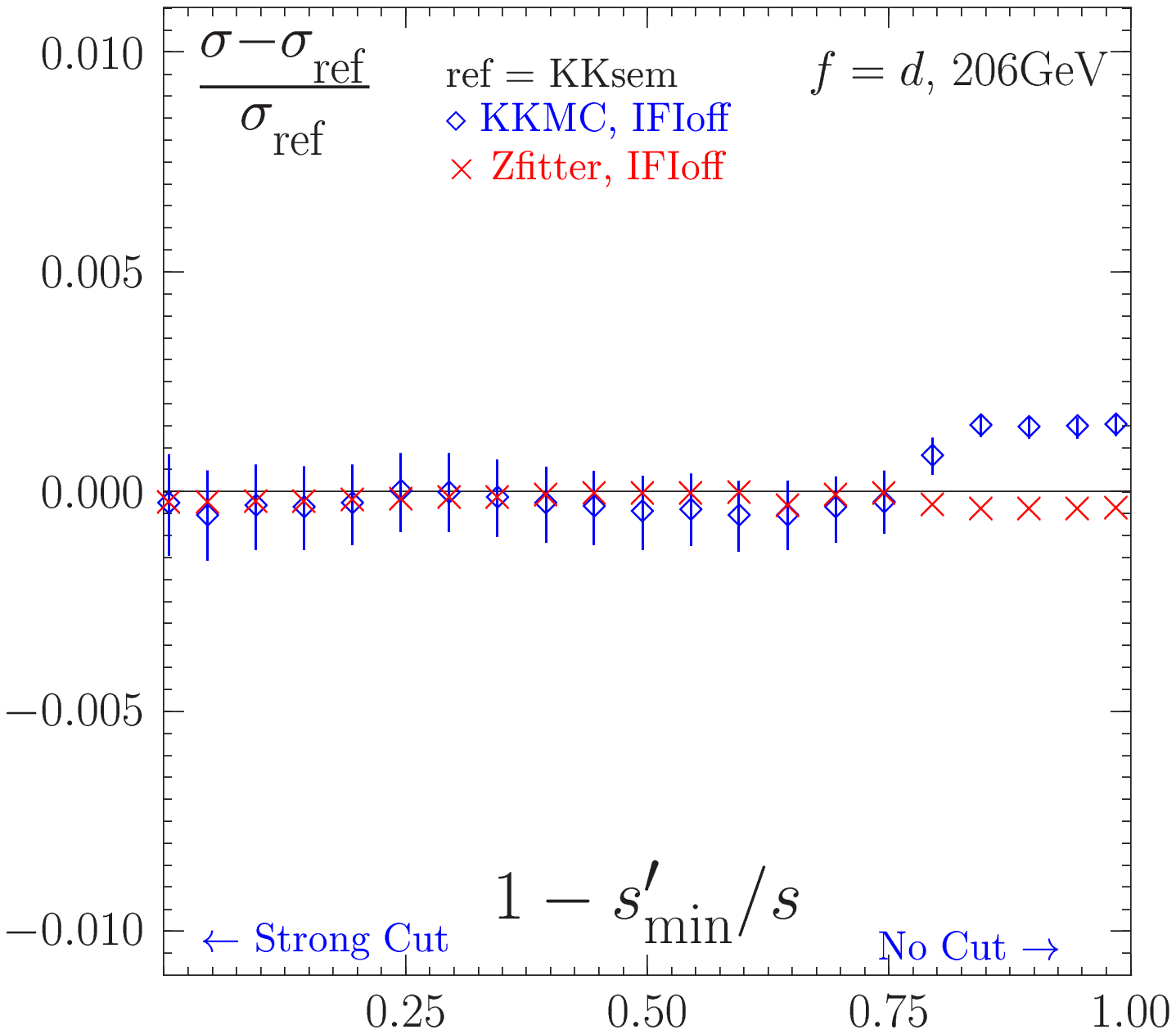,width=55mm,height=55mm}}}
\put(  520,   0){\makebox(0,0)[lb]{\epsfig{file=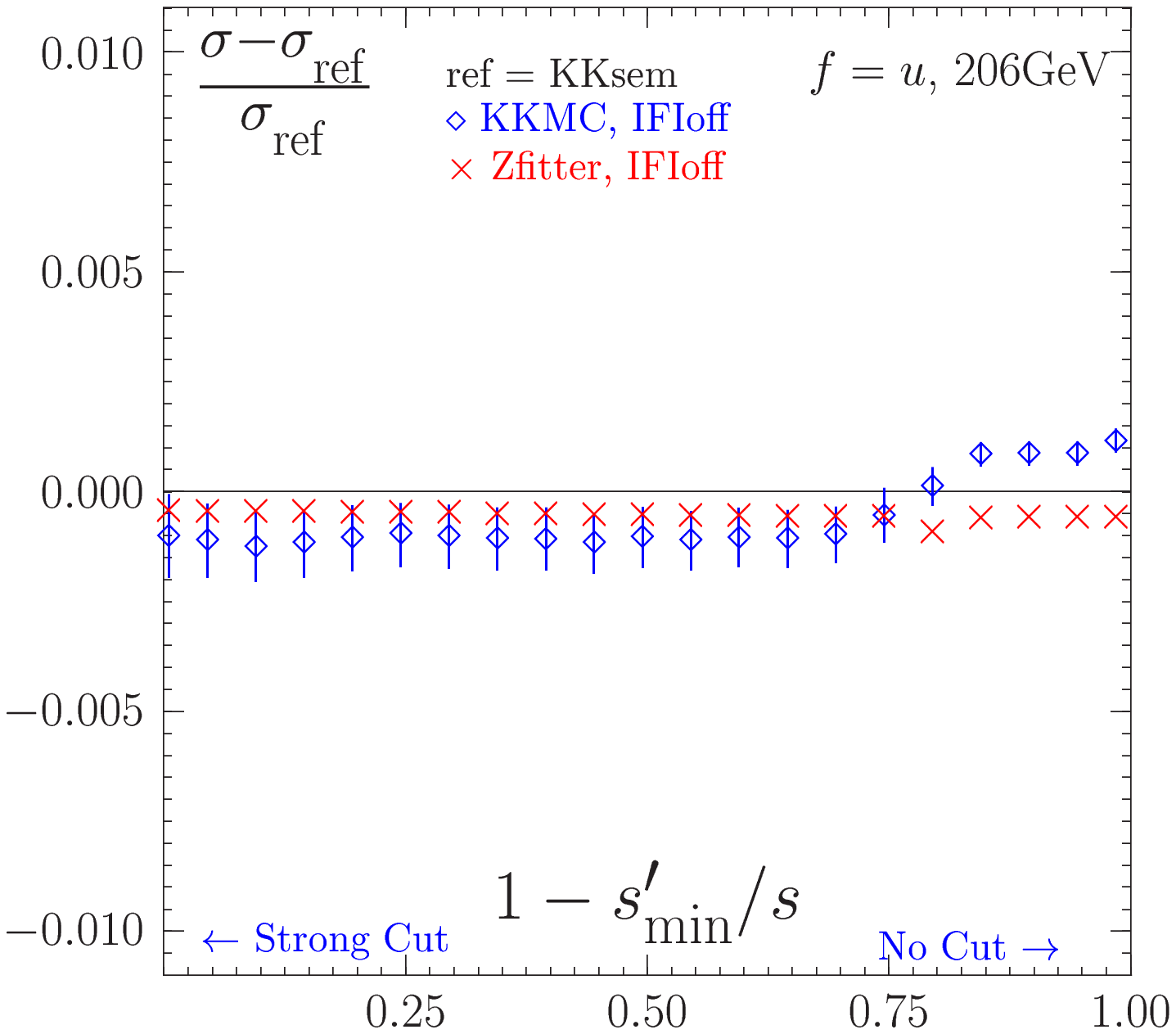,width=55mm,height=55mm}}}
\put( 1070,   0){\makebox(0,0)[lb]{\epsfig{file=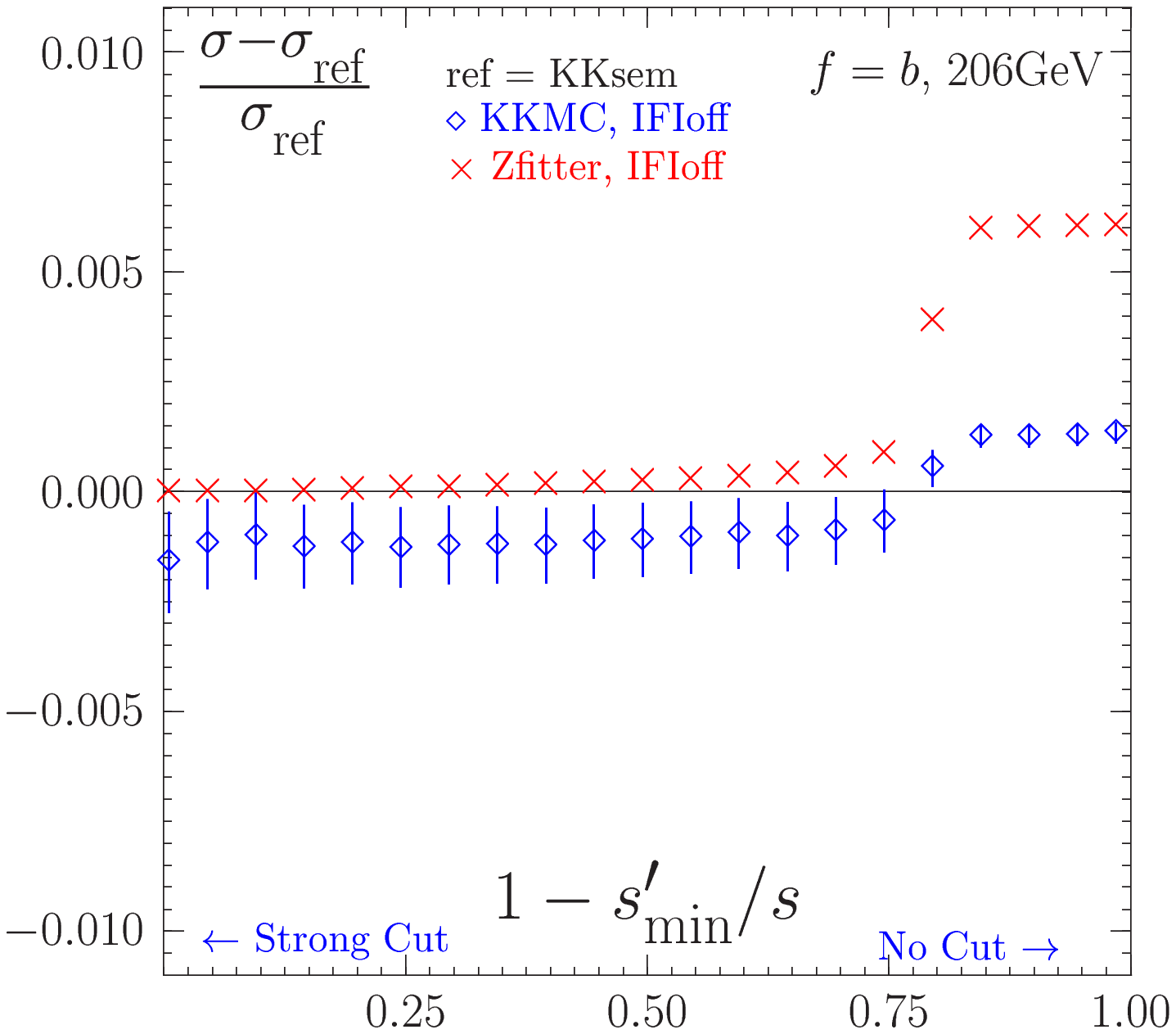,width=55mm,height=55mm}}}
\end{picture}
\caption{\small\sf 
 Total cross section from \zf\ ,  \KKMC\  and ${\cal{KK}}$sem
 at 206GeV for quarks and muon.
 ISR$\otimes$FSR is off.
 Results plotted as a function of the cut on the propagator 
 mass $M_{prop-}=\sqrt{s'}$ relative to ${\cal{KK}}$sem. 
 (The main result of this section.)
}
\label{fig:zf-kk-isr206}
\end{figure}

\begin{table}[!ht]
\centering
\setlength{\unitlength}{0.1mm}
\begin{picture}(1400,1200)
\put(  0, 0){\makebox(0,0)[lb]{\epsfig{file=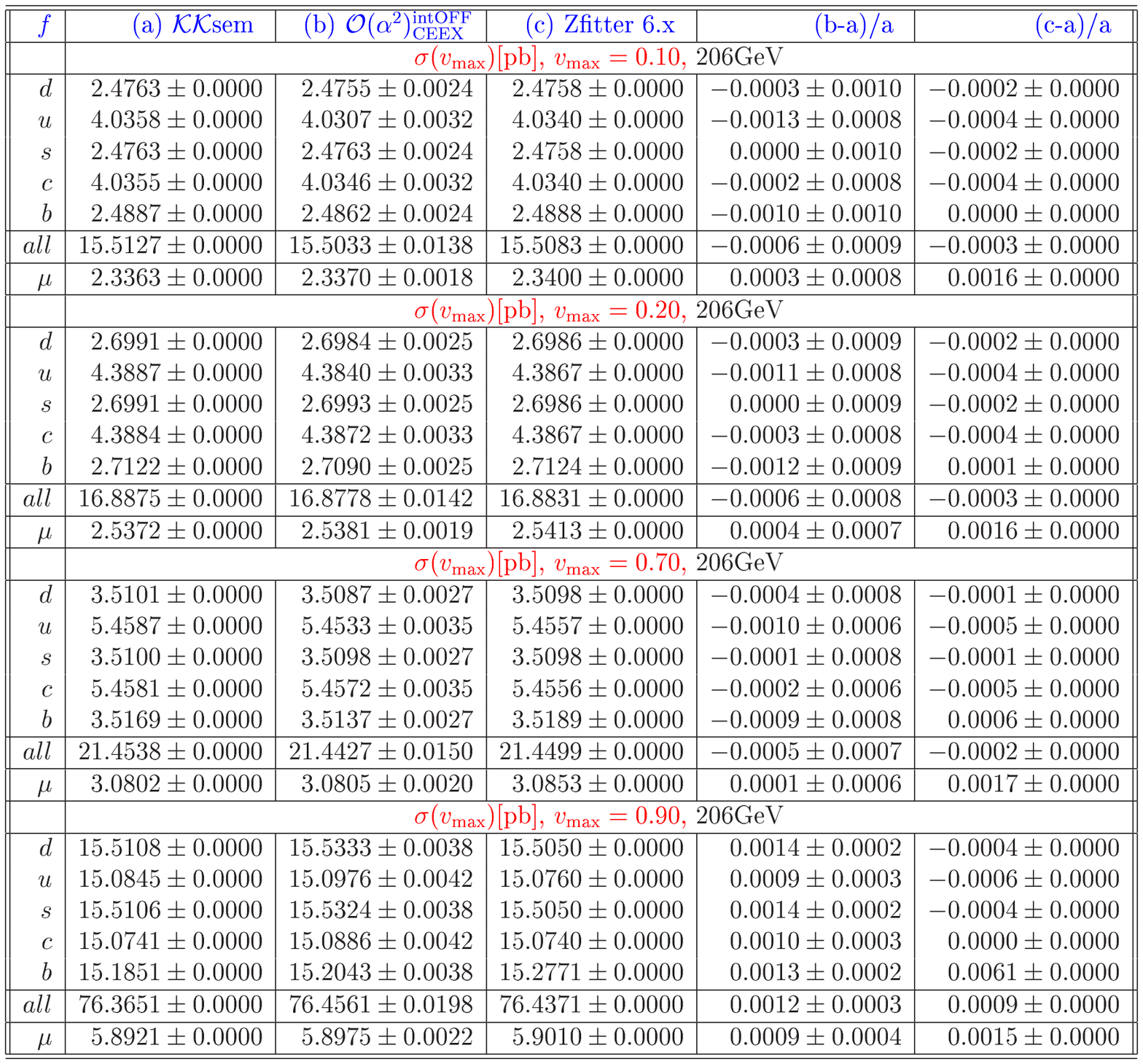,width=140mm,height=120mm}}}
\end{picture}
\caption{\small\sf 
 The same results as in Fig.~\ref{fig:zf-kk-isr206}, 
 for four values of the cut on the propagator mass, $v_{\max}=1-s'_{\min}/s$.
 The QED and QCD FSR corrections are included in both calculations.
}
\label{tab:zf-kk-isr206}
\end{table}

On the other hand, the Fig.~\ref{fig:zf-kk-isr206ibox} represents also a nice
technical cross-check of the implementation of convolution of the QED ISR
structure functions with the effective Born cross section in
\zf\  and ${\cal{KK}}$sem together with the use of DIZET for IBOXF=1.
In both programs we used the SF's of ref.~\cite{jadach:1991} with 
the \Order{L^3\alpha^3} corrections included.
It was also checked, by playing with the input flags of \zf , that
changing from factorized SF's of ref.~\cite{jadach:1991} to the additive
style of ref.~\cite{Kuraev:1985} and ref.~\cite{Montagna:1997jv}
(keeping \Order{L^3\alpha^3}) has very little 
influence (typically $<0.01\%$) on the cross section.
The same exercise was done at $\sqrt{s}$=189GeV and 206GeV, and 
(apart from slight discrepancy for the b-quark for the Z-inclusive cut, 
which seems to be understood)
the agreement was always consistently better than 0.2\%.

\begin{figure}[!ht]
\centering
\setlength{\unitlength}{0.1mm}
\begin{picture}(1200,1200)
\put(  -30, 600){\makebox(0,0)[lb]{\epsfig{file=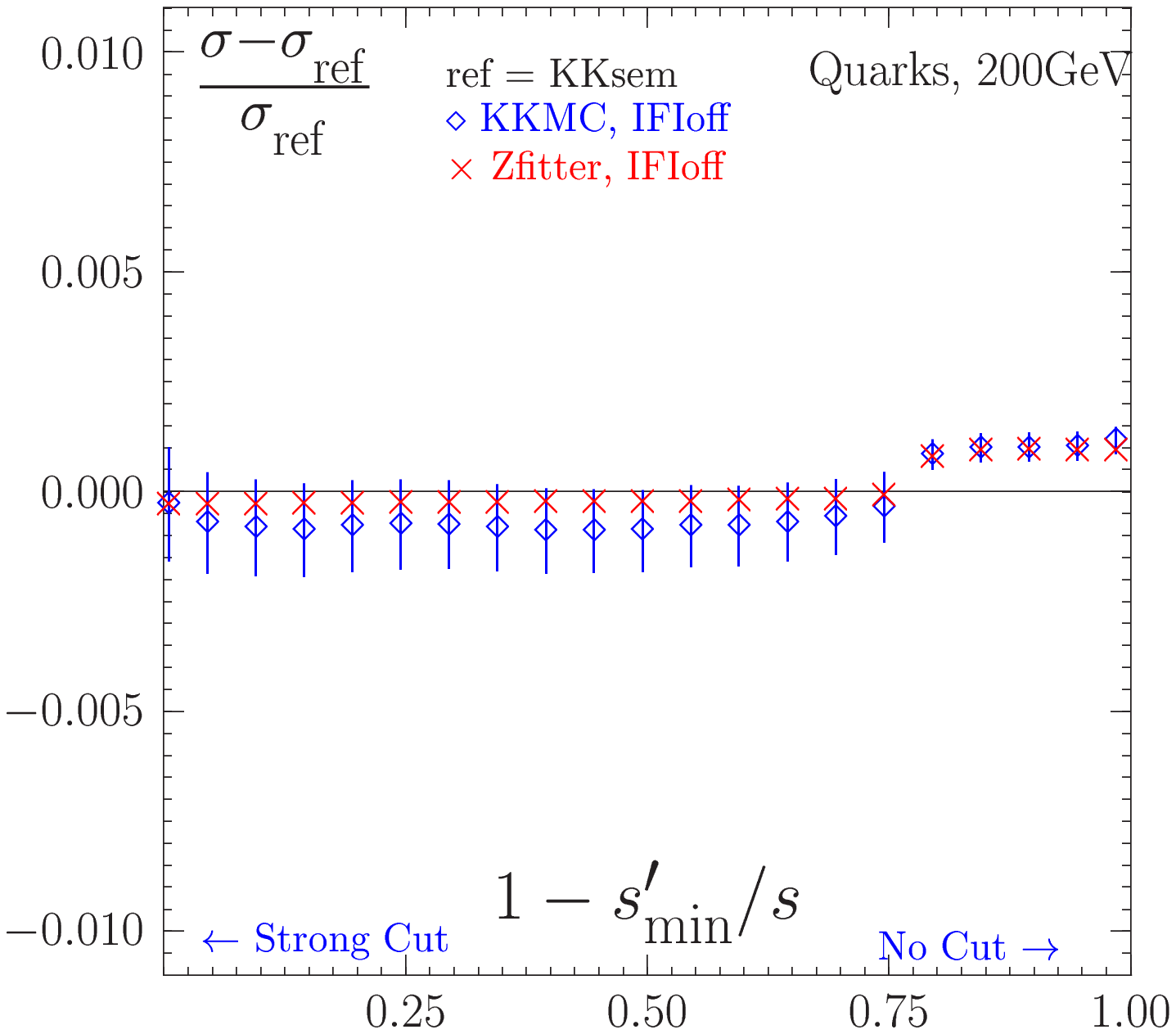,width=60mm,height=60mm}}}
\put(  600, 600){\makebox(0,0)[lb]{\epsfig{file=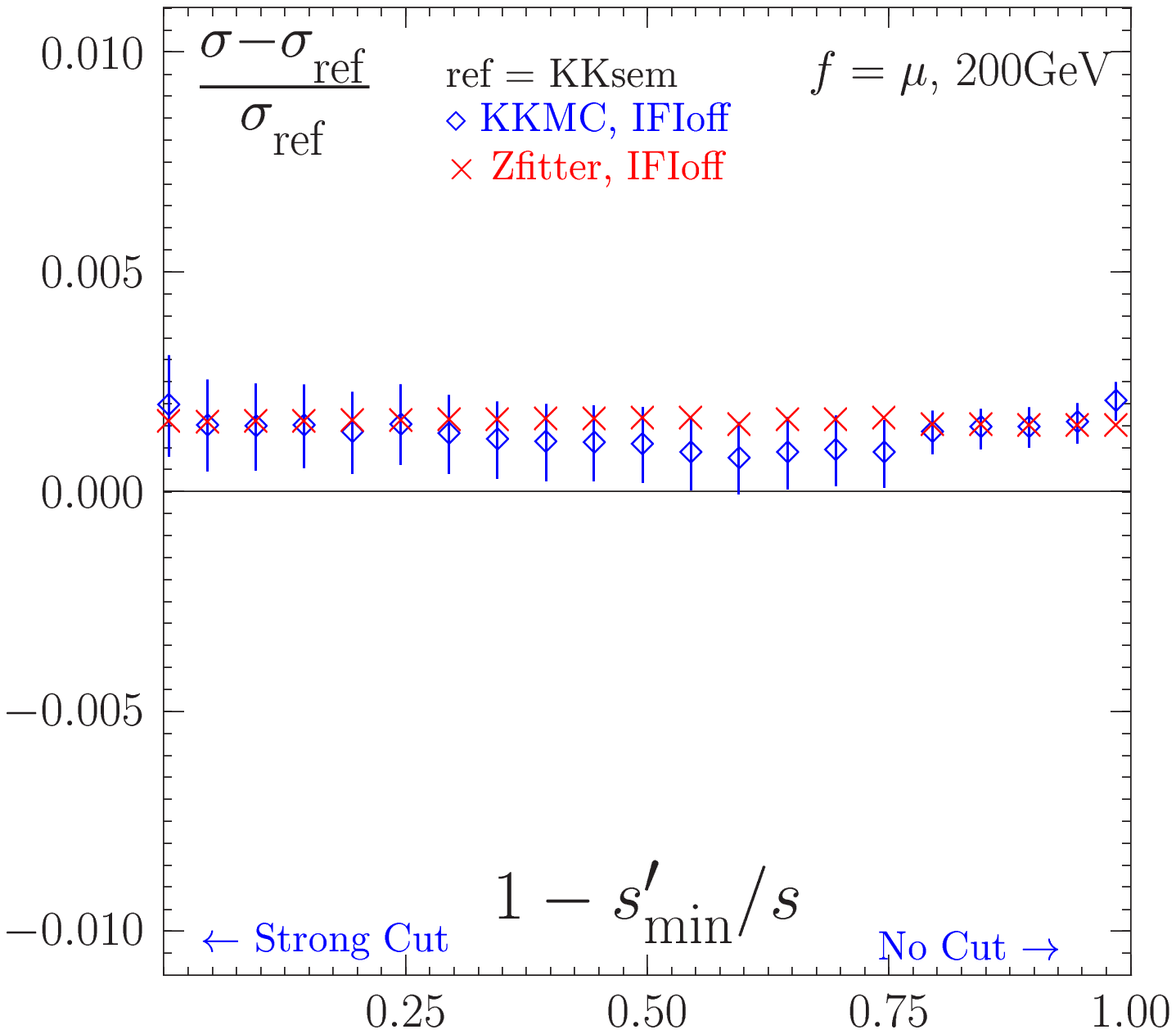,width=60mm,height=60mm}}}
\put(  -30,   0){\makebox(0,0)[lb]{\epsfig{file=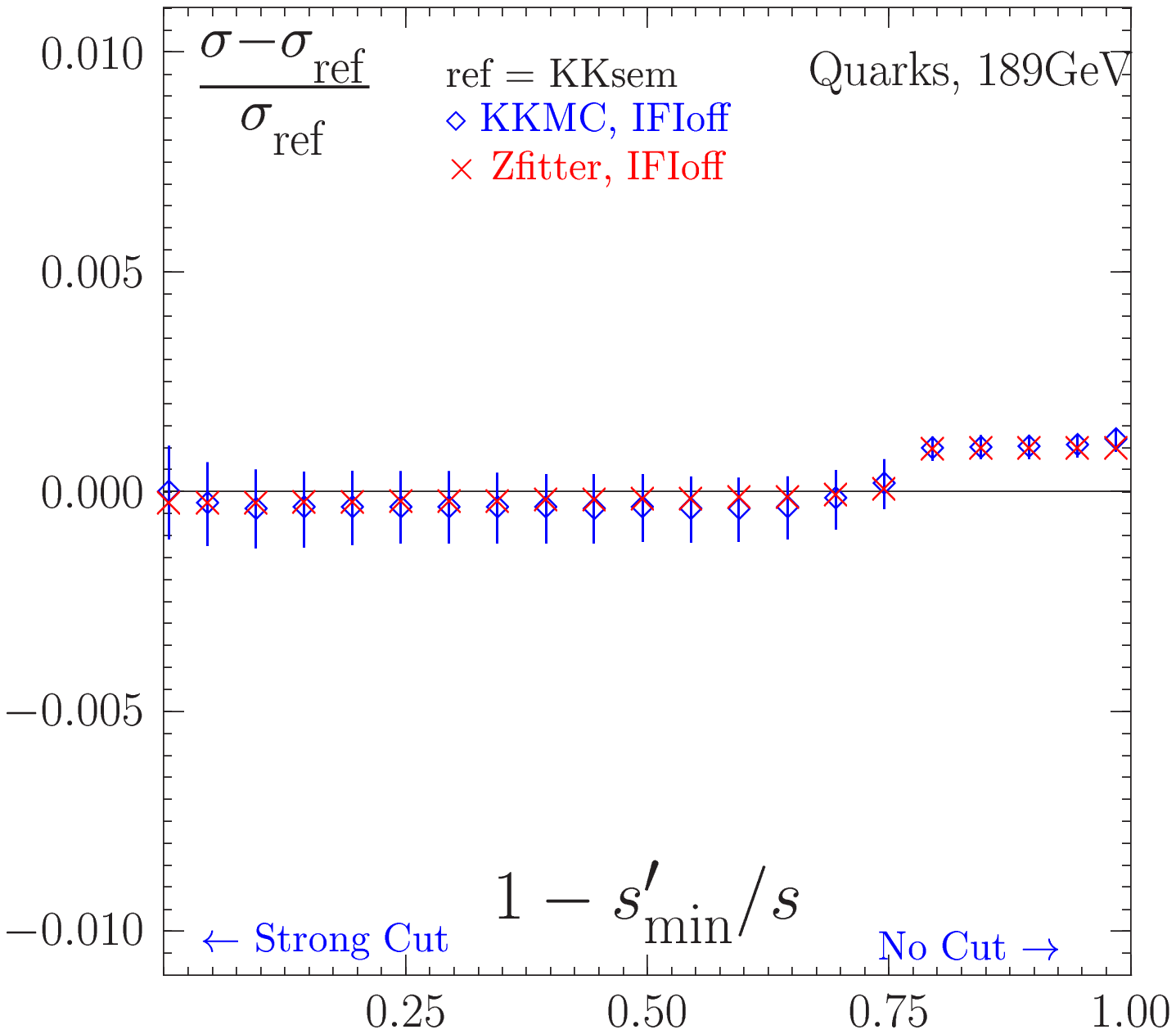,width=60mm,height=60mm}}}
\put(  600,   0){\makebox(0,0)[lb]{\epsfig{file=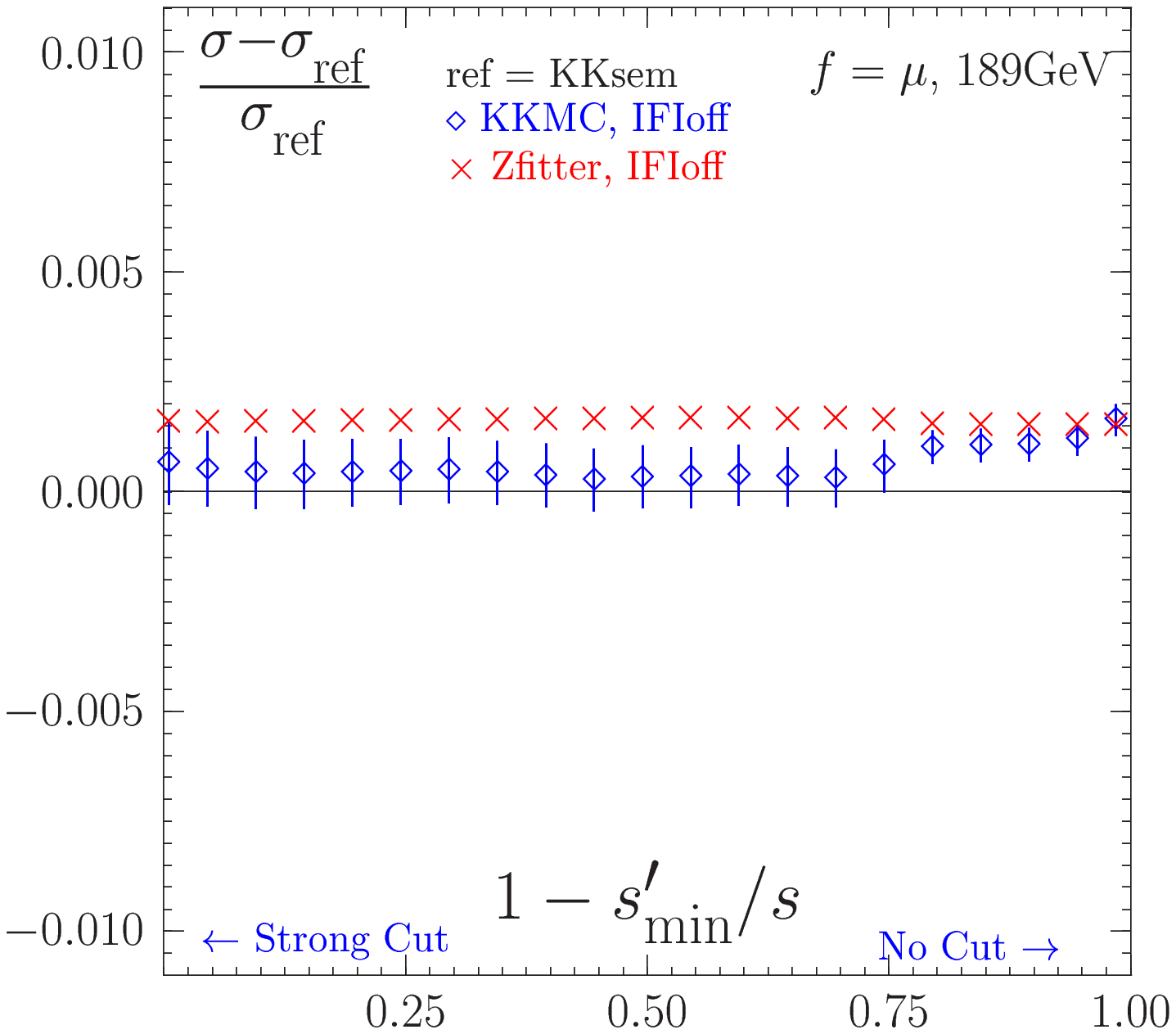,width=60mm,height=60mm}}}
\end{picture}
\caption{\small\sf 
  The same in Fig.~\ref{fig:zf-kk-isr206} for another $\sqrt{s}=$189GeV and 200GeV.
\label{fig:zf-kk-isr189-200}
}
\end{figure}

\subsubsection{All quarks channel by channel and muons, IFI switched off}
In the next exercise we switch on EW-boxes and examine the difference
between \zf\  and \KKMC\  for quarks and muons.
We keep also ${\cal{KK}}$sem all the time in the game -- this has proved to be very 
useful, because at all stages of our comparisons we could often substitute the comparison
among \zf\  and \KKMC\  by faster comparison among \zf\ and ${\cal{KK}}$sem,
profiting from the fact that the CPU time-consuming comparisons
between ${\cal{KK}}$sem and \KKMC\  were already done.
The resulting comparison at $\sqrt{s}=206$GeV is shown in Fig.~\ref{fig:zf-kk-isr206}
and some extract of it also in a numerical form in Tab.~\ref{tab:zf-kk-isr206}.
The agreement is very good $<0.2\%$,
for any value of the cut on propagator mass, for each quark, all quarks and the muon
(except for the $b$-quark, Z-inclusive cut, see remarks above).
The same kind of agreement we observed for 200GeV and 189GeV, 
see Fig.\ref{fig:zf-kk-isr189-200}.
with the maximum discrepancies for hadrons $<0.2\%$ and for muons $<0.3\%$
(a smaller statistical error is needed).

As we have learned during the process of comparisons,
the agreement of Fig.\ref{fig:zf-kk-isr206} was not possible to achieve, without
setting up properly the user options (flags) of \zf\ .  
Flags which were good for LEP1 can not be used at LEP2, 
in particular one should not use the options for approximate treatment of EW boxes
(ZUTHSM interface). One should use instead ZUATSM interface together with {\tt CONV=2}
standing for {\em running} electroweak couplings. 

On the \KKMC\ /${\cal{KK}}$sem part these problem did not arise, 
as they use only one method of implementing EW boxes: through $s$- and $t$-dependent
EWFFs plugged in directly into spin amplitudes, before squaring them.

\begin{figure}[!th]
\centering
\setlength{\unitlength}{0.1mm}
\begin{picture}(1200,600)
\put(  -20, 0){\makebox(0,0)[lb]{\epsfig{file=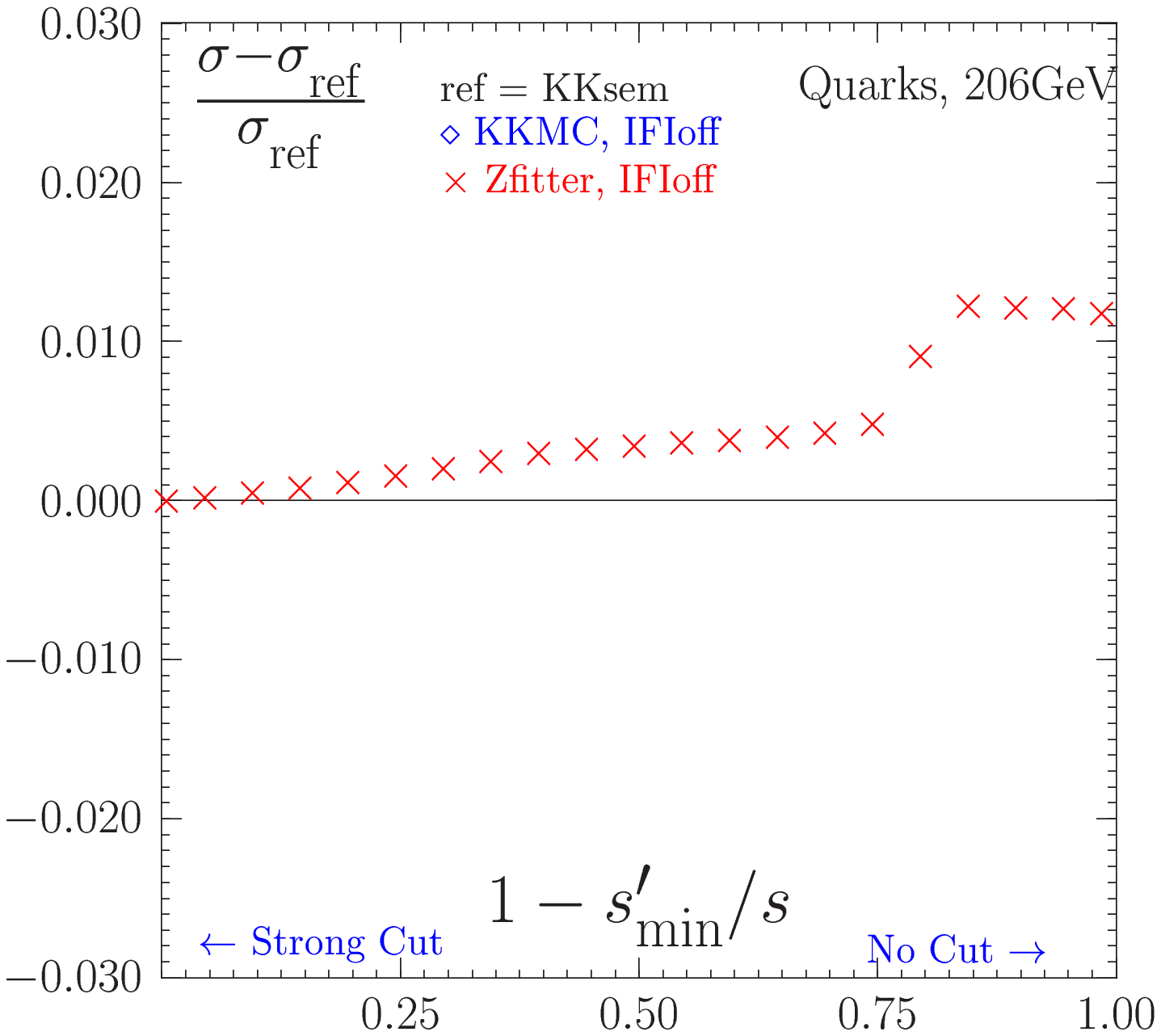,width=60mm,height=60mm}}}
\put(  600, 0){\makebox(0,0)[lb]{\epsfig{file=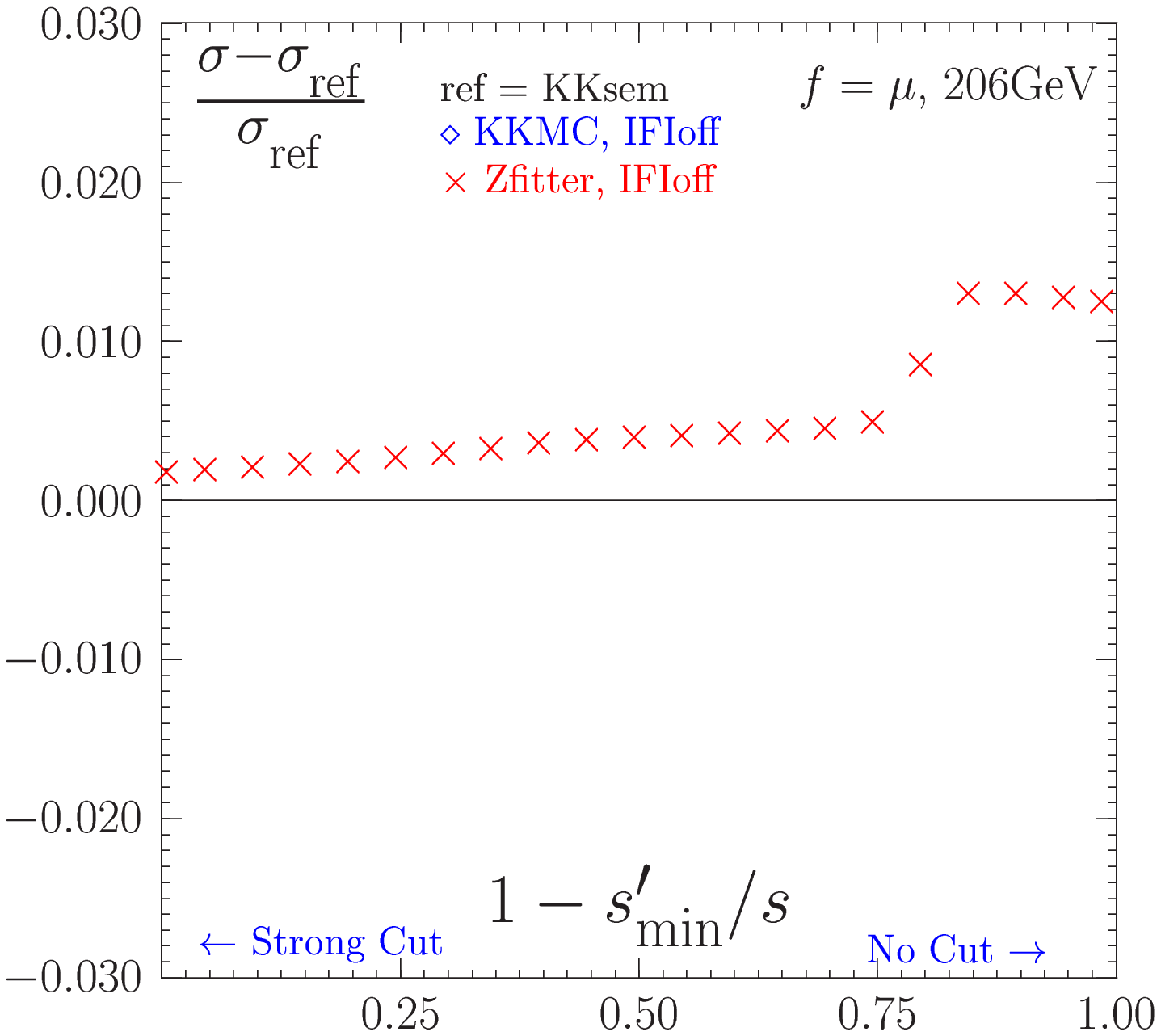,width=60mm,height=60mm}}}
\end{picture}
\caption{\small\sf 
  Illustration of importance of running EWFFs. 
  Plotted is the relative difference of DIZET cross section with the running of EWFFs switched off
  and ${\cal{KK}}$sem with the running  of EWFFs switched on.
}
\label{fig:zf-kk-isr206nrun}
\end{figure}

Finally, in Fig.~\ref{fig:zf-kk-isr206nrun}, we
would like to point out the importance of the running of EWFFs.
The effect is at most 0.1\% for Z-exclusive cuts, 
but is very sizeable $\sim$ 1-2\% at the Z radiative return!
It is a trivial effect but it should not be forgotten. 
Of course, the running of EWFFs was unimportant at LEP1.

Summarizing, the main result of this section is that 
of Figs.~\ref{fig:zf-kk-isr206} and~\ref{fig:zf-kk-isr189-200}.
It was highly nontrivial to get agreement at the level of $\sim$0.2\% of two large codes.
These comparisons test strongly the procedures in which pure EW corrections 
are combined with the QED in both programs and also the reliability of the QED ISR.
In particular results of these test do not invalidate the claim of \KKMC\  
authors that their program controls ISR at the level of 0.2\% for total cross section,
both for Z-exclusive and Z-inclusive acceptances.
The ISR$\otimes$FSR  was excluded from the tests of the present section.
They are done in another section of this report dedicated entirely
to this type of QED correction.

%% file: 2f-Chapt-Part5.tex
\subsection{Pair effects\label{PairEff}}
Let us concentrate in the present section on another class of corrections
which is important for the sub-percent precision tag as demanded by experiments:
the {\it pair corrections}.

Real and virtual secondary fermion pair \fft corrections to primary 2-fermion \ffo final states constitute non-trivial
problems, both experimentally and theoretically. The basis of the problems for real pairs is the existence
of several classes of Feynman diagrams,  all leading to  the final state \ffo\fft,
but  not all suitable of being considered as a radiative correction to \ffo production.
The definition of which part of these \ffo\fft processes should be taken as a radiative
correction to fermion-pair production is ambiguous.  The most useful guidelines for such a
definition are therefore its simplicity and generality, and the achievable accuracy of both 
experimental measurements and theoretical predictions.  The precision aims, as
discussed earlier in this report, are e.g. for hadrons (\ffo=qq) of the order 0.2\% for the "exclusive" high s' selection,
and 0.1\% for the inclusive selection, in order to be negligible with respect to the LEP-combined statistical error
 of these measurements. 
This subsection will  first
discuss basic features of possible 2f+4f signal definitions,  identify the most useful choices,
and  describe their realization in experimental measurements in terms of efficiency determination
and background subtraction and their realization in theoretical predictions. Finally a comparison
between different choices of signal definitions and different theoretical predictions is performed.

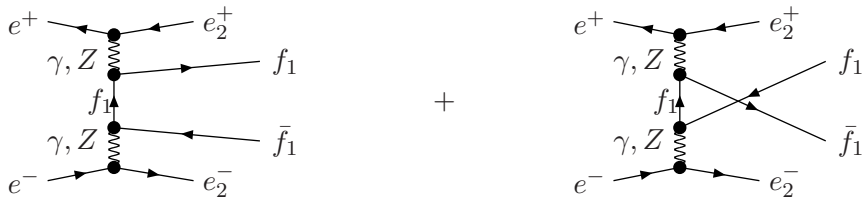
\begin{figure}[h] 
\vspace*{-6mm} 
\begin{eqnarray*} 
\begin{array}{ccc} 
\quad\;\vcenter{\hbox{
   \begin{picture}(150,100)(0,0)
     \ArrowLine(50,75)(25,80)        \Text(10,80)[lc]{$e^{+}$}
     \Photon(50,25)(50,40){2}{5}           \Text(25,30)[lcb]{$ \gamma,Z$}
     \ArrowLine(50,40)(50,60)        \Text(40,50)[lc]{$ f_1$}
     \Photon(50,60)(50,75){2}{5}            \Text(25,70)[lct]{$ \gamma,Z$}
     \ArrowLine(25,20)(50,25)        \Text(10,20)[lc]{$e^{-}$}
     \ArrowLine(80,80)(50,75)     \Text(90,80)[c]{$e^+_2$}
     \ArrowLine(50,25)(80,20)     \Text(90,20)[c]{$e^-_2$}
     \ArrowLine(50,60)(105,65)       \Text(120,65)[rc]{${f}_{1}$}
     \ArrowLine(105,35)(50,40)       \Text(120,35)[rc]{$\bar{f}_{1}$}
     \Vertex(50,75){2.5}
     \Vertex(50,25){2.5}
     \Vertex(50,60){2.5}
     \Vertex(50,40){2.5}
   \end{picture}}} 
&\quad+\qquad &
\vcenter{\hbox{
  \begin{picture}(150,100)(0,0)
     \ArrowLine(50,75)(25,80)        \Text(10,80)[lc]{$e^{+}$}
     \Photon(50,25)(50,40){2}{5}           \Text(25,30)[lcb]{$ \gamma,Z$}
     \ArrowLine(50,40)(50,60)        \Text(40,50)[lc]{$ f_1$}
     \Photon(50,60)(50,75){2}{5}            \Text(25,70)[lct]{$ \gamma,Z$}
     \ArrowLine(25,20)(50,25)        \Text(10,20)[lc]{$e^{-}$}
     \ArrowLine(80,80)(50,75)     \Text(90,80)[c]{$e^+_2$}
     \ArrowLine(50,25)(80,20)     \Text(90,20)[c]{$e^-_2$}
     \ArrowLine(50,60)(105,35)       \Text(120,35)[rc]{$\bar{f}_{1}$}
     \ArrowLine(105,65)(50,40)       \Text(120,65)[rc]{${f}_{1}$}
     \Vertex(50,75){2.5}
     \Vertex(50,25){2.5}
     \Vertex(50,60){2.5}
     \Vertex(50,40){2.5}
   \end{picture}}} 
\end{array} 
\end{eqnarray*} 
\vspace*{-5mm} 
\caption[]{The multi-peripheral (MP) group of diagrams.\label{pairs_MP}} \vspace*{-1mm} 
\end{figure}

\begin{figure}[h] 
\vspace*{-6mm} 
\begin{eqnarray*} 
\begin{array}{cc} 
\quad\;\vcenter{\hbox{
   \begin{picture}(150,100)(0,0)
     \ArrowLine(50,75)(25,80)        \Text(10,80)[lc]{$e^{+}$}
     \ArrowLine(50,25)(50,75)        \Text(40,50)[lc]{$ e$}
     \ArrowLine(25,20)(50,25)        \Text(10,20)[lc]{$e^{-}$}
     \Photon(50,75)(80,80){2}{7}     \Text(65,72.5)[ct]{$\gamma,Z$}
     \Photon(50,25)(80,20){2}{7}     \Text(65,27.5)[cb]{$\gamma$}
     \ArrowLine(105,65)(80,80)       \Text(120,65)[rc]{$\bar{f}_{1}$}
     \ArrowLine(80,80)(105,95)       \Text(120,95)[rc]{$f_{1}$}
     \ArrowLine(105,5)(80,20)        \Text(120,5)[rc]{$\bar{f}_{2}$}
     \ArrowLine(80,20)(105,35)       \Text(120,35)[rc]{$f_{2}$}
     \Vertex(50,75){2.5}
     \Vertex(50,25){2.5}
     \Vertex(80,80){2.5}
     \Vertex(80,20){2.5}
   \end{picture}}} 
&\quad+\qquad 
\vcenter{\hbox{
   \begin{picture}(150,100)(0,0)
     \ArrowLine(50,75)(25,80)        \Text(10,80)[lc]{$e^{+}$}
     \ArrowLine(50,25)(50,75)        \Text(40,50)[lc]{$ e$}
     \ArrowLine(25,20)(50,25)        \Text(10,20)[lc]{$e^{-}$}
     \Photon(50,75)(80,80){2}{7}     \Text(65,72.5)[ct]{$\gamma$}
     \Photon(50,25)(80,20){2}{7}     \Text(65,27.5)[cb]{$\gamma,Z$}
     \ArrowLine(105,65)(80,80)       \Text(120,65)[rc]{$\bar{f}_{2}$}
     \ArrowLine(80,80)(105,95)       \Text(120,95)[rc]{$f_{2}$}
     \ArrowLine(105,5)(80,20)        \Text(120,5)[rc]{$\bar{f}_{1}$}
     \ArrowLine(80,20)(105,35)       \Text(120,35)[rc]{$f_{1}$}
     \Vertex(50,75){2.5} 
     \Vertex(50,25){2.5}
     \Vertex(80,80){2.5}
     \Vertex(80,20){2.5}   
\end{picture}}} 
\end{array} 
\end{eqnarray*} 
\vspace*{-5mm} 
\caption[]{The subgroup ISNS$_\gamma$ of the NC08 sub-family of diagrams.
\label{pairs_ISNSG}} 
\vspace*{-1mm} \end{figure}
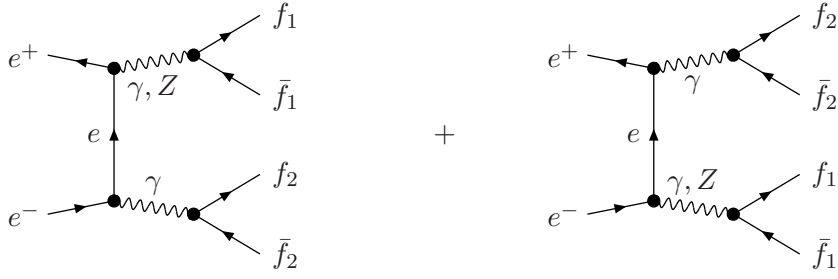

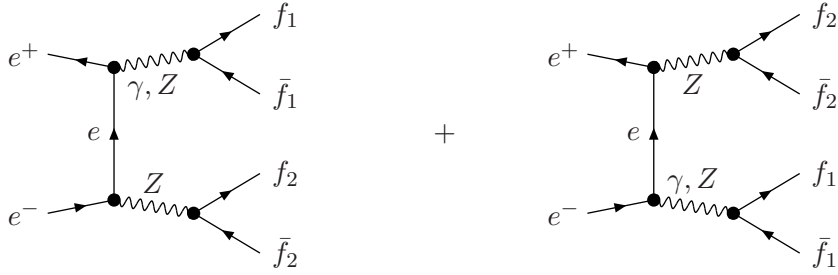
\begin{figure}[h] 
\vspace*{-6mm} 
\begin{eqnarray*} 
\begin{array}{cc} 
\quad\;\vcenter{\hbox{
   \begin{picture}(150,100)(0,0)
     \ArrowLine(50,75)(25,80)        \Text(10,80)[lc]{$e^{+}$}
     \ArrowLine(50,25)(50,75)        \Text(40,50)[lc]{$ e$}
     \ArrowLine(25,20)(50,25)        \Text(10,20)[lc]{$e^{-}$}
     \Photon(50,75)(80,80){2}{7}     \Text(65,72.5)[ct]{$\gamma,Z$}
     \Photon(50,25)(80,20){2}{7}     \Text(65,27.5)[cb]{$Z$}
     \ArrowLine(105,65)(80,80)       \Text(120,65)[rc]{$\bar{f}_{1}$}
     \ArrowLine(80,80)(105,95)       \Text(120,95)[rc]{$f_{1}$}
     \ArrowLine(105,5)(80,20)        \Text(120,5)[rc]{$\bar{f}_{2}$}
     \ArrowLine(80,20)(105,35)       \Text(120,35)[rc]{$f_{2}$}
     \Vertex(50,75){2.5}
     \Vertex(50,25){2.5}
     \Vertex(80,80){2.5}
     \Vertex(80,20){2.5}
   \end{picture}}} 
&\quad+\qquad 
\vcenter{\hbox{
   \begin{picture}(150,100)(0,0)
     \ArrowLine(50,75)(25,80)        \Text(10,80)[lc]{$e^{+}$}
     \ArrowLine(50,25)(50,75)        \Text(40,50)[lc]{$ e$}
     \ArrowLine(25,20)(50,25)        \Text(10,20)[lc]{$e^{-}$}
     \Photon(50,75)(80,80){2}{7}     \Text(65,72.5)[ct]{$Z$}
     \Photon(50,25)(80,20){2}{7}     \Text(65,27.5)[cb]{$\gamma,Z$}
     \ArrowLine(105,65)(80,80)       \Text(120,65)[rc]{$\bar{f}_{2}$}
     \ArrowLine(80,80)(105,95)       \Text(120,95)[rc]{$f_{2}$}
     \ArrowLine(105,5)(80,20)        \Text(120,5)[rc]{$\bar{f}_{1}$}
     \ArrowLine(80,20)(105,35)       \Text(120,35)[rc]{$f_{1}$}
     \Vertex(50,75){2.5} 
     \Vertex(50,25){2.5}
     \Vertex(80,80){2.5}
     \Vertex(80,20){2.5}   
\end{picture}}} 
\end{array} 
\end{eqnarray*} 
\vspace*{-5mm} 
\caption[]{The subgroup ISNS$_Z$ of the NC08 sub-family of diagrams.
\label{pairs_ISNSZ}} 
\vspace*{-1mm} 
\end{figure}

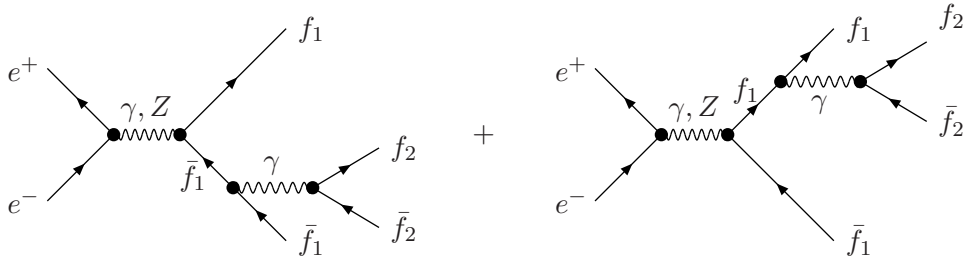
\begin{figure}[h] 
\vspace*{-8mm} 
\begin{eqnarray*} 
\begin{array}{ccc} 
\vcenter{\hbox{
   \begin{picture}(165,100)(0,0)
     \ArrowLine(25,25)(50,50)        \Text(10,25)[lc]{$e^{-}$}
     \ArrowLine(50,50)(25,75)        \Text(10,75)[lc]{$e^{+}$}
     \Photon(50,50)(75,50){2}{7}     \Text(62.5,55)[cb]{$\gamma,Z$}
     \ArrowLine(95,30)(75,50)        \Text(75,40)[lt]{$\bar{f}_{1}$}
     \ArrowLine(115,10)(95,30)       \Text(130,10)[rc]{$\bar{f}_{1}$}
     \ArrowLine(75,50)(115,90)       \Text(130,90)[rc]{$f_{1}$}
     \Photon(95,30)(125,30){2}{7}    \Text(110,35)[cb]{$\gamma$}
     \ArrowLine(150,15)(125,30)      \Text(165,15)[rc]{$\bar{f}_{2}$}
     \ArrowLine(125,30)(150,45)      \Text(165,45)[rc]{$f_{2}$}
     \Vertex(50,50){2.5}
     \Vertex(75,50){2.5}
     \Vertex(95,30){2.5}
     \Vertex(125,30){2.5}
   \end{picture}}} 
&\quad+& 
\vcenter{\hbox{
   \begin{picture}(165,100)(0,0)
     \ArrowLine(25,25)(50,50)        \Text(10,25)[lc]{$e^{-}$}
     \ArrowLine(50,50)(25,75)        \Text(10,75)[lc]{$e^{+}$}
     \Photon(50,50)(75,50){2}{7}     \Text(62.5,55)[bc]{$\gamma,Z$}
     \ArrowLine(115,10)(75,50)       \Text(130,10)[rc]{$\bar{f}_{1}$}
     \ArrowLine(75,50)(95,70)        \Text(77,62)[lb]{$f_{1}$}
     \ArrowLine(95,70)(115,90)       \Text(130,90)[rc]{$f_{1}$}
     \Photon(95,70)(125,70){2}{7}    \Text(110,65)[ct]{$\gamma$}
     \ArrowLine(150,55)(125,70)      \Text(165,55)[rc]{$\bar{f}_{2}$}
     \ArrowLine(125,70)(150,85)      \Text(165,95)[rc]{$f_{2}$}
     \Vertex(50,50){2.5}
     \Vertex(75,50){2.5}
     \Vertex(95,70){2.5}
     \Vertex(125,70){2.5}
   \end{picture}}} 
\end{array} 
\end{eqnarray*} 
\caption[]{The four diagrams of subgroup FS$_\gamma$  belonging to the NC24 process.\label{pairs_FSG}} 
\vspace*{-1mm} 
\end{figure}

\begin{figure}[h] 
\vspace*{-8mm} 
\begin{eqnarray*} 
\begin{array}{ccc} 
\vcenter{\hbox{
   \begin{picture}(165,100)(0,0)
     \ArrowLine(25,25)(50,50)        \Text(10,25)[lc]{$e^{-}$}
     \ArrowLine(50,50)(25,75)        \Text(10,75)[lc]{$e^{+}$}
     \Photon(50,50)(75,50){2}{7}     \Text(62.5,55)[cb]{$\gamma,Z$}
     \ArrowLine(95,30)(75,50)        \Text(75,40)[lt]{$\bar{f}_{1}$}
     \ArrowLine(115,10)(95,30)       \Text(130,10)[rc]{$\bar{f}_{1}$}
     \ArrowLine(75,50)(115,90)       \Text(130,90)[rc]{$f_{1}$}
     \Photon(95,30)(125,30){2}{7}    \Text(110,35)[cb]{$Z$}
     \ArrowLine(150,15)(125,30)      \Text(165,15)[rc]{$\bar{f}_{2}$}
     \ArrowLine(125,30)(150,45)      \Text(165,45)[rc]{$f_{2}$}
     \Vertex(50,50){2.5}
     \Vertex(75,50){2.5}
     \Vertex(95,30){2.5}
     \Vertex(125,30){2.5}
   \end{picture}}} 
&\quad+& 
\vcenter{\hbox{
   \begin{picture}(165,100)(0,0)
     \ArrowLine(25,25)(50,50)        \Text(10,25)[lc]{$e^{-}$}
     \ArrowLine(50,50)(25,75)        \Text(10,75)[lc]{$e^{+}$}
     \Photon(50,50)(75,50){2}{7}     \Text(62.5,55)[bc]{$\gamma,Z$}
     \ArrowLine(115,10)(75,50)       \Text(130,10)[rc]{$\bar{f}_{1}$}
     \ArrowLine(75,50)(95,70)        \Text(77,62)[lb]{$f_{1}$}
     \ArrowLine(95,70)(115,90)       \Text(130,90)[rc]{$f_{1}$}
     \Photon(95,70)(125,70){2}{7}    \Text(110,65)[ct]{$Z$}
     \ArrowLine(150,55)(125,70)      \Text(165,55)[rc]{$\bar{f}_{2}$}
     \ArrowLine(125,70)(150,85)      \Text(165,95)[rc]{$f_{2}$}
     \Vertex(50,50){2.5}
     \Vertex(75,50){2.5}
     \Vertex(95,70){2.5}
     \Vertex(125,70){2.5}
   \end{picture}}} 
\end{array} 
\end{eqnarray*} 
\caption[]{The four diagrams of subgroup FS$_Z$  belonging to the NC24 process.\label{pairs_FSZ}} 
\vspace*{-1mm} 
\end{figure}
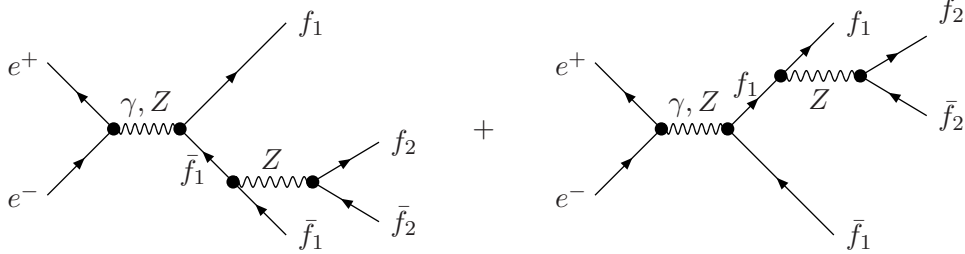

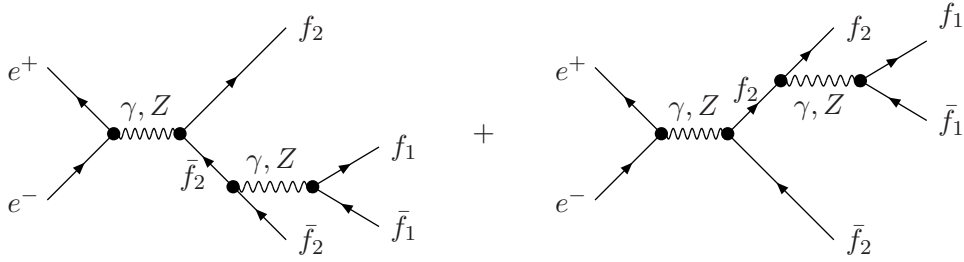
\begin{figure}[h] 
\vspace*{-10mm} 
\begin{eqnarray*} 
\begin{array}{ccc} 
\vcenter{\hbox{
   \begin{picture}(165,100)(0,0)
     \ArrowLine(25,25)(50,50)        \Text(10,25)[lc]{$e^{-}$}
     \ArrowLine(50,50)(25,75)        \Text(10,75)[lc]{$e^{+}$}
     \Photon(50,50)(75,50){2}{7}     \Text(62.5,55)[cb]{$\gamma,Z$}
     \ArrowLine(95,30)(75,50)        \Text(75,40)[lt]{$\bar{f}_{2}$}
     \ArrowLine(115,10)(95,30)       \Text(130,10)[rc]{$\bar{f}_{2}$}
     \ArrowLine(75,50)(115,90)       \Text(130,90)[rc]{$f_{2}$}
     \Photon(95,30)(125,30){2}{7}    \Text(110,35)[cb]{$\gamma,Z$}
     \ArrowLine(150,15)(125,30)      \Text(165,15)[rc]{$\bar{f}_{1}$}
     \ArrowLine(125,30)(150,45)      \Text(165,45)[rc]{$f_{1}$}
     \Vertex(50,50){2.5}
     \Vertex(75,50){2.5}
     \Vertex(95,30){2.5}
     \Vertex(125,30){2.5}
   \end{picture}}} 
&\quad+& 
\vcenter{\hbox{
   \begin{picture}(165,100)(0,0)
     \ArrowLine(25,25)(50,50)
        \Text(10,25)[lc]{$e^{-}$}
     \ArrowLine(50,50)(25,75)        \Text(10,75)[lc]{$e^{+}$}
     \Photon(50,50)(75,50){2}{7}     \Text(62.5,55)[bc]{$\gamma,Z$}
     \ArrowLine(115,10)(75,50)       \Text(130,10)[rc]{$\bar{f}_{2}$}
     \ArrowLine(75,50)(95,70)        \Text(77,62)[lb]{$f_{2}$}
     \ArrowLine(95,70)(115,90)       \Text(130,90)[rc]{$f_{2}$}
     \Photon(95,70)(125,70){2}{7}    \Text(110,65)[ct]{$\gamma,Z$}
     \ArrowLine(150,55)(125,70)      \Text(165,55)[rc]{$\bar{f}_{1}$}
     \ArrowLine(125,70)(150,85)      \Text(165,95)[rc]{$f_{1}$}
     \Vertex(50,50){2.5}
     \Vertex(75,50){2.5}
     \Vertex(95,70){2.5}
     \Vertex(125,70){2.5}   
\end{picture}}} 
\end{array} 
\end{eqnarray*} 
\vspace*{-6mm} 
\caption[]{The eight diagrams of subgroup SF$_\gamma$  and SF$_Z$ belonging to the 
NC24 process.\label{pairs_SFG}} 
\vspace*{-2mm} 
\end{figure}

In order to setup a general framework for the analysis of pair
corrections we largely follow the approach of \cite{GiampPP}.
The key point of the analysis is the separation of pair corrections into
two components: {\em signal} and {\em background}. 
We begin by dividing real 
secondary pair \fft\ contributions 
to all primary pairs \ffo\ except electrons into four groups:
(1) Multi-Peripheral {\bf MP}, (2) Initial State Singlet {\bf ISS}, (3)
Initial State Non-Singlet {\bf ISNS} and (4) Final State {\bf FS}.
We further subdivide groups (3) and (4) into the subgroups
{\bf ISNS$_\gamma$}, {\bf ISNS$_Z$}, {\bf FS$_\gamma$}, {\bf FS$_Z$}, 
where the subscript denotes, if the secondary pair \fft is produced via a
(virtual) $\gamma$ or Z boson. If one drops the condition,
that for the FS diagrams the primary pair \ffo\ has to be that from e$^+$e$^-$
annihilation, there are in addition two interchanged diagrams which
we will denote {\bf SF$_\gamma$} and {\bf SF$_Z$}, 
depending again on the boson 
decaying to \fft.
The group (2) is subdivided into {\bf ISS$_\gamma$}
and {\bf ISS$_Z$}, according to whether the incoming e$^+$ and e$^-$ exchange
a $\gamma$ or a Z. This nomenclature is summarized in Figs.~\ref{pairs_MP} to \ref{pairs_SFG}.

\begin{figure}[h] 
\vspace*{-8mm} 
\begin{eqnarray*} 
\begin{array}{ccc} 
\vcenter{\hbox{
   \begin{picture}(150,100)(0,0)
    \ArrowLine(50,75)(25,80)        \Text(10,80)[lc]{$e^{+}$}
     \Photon(90,25)(70,40){2}{5}           \Text(65,20)[lcb]{$ \gamma,Z$}
     \ArrowLine(70,40)(70,60)        \Text(75,50)[lc]{$ e$}
     \Photon(70,60)(50,75){2}{5}            \Text(40,65)[lct]{$ \gamma,Z$}
     \ArrowLine(25,20)(70,40)        \Text(10,20)[lc]{$e^{-}$}
     \ArrowLine(80,80)(50,75)     \Text(90,80)[c]{$e^+$}
     \ArrowLine(70,60)(100,65)     \Text(110,65)[c]{$e^-$}
     \ArrowLine(120,35)(90,25)       \Text(130,35)[rc]{$\bar{f}_{1}$}
     \ArrowLine(90,25)(120,15)       \Text(130,15)[rc]{$f_{1}$}
     \Vertex(50,75){2.5}
     \Vertex(90,25){2.5}
     \Vertex(70,60){2.5}
     \Vertex(70,40){2.5}
   \end{picture}}} 
&\quad+& 
\vcenter{\hbox{
   \begin{picture}(150,100)(0,0)
     \ArrowLine(50,75)(25,80)        \Text(10,80)[lc]{$e^{+}$}
     \Photon(90,25)(80,50){2}{5}           \Text(65,20)[lcb]{$ \gamma,Z$}
     \ArrowLine(60,50)(80,50)        \Text(70,60)[lc]{$ e^{-}$}
     \Photon(60,50)(50,75){2}{5}            \Text(30,60)[lct]{$ \gamma,Z$}
     \ArrowLine(25,20)(60,50)        \Text(10,20)[lc]{$e^{-}$}
     \ArrowLine(80,80)(50,75)     \Text(90,80)[c]{$e^+$}
     \ArrowLine(80,50)(100,65)     \Text(110,65)[c]{$e^-$}
     \ArrowLine(120,35)(90,25)       \Text(130,35)[rc]{$\bar{f}_{1}$}
     \ArrowLine(90,25)(120,15)       \Text(130,15)[rc]{$f_{1}$}
     \Vertex(50,75){2.5}
     \Vertex(90,25){2.5}
     \Vertex(60,50){2.5}
     \Vertex(80,50){2.5}
\end{picture}}} 
\end{array} 
\end{eqnarray*} 
\vspace*{-5mm} 
\caption[]{The eight diagrams of the singlet group ISS$_\gamma$  and ISS$_Z$.\label{pairs_ISS}} 
\vspace*{-1mm} 
\end{figure}
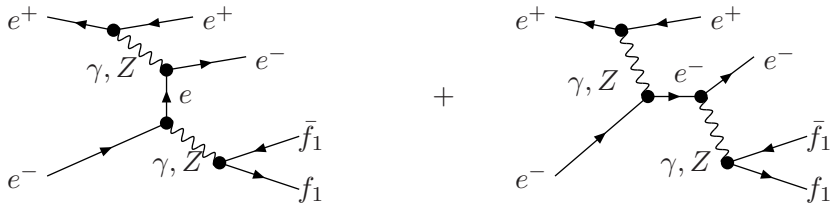

Of course, all of these real pair diagrams come together with their
corresponding virtual pairs in vertex corrections.
For primary electrons \ffo=ee  similar
 sets of diagrams can be plotted (not shown here).
 The main difference to the above diagrams is that the ISNS and FS pairs have now to be attached 
 to both $s$- and $t$-channel e$^+$e$^-$ scattering. 
 The ISNS and FS corrections for $t$-channel Bhabha scattering are thereby identical to the ISS diagrams
 in Figure~\ref{pairs_ISS} with the replacement of \ffo\ by \fft.
 In turn, the singlet diagrams for the $t$-channel Bhabha process are identical to the MP diagrams
 for e$^+$e$^-\to$ e$^+$e$^-$e$^+$e$^-$
  in Figure~\ref{pairs_MP} when the secondary pair is  taken as e$^+_1$e$^-_2$, 
 and  e$^+_2$ forms the primary pair  with the incoming $e^-_1$.
To have the same nomenclature  as for the $s$-channel primary pairs,
pair corrections to $t$-channel electrons will be called 
ISNS for the left-hand  diagram in Fig.~\ref{pairs_ISS}, 
FS for the right-hand  diagram in Fig.~\ref{pairs_ISS}, 
and ISS for the diagrams in Fig.~\ref{pairs_MP}.
The pair corrections to the Bhabha process are further discussed in Section~\ref{LABSMC}.

Appropriate theoretical calculations for ISNS and
ISS pair corrections to primary pairs other than electrons are available in the literature 
\cite{Arbuzov:1999uq,Berends:1988ab, Jadach:1992aa,Kniehl:1988id,Kuraev:1985hb}
with the precision 0.1\% in the LEP2 range \cite{Arbuzov:1999uq},
matching the required experimental precision. 
If no cuts are applied on final state pairs
the real+virtual FS  contribution can  be largely absorbed by evaluating 
 the photonic final state correction $\delta_\gamma$ using $\alpha_{\rm em}(s)$
instead of  $\alpha(0)$, leaving  a tiny residual uncertainty of 0.002\% 
of the 2-fermion cross-section at LEP2 energies~\cite{Hoang:1995ht}.

Concerning the definition of the 4f signal, there are basically two different approaches
\begin{enumerate} 
\item 
{\bf Choosing few (sub)groups of Feynman diagrams as signal definition in such a way that
 cuts on masses and energies of the secondary $\fft$ pairs can be avoided }\\
This is a very useful approach for theoretical predictions, since it avoids 
the calculation
of multi-differential cross-sections for the radiated $\fft$ pair. 
It potentially poses problems 
for experimental measurements since
(sub)groups of Feynman diagrams often cannot easily be extracted from
full 4-vector four-fermion Monte Carlos like KORALW or GRC4f, and  interference of
signal and background is possible.
\item {\bf Choosing  (nearly) all groups of Feynman diagrams and rejecting 
the unwanted part of
phase space by cuts on masses and/or energies of the radiated \fft pair.}\\
If the chosen groups match with those of typical 4-fermion generators, this definition is easy
to implement in experimental measurements, but can mean considerable calculation and programming work
for theoretical predictions.
\end{enumerate}

The discussions in this workshop and in the 2f-LEP2 subgroup of the LEP electroweak working group
converged to a proposal for a LEP-wide definition of a 2f+4f signal, which will be detailed below.
It is made in such a way that it can nearly equivalently be expressed  in both 
approaches, the definition by diagrams (1), and the definition by cuts (2). 
Differences between the approaches are below 0.1\%, and therefore negligible compared to 
the experimental accuracy
This means that experimentalists could perform a measurement within approach (2),
and compare it to a theory prediction with approach (1).   
The proposal is  close to a procedure first used by the OPAL experiment~\cite{MichOPAL},
and will be detailed in the following. 

{\bf \subsubsection{2f+4f signal definition by diagrams}}

\noindent \label{PairEffdiag}
The diagram-based  choice for a 2f+4f signal definition  is\\ 

\begin{itemize}
\item {\bf DEFINITION 1:} ISNS$_\gamma$+FS$_\gamma$\\
No cuts are applied to the mass of the \fft\ pair. For the case of  ISNS$_\gamma$ the
primary pair is required to pass the cut $s^\prime/s > R_{\rm cut}$,
where typical values for $R_{\rm cut}$ at LEP2 are 0.7225 for the "exclusive" 
high $s^\prime$ 
selection and 0.01 for the "inclusive" selection. The treatment of FS$_\gamma$ 
depends on the  $s^\prime$ definition. If the s-channel propagator mass is taken
$s^\prime=M_{\rm prop}^2$, which is possible only in absence
of initial-final state interference (IFI), no cuts are applied to FS$_\gamma$. If one chooses to 
include IFI in the measurement, one has to define $s^\prime=M_{\ffo}^2$, and apply $R_{\rm cut}$
also to FS$_\gamma$.
\end{itemize}

\noindent
The reasoning for the above choice of diagrams is that the bulk of the phase space of all other
diagrams looks kinematically very different from 2-fermion events, and is therefore
rejected in most 2-fermion selections. It would make little sense to re-introduce it via efficiency
corrections, especially since the modeling of  MP and ISS due to poles for forward
electrons or positrons is more  inaccurate than for other \ffo\fft diagrams. 
In addition MP, ISS (comprising Zee) and
ISNS$_{\rm Z}$ (comprising ZZ) cover different types of possible new physics contributions, 
which would  hamper the interpretation of 2-fermion cross-section measurements, if they were
included in the 2f signal\footnote{If an experiment nevertheless chooses to also 
include ISS$_\gamma$
in its signal definition, the measurement can be converted to the above definition
correcting for the contribution of ISS$_\gamma$, which can be obtained from TOPAZ0 \cite{topaz0}
by selecting OSING='SP³ or
from ZFITTER versions 6.21 onwards, calculating the
correction from the  difference between IPSC=3 and IPSC=0 with flag ISPP=2}.

Predictions for the above signal definition with the definition of 
$s^\prime=M_{\rm prop-}$
can be obtained easily from (even old versions) of
the semianalytical programs ZFITTER and TOPAZ0. E.g. in ZFITTER versions up to 5.15 
the corresponding full 2f+4f prediction was obtained setting the flags FOT2=3 and 
INTF=FINR=0.
This old pairs treatment is still available in the actual ZFITTER versions using 
ISPP~=~$-1$.
For ZFITTER 6.21 onwards the flag setting corresponding to the above definition is
INTF = FINR = 0, FSPP = 0, ISPP$\ge$2. 
In TOPAZ0, version 4.4 the recommended flag setting
is ONP=I, for older versions ONP=Y should be used. For both programs the effect of 
real+virtual pair corrections can be
obtained from the difference to flag settings that switch off pairs 
(ISPP=0 in ZFITTER or ONP=N in TOPAZ0).

After convolution of  photon and pair radiation, the $s^\prime$ cut represents
a cut on the combined effect of the two. Different $s^\prime$ cuts for photons and pairs
would require a rather complicated definition of $s^\prime$. 
For the definition  $s^\prime=M_{\rm prop-}$ only initial state photons and pairs have to be
modeled. 
For the definition $s^\prime=M_{\ffo}^2$ the FS$_\gamma$ process is needed explicitely,
which is only available from ZFITTER 6.30 onwards, and not available in TOPAZ0.
The corresponding flag setting in ZFITTER is INTF=2, FINR=1, FSPP=2 (or 1), and 
ISPP$\ge$4.

In the diagram-based signal definition there are no pairing ambiguities for four identical fermions, 
since  the pairing is known.  The only potential problem remains 
 for the ISNS$_\gamma$ subprocess $ee \to \gamma^* \gamma^* \to \ff\ffp$
when both  pairs  fulfill the $s^\prime$ cut.
 Such an event is signal both for a primary pair \ffo=\ff and \ffo=\ffp,
which is {\em per se} not a problem. Only if cross-sections of several channels are
summed up, like for hadronic final states, this can lead to double counting
e.g.\ the same u$\overline{\rm u}$s$\overline{\rm s}$ event could be counted as 
signal for \ffo=u$\overline{\rm u}$ and \ffo=s$\overline{\rm s}$ 
in the theory prediction, while it is counted only once by experimental measurements. 
The amount of such double counting for hadrons depends on the $s^\prime$ cut. 
Obviously there is no double counting for all $R_{\rm cut}\ge 0.25$ 
due to phase space. An estimate using fully simulated GRC4f qqqq events
shows, that at $R_{\rm cut}=0.01$ the double counting is still below
10$^{-4}$ of the qq cross-section, which makes it truly negligible.
Double counting can be fully avoided by imposing  an additional cut of
$M_{\ffo}>M_{\fft}$. It is, however, not possible to apply such a cut
in ZFITTER or TOPAZ0, but only in full 4-fermion generators like GRC4f or KORALW.

{\bf \subsubsection{2f+4f signal definition by cuts} }

\noindent \label{PairEffcut}
The cut-based  choice for a 4f signal definition  is \\

\begin{itemize}
\item {\bf DEFINITION 2: ISNS+FS}, with a cut $M_{\fft}<M_{\rm max}$\\
In this case the meaning of ISNS and FS is
ISNS=ISNS$_\gamma$+ISNS$_{\rm Z}$ and
FS=FS$_\gamma$+FS$_{\rm Z}$+SF$_\gamma$+SF$_{\rm Z}$.
This definition corresponds to using all \ffo\fft\ diagrams with exception
of MP and ISS. Both pairs, if  passing the $s^\prime$ cut, can be
 taken as the primary pair.  
In order to suppress e.g. the unwanted contribution from
ZZ final states, a mass cut on the secondary pair  is added.
It will be shown below, that due to a plateau in the pair cross-section
between the $\gamma*$ peak at low \fft\ masses and the Z peak at high
\fft\ masses, the details of this mass cut don't matter, as long it 
is stays far enough from the Z peak and large enough not to cut
appreciably into the ISNS$_\gamma$ and FS$_\gamma$ processes. 
Suitable choices for a fixed mass cut are $M_{\rm max}=50-80$~GeV, 
while for a fractional mass cut $M^2_{\fft}/s < P_{\rm cut}$ e.g.
the values $P_{\rm cut}=0.10$ and 0.15 are  leading to acceptable ranges of 
$M_{\rm max}=51-65$~GeV and $62-80$~GeV for the
LEP2 centre-of-mass energies between 161 and 206~GeV.
\end{itemize}

\noindent
The advantage of  summing many diagrams is, that experimental  
measurements are able to straight-forward use  full 4-fermion MC generators which include
all these diagrams and their interferences~\footnote{%
Again one could even go for
an additional inclusion of ISS$_\gamma$  and ISS$_Z$
diagrams here,  which the 4-fermion generators KORALW and GRC4f
offer. This question is of no relevance for large $R_{\rm cut}$ values above 0.4 or so,
since the ISS contributions are negligible there. For small $R_{\rm cut}$ values 
of the order 0.01 the ISS contribution is appreciable (some percent of the 2-f
cross-section). While for ISNS and FS+SF  the cut on $M_{\fft}$ 
nearly exclusively selects the ISNS$_\gamma$ and FS$_\gamma$ contributions,
this is not obviously the case for ISS. This 
might lead to non-negligible differences between 
the cut-based definition above and the corresponding diagram-based definition
ISNS$_\gamma$+FS$_\gamma$+ISS$_\gamma$. 
Since none of the LEP experiments included
the ISS  ($\gamma^{*}$/Zee) process in their signal definition so far, this question
has not been quantitatively addressed in this workshop.}

Making no distinction between the various diagrams in 4-fermion generators
and even including the interchanged SF group leaves no  choice for
the  $s^\prime$ definition other than $s^\prime=M_{\ffo}^2$. Only events which
fulfill both the above cut-based definition and the $s^\prime$ cut are counted
as signal. 

Concerning the potential double counting problem the same remarks as for the diagram-based
definition hold. In contrast to the diagram based definition, however, questions arise
for four identical fermions, since the correct pairing is usually not known
(see also the discussion in the footnote of  subsection~\ref{GENTLE1}).
 This effect is still an open problem,
since especially the rejection of ZZ events via the cut on $M_{\fft}$ depends on the chosen pairing.
For estimating the size of the effect, we have  calculated the amount of cut-based real signal pairs using
4 different pairing algorithms for qqqq events, simulated with GRC4f. The pairing was chosen as to maximize 
or minimize certain masses or mass sums  as detailed in Table~\ref{pairing}.
For high $s^\prime$ events the maximum observed difference between any  two algorithms is ranging
from $(0.06\pm0.03)\times10^{-3}$ at 189~GeV to $(0.08\pm0.04)\times10^{-3}$ at 206~GeV, 
 whereas for inclusive events these numbers are
 $(0.4\pm0.1)\times10^{-3}$ at 189~GeV and $(0.8\pm0.2)\times10^{-3}$ at 206~GeV. 
Taking this differences as an estimate for the effect of wrong pairing, it 
increases with centre-of-mass energy as expected from the increasing ZZ cross-section,
but stays below 1 per mil even for inclusive hadrons at the highest energies.
The uncertainty due to pairing ambiguities can in principle be largely reduced 
by correcting for  the difference of a given pairing algorithm to the true pairing,
which can be obtained using  the weights of the REW99 library~\cite{verzocchi:1999} for GRC4f events.
  
\begin{table}[ht]
 \caption[]{Real hadronic pair cross-sections qqq'q' in pb , and relative coirrections in per mill,
 obtained from GRC4f  for the process e$^+$e$^- \to$ hadrons
 at $\sqrt{s}=189$~GeV and 206~geV for four different pairing algorithms applied to the  case of 4 identical quarks  in  the cut-based definition (2).
           } 
 \label{pairing}
 \vspace*{1mm}
 \begin{center}
 \begin{tabular} {|| l || c | c || c | c ||}
 \hline
   qqq'q'     &    $\sigma^{\rm Real} $   &  $\delta^{\rm Real}$    & $\sigma^{\rm Real} $     &  $\delta^{\rm Real}$\\   
\hline
  $R_{\rm cut}$    &      \multicolumn{2}{c|}{0.7225}     &   \multicolumn{2}{c||}{0.01}     \\
 \hline
 algorithm &  \multicolumn{4}{c||}{189 GeV}\\
\hline
 $\min(M_{\fft})$                    &    0.0159    &        0.74        &   0.585   &    6.07    \\
 $\max(M_{\ffo}-M_{\fft})$    &    0.0173    &        0.80       &   0.548   &   5.69    \\                             
 $\max(M_{\ffo})$                  &    0.0173    &        0.80       &   0.564   &    5.85    \\
 $\max(M_{\ffo}+M_{\fft})$     &    0.0173    &        0.80        &   0.589   &    6.11    \\
 \hline
algorithm &  \multicolumn{4}{c||}{206 GeV}\\
\hline
 $\min(M_{\fft})$                    &    0.0117   &        0.68        &   0.568   &    7.30    \\
 $\max(M_{\ffo}-M_{\fft})$    &    0.0130    &        0.76       &   0.509   &   6.54    \\                             
 $\max(M_{\ffo})$                  &    0.0130    &        0.76       &   0.541   &    6.95    \\
 $\max(M_{\ffo}+M_{\fft})$     &    0.0130    &        0.76        &   0.575   &    7.39    \\
 \hline

 \end{tabular}
 \end{center}
 \vspace*{-3mm} 
 \end{table}

As will be shown in the next subsection, for obtaining the correct 2f+4f selection efficiency
and the correct background in experimental measurements it suffices to 
separate the 4-fermion final states, i.e. the {\it real} pairs, into two samples, which form signal
and background, respectively. Virtual pair corrections are signal, but their size 
is irrelevant for the experimental measurements in first order.
Real pair signal samples with the above definition can be obtained e.g. with GRC4f or KORALW. 
In contrast, for obtaining a theoretical prediction the sum of {\it  real and virtual} pair corrections are needed.
Due to mass cuts on $M_{\fft}$, this is not possible with ZFITTER or TOPAZ0 for the above cut-based 
signal definition. A new version 2.11 of GENTLE/4fan  
is able to calculate both real and virtual pair corrections with mass cuts, where the flag setting
corresponding to our above definition is IPPS=6, IGONLY=3, and $P_{\rm cut}=0.10$. 
Another possibility is, to add real pair corrections obtained from KORALW (or GRC4f)
to the virtual pair corrections, which have recently been implemented in the new version 4.14 of KKMC.

\subsubsection{Background subtraction and efficiency determination}

The total 2f+4f signal cross-section has the form
\begin{equation}
    \sigma = \sigma^{\rm Born} + \sigma^{\rm Virt} + \sigma^{\rm Real} 
       \equiv \sigma^{\rm Born}(1 + \delta^{\rm Virt} +  \delta^{\rm Real}),
\end{equation} 
where $\sigma^{\rm Born}$ is the (ISR convoluted) 2-fermion cross-section,
$\sigma^{\rm Real}$ is the (ISR convoluted) cross-section with real pair emission,
and  $\sigma^{\rm Virt}$ is the (negative) correction due to virtual pairs.
The effect of  including a part of the 4f final states as pair emission correction is twofold.
First, obviously only  those 4f events which are not counted as 2f+4f signal contribution 
are to be subtracted as background. 
(Subtracting wrongly all 4-fermion events as background which pass the 2-fermion selection,
can, depending on the $s^\prime$ cut, easily lead to mismeasurements larger  
 than one percent. ) Second, the influence of the real signal pairs on the selection efficiency
has to be taken into account. We call in the following
\begin{eqnarray}
\epsilon_{\rm 2f} &=& \frac{\sigma^{\rm Born}_{\rm vis}}{\sigma^{\rm Born}}\;, 
\\
\epsilon_{\rm 4f} &=& \frac{\sigma^{\rm Real}_{\rm vis}}{\sigma^{\rm Real}}\;,
\end{eqnarray}
where the subscript ``vis'' denotes the part of the cross-section for the respective process
which passes all selection cuts. It is  a very good approximation to assume that the
vertex corrections don't change the selection efficiency, since they lead to the same final state,
so that the efficiency for the "Born+Virt" part of the cross-section is still $\epsilon_{\rm 2f}$. 
This leads to a total selection efficiency for the 2f+4f process of
\begin{eqnarray}
     \epsilon &=& \frac{(1+\delta^{\rm Virt})\epsilon_{\rm 2f} + \delta^{\rm Real}\epsilon_{\rm 4f}}
                                       {1 + \delta^{\rm Virt} +  \delta^{\rm Real}}\\
   & \approx & (1- \delta^{\rm Real} + \delta^{\rm Real}(\delta^{\rm Real}+\delta^{\rm Virt}))\epsilon_{\rm 2f}
                          + (\delta^{\rm Real} - \delta^{\rm Real}(\delta^{\rm Real}+\delta^{\rm Virt}))\epsilon_{\rm 4f}\\
 & \approx & (1- \delta^{\rm Real})\epsilon_{\rm 2f}
                      + \delta^{\rm Real}\epsilon_{\rm 4f},
\end{eqnarray}
where the expression has been expanded up to ${\cal O}(\delta^2)$ in the second line and
to ${\cal O}(\delta)$ in the third line. Since $\delta\sim0.01$ it is fully sufficient to retain the
first order in $\delta$, which means that the experimental measurements need
only to know the fraction $\delta^{\rm Real}$ of real pair emission, and are completely insensitive
to the virtual pair correction $\delta^{\rm Virt}$.
 
Even more transparently one can write the efficiency correction
  $ \Delta\epsilon = \epsilon - \epsilon_{\rm 2f} = 
                                         \delta^{\rm Real} (\epsilon_{\rm 4f} - \epsilon_{\rm 2f})$
which means that the efficiency correction is the product of the real pairs fraction
and the difference in efficiencies between events with and without pairs.
Moreover, if events with very soft or low-mass pairs have identical selection efficiencies
to events without pairs, they need not be explicitly modeled, which  justifies 
cutoffs for soft or low-mass pairs in explicit 4-vector MC generation. Such
cutoffs modify $\delta^{\rm Real}$ and  $\epsilon_{\rm 4f}$ in such a way, that the same 
efficiency correction emerges. 

To give a feeling for the size of the effect of pairs on the selection efficiency, we have listed in Table~\ref{OPALeff} 
 some typical numbers for pair corrections in hadronic and muonic 
selection efficiencies, obtained from a real pair simulation with the GRC4f generator at $\sqrt{s}=189$~GeV
using the cuts based signal definition (2) and the hadronic event selection of the OPAL experiment.
The effect for the other LEP experiments is of similar size.

\begin{table}[ht]
\caption[]{Efficiency corrections $\Delta{\epsilon}$ for the selection of hadrons
and muon pairs due to pair emission corrections for the OPAL experiment at $\sqrt{s}$=189~GeV.
The meaning of the variables is given in the text.
\label{OPALeff}}
\vspace*{1mm}
\begin{center}
\begin{tabular} {|| l || r | r || r | r ||}
\hline
            &  \multicolumn{2}{c|}{ e$^+$e$^- \to$ hadrons}   &   \multicolumn{2}{c||}{ e$^+$e$^- \to\mu^+\mu^-$}       \\     
\hline
$R_{\rm cut}      $      &      0.7225         &         0.01         &     0.7225          & 0.01                   \\
\hline
$\epsilon_{\rm 2f} $     &      87.9\%         &      87.4\%          &      89.8\%         &      79.1\%            \\     
$\epsilon_{\rm 4f} $     &      83.2\%         &      79.7\%          &      86.4\%         &      54.6\%            \\   
$\delta^{\rm{Real}}$     &      0.006          &       0.022          &      0.005          &      0.015             \\
\hline                 
$\Delta{\epsilon}  $     &      $-0.02\%$      &   $-0.17\%$          &   $-0.02\%$         &   $-0.37\%$           \\
\hline
\end{tabular}
\end{center}
\vspace*{-5mm}
\end{table}

For the high $s^\prime$  selection the small fraction $\delta^{\rm{Real}}$     and the small difference between the efficiencies 
$\epsilon_{\rm 4f} $ and  $\epsilon_{\rm 2f} $  results in a very small efficiency correction, well below one per mill. 
Both numbers are larger for the inclusive selections, so that the relative efficiency changes due to pairs  
 for inclusive hadrons is about 2 per mill  ($-0.17\%$ absolute)  and about 5 per mill for muons ($-0.37\%$  absolute). 
Note, that  to obtain these efficiency corrections
both ISNS and FS real pairs have to be generated explicitly,  which is possible e.g. with the GRC4f or KORALW programs.
 
\subsubsection{Pairs in semianalytical and Monte Carlo tools}
For most of the programs, the treatment of pair corrections has been described in 
Section~\ref{sec:programs}
of this report. We  summarize here  the essential points in a comparison of all programs.
The Feynman diagrams included in the programs, and the availability 
of  possible mass cuts  are summarized in Table~\ref{pairs_sumtab}.
It is obvious that with the existing programs a large variety of signal definitions would
in principle be possible, though many of them would be accessible with one program, only.

For  our diagram-based and cut-based signal definitions
we list here the features  needed  for predictions and measurements
of pair corrections. 
\begin{itemize} 
\item theoretical prediction of diagram-based definition 1:\\
Virtual Pairs,  ISNS$_\gamma$, desirably with  common photon-pair exponentiation. 
For all primary pairs , apart from \ffo=~ee,    this 
is available in ZFITTER, TOPAZ0, GENTLE, and in the combination KKMC+KORALW.
For \ffo=~ee only LABSMC has virtual and real pairs for s- and t-channel Bhabhas, yet
without photon-pair convolution.
\item experimental measurement of diagram-based definition 1:\\
 Complete event generation of  ISNS$_\gamma$ and FS$_\gamma$, separable from
 other diagrams. 
For all primary pairs this is possible in KORALW or GRC4f.
\item theoretical prediction of cut-based definition 2:\\
Virtual Pairs and  possibility of mass cuts on secondary pairs for ISNS,  FS, and SF,  
desirably with  photon-pair convolution. 
For all primary pairs , apart from \ffo=~ee, this is possible in GENTLE and KKMC+KORALW.
For \ffo=~ee no program with these features exists.
\item experimental measurement of cut-based definition 2:\\
 Complete event generation of  ISNS, FS,and SF separable from
 other diagrams. 
For all primary pairs this is possible in KORALW or GRC4f.
\end{itemize}
Obviously, ISS is needed for none of the two signal definitions above.
 Initial-final state interference (IFI) of pairs is completely negligible
for any signal definition. Both are nevertheless listed for completeness in Table~\ref{pairs_sumtab}.

\begin{table}
\caption[]{.
\label{pairs_sumtab} Summary of pair corrections available in various programs. For Feynman diagrams 
 the possibility of mass cuts on the secondary  pair is indicated by ``mass''. The convolution of photons and pairs
in a common exponentiation is listed in the row $\gamma$-pair conv. IFI stands for interference of
initial- and final state pairs.  Note, that for LABSMC the singlet contribution listed under
ISS is in fact the multi-peripheral (MP) diagram ee$\to$eeff, and  ISNS, FS and SF refer to both s and t-channel Bhabha
scattering. \\
The last 4 rows indicate for which signal definition theoretical predictions (th) are possible, and for which
signal definition a full 4-fermion signal and background event sample for experimental measurements (exp)
can be obtained. For the 2f+4f 
signal definitions ``real''  means, that only  real pairs can be calculated, and have to be combined with another
program calculating the virtual part ``virt''. }
\vspace*{1mm}
\begin{center}
\begin{tabular}{|l|c|c|c|c|c|c||c|}
\hline
program                             & ZFITTER  & TOPAZ0 & KKMC      & KORALW & GRC4f       &  GENTLE   & LABSMC \\
\hline
virtual pairs        &    yes         &    yes       &   yes          &       no         &    no           &    yes           &   yes         \\
ISNS$_\gamma$&    yes         &    yes       &   no            &      mass      &    mass      &    mass        &   yes         \\
ISNS$_Z$           &      no         &      no       &   no            &      mass      &    mass     &    mass         &   no           \\
FS$_\gamma$    &  mass         &      no      &   no            &       mass      &    mass     &    mass         &   yes         \\
FS$_Z$               &       no        &      no       &   no            &      mass      &    mass     &    mass         &    no        \\
SF$_\gamma$    &       no        &      no       &   no            &      mass     &    mass      &    mass               &    no          \\ 
SF$_Z$               &       no        &     no        &   no           &      mass      &   mass      &     mass              &    no         \\
ISS$_\gamma$   &     yes        &     yes       &  no            &      mass      &   mass      &         no            &     yes (MP) \\
ISS$_Z$              &       no       &       no        & no            &       mass      &   mass      &         no          &   no(MP)    \\
IFI                        &       no       &       no        & no             &       yes        &   yes         &     yes          &   no       \\
$\gamma$-pair conv.&yes     &     yes        & no              &      yes    &     no        &     yes           &  no        \\
\hline
definition 1 (th)   &    yes      &   yes     &         virt       &         real         & real          &   yes   &  yes \\
definition 1 (exp) &      no     &     no     &         no       &         yes         & yes         & no   &  no \\
definition 2 (th)    &     no     &      no    &        virt       &         real         & real           & yes   &  no \\
definition 2 (exp)  &     no     &      no    &        no       &         yes         & yes         & no  &  no\\
\hline 
\end{tabular}
\end{center}
\end{table}

In the following sections we will give a broad variety of
numerical results on pair corrections from several programs.
The programs will then be compared for the  two 4-fermion signal definitions discussed above.

%% file: 2f-Chapt-Part6.tex
\subsubsection{Numerical results and conclusions from GENTLE}
\label{subsec:gentle}

Numerical results obtained with the use of the code GENTLE$\_$4fan v.2.11
\footnote{Accessible from: /afs/cern.ch/user/b/bardindy/public/Gentle2$\_$11\\
also from the Gentle/Zeuthen homepage:\\
http://www.ifh.de/~riemann/doc/Gentle/gentle.html}
are presented in tables and figures shown below. 
They contain $\delta_{\rm{pairs}}$ defined by eq.~(\ref{delta_pairs}) in section~\ref{GENTLE1}
for two processes: $e^+e^-\to\mbox{muons}$ and $e^+e^-\to\mbox{hadrons}$ 
for two cuts on invariant mass of the primary pair
$R_{\rm{cut}}=0.01$ and $0.7225$ and three cms energies: 189, 200 and 206 GeV. 
The wide range of the cut on invariant mass of the secondary pair was studied,
$P_{\rm{cut}}=10^{-4} - 1$.
Results for several typical selections of groups of Feynman diagrams 
({ IPPS, IGONLY}) are shown. 

In Tables~\ref{tab_mu} and Fig.~\ref{fig_mu}
we show $\delta_{\rm{pairs}}(P_{\rm{cut}})$ for the process $e^+e^-\to\mbox{muons}$ for
two $R_{\rm{cut}}$ and three c.m.s. energies. We note drastic dependence on 
$R_{\rm{cut}}$,
moderate energy dependence and plateau-like $P_{\rm{cut}}$ dependence in cases when
$Z$ exchange is not included. Solid lines show $\delta_{\rm{pairs}}(P_{\rm{cut}})$ when 
only ISPP mediated by $\gamma$ exchange is taken into account. Adding on top of it
the FSPP, has practically no influence for $R_{\rm{cut}}=0.01$ and gives almost constant
negative shift for $R_{\rm{cut}}=0.7225$. For the latter case 
$\delta_{\rm{pairs}}(P_{\rm{cut}})$
are very flat, practically no $P_{\rm{cut}}$ dependence is seen. For $R_{\rm{cut}}=0.01$,
$\delta_{\rm{pairs}}(P_{\rm{cut}})$ exhibits some $P_{\rm{cut}}$ dependence.
Allowing $Z$ exchange we observe an interesting phenomenon which we called
{\em $Z$ opening}. It occurs when two cuts allow resonance production of the $Z$ boson. 
We emphasize, that $Z$ opening has nothing to do with $Z$ radiative return (ZRR).
As seen from the last Fig.~\ref{fig_mu} it takes place in the case when the ISR 
convolution is ignored. $Z$ opening is expected qualitatively, and from the figures
we may easily see its quantitative size. $Z$ opening rapidly grows with energy,
reaching half a per cent at 206 GeV. Therefore, it is relatively important and one
has to bother about cutting of such events. (Blind use of $P_{\rm{cut}}=1$ may be 
dangerous.)

In Fig.~\ref{fig_mu} we also show by dots ZFITTER v.6.30 results, which are shown
only at $P_{\rm{cut}}=1$ since \zf\  doesn't allow for cutting of ISPP.
The agreement between GENTLE and \zf\  is at the level half a per mill
for $R_{\rm{cut}}=0.7225$ and --- one per mill for $R_{\rm{cut}}=0.01$. 
(Note much better agreement for the case when the ISR is ignored.) 
It is not surprising since GENTLE and 
\zf\  exploit very different approach for the ISR convolution.
\zf\  uses fully expanded, order-by-order additive approach 
(see subsection~\ref{pairZF}). As for pairs is concerned this means:
\begin{equation}
\sigma(QED)=\sigma(photonic)+\sigma(pairs~{\cal{O}}(\alpha^2))
        +\sigma({\cal{O}}(\alpha^3))+\sigma({\cal{O}}(\alpha^4)),
\end{equation}
with the two last terms computed in~Ref.~\cite{Arbuzov:1999uq} in the LLA 
(Leading Logarithmic Approximation).
The term $\sigma({\cal{O}}(\alpha^3))$ represents the lowest order QED correction 
to the Born ${\cal{O}}(\alpha^2)$ pair production.

In the framework of GENTLE-like "multiplicative" approach 
one computes a convolution integral from a $4f$ kernel,
which takes into account multiple photon emission. So, the 1 per mill or better 
agreement between GENTLE and \zf\  is far from being trivial.
Given completely different treatments of ISR convolution we may trade the difference,
which arises after convolution, for a measure of the theoretical uncertainty which 
is due to ISR convolution.

One should note that ISR convolution is very important. Even for $R_{\rm{cut}}=0.7225$
it reaches $10\%$ while for $R_{\rm{cut}}=0.01$ it changes the result by factor of three.
Here, however, the bulk of the effect is due to ZRR which enhances the denominator 
of~eq.~(\ref{delta_pairs}) reducing thereby the factor $\delta_{\rm{pairs}}(P_{\rm{cut}})$
drastically. 
\begin{table}[!h]
\begin{center}
\setlength{\unitlength}{1cm}
\begin{picture}(16,20)
\put( 0,10.1){\makebox(0,0)[lb]{\epsfig{file=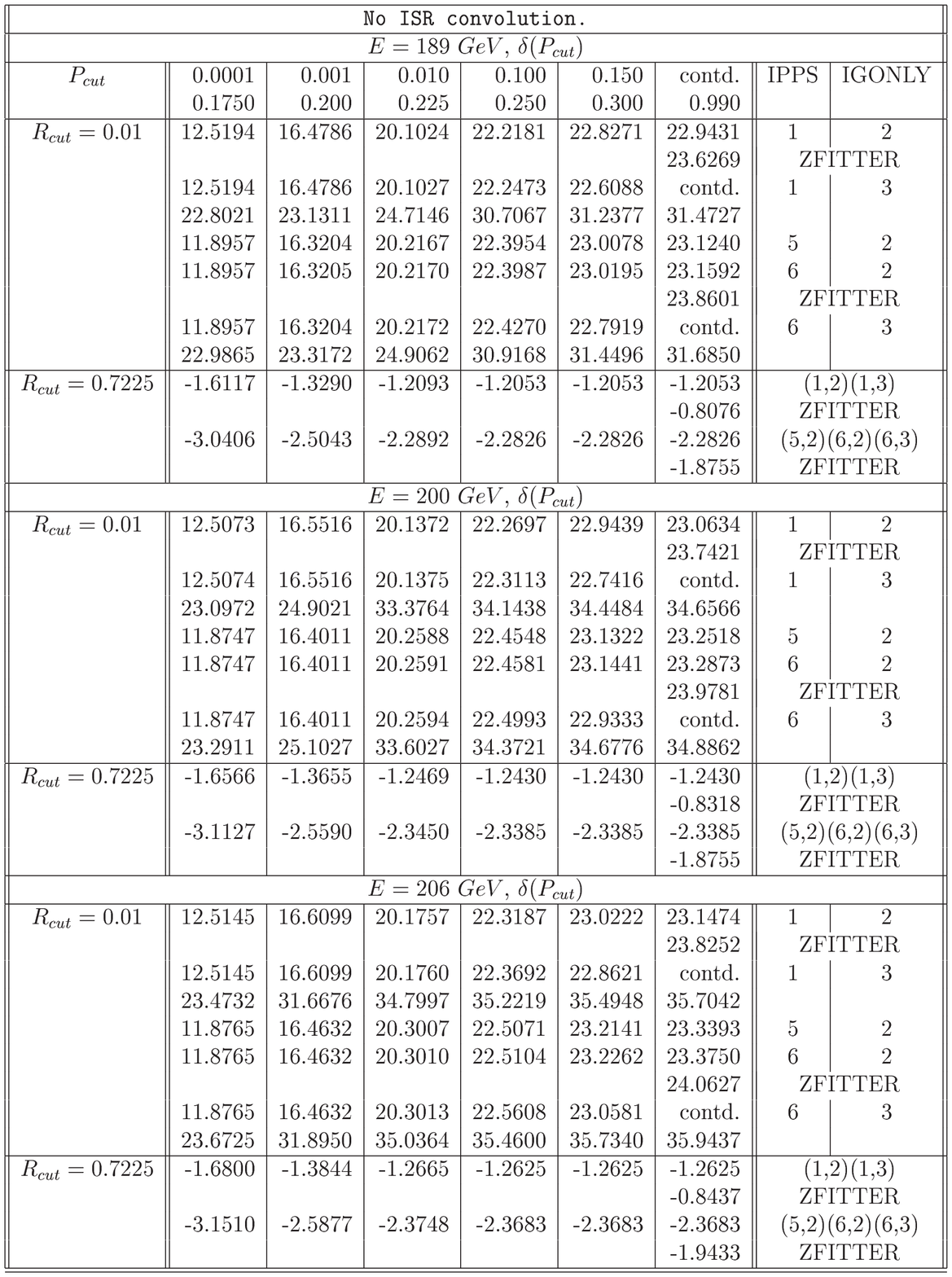,width=10cm,height=16cm,angle=90}}}
\put( 0, 0  ){\makebox(0,0)[lb]{\epsfig{file=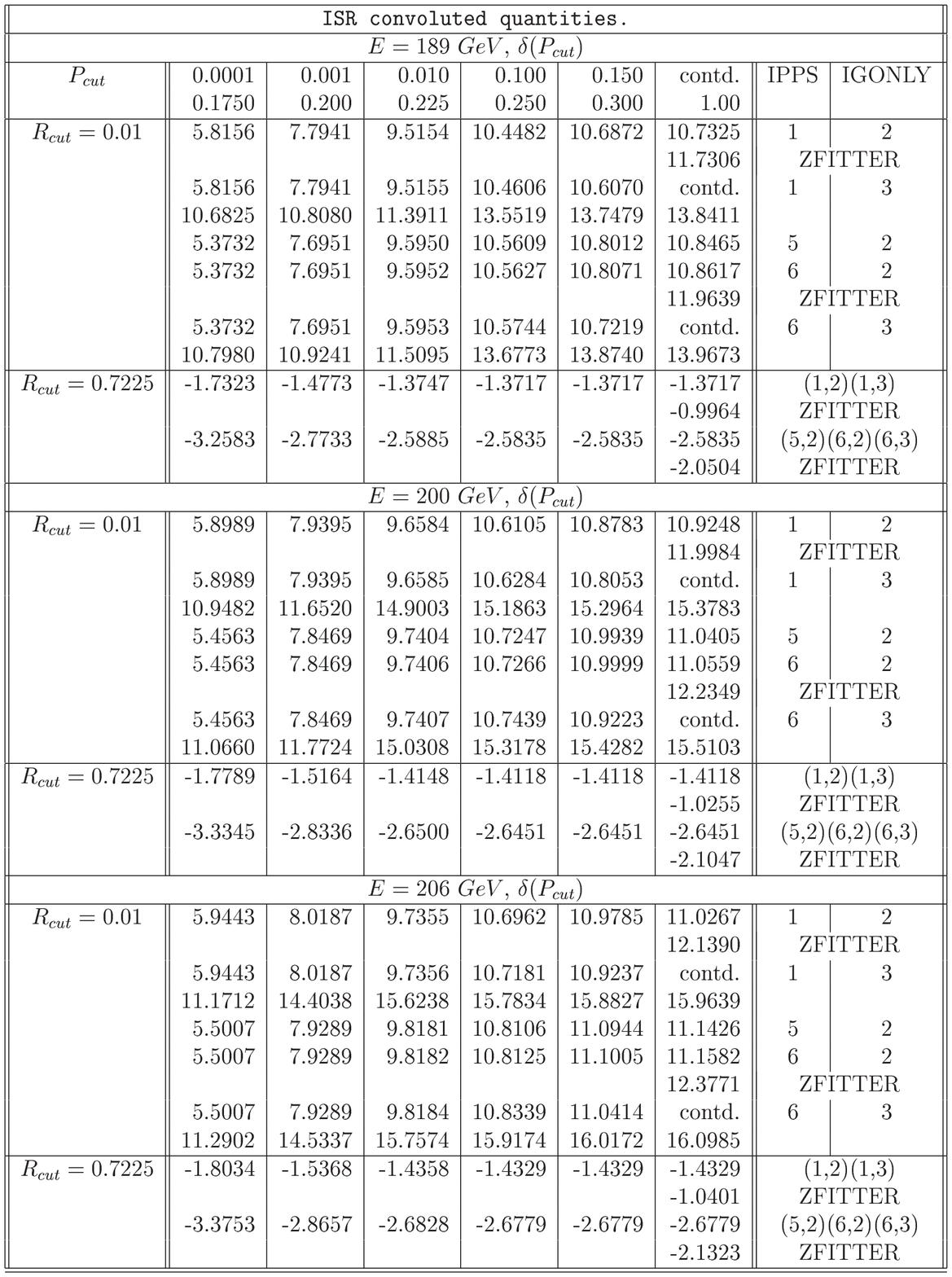,width=10cm,height=16cm,angle=90}}}
\end{picture}
\caption{GENTLE/4$\_$fan v.2.11. Process $e^{+}e^{-}\to\mu^{+}\mu^{-}$.        
{\tt IPPS - IGONLY} rows. Hadronic language.       
}
\label{tab_mu}
\end{center}
\end{table}


\begin{figure}[t]
\centering
\setlength{\unitlength}{0.1mm}
\begin{picture}(1600,1500)
\put(-50,700){\makebox(0,0)[lb]{\epsfig{file=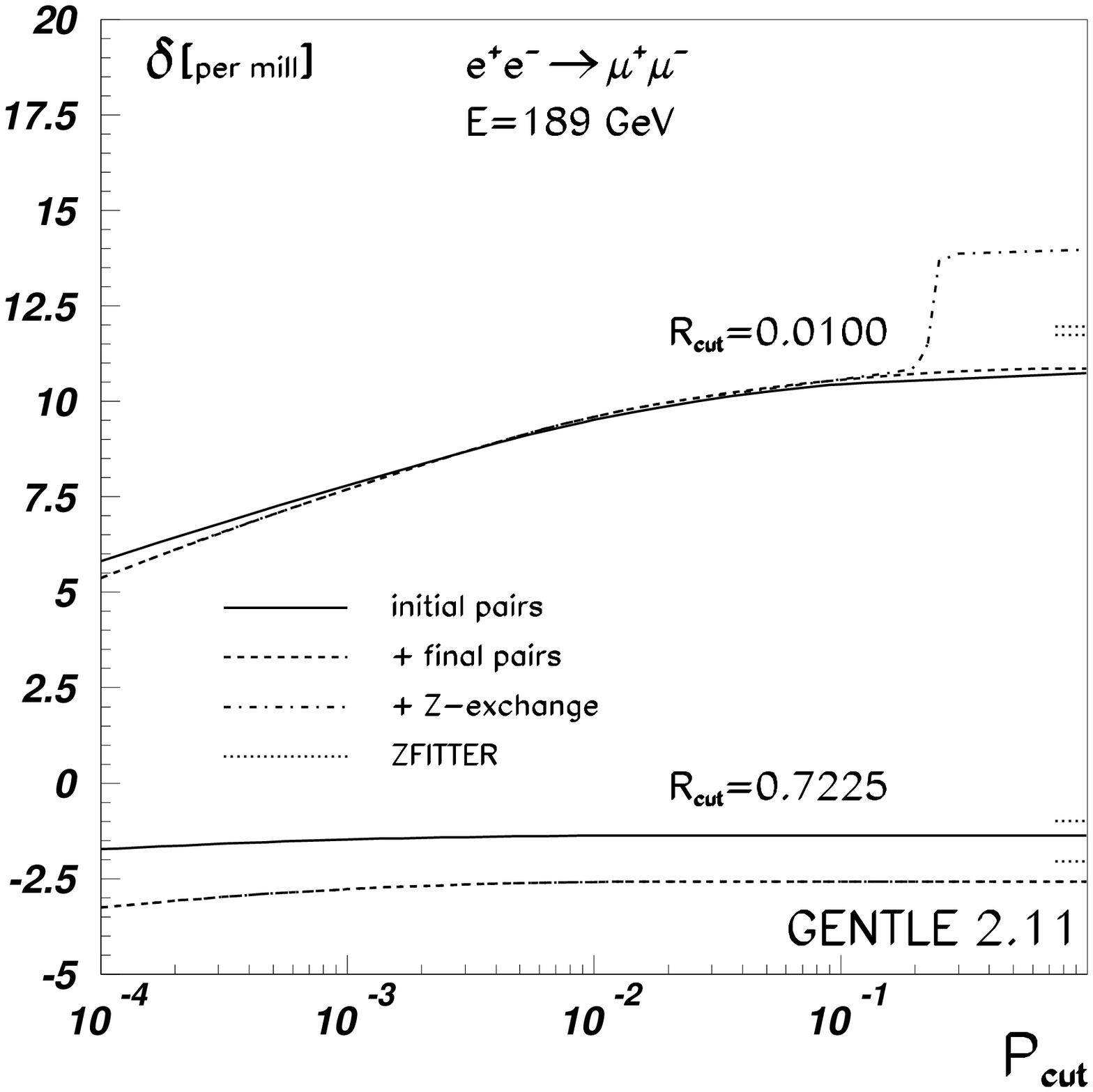,width=85mm,height=85mm}}}
\put(750,700){\makebox(0,0)[lb]{\epsfig{file=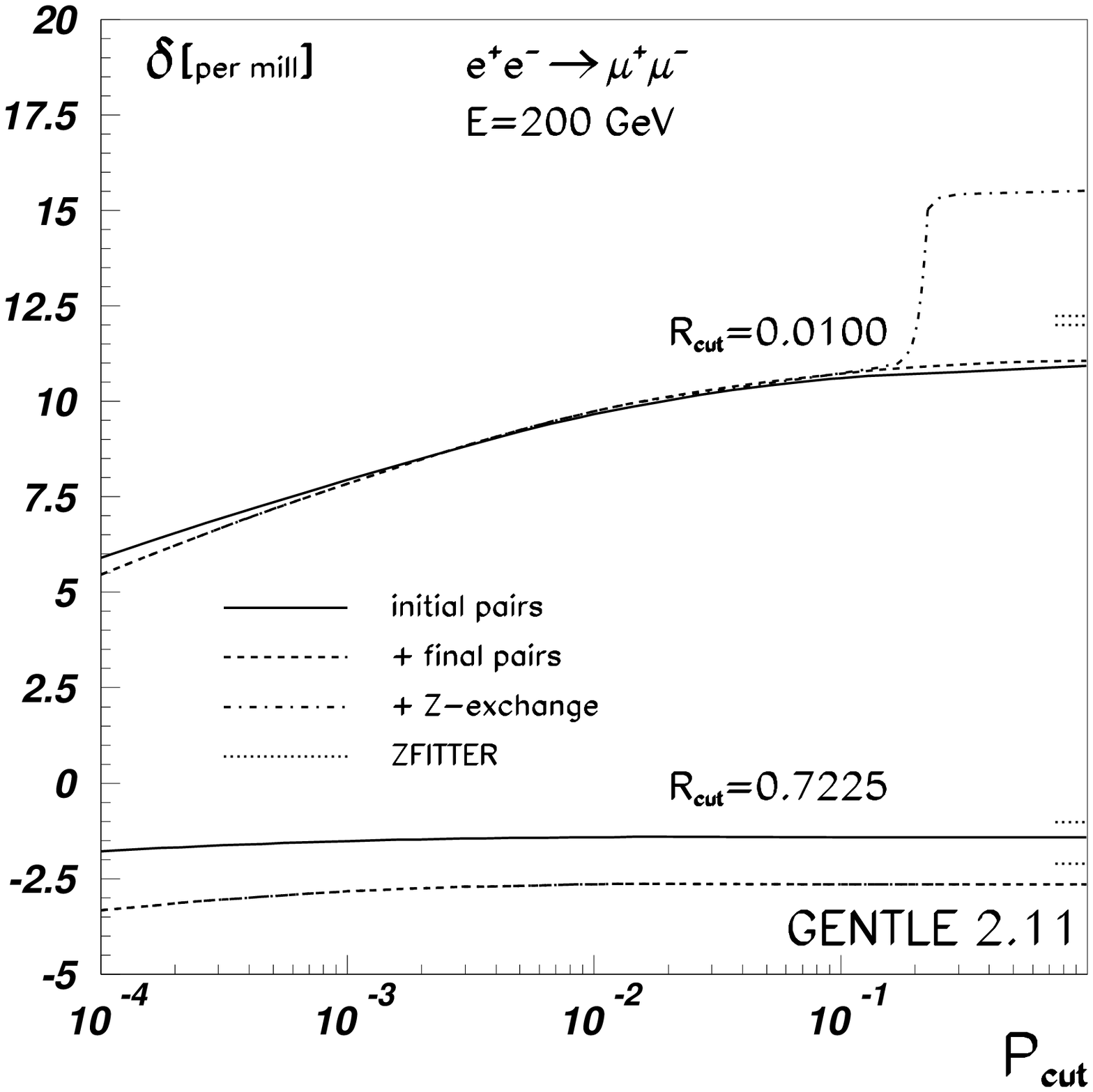,width=85mm,height=85mm}}}
\put(-50,-50){\makebox(0,0)[lb]{\epsfig{file=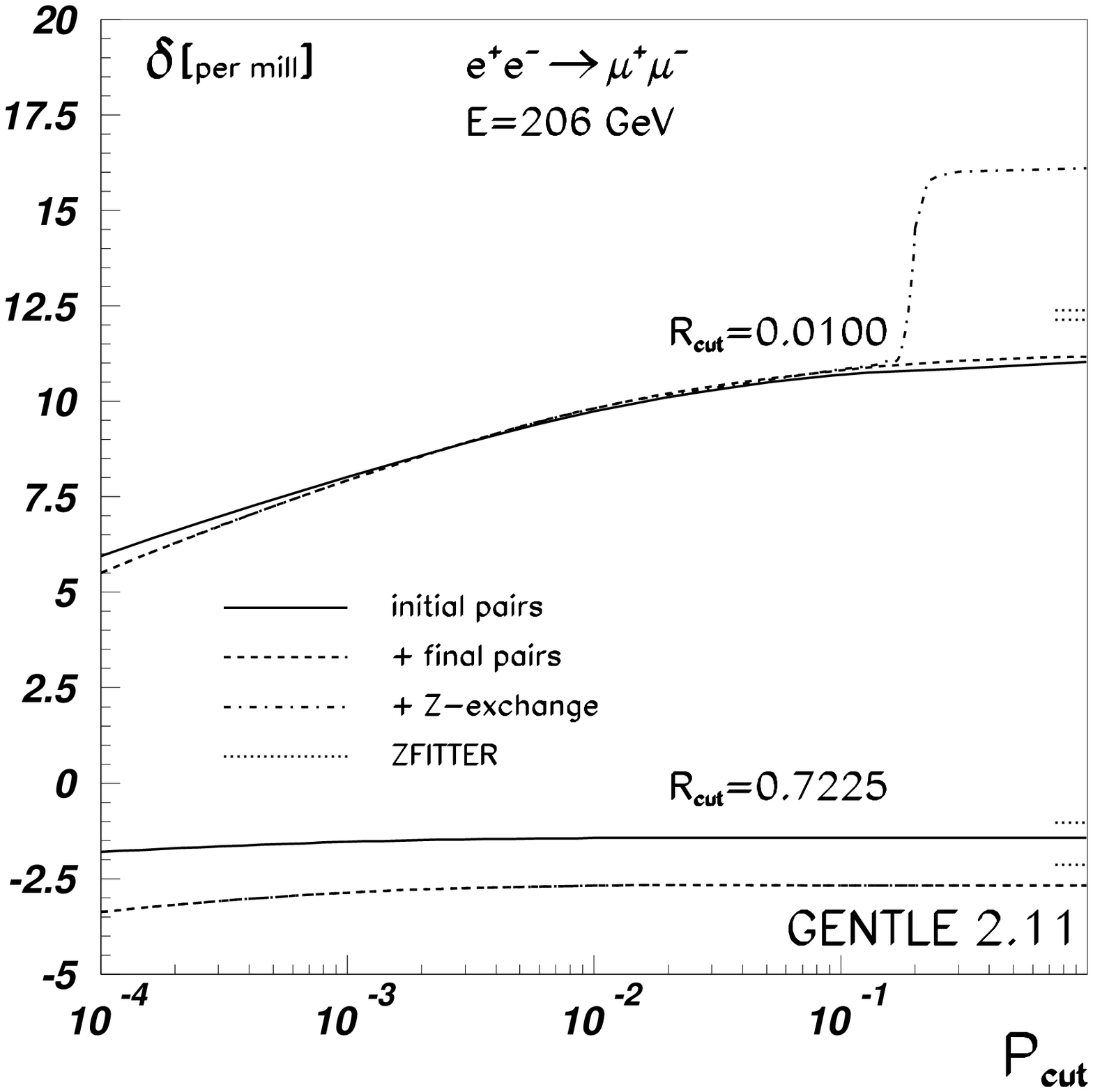,width=85mm,height=85mm}}}
\put(750,-50){\makebox(0,0)[lb]{\epsfig{file=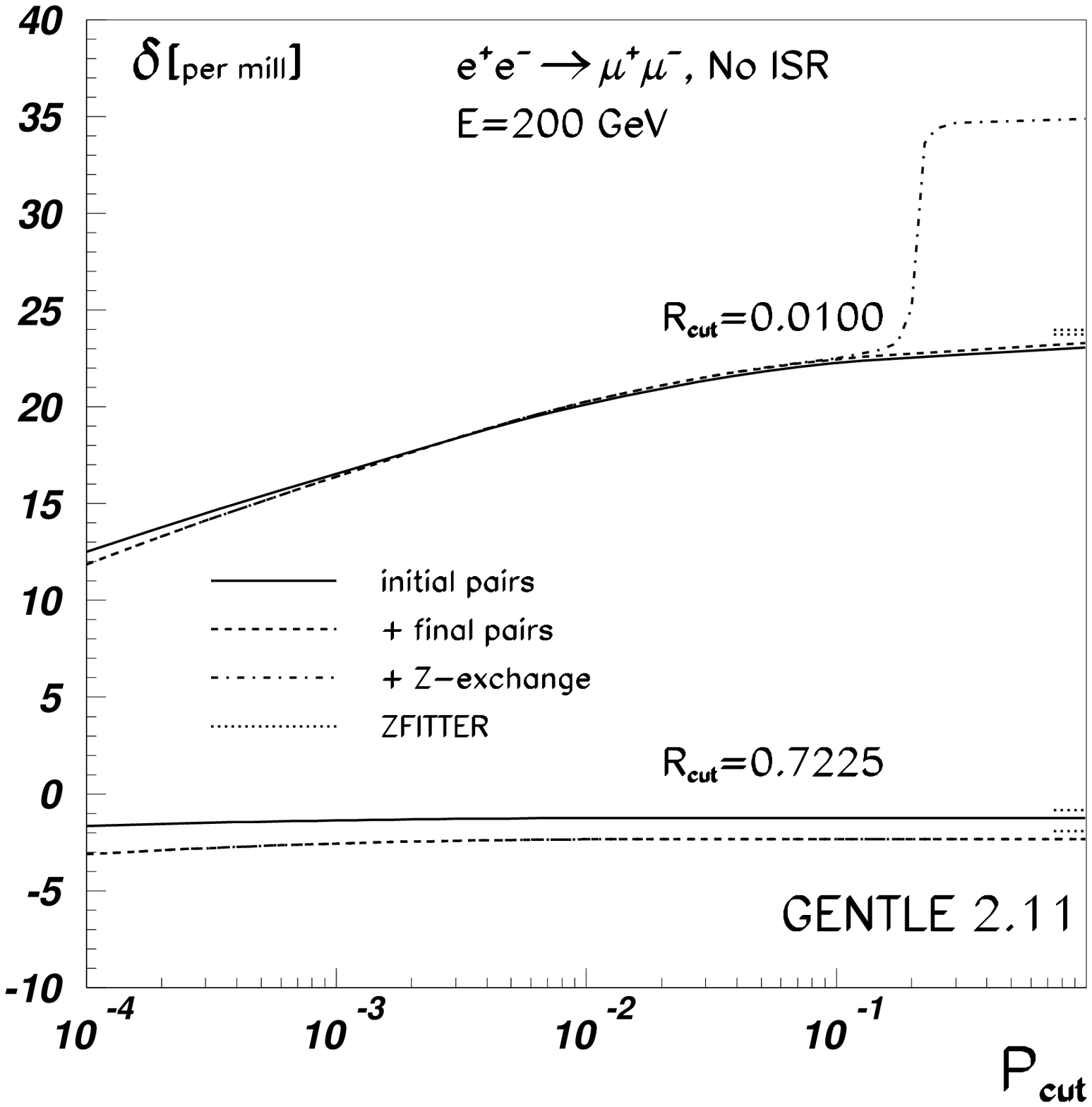,width=85mm,height=85mm}}}
\end{picture}
\caption{\small\sf 
$\delta_{\rm{pairs}}(P_{\rm{cut}})$ for the process $e^+e^-\to\mbox{muons}$ for two $R_{\rm{cut}}$
and three c.m.s energies. In the fourth figure we show $\delta_{\rm{pairs}}(P_{\rm{cut}})$
computed without ISR convolution.\label{fig_mu}}
\end{figure}

The quantities $\delta_{\rm{pairs}}(P_{\rm{cut}})$ for the process $e^+e^-\to\mbox{hadrons}$ are
shown in tables~\ref{tab_qu}  and Fig.~\ref{fig_qu}.
They exhibit very similar to the case of process $e^+e^-\to\mbox{muons}$ behavior and
actually the same discussion applies for them. We note that the size of the effect is
nearly two times bigger as compared to muon case.
This is due to the fact that pair emission contributes strongly to the return to the Z,
which has a much larger hadronic branching fraction than the mixture of 
virtual  photon and Z in the $s$-channel propagator at the full centre-of-mass energy.  
Another funny feature of $\delta_{\rm{pairs}}(P_{\rm{cut}})$'s 
for the process $e^+e^-\to\mbox{hadrons}$ is much better agreement between 
GENTLE and \zf\  leading to an impossibility to see the difference
in the case if ISR is ignored. 
Therefore, in this case the difference seen in the first three figures of
Fig.~\ref{fig_qu} is totally due to ISR convolution.

\begin{table}[!h]
\begin{center}
\setlength{\unitlength}{1cm}
\begin{picture}(16,20)
\put( 0,10.1){\makebox(0,0)[lb]{\epsfig{file=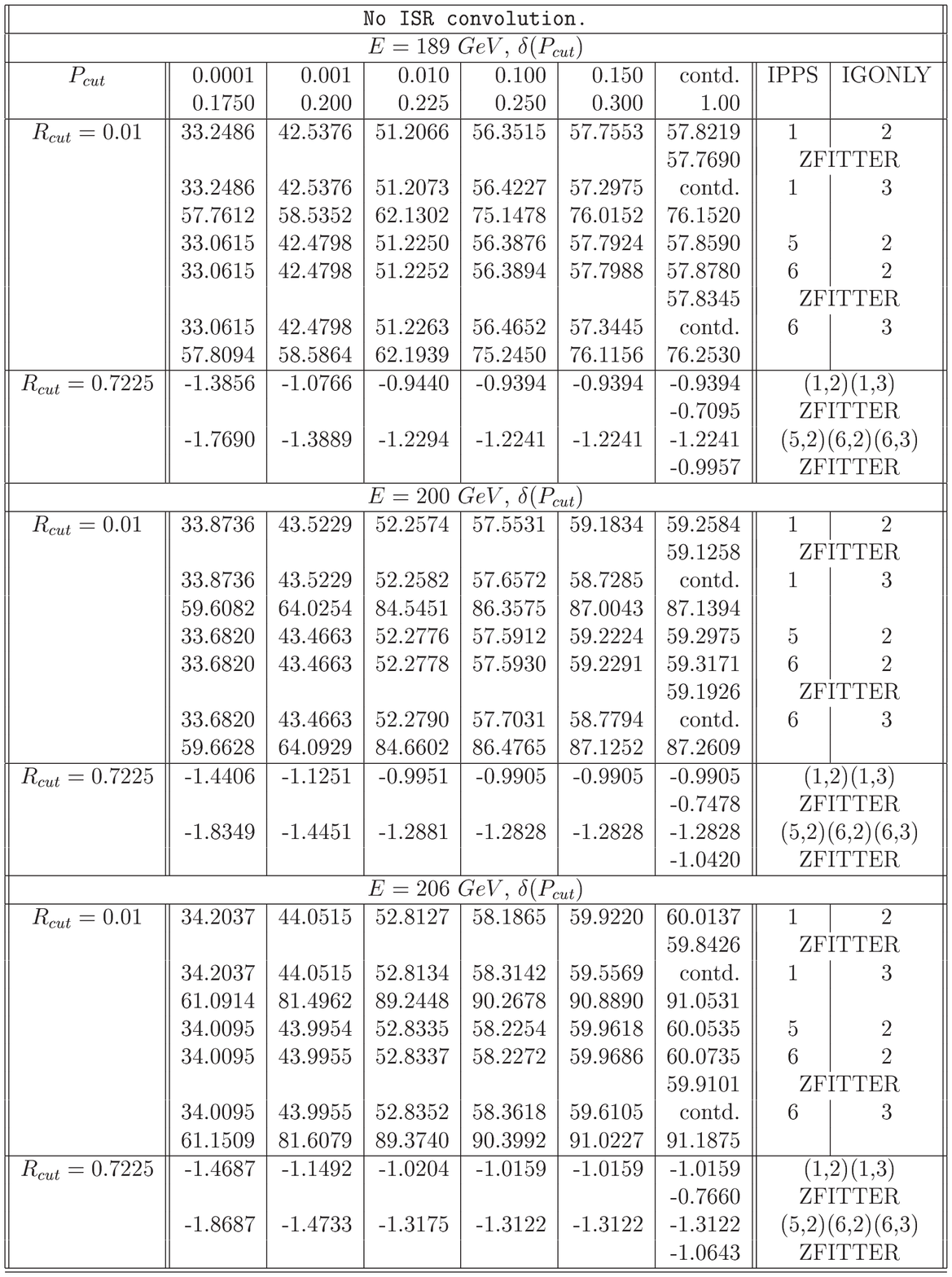,width=10cm,height=16cm,angle=90}}}
\put( 0, 0  ){\makebox(0,0)[lb]{\epsfig{file=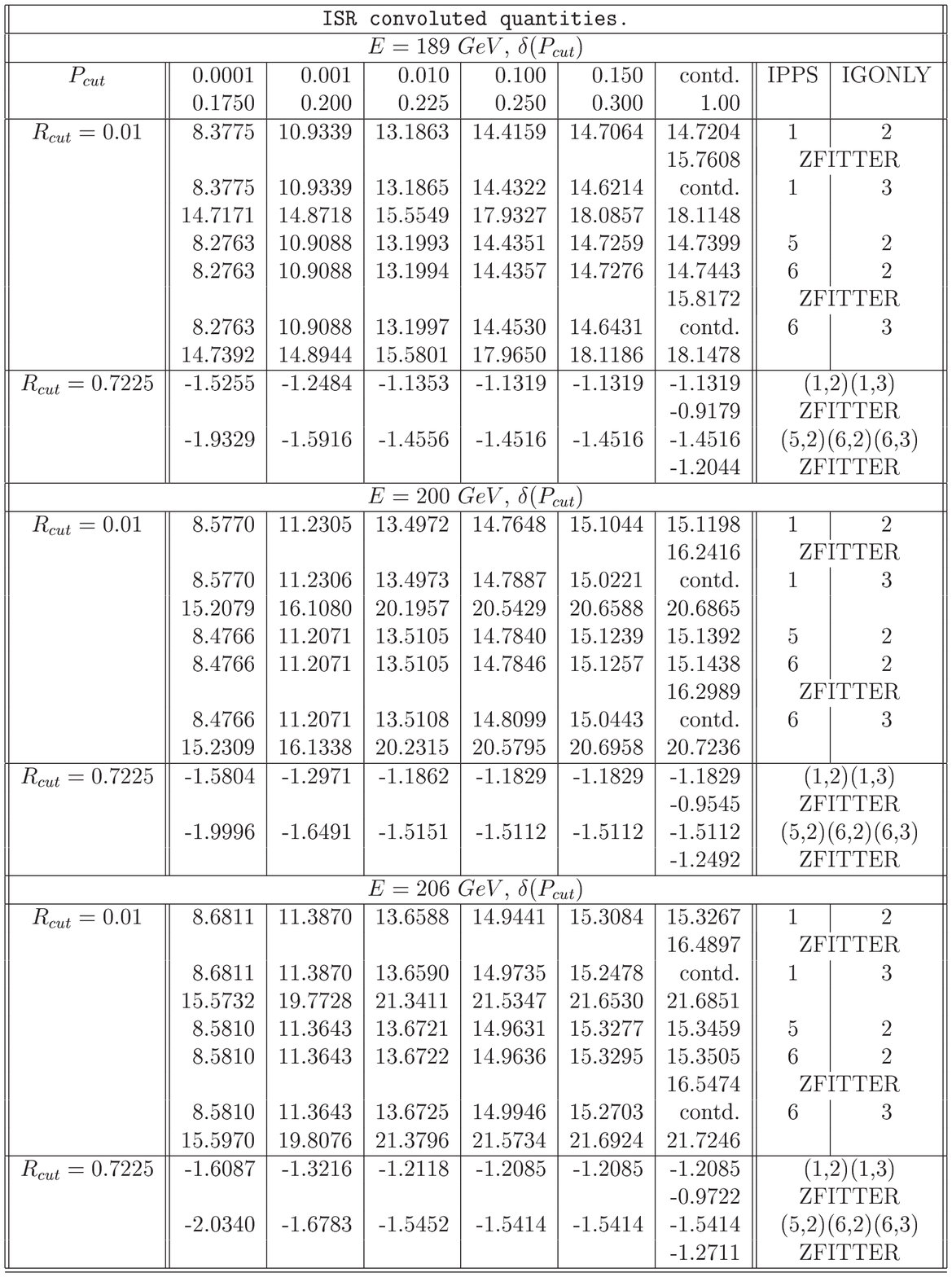,width=10cm,height=16cm,angle=90}}}
\end{picture}
\caption{GENTLE/4$\_$fan v.2.11. Process $e^{+}e^{-}\to hadrons$.        
{\tt IPPS - IGONLY} rows. Hadronic language.       
}
\label{tab_qu}
\end{center}
\end{table}

\newpage


The eight  tables~\ref{kobel1}-\ref{kobel2} contain the partial contributions  
to $\delta_{\rm{pairs}}$,  i.e.
separately  20 channels for
five primary $\otimes$ four secondary pairs for the process $e^+e^-\to\mbox{hadrons}$: 
$d,u,s,c,b\;\otimes\;\mbox{hadrons},e,\mu,\tau$,\\
for { IPPS=5, IGONLY=2} --
$[E_{\rm{cm}}=189,\;206\;\mbox{GeV}]\otimes[P_{\rm{cut}}=1.0,\;0.7225]$; \\
for { IPPS=6, IGONLY=3} --
$[E_{\rm{cm}}=189,\;206\;\mbox{GeV}]\otimes[P_{\rm{cut}}=0.1,\;0.7225]$.\\
The difference between two sets with { IPPS=5, IGONLY=2} and { IPPS=6, 
IGONLY=3}
is due to $P_{\rm{cut}}$, which is small owing to the plateau-like dependence, and 
due to $Z$ exchange, which is also small  since $Z$ doesn't open yet for 
a $P_{\rm{cut}}$ at 0.10.

Finally, two tables~\ref{Maciek} contain four partial contributions to 
$\delta_{\rm{pairs}}$ for the process $e^+e^-\to\mbox{muons}$
for the same set of input parameters. 
However, we show here both ``ISR off'' and ``ISR on'' cases and only {IPPS=5, IGONLY=2}
selection. 
The ``ISR on'' exhibits similar to the process $e^+e^-\to\mbox{hadrons}$
properties and the same discussion applies in this case.

\begin{figure}[t]
\centering
\setlength{\unitlength}{0.1mm}
\begin{picture}(1600,1500)
\put(-50,700){\makebox(0,0)[lb]{\epsfig{file=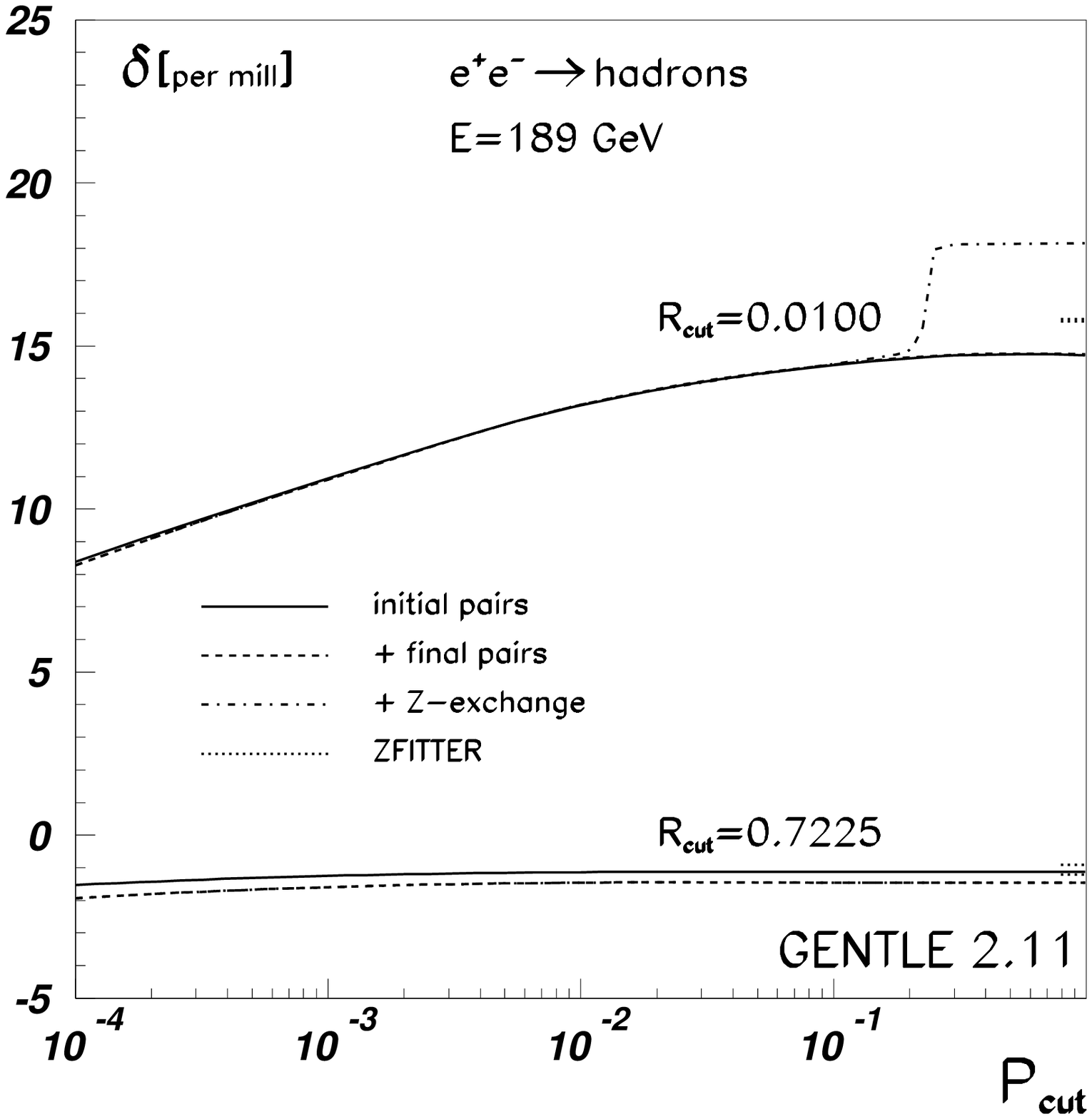,width=85mm,height=85mm}}}
\put(750,700){\makebox(0,0)[lb]{\epsfig{file=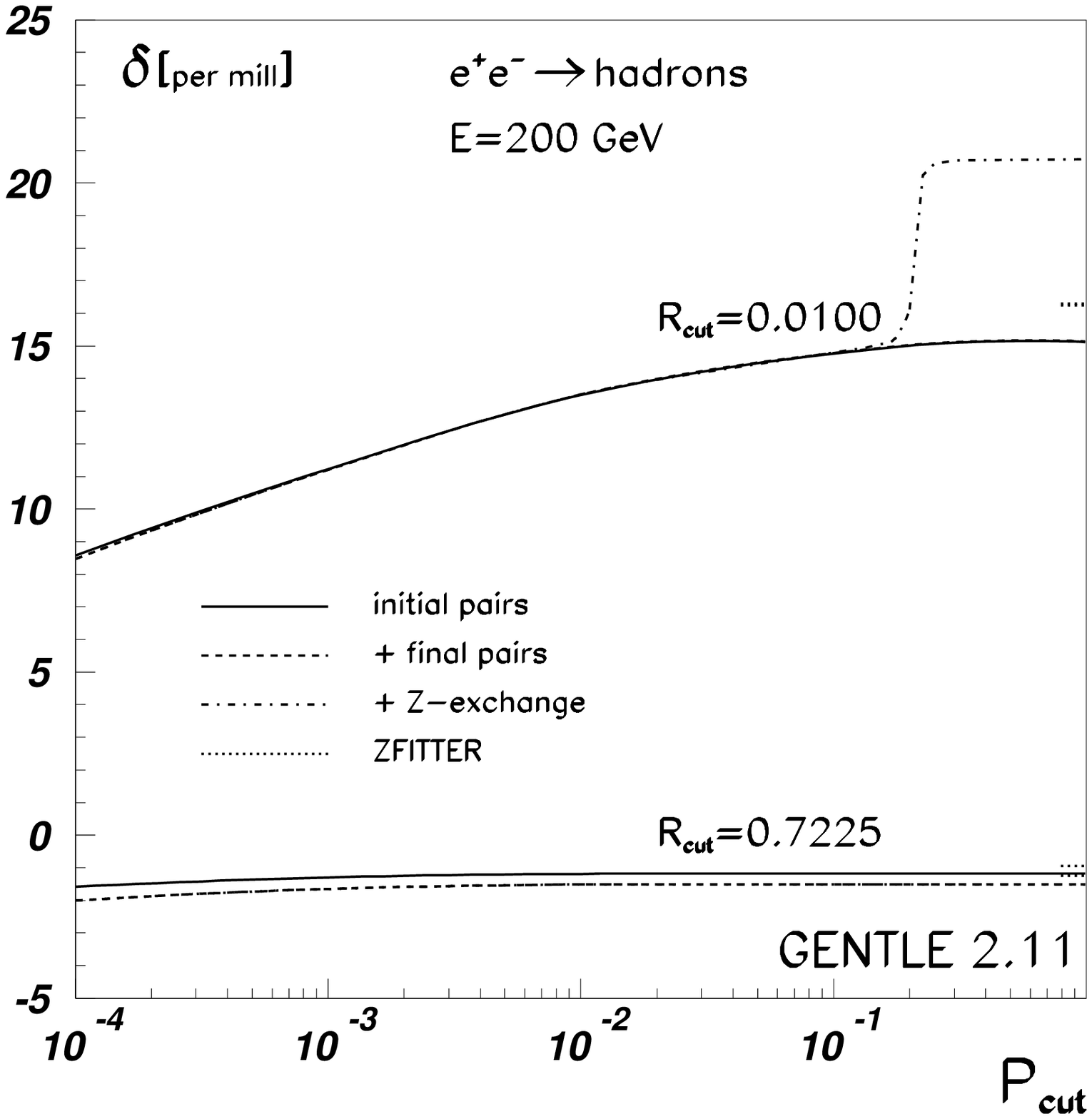,width=85mm,height=85mm}}}
\put(-50,-50){\makebox(0,0)[lb]{\epsfig{file=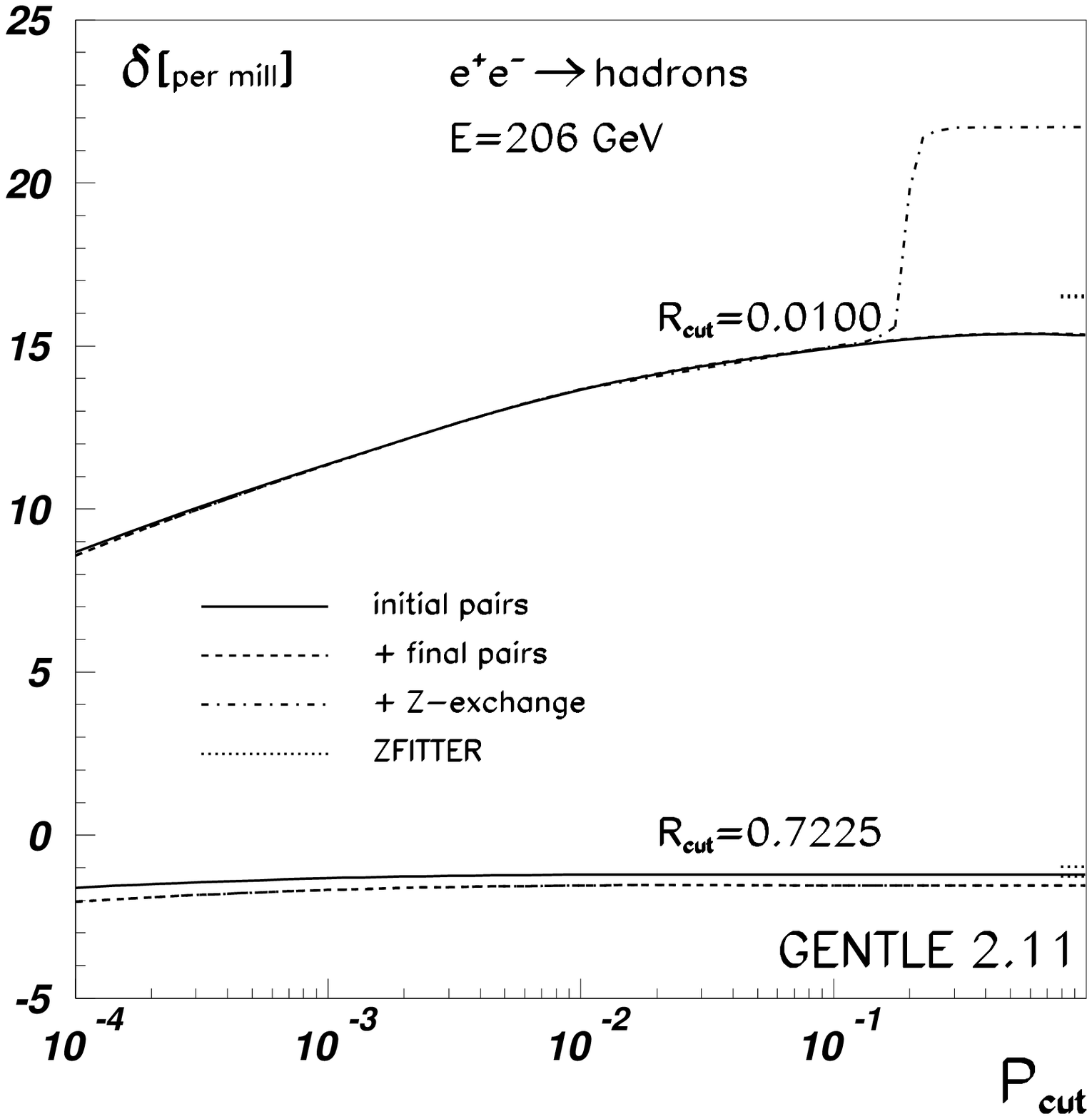,width=85mm,height=85mm}}}
\put(750,-50){\makebox(0,0)[lb]{\epsfig{file=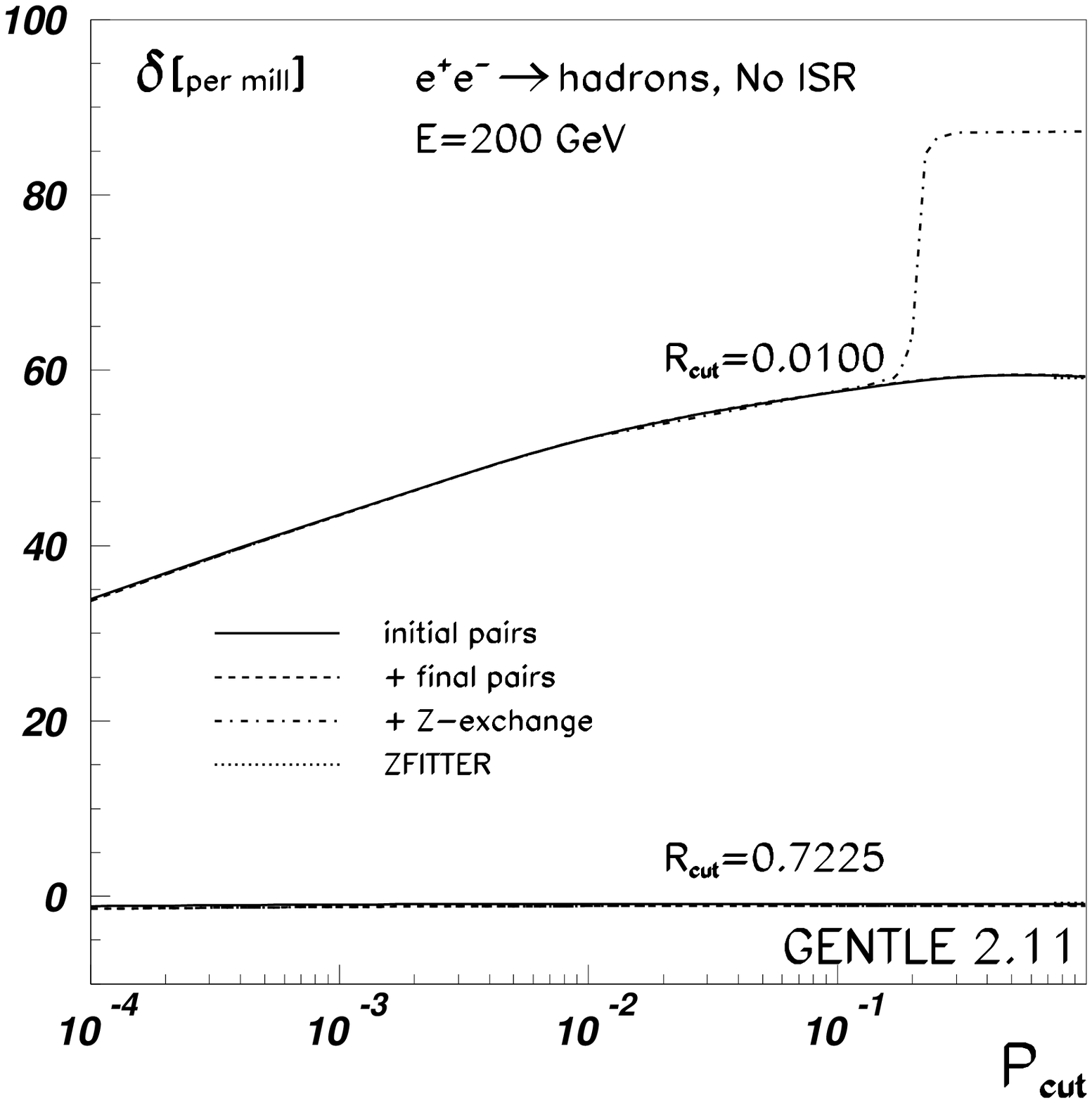,width=85mm,height=85mm}}}
\end{picture}
\caption{\small\sf 
$\delta_{\rm{pairs}}(P_{\rm{cut}})$ for the process $e^+e^-\to\mbox{hadrons}$ for two $R_{\rm{cut}}$
and three c.m.s energies. In the fourth figure we show $\delta_{\rm{pairs}}(P_{\rm{cut}})$
computed without ISR convolution.\label{fig_qu}}
\end{figure}

\begin{table}[!h]
\begin{center}
\setlength{\unitlength}{1cm}
\begin{picture}(16,20)
\put( 0,10.1){\makebox(0,0)[lb]{\epsfig{file=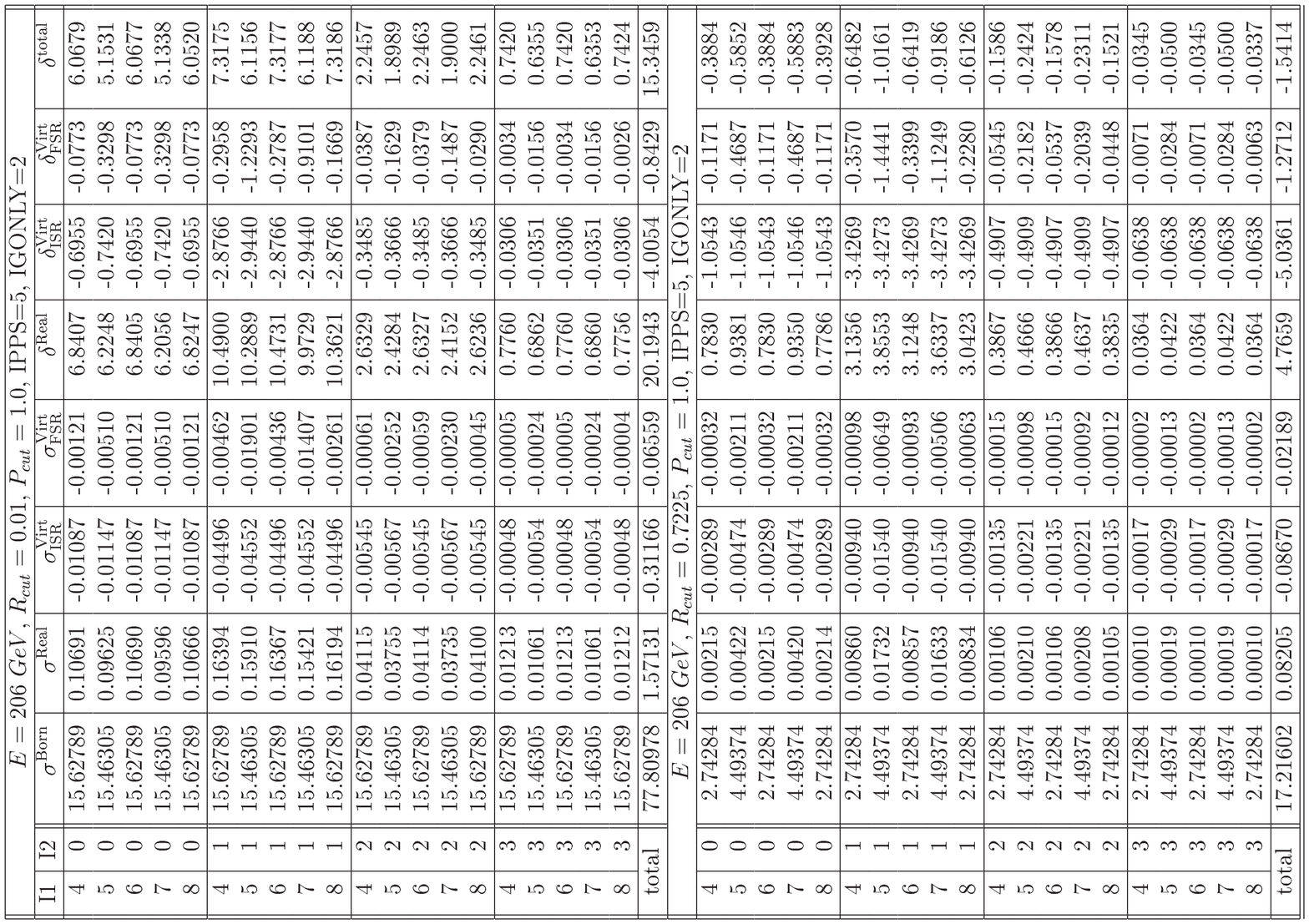,width=10cm,height=16cm,angle=90}}}
\put( 0, 0  ){\makebox(0,0)[lb]{\epsfig{file=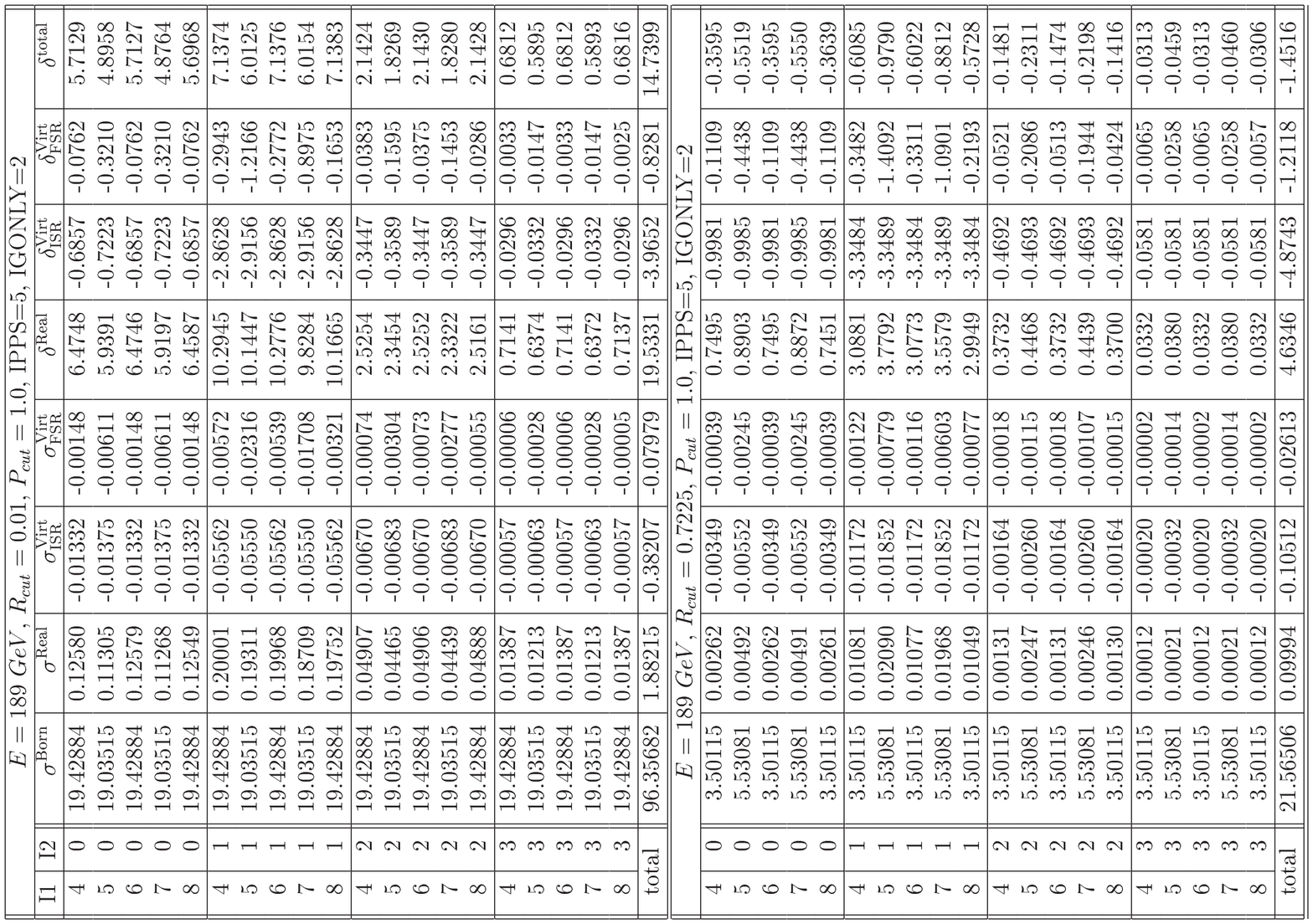,width=10cm,height=16cm,angle=90}}}
\end{picture}
\caption{LEP2: GENTLE$\_$4fan v.2.11, $e^+e^-\to\mbox{hadrons}$.
                 Partial contributions. I1-primary pair, I2-secondary pair.
           I1,I2$\;=\;0-hadrons,\;1-e,\;2-\mu,\;3-\tau,\;4-d,\;5-u,\;6-s,\;7-c,\;8-b$. 
                 Cross-sections in pb, $\delta$'s in per mill.          
The parameters (IPPS,IGONLY)=(5,2) correspond to the diagram-based signal definition with 
$s^\prime=M_{\rm inv}^2$.      
}
\label{kobel1}
\end{center}
\end{table}

\begin{table}[!h]
\begin{center}
\setlength{\unitlength}{1cm}
\begin{picture}(16,20)
\put( 0,10.1){\makebox(0,0)[lb]{\epsfig{file=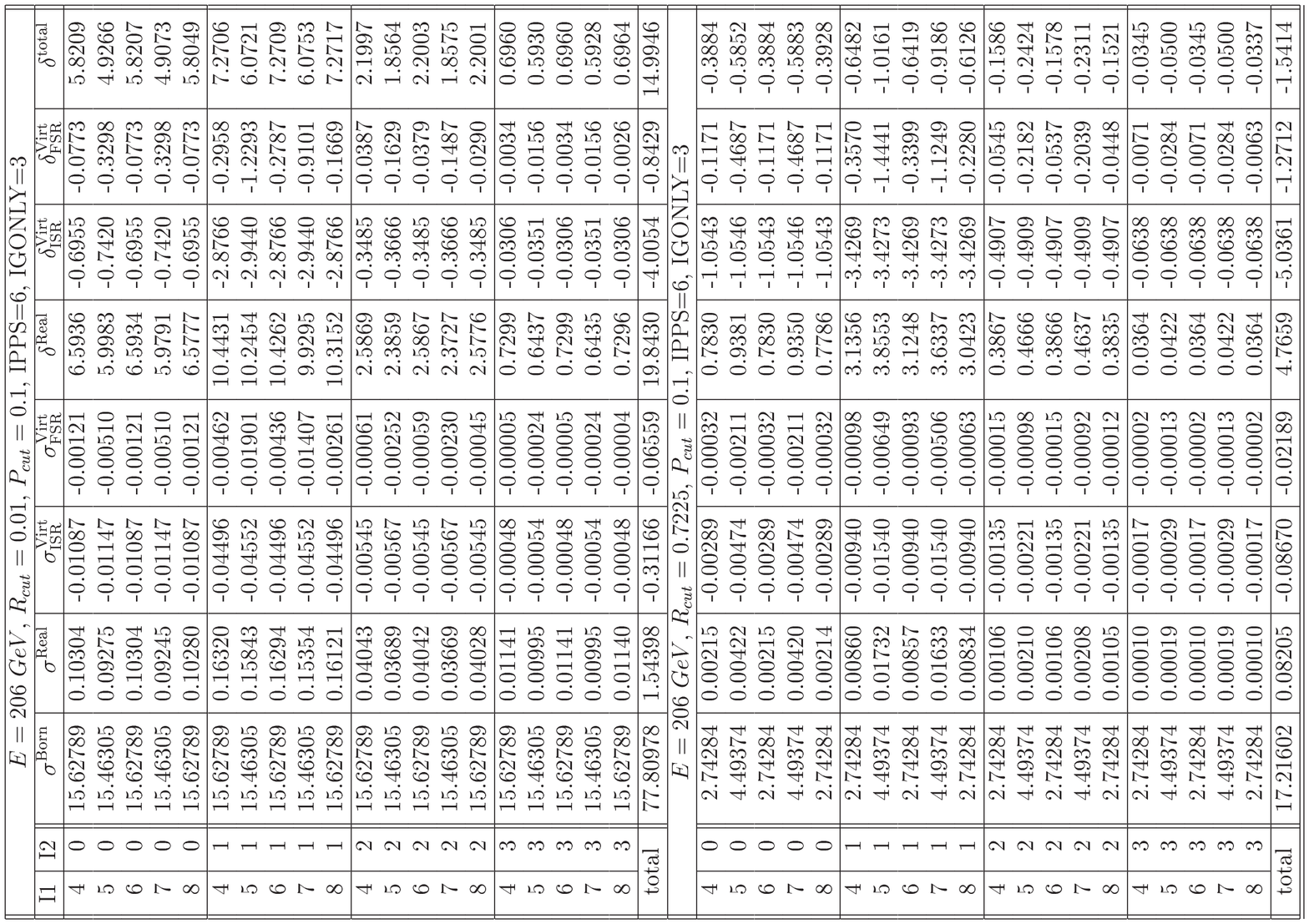,width=10cm,height=16cm,angle=90}}}
\put( 0, 0  ){\makebox(0,0)[lb]{\epsfig{file=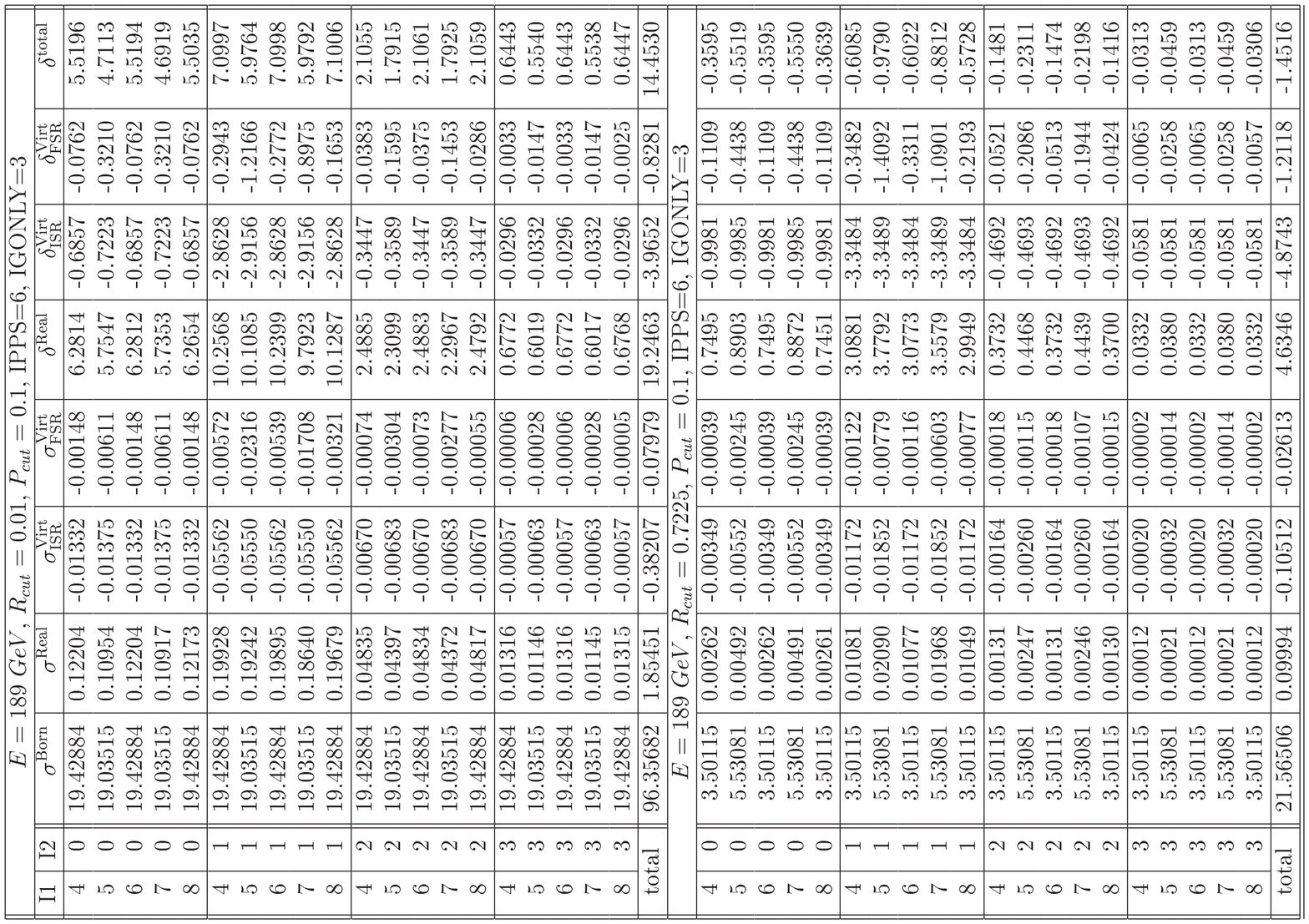,width=10cm,height=16cm,angle=90}}}
\end{picture}
\caption{LEP2: GENTLE$\_$4fan v.2.11, $e^+e^-\to\mbox{hadrons}$.
                 Partial contributions. I1-primary pair, I2-secondary pair.
           I1,I2$\;=\;0-hadrons,\;1-e,\;2-\mu,\;3-\tau,\;4-d,\;5-u,\;6-s,\;7-c,\;8-b$. 
                 Cross-sections in pb, $\delta$'s in per mill.          
The parameters (IPPS,IGONLY)=(6,3) together with $P_{\rm cut}$=0.10  correspond to the cut-based signal definition.  
}
\label{kobel2}
\end{center}
\end{table}



\begin{table}[!h]
\begin{center}
\epsfig{file=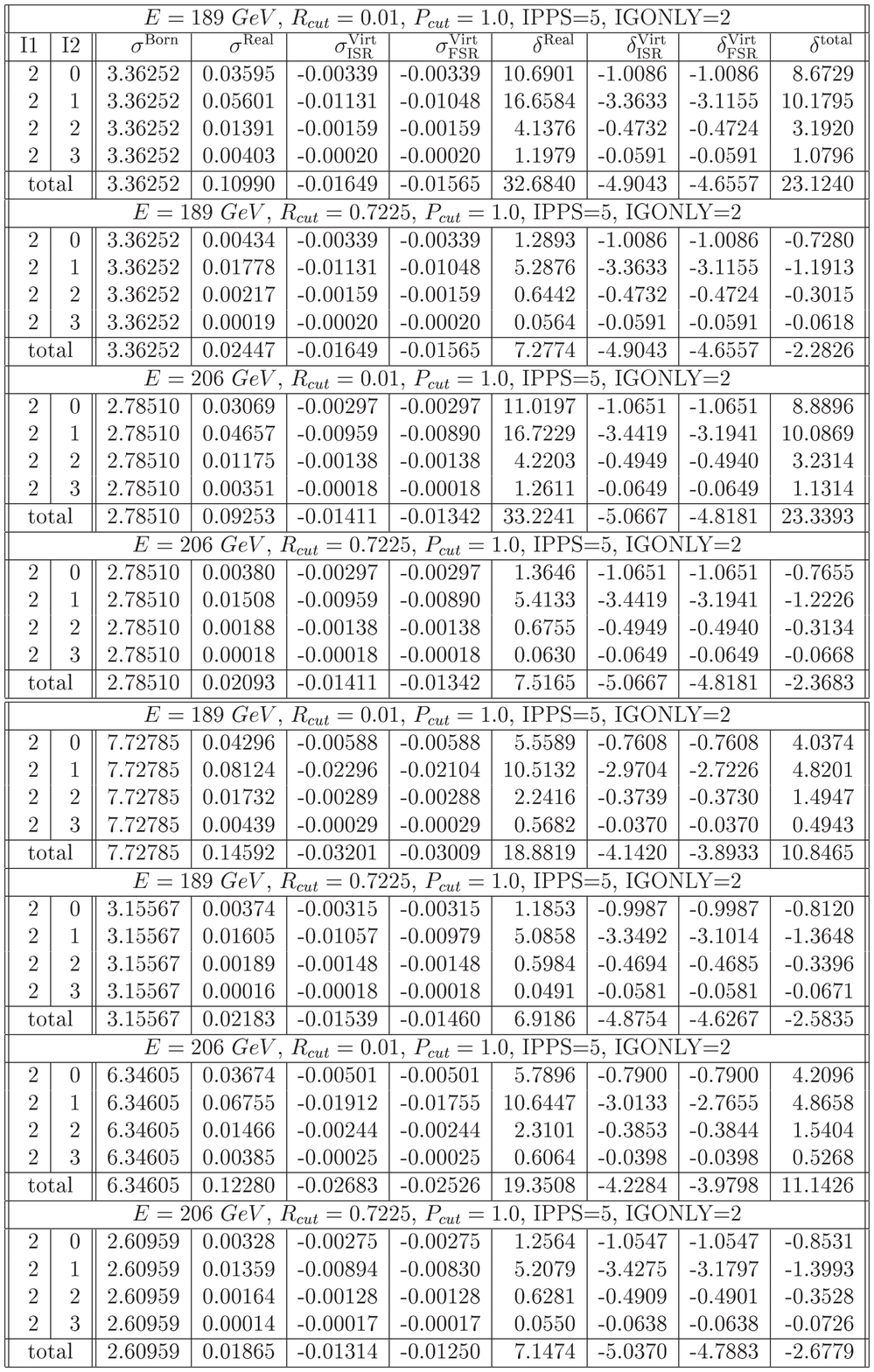,width=14cm,height=21cm,angle=0}
\caption[]{LEP2: GENTLE\_4fan v.2.11, $e^{+}e^{-}\to\mbox{muons}$.
                 Partial contributions. I1-primary pair, I2-secondary pair.\\
                 Cross-sections in pb, $\delta$'s in per mill. 
                 Upper part ISR off, lower part ISR on.
\label{Maciek}}
\end{center}
\end{table}

%% file: 2f-Chapt-Part7.tex
\subsubsection{Results on Pairs from \KKMC\  and KORALW}

There is an intriguing possibility to define and realize with
\KKMC+KORALW an alternative definition of the $2f$ signal:
{\fbox{ $2f$ Experimental Signal $\equiv$ $2f$ signal without any pairs}} 
realized by:
\begin{itemize} 
\item
Eliminating completely all $4f$ background together with other backgrounds and
detector efficiency using KORALW. 
\item 
Eliminating virtual par
contributions $\sim 1\%$ together with the ISR*FSR
interference using \KKMC.
\item 
Switching off pairs in ZFITTER or TOPAZ0.
\end{itemize}

The above scenario was usually not emphasized in the past, 
because of the potential technical difficulties 
with the MC integration, and the cancellation of the mass singularities.
On the other hand, while constructing the KORALW
program this application was kept in mind \cite{koralw:1998}
and the appropriate coverage of phase space integration was ensured.
Together with the recent upgrade of \KKMC\ with the virtual pair form-factors
this opens the way to the first exercises in this direction.
The cancellation of fermion masses in KORALW+\KKMC\ is expected
to be technically and physically as good as in the semi-analytical programs 
that are compared with them%
\footnote{
 This approach to pairs was already
 realized in the case of Bhabha scattering in the BHLUMI 2.30 of 
 ref.~\cite{bhlumi-pairs:1997}, where multiple real pairs 
 are generated within the full exclusive phase space and
 the virtual and real pair corrections cancel numerically.
}.

\begin{figure}[!ht]
\centering
\setlength{\unitlength}{0.1mm}
\begin{picture}(1600,1020)
\put(300,250){\begin{picture}( 1200,720)
\put(0,0){\framebox( 1200,720){ }}
\multiput(  300.00,0)(  300.00,0){   4}{\line(0,1){25}}
\multiput(     .00,0)(   30.00,0){  41}{\line(0,1){10}}
\multiput(  300.00,720)(  300.00,0){   4}{\line(0,-1){25}}
\multiput(     .00,720)(   30.00,0){  41}{\line(0,-1){10}}
\put( 300,-25){\makebox(0,0)[t]{\large $    0.25 $}}
\put( 600,-25){\makebox(0,0)[t]{\large $    0.5 $}}
\put( 900,-25){\makebox(0,0)[t]{\large $    0.75 $}}
\put(1200,-25){\makebox(0,0)[t]{\large $    1 $}}
\put( 750,-125){\makebox(0,0)[r]{\Large $v_{cut}=1-R_{cut}$}}
\multiput(0,     .00)(0,  240.00){   4}{\line(1,0){25}}
\multiput(0,   24.00)(0,   24.00){  30}{\line(1,0){10}}
\multiput(1200,     .00)(0,  240.00){   4}{\line(-1,0){25}}
\multiput(1200,   24.00)(0,   24.00){  30}{\line(-1,0){10}}
\put(-25,   0){\makebox(0,0)[r]{\large $     0 $}}
\put(-25, 240){\makebox(0,0)[r]{\large $     0.005 $}}
\put(-25, 480){\makebox(0,0)[r]{\large $    0.01 $}}
\put(-25, 720){\makebox(0,0)[r]{\large $    0.015 $}}
\put(-125, 600){\makebox(0,0)[r] {\Large $\sigma$ [pb]}}
\end{picture}}
\put(300,250){\begin{picture}( 1200,1200)
\newcommand{\R}[2]{\put(#1,#2){\circle*{  3}}}
\R{   0}{  -1} \R{   9}{   1} \R{  19}{   3} \R{  28}{   6}
\R{  38}{   9} \R{  48}{  11} \R{  58}{  14} \R{  67}{  16}
\R{  77}{  18} \R{  87}{  20} \R{  97}{  21} \R{ 107}{  23}
\R{ 116}{  25} \R{ 126}{  26} \R{ 136}{  28} \R{ 146}{  29}
\R{ 156}{  31} \R{ 166}{  32} \R{ 176}{  34} \R{ 186}{  35}
\R{ 196}{  36} \R{ 206}{  37} \R{ 216}{  39} \R{ 225}{  40}
\R{ 235}{  41} \R{ 245}{  42} \R{ 255}{  43} \R{ 265}{  44}
\R{ 275}{  46} \R{ 285}{  47} \R{ 295}{  48} \R{ 305}{  49}
\R{ 315}{  50} \R{ 325}{  51} \R{ 335}{  52} \R{ 345}{  53}
\R{ 355}{  54} \R{ 365}{  55} \R{ 375}{  56} \R{ 385}{  57}
\R{ 395}{  58} \R{ 404}{  59} \R{ 414}{  59} \R{ 424}{  60}
\R{ 434}{  61} \R{ 444}{  62} \R{ 454}{  63} \R{ 464}{  64}
\R{ 474}{  65} \R{ 484}{  66} \R{ 494}{  67} \R{ 504}{  68}
\R{ 514}{  69} \R{ 524}{  70} \R{ 534}{  70} \R{ 544}{  71}
\R{ 554}{  72} \R{ 564}{  73} \R{ 574}{  74} \R{ 584}{  75}
\R{ 594}{  76} \R{ 604}{  77} \R{ 614}{  78} \R{ 624}{  79}
\R{ 633}{  80} \R{ 643}{  81} \R{ 653}{  82} \R{ 663}{  83}
\R{ 673}{  84} \R{ 683}{  85} \R{ 693}{  86} \R{ 703}{  87}
\R{ 713}{  88} \R{ 723}{  89} \R{ 733}{  90} \R{ 743}{  92}
\R{ 753}{  93} \R{ 763}{  94} \R{ 773}{  96} \R{ 782}{  97}
\R{ 792}{  99} \R{ 802}{ 100} \R{ 812}{ 102} \R{ 822}{ 104}
\R{ 832}{ 107} \R{ 841}{ 109} \R{ 851}{ 112} \R{ 860}{ 115}
\R{ 869}{ 119} \R{ 878}{ 125} \R{ 886}{ 131} \R{ 892}{ 138}
\R{ 898}{ 146} \R{ 902}{ 156} \R{ 905}{ 166} \R{ 907}{ 176}
\R{ 908}{ 186} \R{ 909}{ 195} \R{ 910}{ 205} \R{ 912}{ 215}
\R{ 913}{ 225} \R{ 914}{ 235} \R{ 915}{ 245} \R{ 916}{ 255}
\R{ 917}{ 265} \R{ 918}{ 275} \R{ 918}{ 285} \R{ 919}{ 295}
\R{ 920}{ 305} \R{ 921}{ 315} \R{ 922}{ 325} \R{ 924}{ 335}
\R{ 925}{ 345} \R{ 926}{ 355} \R{ 927}{ 365} \R{ 928}{ 374}
\R{ 930}{ 384} \R{ 931}{ 394} \R{ 932}{ 404} \R{ 934}{ 414}
\R{ 936}{ 424} \R{ 938}{ 434} \R{ 941}{ 443} \R{ 946}{ 451}
\R{ 954}{ 457} \R{ 963}{ 462} \R{ 972}{ 466} \R{ 982}{ 469}
\R{ 991}{ 472} \R{1001}{ 474} \R{1011}{ 476} \R{1021}{ 478}
\R{1031}{ 479} \R{1040}{ 481} \R{1050}{ 483} \R{1060}{ 485}
\R{1070}{ 487} \R{1080}{ 489} \R{1089}{ 491} \R{1099}{ 493}
\R{1109}{ 495} \R{1119}{ 498} \R{1128}{ 501} \R{1138}{ 504}
\R{1147}{ 508} \R{1156}{ 512} \R{1165}{ 516} \R{1173}{ 522}
\R{1178}{ 532} \R{1181}{ 542} \R{1183}{ 551} \R{1185}{ 561}
\R{1186}{ 571} \R{1188}{ 581} \R{1189}{ 591} \R{1190}{ 601}
\R{1191}{ 611} \R{1192}{ 621} \R{1193}{ 631} \R{1194}{ 641}
\R{1195}{ 651} \R{1196}{ 661} \R{1196}{ 671} \R{1197}{ 681}
\R{1198}{ 691} \R{1198}{ 701} \R{1199}{ 711} \R{1200}{ 721}
\R{1200}{ 731}
\end{picture}} 
\put(300,250){\begin{picture}( 1200,1200)
\newcommand{\R}[2]{\put(#1,#2){\circle*{  6}}}
\newcommand{\E}[3]{\put(#1,#2){\line(0,1){#3}}}
\R{   8}{   3}
\E{   8}{    3}{   2}
\R{  23}{   9}
\E{  23}{    8}{   2}
\R{  38}{  15}
\E{  38}{   13}{   2}
\R{  53}{  19}
\E{  53}{   18}{   4}
\R{  68}{  23}
\E{  68}{   21}{   4}
\R{  83}{  26}
\E{  83}{   24}{   4}
\R{  98}{  29}
\E{  98}{   27}{   4}
\R{ 113}{  33}
\E{ 113}{   30}{   4}
\R{ 128}{  35}
\E{ 128}{   33}{   4}
\R{ 143}{  39}
\E{ 143}{   36}{   4}
\R{ 158}{  43}
\E{ 158}{   41}{   4}
\R{ 173}{  45}
\E{ 173}{   43}{   6}
\R{ 188}{  48}
\E{ 188}{   46}{   6}
\R{ 203}{  50}
\E{ 203}{   48}{   6}
\R{ 218}{  53}
\E{ 218}{   51}{   6}
\R{ 233}{  55}
\E{ 233}{   53}{   6}
\R{ 248}{  58}
\E{ 248}{   55}{   6}
\R{ 263}{  60}
\E{ 263}{   57}{   6}
\R{ 278}{  63}
\E{ 278}{   60}{   6}
\R{ 293}{  65}
\E{ 293}{   62}{   6}
\R{ 308}{  67}
\E{ 308}{   64}{   6}
\R{ 323}{  69}
\E{ 323}{   66}{   6}
\R{ 338}{  72}
\E{ 338}{   69}{   6}
\R{ 353}{  74}
\E{ 353}{   71}{   6}
\R{ 368}{  75}
\E{ 368}{   72}{   6}
\R{ 383}{  77}
\E{ 383}{   74}{   6}
\R{ 398}{  79}
\E{ 398}{   76}{   6}
\R{ 413}{  80}
\E{ 413}{   77}{   6}
\R{ 428}{  82}
\E{ 428}{   79}{   6}
\R{ 443}{  83}
\E{ 443}{   80}{   6}
\R{ 458}{  85}
\E{ 458}{   82}{   6}
\R{ 473}{  86}
\E{ 473}{   83}{   6}
\R{ 488}{  88}
\E{ 488}{   84}{   6}
\R{ 503}{  89}
\E{ 503}{   86}{   6}
\R{ 518}{  91}
\E{ 518}{   88}{   6}
\R{ 533}{  93}
\E{ 533}{   90}{   6}
\R{ 548}{  95}
\E{ 548}{   91}{   6}
\R{ 563}{  96}
\E{ 563}{   93}{   6}
\R{ 578}{  98}
\E{ 578}{   95}{   6}
\R{ 593}{ 100}
\E{ 593}{   96}{   6}
\R{ 608}{ 101}
\E{ 608}{   98}{   6}
\R{ 623}{ 103}
\E{ 623}{  100}{   6}
\R{ 638}{ 104}
\E{ 638}{  101}{   6}
\R{ 653}{ 106}
\E{ 653}{  103}{   6}
\R{ 668}{ 108}
\E{ 668}{  105}{   6}
\R{ 683}{ 109}
\E{ 683}{  106}{   6}
\R{ 698}{ 111}
\E{ 698}{  108}{   6}
\R{ 713}{ 113}
\E{ 713}{  110}{   6}
\R{ 728}{ 114}
\E{ 728}{  111}{   6}
\R{ 743}{ 117}
\E{ 743}{  113}{   6}
\R{ 758}{ 118}
\E{ 758}{  115}{   6}
\R{ 773}{ 121}
\E{ 773}{  117}{   6}
\R{ 788}{ 123}
\E{ 788}{  120}{   8}
\R{ 803}{ 126}
\E{ 803}{  123}{   8}
\R{ 818}{ 128}
\E{ 818}{  125}{   8}
\R{ 833}{ 133}
\E{ 833}{  129}{   8}
\R{ 848}{ 138}
\E{ 848}{  134}{   8}
\R{ 863}{ 143}
\E{ 863}{  139}{   8}
\R{ 878}{ 157}
\E{ 878}{  152}{   8}
\R{ 893}{ 180}
\E{ 893}{  175}{  10}
\R{ 908}{ 274}
\E{ 908}{  268}{  14}
\R{ 923}{ 499}
\E{ 923}{  490}{  20}
\R{ 938}{ 556}
\E{ 938}{  545}{  20}
\R{ 953}{ 576}
\E{ 953}{  566}{  20}
\R{ 968}{ 585}
\E{ 968}{  575}{  20}
\R{ 983}{ 591}
\E{ 983}{  580}{  20}
\R{ 998}{ 595}
\E{ 998}{  585}{  20}
\R{1013}{ 600}
\E{1013}{  590}{  20}
\R{1028}{ 603}
\E{1028}{  593}{  20}
\R{1043}{ 607}
\E{1043}{  597}{  20}
\R{1058}{ 610}
\E{1058}{  600}{  20}
\R{1073}{ 615}
\E{1073}{  605}{  20}
\R{1088}{ 620}
\E{1088}{  609}{  20}
\R{1103}{ 625}
\E{1103}{  615}{  20}
\R{1118}{ 632}
\E{1118}{  621}{  20}
\R{1133}{ 640}
\E{1133}{  629}{  20}
\R{1148}{ 650}
\E{1148}{  640}{  20}
\R{1163}{ 667}
\E{1163}{  657}{  22}
\R{1178}{ 697}
\E{1178}{  686}{  22}
\end{picture}} 
\put(300,250){\begin{picture}( 1200,1200)
\newcommand{\R}[2]{\put(#1,#2){\circle{ 10}}}
\newcommand{\E}[3]{\put(#1,#2){\line(0,1){#3}}}
\R{   8}{   3}
\E{   8}{    2}{   2}
\R{  23}{   8}
\E{  23}{    7}{   2}
\R{  38}{  11}
\E{  38}{    9}{   2}
\R{  53}{  13}
\E{  53}{   12}{   2}
\R{  68}{  17}
\E{  68}{   15}{   4}
\R{  83}{  18}
\E{  83}{   17}{   4}
\R{  98}{  21}
\E{  98}{   19}{   4}
\R{ 113}{  23}
\E{ 113}{   21}{   4}
\R{ 128}{  25}
\E{ 128}{   23}{   4}
\R{ 143}{  27}
\E{ 143}{   25}{   4}
\R{ 158}{  30}
\E{ 158}{   28}{   4}
\R{ 173}{  31}
\E{ 173}{   29}{   4}
\R{ 188}{  33}
\E{ 188}{   31}{   4}
\R{ 203}{  35}
\E{ 203}{   33}{   4}
\R{ 218}{  36}
\E{ 218}{   34}{   4}
\R{ 233}{  37}
\E{ 233}{   35}{   4}
\R{ 248}{  39}
\E{ 248}{   37}{   4}
\R{ 263}{  40}
\E{ 263}{   38}{   4}
\R{ 278}{  42}
\E{ 278}{   40}{   4}
\R{ 293}{  44}
\E{ 293}{   42}{   4}
\R{ 308}{  46}
\E{ 308}{   43}{   4}
\R{ 323}{  48}
\E{ 323}{   45}{   4}
\R{ 338}{  49}
\E{ 338}{   47}{   4}
\R{ 353}{  51}
\E{ 353}{   48}{   4}
\R{ 368}{  52}
\E{ 368}{   49}{   4}
\R{ 383}{  53}
\E{ 383}{   51}{   4}
\R{ 398}{  55}
\E{ 398}{   52}{   4}
\R{ 413}{  56}
\E{ 413}{   53}{   6}
\R{ 428}{  57}
\E{ 428}{   55}{   6}
\R{ 443}{  58}
\E{ 443}{   56}{   6}
\R{ 458}{  59}
\E{ 458}{   57}{   6}
\R{ 473}{  61}
\E{ 473}{   58}{   6}
\R{ 488}{  62}
\E{ 488}{   59}{   6}
\R{ 503}{  63}
\E{ 503}{   61}{   6}
\R{ 518}{  65}
\E{ 518}{   62}{   6}
\R{ 533}{  66}
\E{ 533}{   64}{   6}
\R{ 548}{  68}
\E{ 548}{   65}{   6}
\R{ 563}{  69}
\E{ 563}{   67}{   6}
\R{ 578}{  71}
\E{ 578}{   68}{   6}
\R{ 593}{  72}
\E{ 593}{   70}{   6}
\R{ 608}{  74}
\E{ 608}{   71}{   6}
\R{ 623}{  76}
\E{ 623}{   73}{   6}
\R{ 638}{  77}
\E{ 638}{   74}{   6}
\R{ 653}{  78}
\E{ 653}{   75}{   6}
\R{ 668}{  80}
\E{ 668}{   77}{   6}
\R{ 683}{  82}
\E{ 683}{   79}{   6}
\R{ 698}{  83}
\E{ 698}{   80}{   6}
\R{ 713}{  85}
\E{ 713}{   82}{   6}
\R{ 728}{  86}
\E{ 728}{   83}{   6}
\R{ 743}{  88}
\E{ 743}{   85}{   6}
\R{ 758}{  90}
\E{ 758}{   87}{   6}
\R{ 773}{  92}
\E{ 773}{   89}{   6}
\R{ 788}{  94}
\E{ 788}{   91}{   6}
\R{ 803}{  98}
\E{ 803}{   94}{   6}
\R{ 818}{  99}
\E{ 818}{   96}{   6}
\R{ 833}{ 103}
\E{ 833}{  100}{   6}
\R{ 848}{ 108}
\E{ 848}{  105}{   6}
\R{ 863}{ 114}
\E{ 863}{  111}{   6}
\R{ 878}{ 128}
\E{ 878}{  124}{   8}
\R{ 893}{ 151}
\E{ 893}{  147}{   8}
\R{ 908}{ 245}
\E{ 908}{  239}{  12}
\R{ 923}{ 470}
\E{ 923}{  460}{  18}
\R{ 938}{ 525}
\E{ 938}{  515}{  20}
\R{ 953}{ 545}
\E{ 953}{  534}{  20}
\R{ 968}{ 553}
\E{ 968}{  543}{  20}
\R{ 983}{ 558}
\E{ 983}{  548}{  20}
\R{ 998}{ 562}
\E{ 998}{  552}{  20}
\R{1013}{ 567}
\E{1013}{  557}{  20}
\R{1028}{ 570}
\E{1028}{  560}{  20}
\R{1043}{ 574}
\E{1043}{  564}{  20}
\R{1058}{ 577}
\E{1058}{  567}{  20}
\R{1073}{ 581}
\E{1073}{  571}{  20}
\R{1088}{ 585}
\E{1088}{  575}{  20}
\R{1103}{ 591}
\E{1103}{  580}{  20}
\R{1118}{ 597}
\E{1118}{  587}{  20}
\R{1133}{ 605}
\E{1133}{  595}{  20}
\R{1148}{ 615}
\E{1148}{  605}{  20}
\R{1163}{ 632}
\E{1163}{  622}{  20}
\R{1178}{ 661}
\E{1178}{  651}{  20}
\end{picture}} 
\put(300,250){\begin{picture}( 1200,1200)
\newcommand{\R}[2]{\put(#1,#2){\circle{  6}}}
\newcommand{\E}[3]{\put(#1,#2){\line(0,1){#3}}}
\R{   8}{   3}
\E{   8}{    2}{   2}
\R{  23}{   8}
\E{  23}{    7}{   2}
\R{  38}{  11}
\E{  38}{    9}{   2}
\R{  53}{  13}
\E{  53}{   12}{   2}
\R{  68}{  17}
\E{  68}{   15}{   4}
\R{  83}{  18}
\E{  83}{   17}{   4}
\R{  98}{  21}
\E{  98}{   19}{   4}
\R{ 113}{  23}
\E{ 113}{   21}{   4}
\R{ 128}{  25}
\E{ 128}{   23}{   4}
\R{ 143}{  27}
\E{ 143}{   25}{   4}
\R{ 158}{  30}
\E{ 158}{   28}{   4}
\R{ 173}{  31}
\E{ 173}{   29}{   4}
\R{ 188}{  33}
\E{ 188}{   31}{   4}
\R{ 203}{  35}
\E{ 203}{   33}{   4}
\R{ 218}{  36}
\E{ 218}{   34}{   4}
\R{ 233}{  37}
\E{ 233}{   35}{   4}
\R{ 248}{  39}
\E{ 248}{   37}{   4}
\R{ 263}{  40}
\E{ 263}{   38}{   4}
\R{ 278}{  42}
\E{ 278}{   40}{   4}
\R{ 293}{  44}
\E{ 293}{   42}{   4}
\R{ 308}{  46}
\E{ 308}{   43}{   4}
\R{ 323}{  48}
\E{ 323}{   45}{   4}
\R{ 338}{  49}
\E{ 338}{   47}{   4}
\R{ 353}{  51}
\E{ 353}{   48}{   4}
\R{ 368}{  52}
\E{ 368}{   49}{   4}
\R{ 383}{  53}
\E{ 383}{   51}{   4}
\R{ 398}{  55}
\E{ 398}{   52}{   4}
\R{ 413}{  56}
\E{ 413}{   53}{   6}
\R{ 428}{  57}
\E{ 428}{   55}{   6}
\R{ 443}{  58}
\E{ 443}{   56}{   6}
\R{ 458}{  59}
\E{ 458}{   57}{   6}
\R{ 473}{  61}
\E{ 473}{   58}{   6}
\R{ 488}{  62}
\E{ 488}{   59}{   6}
\R{ 503}{  63}
\E{ 503}{   61}{   6}
\R{ 518}{  65}
\E{ 518}{   62}{   6}
\R{ 533}{  66}
\E{ 533}{   64}{   6}
\R{ 548}{  68}
\E{ 548}{   65}{   6}
\R{ 563}{  69}
\E{ 563}{   67}{   6}
\R{ 578}{  71}
\E{ 578}{   68}{   6}
\R{ 593}{  72}
\E{ 593}{   70}{   6}
\R{ 608}{  74}
\E{ 608}{   71}{   6}
\R{ 623}{  76}
\E{ 623}{   73}{   6}
\R{ 638}{  77}
\E{ 638}{   74}{   6}
\R{ 653}{  78}
\E{ 653}{   75}{   6}
\R{ 668}{  80}
\E{ 668}{   77}{   6}
\R{ 683}{  82}
\E{ 683}{   79}{   6}
\R{ 698}{  83}
\E{ 698}{   80}{   6}
\R{ 713}{  85}
\E{ 713}{   82}{   6}
\R{ 728}{  86}
\E{ 728}{   83}{   6}
\R{ 743}{  88}
\E{ 743}{   85}{   6}
\R{ 758}{  90}
\E{ 758}{   87}{   6}
\R{ 773}{  92}
\E{ 773}{   89}{   6}
\R{ 788}{  94}
\E{ 788}{   91}{   6}
\R{ 803}{  97}
\E{ 803}{   94}{   6}
\R{ 818}{  99}
\E{ 818}{   96}{   6}
\R{ 833}{ 103}
\E{ 833}{  100}{   6}
\R{ 848}{ 108}
\E{ 848}{  105}{   6}
\R{ 863}{ 114}
\E{ 863}{  110}{   6}
\R{ 878}{ 127}
\E{ 878}{  123}{   8}
\R{ 893}{ 147}
\E{ 893}{  143}{   8}
\R{ 908}{ 227}
\E{ 908}{  221}{  12}
\R{ 923}{ 411}
\E{ 923}{  401}{  18}
\R{ 938}{ 454}
\E{ 938}{  444}{  20}
\R{ 953}{ 469}
\E{ 953}{  459}{  20}
\R{ 968}{ 475}
\E{ 968}{  465}{  20}
\R{ 983}{ 479}
\E{ 983}{  469}{  20}
\R{ 998}{ 482}
\E{ 998}{  472}{  20}
\R{1013}{ 486}
\E{1013}{  476}{  20}
\R{1028}{ 488}
\E{1028}{  478}{  20}
\R{1043}{ 490}
\E{1043}{  480}{  20}
\R{1058}{ 492}
\E{1058}{  482}{  20}
\R{1073}{ 495}
\E{1073}{  485}{  20}
\R{1088}{ 498}
\E{1088}{  487}{  20}
\R{1103}{ 501}
\E{1103}{  491}{  20}
\R{1118}{ 505}
\E{1118}{  495}{  20}
\R{1133}{ 510}
\E{1133}{  500}{  20}
\R{1148}{ 516}
\E{1148}{  506}{  20}
\R{1163}{ 527}
\E{1163}{  517}{  20}
\R{1178}{ 545}
\E{1178}{  535}{  20}
\end{picture}} 
\end{picture} 
\vspace*{-10mm}
\caption{\small \sf
Total cross sections [pb] for $e\bar e\to\mu\bar\mu\tau\bar\tau$ from KORALW
MC in various
approximations of matrix element: ISNS$_{\gamma+Z}$ (big open circles),
ISNS$_{\gamma}$ (small open circles), complete 4f (big dots) and
semianalytical ISNS$_{\gamma}$ of Ref.\
\cite{Kniehl:1988id} (small dots)
as a function of $v_{cut}=1-R_{cut}$ for $\sqrt{s}=189$ GeV. Note that
$m_\tau$ is set equal to $m_\mu$.
}
\label{kw-kkks}
\end{figure}
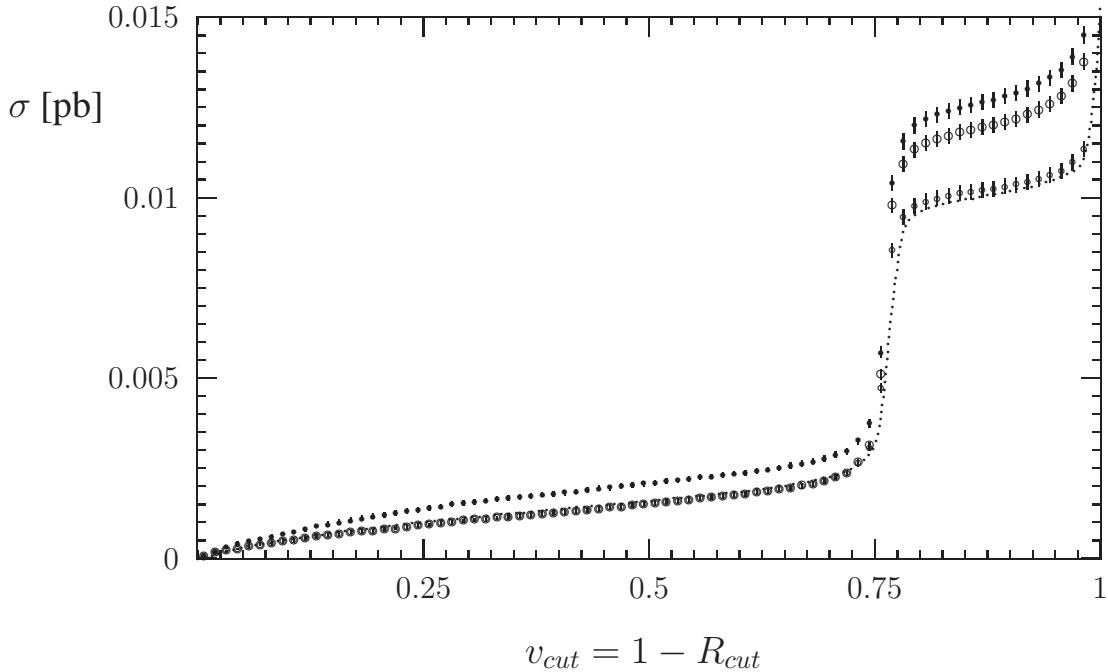
Note that that the evaluation of the
{\em experimental efficiency} and elimination of the {\em background}
requires running \KKMC\ and KORALW anyway, so complete elimination
of the secondary pair effects would come essentially as a byproduct
of the above procedure, with little theoretical uncertainty.
Before the above scenario could be realized several technical points need to be
checked:
\begin{itemize}
\item
  Even though a lot of technical 
  tests were already performed on KORALW in all corners of $4f$ phase space, 
  also with untagged electrons, additional tests need to be (re)done.
  As an example of such technical tests we show in 
  Fig.~\ref{kw-kkks} the comparison of KORALW with analytical result of 
  \cite{Kniehl:1988id} for the $\mu\bar\mu\tau\bar\tau$ final state as a
  function of $v_{cut}=1-R_{cut}$ for $\sqrt{s}=189$GeV. 
  The $\tau$ mass is set equal to $\mu$ mass. 
  The three histograms correspond to different approximations of matrix
  element in KORALW: ISNS$_{\gamma}$ ({\tt ISWITCH}=5), ISNS$_{\gamma+Z}$
  ({\tt ISWITCH}=2) and complete 4f ({\tt ISWITCH}=1). 
  Apart from the discrepancy at the $Z$ peak due to finite binning
  size the semianalytical  and corresponding Monte Carlo results agree within the statistical
  errors. 
  Another possible test is the comparison with semianalytical program
  GENTLE. This comparison is currently under study.
\item
 For the sake of comparisons the option of reducing matrix element
 of KORALW to ISNS$_{\gamma}$, FSNS$_{\gamma}$ etc.\ has been introduced,
 as described in Sect.~\ref{KORALW1} of this Report.
\item
 The virtual ISNS$_{\gamma}$ and FSNS$_{\gamma}$ terms have been 
 incorporated into \KKMC\ as described in Sect.~\ref{subsec:kkmc-virt} 
 of this Report.
\end{itemize}

\noindent
With both programs updated, as an example of the numerical results, the correction to the process 
$e\bar e\to\mu\bar\mu$ due to emission of one real pair has been calculated by {\tt KORALW} with
the following cuts:
\begin{enumerate}
\item
   mass of $\mu\bar\mu$ pair with highest mass bigger than (A)
   $0.9\sqrt{s}$ or (B) $0.4\sqrt{s}$ (two cuts);
\item
   angle of muon from $\mu\bar\mu$ pair with highest mass with respect
   to the beam: $\vert\cos\theta_\mu\vert \leq 0.95$;
\item
   sum of transverse momenta of neutrina less than $0.3(\sqrt{s}-\sum
   E_\nu)$.
\end{enumerate}

\begin{table}[!h]
\begin{center}
\begin{tabular}{||c|c|c|c||}
\hline\hline
Cut   & $\sigma_{tot} [pb]$   &  \KKMC\ $\sigma_{virt}$ [pb]   & KORALW
$\sigma_{real}$ [pb] \\
\hline
(A)   & 2.67             &  $-0.025\pm.001$  & $+0.020\pm.001$ \\ 
(B)   & 6.70             &  $-0.070\pm.001$  & $+0.497\pm.006$ \\
\hline\hline
\end{tabular}
\end{center}
\caption{\small \sf
Pair corrections to $e\bar e\to\mu\bar\mu$ calculated by 
KORALW (real) and \KKMC\ (virtual) for $\sqrt{s}=189$GeV. 
All quark and lepton pairs are included. Cuts (A) and (B) are defined in 
the text.
}
\label{kw-kk}
\end{table}
The calculation is quite fast and numerically stable. 
The cuts (A,B) correspond for example to (roughly) \citobs{IAleph5}, \citobs{IAleph6}.  
Additional cut on neutrino is based on L3 realistic
cut on secondary pair (Sect.\ 2.43.3). Its aim is to reduce $W$-pair
production background by requiring transverse energy imbalance to be
smaller than 0.3E$_{vis}$.
Together with the virtual component calculated by \KKMC\ the results are
summarized in Table~\ref{kw-kk}.

The following comments are in order here:
\begin{itemize}
\item[(a)]
  The $\sigma_{real}$ of KORALW is with {\em complete} $4f$
  matrix element for all fermions $f=d,u,s,c,b,\mu,\tau,\nu's$
  and {\em not} the ISNS signal of the signal definition proposal outlined 
  in sections~\ref{PairEffdiag} and  ~\ref{PairEffcut}.
\item[(b)]
  Masses $m_f$ are taken 0.2GeV for $f=d,u,s$ and PDG values for the rest.
  Changing $m_f$ of light quarks by factor two induces only
  $\delta\sigma_{virt}/\sigma=0.04\%$!  
\item[(c)]
  ISR was switched {\em off} in KORALW and {\em on} in \KKMC.
\item[(d)]
  The virtual pair correction is $-0.9\%$ with
  little dependence on the cut on $M_{\mu\bar\mu}$. 
\item[(e)]
  The electron channel dominates in real pair contributions.
\end{itemize}

This project is at the moment unfinished. 
In particular we have not discussed here the issues related to bremsstrahlung.
It is instructive to note in this context that one of the reasons that
make possible numerical cancellation amongst two separate Monte Carlo
programs is the fact that bremsstrahlung in both KORALW and \KKMC\ 
is implemented in the same way, based on YFS principle. 
As there are number of different ways of simulating photonic cascades 
in different Monte Carlo codes this issue may be a nontrivial one in
some cases.

The merits of the above definition of the $2f$ signal are to be
judged by LEP experimentalists. 
The authors of \KKMC\ and KORALW will provide the tools if there
is an interest.
Finally let us also stress that the tandem KORALW+\KKMC\
can be usefull not only for implementing the scenario ``without pairs''
decribed in the beginning of this section, but also for implementing
any other two-fermion signal and for comparisons with
any other semianalytical or Monte Carlo programs.
It is an equally important role.

\subsubsection{Results on Pairs from TOPAZ0}%

Here we shortly describe the implementation of pairs in 
TOPAZ0~\footnote{
  Authors of the report thank G.~Passarino for providing numerical results.}.
Since version 4.4 (April 1999) TOPAZ0~\cite{topaz0} allows the additional 
value {\bf ONP = 'I'}, where an extension of the Kuraev-Fadin (KF)
approach~\cite{Kuraev:1985hb} for virtual pairs and for soft and exponentiated 
ISNS$_{\gamma}$ pairs is used; the extension is also applicable to hadron 
pairs, because one uses KKKS results~\cite{Kniehl:1988id} for $\ord{\alpha^2}$ 
and write it in terms of moments. Then one matches it to KF and generalizes KF 
to soft and exponentiated hadrons pairs~\cite{TOPAZ0_unpu}.

Next, one uses the generalized KF-approach for virtual $\,+\,$ ISNS soft pairs
and cut to the same $s'$ value of IS QED radiation,
$s'$ being the centre-of-mass energy of the $e^+e^-$ system after initial state
radiation of photons and pairs. This choice, however, is not strictly needed.

Finally, one includes soft pairs only up to some cut $\Delta$ that is 
compatible with $E \gg \Delta \gg 2\,m$, where $E$ is the energy of the 
incoming electron(positron). Above it one uses ISNS$_{\gamma}$ hard
pairs according to KKKS formulation but not added linearly, KKKS in 
convolution with IS QED radiation. The radiator used here is a LL one, with 
options {\tt ORISPP = 'S'} (second order) and {\tt ORISPP = 'T'} (third order). 

An old comparison with {\tt MIZA} in the JMS~\cite{Jadach:1992aa} approach gave a 
nice agreement for energies around the $Z$ peak, below $0.03$ per mill  from 
$88\,$GeV to $94\,$GeV.

ISS$_{\gamma}$ pairs~\cite{Berends:1988ab}, i.e.\ ISS-pairs where the 
$t$-channel exchange is only via a photon, can be included by selecting 
{\bf OSING = 'SP'}.

After some tuned comparison with GENTLE/ZFITTER version $4.4$ of
TOPAZ0 has been slightly upgraded%
\footnote{
  ({\sl http://www.to.infn/\~{}giampier/topaz0$_{\_}$v44$_{\_}$rs8$_{\_}$220600.f)}}
 to cure instability problems in 
virtual pairs and in real $\tau$-pair production for very low values of the 
$s'$-cut. 

A sample of results is shown in Tabs.~\ref{tabppT_1}--\ref{tabppT_4} 
where $\delta$ (in per mil) is the relative effect of pair production,
$\sigma(\rm pairs)/\sigma$.
Pair corrections are shown in Tabs.~\ref{tabppT_1}--\ref{tabppT_2} for 
$e^+e^- \to\,$hadrons and in Tabs.~\ref{tabppT_3}--\ref{tabppT_4} for 
$e^+e^- \to \mu^+\mu^-$ for the two values of $s'/s$ and for $\sqrt{s} =
189\,$GeV. Results are shown for all secondary pairs both virtual and real. 
When compared with GENTLE's predictions in {\em hadronic} language
in Table~\ref{kobel1} for hadrons and Table~\ref{Maciek} (IPPS,IGONLY)=(5,2),
ISR on, for muons 
we observe a nice agreement everywhere for virtual pairs, with a maximum 
deviation of $0.1$ per mil.
When we neglect FS$_{\gamma}$ pairs, not implemented in TOPAZ0 
and compare the real pairs for hadrons with  Table~\ref{kobel1} 
it follows that for the total contribution to the hadronic cross-section the 
differences range from $0.9$ per mill  when the radiative return is allowed 
($s'/s \ge 0.01$) to $0.6$ per mill  when the radiative return is inhibited 
($s'/s \ge 0.7225$). 
Inclusion of  FS$_\gamma$ would increase this difference somewhat.
A comparison of the total corrections to the  options without FS pairs in GENTLE
and ZFITTER in Table~\ref{pairs_hadrons} shows a maximal difference
of 1.7 per mill  for hadrons. 
For muons we have  a  similar $1.5$ per mill  maximal difference in Table~\ref{pairs_muons}.
Note also that GENTLE -- ZFITTER agreement is $0.5 \div 1$
per mill with ZFITTER closer to TOPAZ0, so that  we see a $0.7$
per mill maximal difference between TOPAZ0 and ZFITTER.  
\begin{table}[ht]
\begin{center}
\begin{tabular}{|c||c|c|c|}
  \hline
  & \multicolumn{3}{c|}{$\sqrt{s} = 189\,$GeV $e^+e^- \to\,$hadrons,
$s'/s = 0.01$} \\
  \cline{2-4}
  & $\delta^{\rm V,ISR}_{\rm T}$  
  & $\delta^{\rm R,ISR}_{\rm T}$  
  & $\delta^{\rm T,ISR}_{\rm T}$  \\
\hline 
\hline
$e^+e^-$       & -2.91 & +10.37 &  +7.46  \\ 
$\mu^+\mu^-$   & -0.35 &  +2.57 &  +2.22  \\ 
$\tau^+\tau^-$ & -0.03 &  +0.72 &  +0.69  \\ 
hadrons        & -0.71 &  +6.76 &  +6.05  \\
all            & -4.00 & +20.42 & +16.42  \\
\hline
\hline 
\end{tabular}
\end{center}
\caption[]{TOPAZ0 predictions for virtual (V), real (R) and total
(T) pair-production corrections to $e^+e^- \to\,$hadrons at $\sqrt{s} =
189\,$GeV and for $s'/s \ge 0.01$. Results are shown for all {\em secondary}
pairs. All $\delta$'s are in per mill.}
\label{tabppT_1}
\end{table}
\begin{table}[!t]
\begin{center}
\begin{tabular}{|c||c|c|c|}
  \hline
  & \multicolumn{3}{c|}{$\sqrt{s} = 189\,$GeV $e^+e^- \to\,$hadrons,
$s'/s = 0.7225$} \\
  \cline{2-4}
  & $\delta^{\rm V,ISR}_{\rm T}$  
  & $\delta^{\rm R,ISR}_{\rm T}$  
  & $\delta^{\rm T,ISR}_{\rm T}$  \\
\hline 
\hline
$e^+e^-$       & -3.42 & +2.93  & -0.49  \\ 
$\mu^+\mu^-$   & -0.48 & +0.37  & -0.11  \\ 
$\tau^+\tau^-$ & -0.06 & +0.05  & -0.01  \\ 
hadrons        & -1.03 & +0.66  & -0.37  \\
all            & -4.99 & +4.01  & -0.98  \\
\hline
\hline 
\end{tabular}
\end{center}
\vspace*{-3mm}
\caption[]{TOPAZ0 predictions for virtual (V), real (R) and total
(T) pair-production corrections to $e^+e^- \to\,$hadrons at $\sqrt{s} =
189\,$GeV and for $s'/s \ge 0.7225$. Results are shown for all {\em secondary}
pairs. All $\delta$'s are in per mill.}
\label{tabppT_2}
\end{table}
\begin{table}[!h]
\begin{center}
\begin{tabular}{|c||c|c|c|}
  \hline
  & \multicolumn{3}{c|}{$\sqrt{s} = 189\,$GeV $e^+e^- \to \mu^+\mu^-$,
$s'/s = 0.01$} \\
  \cline{2-4}
  & $\delta^{\rm V,ISR}_{\rm T}$  
  & $\delta^{\rm R,ISR}_{\rm T}$  
  & $\delta^{\rm T,ISR}_{\rm T}$  \\
\hline 
\hline
$e^+e^-$       & -2.98 &  +8.52 &  +5.54  \\ 
$\mu^+\mu^-$   & -0.38 &  +2.04 &  +1.66  \\ 
$\tau^+\tau^-$ & -0.04 &  +0.57 &  +0.53  \\ 
hadrons        & -0.77 &  +5.30 &  +4.53  \\
all            & -4.17 & +16.43 & +12.26  \\
\hline
\hline 
\end{tabular}
\end{center}
\vspace*{-3mm}
\caption[]{TOPAZ0 predictions for virtual (V), real (R) and total
(T) pair-production corrections to $e^+e^- \to \mu^+\mu^-$ at $\sqrt{s} =
189\,$GeV and for $s'/s \ge 0.01$. Results are shown for all {\em secondary}
pairs. All $\delta$'s are in per mil.}
\label{tabppT_3}
\end{table}
\begin{table}[!h]
\begin{center}
\begin{tabular}{|c||c|c|c|}
  \hline
  & \multicolumn{3}{c|}{$\sqrt{s} = 189\,$GeV $e^+e^- \to \mu^+\mu^-$,
$s'/s = 0.7225$} \\
  \cline{2-4}
  & $\delta^{\rm V,ISR}_{\rm T}$  
  & $\delta^{\rm R,ISR}_{\rm T}$  
  & $\delta^{\rm T,ISR}_{\rm T}$  \\
\hline 
\hline
$e^+e^-$       & -3.31 & +2.79  & -0.52  \\ 
$\mu^+\mu^-$   & -0.46 & +0.34  & -0.12  \\ 
$\tau^+\tau^-$ & -0.06 & +0.05  & -0.01  \\ 
hadrons        & -0.99 & +0.61  & -0.38  \\
all            & -4.83 & +3.80  & -1.03  \\
\hline
\hline 
\end{tabular}
\end{center}
\vspace*{-3mm}
\caption[]{TOPAZ0 predictions for virtual (V), real (R) and total
(T) pair-production corrections to $e^+e^- \to \mu^+\mu^-$ at $\sqrt{s} =
189\,$GeV and for $s'/s \ge 0.7225$. Results are shown for all {\em secondary}
pairs. All $\delta$'s are in per mil.}
\label{tabppT_4}
\end{table}
There are unsolved problems that will constitute the bulk of next TOPAZ0
upgrading. They are:

\begin{enumerate}

\item Double-counting of real pairs and pairing ambiguities in the realistic
language of hadrons, i.e.\ not at the parton level;

\item Identical particles in primary and secondary pairs;

\item Splitting of real pairs into different channels, i.e.\ 
how to define $e^+e^- \to e^+e^- \barb b$, $\barb b$ pair-correction to Bhabha? 
$e^+e^-$ pair correction to $\sigma_{\rm had}$? Background?

\item Flavor misinterpretation;

\item extension to Bhabha scattering, i.e.\ implementation of pair corrections
with realistic cuts, collinearity and energy thresholds, instead of simple
$s'$-cuts.
\end{enumerate}

Cutting on secondary pairs is not a real problem once pairing -- 
double-counting ambiguities are solved in hadronic language via 
$R_{\rm had}$. The reason why this cut was never implemented is that
hadron pairs are easily constructed in the language of 
moments~\cite{Kniehl:1988id} which requires integrating over the {\em defined} 
secondary pair. If one cuts on it the answer is at parton level and should be 
folded with $R_{\rm had}$ and, presently, there is no routine capable of 
giving $R_{\rm had}(s)$ for $0 < s < 200\,$GeV  without doing something extra 
work at very low $s$ and around the thresholds~\cite{TOPAZ0_tt_pc}.

\subsubsection{Results on Pairs for Bhabhas from LABSMC}%
\label{LABSMC_pair}
With the default version of LABSMC (i.e. without the multi-peripheral contribution)
the sum of virtual+real pair effects was determined for the  Bhabha observables
listed in Table~\ref{LABS_sp}.
The corrections were calculated in respect to the cross sections,
where all other types of RC have been already applied.
There is a simple dependence of the size of corrections
on the applied cuts. The most strong cuts on real emission are there, 
the most large (and negative) effect is coming out. 
The largest corrections are found for some idealized observables, where
also the final state corrections do give a lot.

\begin{table}[!ht]
\scriptsize
\centering
\caption{\small LABSMC corrections due to pairs in per mill of the  cross-sections
for the respective observables.\label{LABS_sp}}
\begin{tabular} {||l|l|l|l||}
\hline\hline
      obs.              &            189   &     200    &    206  [GeV] 
\\
\hline
\multicolumn{4}{||c||}{realistic observables}\\
\hline
   \citobs{Aleph3}      &      -2.133   &  -2.180   &  -2.130  \\
   \citobs{Aleph4}       &     -2.281    & -2.286    & -2.287  \\
   \citobs{Delphi3}       &    -2.197    & -2.228    & -2.257  \\
   \citobs{LT4}           &    -2.618     &-2.660     &-2.684  \\
   \citobs{LT5}            &   -1.871     &-1.894     &-1.924  \\
   \citobs{LT6}               &-0.887     &-0.920     &-0.889  \\
   \citobs{LT7}               &-0.482     &-0.668     &-0.728  \\
   \citobs{LT8}               & 0.206     & 0.150     & 0.184  \\
   \citobs{Opal3}             & 0.131     & 0.014     & 0.072  \\
   \citobs{Opal4}             &-2.626     &-2.699     &-2.706  \\
   \citobs{Opal5}             &-1.635     &-1.669     &-1.669  \\
\hline
\multicolumn{4}{||c||}{idealized observables}\\
\hline
   \citobs{IAleph3}           &-5.846     &-6.009     &-5.994  \\
   \citobs{IAleph4}           &-6.774     &-6.925     &-6.967  \\
   \citobs{ILT4}              &-5.322     &-5.427     &-5.451  \\
   \citobs{ILT5}              &-1.867     &-1.890     &-1.920  \\
   \citobs{ILT6}              &-0.885     &-0.918     &-0.887  \\
   \citobs{ILT7}              &-0.481     &-0.666     &-0.725  \\
   \citobs{ILT8}              & 0.206     & 0.150     & 0.184  \\
   \citobs{IOpal3}            & 0.131     & 0.014     & 0.072  \\
   \citobs{IOpal4}            &-2.716     &-2.691     &-2.699  \\
   \citobs{IOpal5}            &-1.628     &-1.662     &-1.662  \\
\hline\hline
\end{tabular}
\end{table}

Concerning the multi-peripheral two--photon corrections, there are visible contributions
only for a few observables. The only large correction is for  \citobs{IOpal3} because of 
the wide range of allowed
collinearity and a very low energy threshold for electrons (1 GeV).
For all other observables, not listed in Table~\ref{LABS_MP}, the multi-peripheral reaction is cut away.

\begin{table}[!ht]
\scriptsize
\centering
\caption{\small LABSMC pair corrections due to multi-peripheral two-photon processes. \label{LABS_MP}}
\begin{tabular} {||l|l|l|l||}
\hline\hline
      obs.              &            189   &     200    &    206  [GeV] 
\\
\hline
\hline
   \citobs{Delphi3}           &0.105     &0.106     &0.107 \\
   \citobs{IOpal3}            &2.336     &2.359     &2.370 \\
   \citobs{IOpal4}            &0.070     &0.069     &0.069 \\
   \citobs{IOpal5}            &0.423     &0.426     &0.428 \\
\hline\hline
\end{tabular}
\end{table}

The accuracy on the above numbers can be estimated to be about 20\%,
which is mainly coming from the uncertainty in the  description of
secondary hadronic pairs.

\subsubsection{Comparison of Results for Hadrons and Muons}

In the following we will compare the results of the different signal definition obtained
from various programs for primary hadrons and primary muons. The Bhabha process
will be discussed in the next section. 

\subsubsection*{Real pairs}

As explained above, only the amount of {\it  real} secondary ISNS plus FS pair \fft\ emission 
enters the  experimental measurements of 2-fermion cross-sections. For all primary pairs 
apart from \ffo = ee this contribution  can be calculated  in GENTLE, KORALW, and GRC4f.
 Since the checks of the new KORALW code were not completely finished at the time of writing
this report, we compare in the following GRC4f and GENTLE, using the cut-based 
signal definition (2), which is realized in GENTLE via IPPS=6, IGONLY=3, and $P_{\rm cut}=0.10$.
The GRC4f prediction has  been obtained with method (A) described in Section~\ref{GRC4f1}. 
Using method (B)  gives consistent results. The result of the comparison for qq\fft\ corrections to the process e$^+$e$^- \to$ hadrons
at $\sqrt{s}$=189~GeV is listed in Table~\ref{realpairs} and  \ref{realpairs1}.
Comparisons at other centre-of-mass energies result in similar numbers.

\begin{table}[ht]
\caption[]{Real pair cross-sections in pb , and relative corrections in per mill,
obtained from GRC4f and GENTLE for the process e$^+$e$^- \to$ hadrons
at $\sqrt{s}=189$~GeV for {\bf high $s^\prime$ events of ${\boldmath R_{\rm cut}=0.7225}$}
according to the cut-based definition (2).
The last column lists the difference between  GRC4f and GENTLE in per mill of the
hadronic cross-section. The errors given are statistical, only. 
          }
\label{realpairs}
\vspace*{3mm}
\begin{center}
\begin{tabular} {|| l || c | c || c | c || c ||}
\hline
\ffo\fft & $\sigma^{\rm Real}_{\rm GRC4f} $  &  $\sigma^{\rm Real}_{\rm GENTLE}  $      &
$\delta^{\rm Real}_{\rm GRC4f}$&$\delta^{\rm Real}_{\rm GENTLE}$&$\Delta\delta^{\rm Real}$\\   
\hline
qqee               & $0.090\pm0.002$ & 0.073   &  $ 4.2\pm0.1$   &  3.4     &  $+0.8$  \\
qq$\mu\mu+\tau\tau$& $0.013\pm0.002$ & 0.010   &  $ 0.6\pm0.1$   &  0.4     &  $+0.2$  \\
qqqq               & $0.016\pm0.002$ & 0.018   &  $ 0.7\pm0.1$   &  0.8     &  $-0.1$  \\
\hline
total qq\fft       & $0.119\pm0.003$ & 0.100   &  $ 5.5\pm0.2$   &  4.6     &  $+0.9$  \\
\hline
\end{tabular}
\end{center}
\end{table}

\begin{table}[ht]
\caption[]{Real pair cross-sections in pb , and relative corrections in per mill,
obtained from GRC4f and GENTLE for the process e$^+$e$^- \to$ hadrons
at $\sqrt{s}=189$~GeV
for {\bf inclusive events} of  $R_{\rm  cut}=0.01$. 
The last column lists the difference between  GRC4f and GENTLE in per mill of the
hadronic cross-section. The errors given are statistical, only.
        }
\label{realpairs1}
\vspace*{3mm}
\begin{center}
\begin{tabular} {|| l || c | c || c | c || c ||}
\hline
\ffo\fft   &   $\sigma^{\rm Real}_{\rm GRC4f} $   &  $\sigma^{\rm Real}_{\rm GENTLE} $ &
$\delta^{\rm Real}_{\rm GRC4f}$&$\delta^{\rm Real}_{\rm GENTLE}$&$\Delta\delta^{\rm Real}$\\   
\hline
qqee               & $1.16\pm0.01$ & 0.97   &  $ 12.1\pm0.1$    &  10.1    &  $+2.0$  \\
qq$\mu\mu+\tau\tau$& $0.35\pm0.01$ & 0.29   &  $  3.6\pm0.1$    &   3.0    &  $+0.6$  \\
qqqq               & $0.58\pm0.03$ & 0.59   &  $  6.0\pm0.3$    &   6.1    &  $-0.1$  \\
\hline
total qq\fft       & $2.09\pm0.03$ & 1.85   &  $  21.7\pm0.3$   &  19.2    &  $+2.5$  \\
\hline
\end{tabular}
\end{center}
\end{table}

The results show, that for secondary lepton pairs
GRC4f has  about 20\% more   real pairs than GENTLE, though the 
difference is not statistically significant for $\mu\mu$ and $\tau\tau$ in the high $s^\prime$
sample. For hadronic pairs there is perfect agreement, despite the fact that GENTLE 
uses a pure hadronic approach (in terms of $R_{\rm had})$ to obtain qqqq, while GRC4f
calculates partonic corrections (where we have used quark masses of $m_u = m_d = 0.14$~GeV)
which have been  a posteriori corrected  for  the effect 
of the low-mass hadronic $R_{\rm had}$ ratio and of the running $\alpha_{\rm em}$  
via re-weighting of the GRC4f events (not available in default GRC4f). 
This latter  corrections have only a small impact on the comparison, since they tend to
cancel each other, if   $\alpha_{\rm em}(s)$  has been used in the generation of the events,
resulting in a total correction of $-0.001$~pb and $-0.01$~pb for $\sigma^{\rm Real}_{\rm GRC4f}$(qqqq)
for $R_{\rm cut}$ = 0.7225 and 0.01, respectively.   
       
 The total real pairs difference between GRC4f and GENTLE 
in terms of the corresponding hadronic cross-section
  is 0.9 per mill  for the high $s^\prime$
 selection and  2.5 per mill  for the inclusive selection.
A possible source of this difference is the more sophisticated treatment of common
photon and pairs emission in GENTLE while GRC4f simply attaches a photon radiator 
function to the 4-fermion matrix element. Note that  the agreement for qq pairs 
depends on  the choice of quark masses in GRC4f.

Comparing with Table~\ref{OPALeff} is it evident that even in the worst case (inclusive muons)
a 20\% error on $\delta_{\rm Real}$ means a relative error of $0.1\%$ for the combined 2f+4f
efficiency. 
One can therefore conclude that GRC4f is adequate to calculate the influence of pair emission
on the efficiency to better than 0.1\%.

\subsubsection*{Comparison between signal definitions}
 
Since the virtual pair corrections are identical for the above diagram based (1) , and cut-based (2)
signal definitions,  the total difference between the definitions is given by the   difference in the amount of real pairs.
Repeating the above calculations with the diagram-based signal definition (1), using the same $s^\prime$ definition
(i.e. IPPS=5, IGONLY=2, $P_{\rm cut}=1.0$ (no cut on secondary pairs) in GENTLE) results in 
Table~\ref{pairs_sigdiff} of differences in real pair cross-sections. 

\begin{table}[ht]
\caption[]{ Differences between diagram-based definition (1) and cut-based definition (2)
for real pairs in per mill of the hadronic cross-section,
obtained from GRC4f and GENTLE for the process e$^+$e$^- \to$ hadrons
at $\sqrt{s}=189$~GeV for  high $s^\prime$ events with $R_{\rm cut}=0.7225$
and   inclusive events with $R_{\rm cut}=0.01$. 
For $R_{\rm cut}=0.7225$ no significant difference was observed within the statistical errors
of the comparison in both programs. For qqee the diagram-based definition 
was not available for GRC4f.  
          }
\label{pairs_sigdiff}
\vspace*{3mm}
\begin{center}
\begin{tabular} {|| l || c | c | c | c ||}
\hline
\ffo\fft  &  \multicolumn{2}{c|}{$\Delta\delta^{\rm Real}_{\rm GRC4f} $}  
            & \multicolumn{2}{c||}{$\Delta\delta^{\rm Real}_{\rm GENTLE} $} \\
\hline
$R_{\rm cut}$& 0.7225  &  0.01            &    0.7225        &  0.01        \\
\hline
qqee         & $ - $   &  $ - $       &  $ < 0.0001 $   &  0.04       \\
qq$\mu\mu+\tau\tau$ & $<0.1 $  &  $0.02\pm0.12$       &  $ < 0.0001 $   &  0.07       \\
qqqq                & $<0.1 $  &  $0.31\pm0.33$       &  $ < 0.0001 $   &  0.18       \\
\hline
total qq\fft      & $ - $   &  $ - $       &  $ < 0.0001 $    &  0.29       \\
\hline
\end{tabular}
\end{center}
\end{table}

This comparison shows that for inclusive hadrons  the difference between the two definitions,
as predicted by GENTLE, is of order $10^{-4}$ for each class of pairs listed,
 amounting to a total of $2.9\times 10^{-4}$. 
The results of GRC4f are consistent, though their sensitivity is limited by the 
statistical error of some $10^{-4}$. For high $s^\prime$ hadrons no difference between
definitions 1 and 2 is visible even on the level of $10^{-7}$ in GENTLE.

\subsubsection*{Comparison of real+virtual pairs}

In the following we compare as a typical example the sum of virtual and real pair corrections
for primary hadrons and primary muons at $\sqrt(s)=189$~GeV for four different definitions
of the secondary pair signal:
\begin{itemize}
\item[A] Diagram-based definition (1) with $s^\prime=M_{\rm prop}^2$
\item[B] Diagram-based definition (1) with $s^\prime=M_{\rm inv}^2$
\item[C] Cut-based definition (2) with $s^\prime=M_{\rm inv}^2$ and $P_{\rm cut}=0.15$
\item[D] Cut-based definition (2) with $s^\prime=M_{\rm inv}^2$ and $P_{\rm cut}=0.10$
\end{itemize}

\begin{table}[ht]
\caption[]{Relative virtual + real pair corrections in per mill of the total cross-section for the process
e$^+$e$^-\to$hadrons at $\sqrt{s}=189$~GeV for different signal definitions. A $-$ means that this definition
is not accessible in the respective program.
          }
\label{pairs_hadrons}
\vspace*{3mm}
\begin{tabular} {|| l || c | c | c || c | c | c || }
\hline
  & GENTLE & ZFITTER & TOPAZ0 & GENTLE & ZFITTER & TOPAZ0 \\
\hline 
$R_{\rm cut}$    & \multicolumn{3}{c||}{0.01} & \multicolumn{3}{c||}{0.7225}            \\
\hline
A) diagram,  $s^\prime=M_{\rm prop}^2$   & 14.72  & 15.76   &  16.42 & $-1.13$   & $-0.92$ &  $-0.98$    \\
B) diagram,  $s^\prime=M_{\rm inv}^2 $   & 14.74  & 15.82   &  $-$    & $-1.45$   & $-1.20$ &  $-$       \\
C) cuts,     $ P_{\rm cut}=0.15 $        & 14.64  & $-$     &  $-$    & $-1.45$   & $-$        &  $-$    \\
D) cuts,     $ P_{\rm cut}=0.10 $        & 14.45  & $-$     &  $-$    & $-1.45$   & $-$        &  $-$    \\
\hline
$\delta_{\rm B}-\delta_{\rm A}$          &  0.02  &  0.06   &  $-$    &  $-0.32$   & $-0.29$ & $-$       \\
$\delta_{\rm B}-\delta_{\rm C}$          &  0.10  &  $-$    &  $-$    &  $ < 0.0001$   & $-$ & $-$       \\
$\delta_{\rm B}-\delta_{\rm D}$          &  0.29  &  $-$    &  $-$    &  $ < 0.0001$   & $-$ & $-$       \\
\hline
\end{tabular}
\end{table}

\begin{table}[ht]
\caption[]{Relative virtual + real pair corrections in per mill of the total cross-section for the process 
e$^+$e$^-\to\mu^+\mu^-$ at $\sqrt{s}=189$~GeV for different signal definitions. A $-$ means that this definition
is not accessible in the respective program.
         }
\label{pairs_muons}
\vspace*{3mm}
\begin{tabular} {|| l || c | c | c || c | c | c || }
\hline
 & GENTLE & ZFITTER & TOPAZ0 & GENTLE & ZFITTER & TOPAZ0 \\
\hline 
$R_{\rm cut}$        & \multicolumn{3}{c||}{0.01} & \multicolumn{3}{c||}{0.7225}            \\
\hline
A) diagram,  $s^\prime=M_{\rm prop}^2$   & 10.73  & 11.73   &  12.26 & $-1.37$   & $-1.00$ &  $-0.98$       \\
B) diagram,  $s^\prime=M_{\rm inv}^2 $   & 10.84  & 11.96   &  $-$    & $-2.58$  & $-2.05$ &  $-$           \\
C) cuts,     $ P_{\rm cut}=0.15 $        & 10.72  & $-$     &  $-$    & $-2.58$  & $-$     &  $-$           \\
D) cuts,     $ P_{\rm cut}=0.10 $        & 10.57  & $-$     &  $-$    & $-2.58$  & $-$     &  $-$           \\
\hline
$\delta_{\rm B}-\delta_{\rm A}$          &  0.11  &  0.23   &  $-$    &  $-1.21$   & $-1.05$ & $-$          \\
$\delta_{\rm B}-\delta_{\rm C}$          &  0.12  &  $-$    &  $-$    &  $ < 0.0001$   & $-$ & $-$          \\
$\delta_{\rm B}-\delta_{\rm D}$          &  0.27  &  $-$    &  $-$    &  $ < 0.0001$   & $-$ & $-$          \\
\hline 
\end{tabular}
\end{table}

From these tables several conclusions can be drawn.
The comparison between GENTLE, ZFITTER, and TOPAZ0 for
the diagram-based definition  with $s^\prime=M_{\rm prop}^2$  (A)
reveals maximum differences of 1.7 (1.5) per mill for inclusive hadrons (muons) and
0.2 (0.4) per mill for   high $s^\prime$ hadrons (muons) between any two of the programs.
Differences between cut-based and diagram-based signal definitions are between 1 and $3\times10^{-4}$
for inclusive selections for $P_{\rm cut}$ in the range from 0.10 to 0.15   (compare also Table~\ref{pairs_sigdiff}),
and below $10^{-7}$ for the high $s^\prime$ selection, as long $s^\prime$ is defined as $M_{\rm inv}^2$  everywhere.    
Whereas for the inclusive selections the difference between $s^\prime=M_{\rm prop}^2$   and
$s^\prime=M_{\rm inv}^2  $   is at most  0.2 per mill, it is about 0.3 (1.1) per mill for high $s^\prime$ hadrons (muons).
Compared to the LEP-combined statistical precision of the measurements all these differences are  small.
Even the 1.7 per mill difference between GENTLE and TOPAZ0 for inclusive hadrons is only about  half of the 
expected LEP-combined statistical error, summed over all centre-off mass energies,and is thus not far from the
precision tag of 1.1 per mill.     

\subsubsection{Results for Bhabhas}

There is only one program, LABSMC, which is able to calculate virtual+real pair effects for s+t channel Bhabha scattering
for the signal definition given above. Therefore a comparisons like the one performed above for hadrons and muons
cannot be done for primary electron pairs. 
We just state here that the pair corrections for idealized observables range from +0.2 per mill for \citobs{ILT8}
to $-7.0$ per mill for \citobs{IAleph4}. Largish corrections between 5 and 7 per mill occur only for observables which
cut hard on $M_{\rm inv}$. For them a relative accuracy of 20-30\% would be needed to meet the experimental precision
tags, which are between 0.13 and 0.21\% for the cross-sections. 
All other corrections are below 3 per mill so that for them a 50\% pair correction accuracy would suffice. 
The author of LABSMC estimates a relative accuracy of 20\% for pair corrections, which would mean
that all experimental precision requirements are met. Yet, it would be very valuable if the pair corrections in 
LABSMC could be cross-checked against another code.

\subsubsection{Conclusions for pair effects}
\label{pairconclusions}

Shortly before and during this workshop a lot of new code for pair corrections at LEP2 were developed.
Before 1999, essentially only the diagram-based pair correction with $s^\prime = M_{\rm prop}^2$, i.e.
inclusive FS pairs, could be calculated by ZFITTER and TOPAZ0 for all primary pairs
apart from electrons. 
Common exponentiation of initial state photons and
ISNS$_{\gamma}$ pairs for energies away from the $Z$-peak as well as 
optional ISS$_{\gamma}$ pairs were implemented in both codes in 1999.
ZFITTER has now been upgraded to include  explicit FS$_\gamma$ with the possibility of mass cuts.  
The new GENTLE/4fan  offers even
more options with mass cuts on all pairs and inclusion of pairs from virtual Z's and swapped FS diagrams.
Finally a new powerful combination of \KKMC\ and KORALW is being developed which contributed
first  numbers to this document.  
This makes  a whole variety of options for pair treatment available.

The main achievement of the pair study described above are:
 \begin{itemize}
\item A proposal for a signal definition which can be, to better than 0.1\% accuracy defined either
based on  (experiment oriented) cuts or on  (theory oriented) diagrams.
\item The determination of efficiency corrections using full event generators has been checked
 for GRC4f to a precision of 0.1\%, from a comparison of real pair cross-sections with GENTLE. 
\item Double counting of hadronic events  has been studied
with GRC4f and found to be smaller than 10$^{-4}$.
\item Problems of pairing ambiguities for 4 identical fermions become increasingly important with
the larger ZZ cross-sections at high energies, especially for inclusive measurements
with the signal definitions adopted here. From varying pairing algorithms, 
a worst-case diffrence of  0.8 
 per mill was found for inclusive hadrons  at 206 GeV. 
\item Differences for pair corrections
between $s^\prime$ definitions via the propagator or primary pair mass
in the diagram-based approach have been determined.
 GENTLE and ZFITTER both find them  to be about 0.3 (1.1) per mill for high $s^\prime$ hadrons (muons). 
\item Maximum differences for the  diagram-based pair correction of 1.7 (1.5) per mill for inclusive hadrons (muons) and
0.2 (0.4) per mill for   high $s^\prime$ hadrons (muons) between any two of the programs GENTLE, ZFITTER
and TOPAZ0 have been found.
\item A first complete calculation of pair corrections for Bhabhas has been done by LABSMC.
\end{itemize}

We conclude, that for the proposed signal definition sufficient comparisons have been performed
to be confident that the above numbers are limitations of the theoretical uncertainties. 
With the exception of the 
1.7 per mill difference for inclusive hadrons, which is slightly above  the respective precision tag of 1.1 per mill,
all theoretical uncertainties  are well below   the experimental precision tags.
For  other signal definitions, however, uncertainties have not been checked systematically.
Some more information is present from the tables presented in this section. It is in any case
not advisable to chose a signal definition for pairs that can be calculated by one program, only. 
Especially for the case of Bhabha scattering it would be highly desirable to have more than one code
predicting the effects of secondary pairs. 
Improvements are still expected in GENTLE, TOPAZ0 and \KKMC\ + KORALW.

%% file: 2f-Chapt-Summary.tex
\section{Summary }
\label{conclusion}

In this report we have addressed the question of 
uncertainties of the predictions of {\em theoretical} calculations
for quantities measured in LEP2 experiments (LEP2 observables).
We also have studied uncertainties in relating measured quantities
to theoretical predictions, via signal definitions and acceptance corrections,
especially for the effect of secondary pair radiation, and to the lesser 
degree wide angle Bhabha scattering.
The calculations were  contributed by several theoretical groups and are
implemented using various techniques and approaches embodied 
in most cases in Monte Carlo's, but also in semi-analytical
codes -- most of them are already in use in all LEP2 collaborations.
We tried to cover all two-fermion processes:
production of quark-, muon- and tau-pairs, the Bhabha process.
The considerable effort was also invested in the processes with
(additional) tagged single and double photons. 
The neutrino channel and Bhabha channel are of course the most important in this class.
The whole analysis of the theoretical prediction was done for
the rather complete list of the LEP2 observables,
which we tried to have as close as possible to the ones used by LEP experiments.
Most of the collected theoretical predictions are for simplified experimental
acceptances (so called idealized observables), 
but we also tried, not without some success,
to produce and discuss theoretical predictions for {\em realistic}
LEP2 observables.

Already from the beginning of the work of our two-fermion 
group it was expected  that for the full LEP2 statistics the
most important issues in the evaluation of the theoretical uncertainties would be:
\begin{enumerate}
\item
 Production of  secondary 
pairs,
\item
 QED initial-final state interference,
\item
 Precise cross-checks of QED initial state radiation and multi-photon effects,
\item
 Numerical cross-checks of the electroweak corrections.
\end{enumerate}

The QED parts of the calculations have turned out to be under very good control,
the biggest and most important contributions from the initial state
radiation (which is up to $200\%$)
is now under control down to $0.2-0.4\%$ (final state radiation included)
in most cases.

The initial-final state interference, which is $\sim 2\%$,
is now controlled down to $0.2\%$.

The above is true for the ordinary two-fermion LEP2 observables,
which do not require the presence of one or two visible (tagged) photons.
In the case of tagged photons the achieved precision varies
with the type of the cut and 
is  typically 4\%for  the $\nu \bar\nu \gamma$ observables,
3 \% for Bhabha and 1 \% for $\mu^+ \mu^- \gamma$,
or $\tau^+ \tau^- \gamma$.
The initial-final state interference has turned out to be rather important
for the observables with tagged photons.

For Bhabha process the  QED bremsstrahlung was found to be
one of the main sources of uncertainties for the end-cap observables.
For the LEP2 observables relying on the data from the barrel (wide angle) detector
only, there is a sizable contribution from the electroweak part of 
the calculation. 
We were unable to explore this subject before the end of  the workshop.

In the work of our group we did not have a chance to scrutinize
once again the pure electroweak corrections.
To some extent it was already done in the earlier LEP workshops,
so one could say that it is not really necessary, 
on the other hand it is always necessary
to cross-check the codes like ZFITTER and \KKMC\  used by LEP2 experiments once again,
on every possible aspect.
We clearly recommend that it should be done in the near future.

As a collorary of the tests of ZFITTER and \KKMC\  we have noticed that
the numerical contribution from the so-called electroweak boxes,
which is negligible at LEP1,
and is still rather small at lower CMS energies of LEP2 like 189GeV, is, 
however, already quite large $\sim 2\%$ at higher LEP2 energies like 206 GeV. 
In other words, as far as electroweak phenomena are concerned,
the LEP experiments at the LEP2 top energies start to be in the same situation
as the future Linear Colliders!

The  production of the secondary  pairs was on the top
of the list of priorities of this workshop. We have come up with 
a proposal for a 2-fermion signal definition including pair radiation, which is 
simple and equally applicable to all final state fermions, from electrons to hadrons,
and to all $s^\prime$ cuts. This signal definition comes in fact optionally
as a diagram-based or as a cut-based definition, where the one best suited 
for the given experimental or theoretical setup can be chosen. 
All definitions  are numerically identical
within 0.3 per mill for hadrons and 1.1 per mill for muons for any $s^\prime$ cut.  

For this set of signal definitions we have performed
a broad variety of comparisons and tests, starting from effects like
pairing ambiguities, double counting, and efficiency corrections,  
eventually  leading to a  comparison between different theoretical codes
for computing real and virtual pair corrections.    
Uncertainties and biases resulting from these studies were  found to
range from below 0.1 per mill to at most 1.7 per mill of the 2-fermion cross-sections.
This corresponds to up to 20\% relative uncertainties on the pair corrections.
Some theoretical codes have provided results also for other definitions
of pair signals, which are however often  calculable by one code, only.
This situation is expected to improve with forthcoming upgrades of GENTLE, TOPAZ0
and \KKMC+KORALW.
For the Bhabha scattering process only one theoretical calculation of pair effects
is available for the proposed signal definition.   


For the nearest future after the present workshop we would most urgently recommend
cross-checks for  the purely electroweak corrections, especially
for wide angle Bhabha, and 
for secondary pair corrections  for the Bhabha process. 


There seem to be little unresolved problems as far as QED is concerned,
except for certain observables with the tagged photons $e^+e^- \to \nu \bar \nu \gamma $
and (to a lesser degree) for wide angle Bhabhas, 
where further improvements  of the theoretical precision
by factor $4-10$ are necessary.
\begin{table}
\caption[]{ Comparison of the typical theoretical uncertainties with the
typical experimental precision tags}
\label{fintab}
\begin{center}
\begin{tabular}{|l|c|c|}
\hline
class of observables & theoretical uncertainty & experimental precision tag
\\
\hline
$e^+e^- \to q \bar q (\gamma)$      & 0.26 \%            & 0.1 \% -0.2 \%
\\
$e^+e^- \to \mu^+ \mu^- (\gamma)$   & 0.4 \%             & 0.4 \% -0.5 \%
\\
$e^+e^- \to \tau^+ \tau^- (\gamma)$ & 0.4 \%             & 0.4 \% -0.6 \%
\\
$e^+e^- \to e^+ e^- (\gamma)$  (endcap)      & 0.5 \%    & 0.13 \% 
\\
$e^+e^- \to e^+ e^- (\gamma)$  (barrel)     & 2.0 \%     & 0.21 \%
\\
$e^+e^- \to e^+ e^- \gamma$         & 3 \%               & 1.5 \%
\\
$e^+e^- \to \l^+ \l^- \gamma$       & 1 \%               & 1.5 \%
\\
$e^+e^- \to \nu \bar \nu \gamma$   &  4 \%               & 0.5 \%
\\ 
\hline 
\end{tabular}
\end{center}
\end{table}

All of the above information we tried to summarize once again
in table \ref{fintab}, where we list
the theoretical precisions attained in our study, 
for the LEP2 observables defined in our Section~\ref{list}
in comparison to typical ultimate experimental requirements for combined  LEP2 data.

%% file: 2f-Chapt-Biblio.tex
\providecommand{\href}[2]{#2}